\documentclass[a4paper,floatfix,12pt,apl]{revtex4-2}

\usepackage{amssymb}
\usepackage{amsmath}
\usepackage{epsfig}
\usepackage{graphicx}
\usepackage{multirow}
\usepackage{xcolor}
\usepackage{hyperref}

\begin{document}

\author{R. J. Hudspith}
\email{renwick.james.hudspith@googlemail.com}
\affiliation{GSI Helmholtzzentrum f\"ur Schwerionenforschung, 64291 Darmstadt, Germany}
\author{D. Mohler}
\email{d.mohler@gsi.de}
\affiliation{GSI Helmholtzzentrum f\"ur Schwerionenforschung, 64291 Darmstadt, Germany}
\affiliation{Institut f\"ur Kernphysik, Technische Universit\"at Darmstadt,
Schlossgartenstrasse 2, 64289 Darmstadt, Germany}

\title{Exotic Tetraquark states with two $\bar{b}$-quarks and $J^P=0^+$ and
  $1^+$ $B_s$ states in a nonperturbatively-tuned Lattice NRQCD setup}

\begin{abstract}
We use $n_f=2+1$ Wilson-clover gauge-field ensembles from the CLS consortium in a Lattice NRQCD
setup to predict the binding energy of a $I(J^P)=0(1^+)$
$ud\bar{b}\bar{b}$ tetraquark and a $\frac{1}{2}(1^+)$ $\ell s\bar{b}\bar{b}$
tetraquark. We determine the binding energies with
respect to the relevant $BB^*$ and $B_sB^*$ thresholds respectively to be $112.0(13.2)$~MeV for the
$ud\bar{b}\bar{b}$, and $46.4(12.3)$~MeV for the $\ell s\bar{b}\bar{b}$. We
also determine the ground-state $J^P=0^+$ $B_{s0}^*$ and $1^+$ $B_{s1}$ mesons to lie $75.4(14.0)$ and $78.7(13.9)$~MeV below the $BK$ and $B^*K$ thresholds respectively. Our errors are entirely dominated by systematics due to discretisation effects. To achieve these measurements, we performed a neural network based nonperturbative tuning of the Lattice NRQCD Hamiltonian's parameters against the basic bottomonium spectrum.
For all
lattice spacings considered we can reproduce the continuum splittings of low-lying
bottomonia. It is worth remarking that our nonperturbative tuning parameters deviate
from 1 by significant amounts, particularly the term $c_2$.

\end{abstract}
\maketitle
\newpage

\section{Introduction}


The study of doubly-heavy tetraquarks with (anti)bottom quarks is currently an
area of considerable interest, both on the lattice and in phenomenology. These
states are of an explicitly exotic nature, and initial
studies of  doubly-heavy $ud\bar{b}\bar{b}$-tetraquarks, e.g. \cite{Bicudo:2015vta,Francis:2016hui} and
references therein, suggest rather large binding energies.

Early lattice studies found an attractive heavy-light meson-meson potential
\cite{Richards:1990xf,Mihaly:1996ue,Green:1998nt,Stewart:1998hk,Pennanen:1999xi,Detmold:2007wk,Brown:2012tm,Wagner:2011ev}
indicative of the possibility to admit a bound $ud\bar{b}\bar{b}$ tetraquark
($T_{bb}$) in nature. More recently, dynamical light-quark simulations with
static b-quarks
\cite{Bicudo:2012qt,Bicudo:2015kna,Bicudo:2015vta,Bicudo:2016ooe}, and with Lattice non-relativistic QCD (NRQCD) b-quarks
\cite{Francis:2016hui,Junnarkar:2018twb,Leskovec:2019ioa,Mohanta:2020eed} have
predicted a strong-interaction-stable $I(J^P)=0(1^+)$, $ud\bar{b}\bar{b}$
tetraquark. Phenomenologically this state is almost unequivocally expected to be deeply bound \cite{Zouzou:1986qh,Carlson:1987hh,Semay:1994ht,Pepin:1996id,Brink:1998as,Vijande:2003ki,Ebert:2007rn,Vijande:2007rf,Zhang:2007mu,Lee:2009rt,Yang:2009zzp,Wang:2017uld,Karliner:2017qjm,Eichten:2017ffp,Czarnecki:2017vco,Mehen:2017nrh,Maiani:2019lpu,Hernandez:2019eox,Lu:2020rog,Braaten:2020nwp,Cheng:2020wxa,Meng:2020knc,Weng:2021hje,Faustov:2021hjs,Deng:2021gnb,Kim:2022mpa} with respect to the lowest-lying non-interacting $BB^*$ threshold.

From a diquark perspective, if a $ud\bar{b}\bar{b}$ tetraquark is deeply bound
the next logical candidate with a slightly less attractive light good-diquark configuration has flavor $\ell s\bar{b}\bar{b}$ and quantum numbers $I(J^P)=\frac{1}{2}(1^+)$. Such a state has been measured on the lattice to lie below the $B_sB^*$ threshold in \cite{Francis:2016hui,Junnarkar:2018twb,Meinel:2022lzo}, and is somewhat more shallowly bound than the $ud\bar{b}\bar{b}$. Phenomenologically this state is expected to lie quite close to threshold \cite{Silvestre-Brac:1993zem,Semay:1994ht,Ebert:2007rn,Lee:2009rt,Du:2012wp,Eichten:2017ffp,Wang:2017uld,Park:2018wjk,deng:2018kly,Braaten:2020nwp,Cheng:2020wxa,Meng:2020knc,Weng:2021hje,Dai:2022ulk,Deng:2021gnb,Kim:2022mpa}, with the majority of relativistic quark models \cite{Godfrey:1985xj} suggesting it is in fact unbound \cite{Ebert:2007rn,Lu:2020rog,Faustov:2021hjs}. 

Further states of interest are the  $J^P=0^+$ $B_{s0}^*$ and $1^+$ $B_{s1}$ mesons. Their lighter $D^*_{s0}(2317)^{\pm}$ and $D_{s1}(2460)^{\pm}$
counterparts show properties not expected in quark-model calculations and were
among the first exotic states discovered in the era of the b-factories \cite{BaBar:2003oey,CLEO:2003ggt}. While a
modern Lattice QCD scattering calculation obtains these states significantly below the
respective threshold \cite{Lang:2015hza} (using a relativistic formalism),
other approaches obtain results consistent with
\cite{Gregory:2010gm,Wurtz:2015mqa} (using Lattice
NRQCD) \cite{Koponen:2007nr} (using static b-quarks) the
lowest-lying two-meson thresholds $BK$ and $B^*K$ respectively.  Similar
statements can be made about various phenomenological models \cite{Orsland:1998de,Bardeen:2003kt,Kolomeitsev:2003ac,Matsuki:2004db,Guo:2006fu,Guo:2006rp,Badalian:2007yr,Vijande:2007ke,Cleven:2010aw,Dmitrasinovic:2012zz,Colangelo:2012xi,Altenbuchinger:2013vwa,Torres-Rincon:2014ffa,Wang:2015mxa,Ortega:2016pgg,Albaladejo:2016ztm,Cheng:2017oqh,Du:2017zvv,Alhakami:2020vil,Fu:2021wde,Yang:2022vdb,Godfrey:1986wj,Lahde:1999ih,DiPierro:2001dwf,Lakhina:2006fy,Ebert:2009ua,Sun:2014wea,Godfrey:2016nwn}. Typically, relativistic quark model calculations predict these states to lie at or above threshold, whereas other approaches usually predict they lie below. It has been a motivation of our work to investigate the impact of the Lattice NRQCD tuning on these states and provide an approach to determining these states differing from previous lattice studies. 

NRQCD measurements have a long and storied history
within the field of Lattice QCD. They are numerically very cheap to perform
and can be statistically precise. NRQCD is an effective field theory
approach to describing heavy bottom-quark (b-quark) dynamics, which however has some
difficult-to-quantify systematics. Nevertheless, Lattice NRQCD studies have
made important contributions to standard model quantities such as $\alpha_s$
\cite{McNeile:2010ji}, the b-quark mass \cite{Davies:1994pz,Gray:2005ur}, and
b-meson decays and $|V_{cb}|$ \cite{Na:2015kha}, to name but a few. NRQCD is but one approach to b-physics. Others include: relativistic heavy quark actions \cite{El-Khadra:1996wdx,Christ:2006us,Oktay:2008ex}, heavy quark effective theory \cite{Heitger:2003nj}, and various extrapolation approaches \cite{ETM:2009sed,Parrott:2020vbe,Colquhoun:2022atw}. Resolving the physical b as a valence quark with the same action as the light quarks requires a very fine lattice spacing and a large box; with current technology generating such ensembles is very expensive.

Tuning the Lattice NRQCD action has periodically been of interest
\cite{Lepage:1992tx,Meinel:2010pv,Davies:2018fwg}; particularly when it comes
to choices of different tadpole improvement factors
\cite{Lepage:1992xa,Lepage:1997id}, and to whether the action's parameters
should be perturbatively improved
\cite{Morningstar:1994qe,Morningstar:1993de}. We will show that all
coefficients of a simple truncation of the NRQCD action can be tuned entirely
non-perturbatively using machine learning to reproduce the simple low-lying
spectrum of bottomonia to a level compatible with the uncertainty on our lattice
spacing (roughly 1\%). The results for the simple spectrum including excitations from our approach will be compared to those using tree-level coefficients.

In Section \ref{methodology} the lattice methodology for our calculation is
described with a particular focus on the nonperturbative Lattice NRQCD
tuning. The results of this tuning are summarized in Section \ref{results},
where we also present the impact of our tuning on the spectrum of low-lying
bottomonium excitations. In Sections \ref{sec:udbb}, \ref{sec:lsbb}, and \ref{sec:bs} we
present our main physics analysis for the doubly heavy tetraquarks and $B_s$
states respectively. We proceed with a discussion of systematic uncertainties
in Section \ref{sec:higher_order_syst} and conclude in Section
\ref{conclusions}.

\section{Methodology\label{methodology}}

\subsection{NRQCD Hamiltonian and evolution}

For the implementation of the $O(v^4)$ NRQCD Hamiltonian (with higher-order discretisation-correction terms $c_5$ and $c_6$) we follow \cite{Lepage:1992tx} and \cite{Mathur:2002ce}. To obtain our heavy-quark propagator we apply the symmetric evolution to some prepared source $G(x,t)$ (technical aspects of the lattice NRQCD implementation can be found in App.~\ref{app:NRQCD_tech} for the adventurous reader),
\begin{equation}\label{eq:evolutionNRQCD}
\begin{aligned}
G(x,t+1) &= \left( 1 - \frac{\delta H}{2} \right) \left( 1 - \frac{H_0}{2n}\right)^n \tilde{U}_t(x,t)^\dagger \left( 1 - \frac{H_0}{2n} \right)^n \left( 1 - \frac{\delta H}{2} \right)  G(x,t),
\end{aligned}
\end{equation}
with
\begin{equation}\label{eq:nrqcd_ac}
\begin{aligned}
H_0 = &-\frac{1}{2 aM_0} \Delta^2,\\
H_I = &\left(-c_1\frac{1}{8 (aM_0)^3}-c_6 \frac{1}{16 n (aM_0)^2}\right)\left(\Delta^2\right)^2 \\
&+c_2 \frac{i}{8 (aM_0)^2}( \tilde\Delta\cdot \tilde{E} - \tilde{E} \cdot \tilde\Delta) + c_5 \frac{\Delta^4}{24 (aM_0)}\\
H_D = &-c_3 \frac{1}{8 (aM_0)^2}\sigma\cdot\left( \tilde\Delta\times \tilde{E} - \tilde{E} \times \tilde\Delta \right) -c_4 \frac{1}{2 (aM_0)} \sigma\cdot \tilde{B}\\
 \delta H = &H_I + H_D.
\end{aligned}
\end{equation}
Where for readibility we have separated the contributions to $\delta_H$ into the spin-dependent $H_D$ and spin-independent $H_I$ terms. Here $aM_0$ is the bare b-quark mass and $n$ is the stability
parameter (which we will always set to 4 throughout this work). The tilde
indicates a higher-order improved version of the derivative, tadpole improvement of the links, or improvement of the traceless clover field-strength tensor, of which $E$ and $B$ are the usual
components \cite{Meinel:2010pv}. Two typical choices for the
tadpole-improvement factor $U_0$ exist in the literature: the fourth-root
of the plaquette $U_{0P}$, and the mean Landau link $U_{0L}$. One of the facets of this work will be to investigate the dependence of our tuning on these choices \footnote{As we will use a mix of periodic
  and open boundary condition ensembles in time, we will look at a single, coarse, periodic
  box to do this comparison.}. As the Hamiltonian is derived from a Taylor expansion in
$aM_0$, results from measurements at $aM_0 \approx 1$ are not likely to be
trustworthy, and this puts a constraint on how fine a lattice spacing can be
used. 

We will later propose a strategy to allow all of the coefficients $c_i$ to be tuned nonperturbatively, but for now we discuss some expectations for these coefficients from the literature: the coefficient $c_3$ is expected to be close to 1 as there is a 
spin-average combination that is approximately proportional to $c_3$ \cite{Gray:2005ur} and it turns out to match experiment well when $c_3\approx 1$. The other spin-dependent term
$c_4$ is expected to affect most-strongly the $\Upsilon-\eta_b$ hyperfine splitting \cite{Dowdall:2012ab}
and plays a r\^ole in the spin-orbit splitting, typically it has a value greater than 1. The coefficient $c_2$ is commonly set to its tree-level value of 1. The terms $c_5$ and $c_6$ are pure discretisation-effect canceling terms for the spatial and temporal applications of $H_0$ respectively, and if these effects are small, should be $O(1)$. $c_1$ is suppressed by a further power of the mass and is expected to be close to 1 as well.

The NRQCD action is known to the next higher order ($O(v^6)$) \cite{Lepage:1992tx,Manohar:1997qy} with coefficients $c_7\dots c_{11}$,
\begin{equation}\label{eq:nrqcd_hospin}
\begin{aligned}
H_I^{v6} = &-c_{10}\frac{1}{8(aM_0)^3}\left(\tilde{E}\cdot \tilde{E}+\tilde{B}\cdot \tilde{B}\right) 
            - c_{11}\frac{1}{192(n)^2(aM_0)^3}(\Delta^2)^3 \\
H_D^{v6} = 
&-c_7\frac{1}{8(aM_0)^3}\left\{ \tilde{\Delta}^2,\sigma\cdot \tilde{B}\right\} 
-c_8\frac{3}{64(aM_0)^4}\tilde\Delta^2 \sigma\cdot\left( \tilde\Delta\times \tilde{E} - \tilde{E} \times \tilde\Delta \right)  \\
&- c_9\frac{i}{8(aM_0)^3}\sigma\cdot \left( \tilde{E}\times \tilde{E} + \tilde{B}\times \tilde{B} \right)\;.
\end{aligned}
\end{equation}
Inclusion of these higher-order terms has been argued in the literature to
have some small effect \cite{Lewis:1998ka,Meinel:2010pv,Hughes:2015dba} at typical
lattice spacings for bottomonia, similarly for the case of some
$bbb$-baryon splittings they do have some significant impact \cite{Meinel:2012qz}. We intend to see how well we can reproduce the bottomonium spectrum
with just the terms listed in Eq.~\ref{eq:nrqcd_ac}, as this will allow for a more quantitative comparison against previous works which measure related quantities. There is
nothing inherently stopping us from including higher
orders of the NRQCD Hamiltonian in our tuning, provided there are enough states that can be used as inputs. We address this point later in the paper in Sec.~\ref{sec:higher_order_syst}, where we consider adding only the tree-level spin-dependent higher-order coefficients $c_7=c_8=c_9=1$ to our tuning to probe their impact, and we will find that their inclusion does improve some of our heavy-light spin-splittings. We also consider, conversely, the implications of reducing the number of tuneable parameters in App.~\ref{sec:app_fixed} and find that it worsens the quality of our tuning.

As we see from Eq.~\ref{eq:evolutionNRQCD}, Lattice NRQCD incorporates corrections to
the static Wilson-line propagator, so working at very coarse lattice spacings
will likely include strong discretisation effects from the gluon field itself,
suggesting that there is an appropriate window for expecting accurate results. When investigating heavy-light B-meson physics, as we will do here, we are further
constrained from additional light-quark discretisation effects. As any
measurement with Lattice NRQCD inherently means we do not have a formal continuum
limit, serious care is needed in working within an appropriate range of
lattice spacings and in conservatively estimating discretisation effects.

\subsection{Nonperturbative tuning}

Lattice NRQCD calculations will often either use tree-level
parameters ($c_i=1$) \cite{Mathur:2002ce,Wurtz:2015mqa,Francis:2016hui},
or include a few determined at
$O(\alpha_s)$ using lattice perturbation theory
\cite{Morningstar:1994qe,Hammant:2011bt,Hammant:2013sca}. Occasionally, some \cite{Gray:2005ur,Dowdall:2012ab,Meinel:2012qz} have 
individually-tuned $c_4$ and/or $c_3$ by a non-perturbative prescription. Instead of any of these approaches we
will see how precisely we can determine the simple ground-state bottomonium
spectrum by allowing all of the parameters to vary and tuning them
simultaneously via a neural network. It is important to note that we will be
forcing a truncated series to approximate the continuum spectrum and the
parameters we determine will therefore be absorbing cut-off effects from many sources.

As Lattice NRQCD has an additive mass renormalisation, we cannot directly determine the masses of the pseudoscalar ($\eta_b$) or vector ($\Upsilon$) mesons. Instead, the overall b-quark mass $aM_0$ is tuned from the non-relativistic dispersion relation of the $\eta_b$ and $\Upsilon$ via the expansion
\begin{equation}
aE(p) = aM_1 + \frac{a^2p^2}{2aM_2} + \cdots,
\end{equation}
to match the kinetic mass, $aM_2$, to the spin-averaged continuum $\eta_b$ and $\Upsilon$ masses. We will use the value 9.445(2) GeV for this spin-average from \cite{Dowdall:2012ab}.
In practice this is done by us using partially-twisted boundary conditions \cite{Bedaque:2004kc,Sachrajda:2004mi}.

The operators and their continuum analogs used in our tuning can be found in Tab.~\ref{tab:optab}; these are all taken from \cite{Davies:1994mp} and use the quark-line-connected contractions of the simple operators
\begin{equation}
O(x) = \bar{\psi} \Gamma(x) \psi,
\end{equation}
with Coulomb gauge-fixed \footnote{Fixed to a precision of $10^{-14}$ using the FACG algorithm \cite{Hudspith:2014oja}.} wall-source, sink-smeared propagators.

\begin{table}
  \begin{tabular}{cc|c}
    \toprule
    State & PDG mass [GeV] \cite{Zyla:2020zbs} & $\Gamma(x)$\\
    \hline
    $\eta_b(1S)$   & 9.3987(20) & $\gamma_5$ \\
    $\Upsilon(1S)$ & 9.4603(3) & $\gamma_i$ \\
    \hline
    $\chi_{b0}(1P)$ & 9.8594(5) & $\sigma\cdot \Delta$ \\
    $\chi_{b1}(1P)$ & 9.8928(4) & $\sigma_j\Delta_i-\sigma_i\Delta_j$ $(i \neq j)$ \\
    $\chi_{b2}(1P)$ & 9.9122(4) & $\sigma_j\Delta_i+\sigma_i\Delta_j$ $(i \neq j)$ \\
    $h_b(1P)$      & 9.8993(8) & $\Delta_i$ \\
    \botrule
  \end{tabular}
  \caption{Table of lattice operators used with their continuum-state analogs. Here $\sigma_i$ are the usual Pauli matrices and $\Delta_i$ is the symmetric lattice finite difference.}\label{tab:optab}
\end{table}

We have chosen the simple ground-state operators of Tab.~\ref{tab:optab} for
our tuning data as there should be little ambiguity in determining their
masses precisely from a correlated single-exponential fit to the resulting correlator. This allows for higher-order excitations of bottomonia to become predictions/indications of the quality our tuning as discussed in Sec.~\ref{sec:excited_b}. We chose to tune our parameters against the splittings with regard to the
$\eta_b$ of the states listed in Tab.~\ref{tab:optab}. For the continuum splittings we will use the most-recent PDG values \cite{Zyla:2020zbs}, while noting that there is some tension between the experimental determinations of the $\Upsilon-\eta_b$ hyperfine splitting.

Often studies will focus on splittings within the S- and P-wave states
\cite{Davies:1994mp} separately, because the measured S-wave -- P-wave splitting is usually quite
far from its continuum value due to unknown radiative corrections and
higher-order effects not accounted for. An example of this can be seen in e.g. \cite{Wurtz:2015mqa}, where a significant difference between the PACS-CS lattice spacing and one derived from the $1P-1S$ splitting was seen. The procedure of remedying this by re-defining the lattice spacing after tuning $aM_0$ by this splitting is somewhat common in practice \cite{HPQCD:2003rsu,Meinel:2009rd}. For us, 
it is hard to justify re-determining the lattice spacing to match a physical
splitting from Lattice NRQCD, and we will show with our tuning that it is possible to absorb these differences into the coefficients themselves.

\begin{figure}
  \includegraphics[scale=0.6]{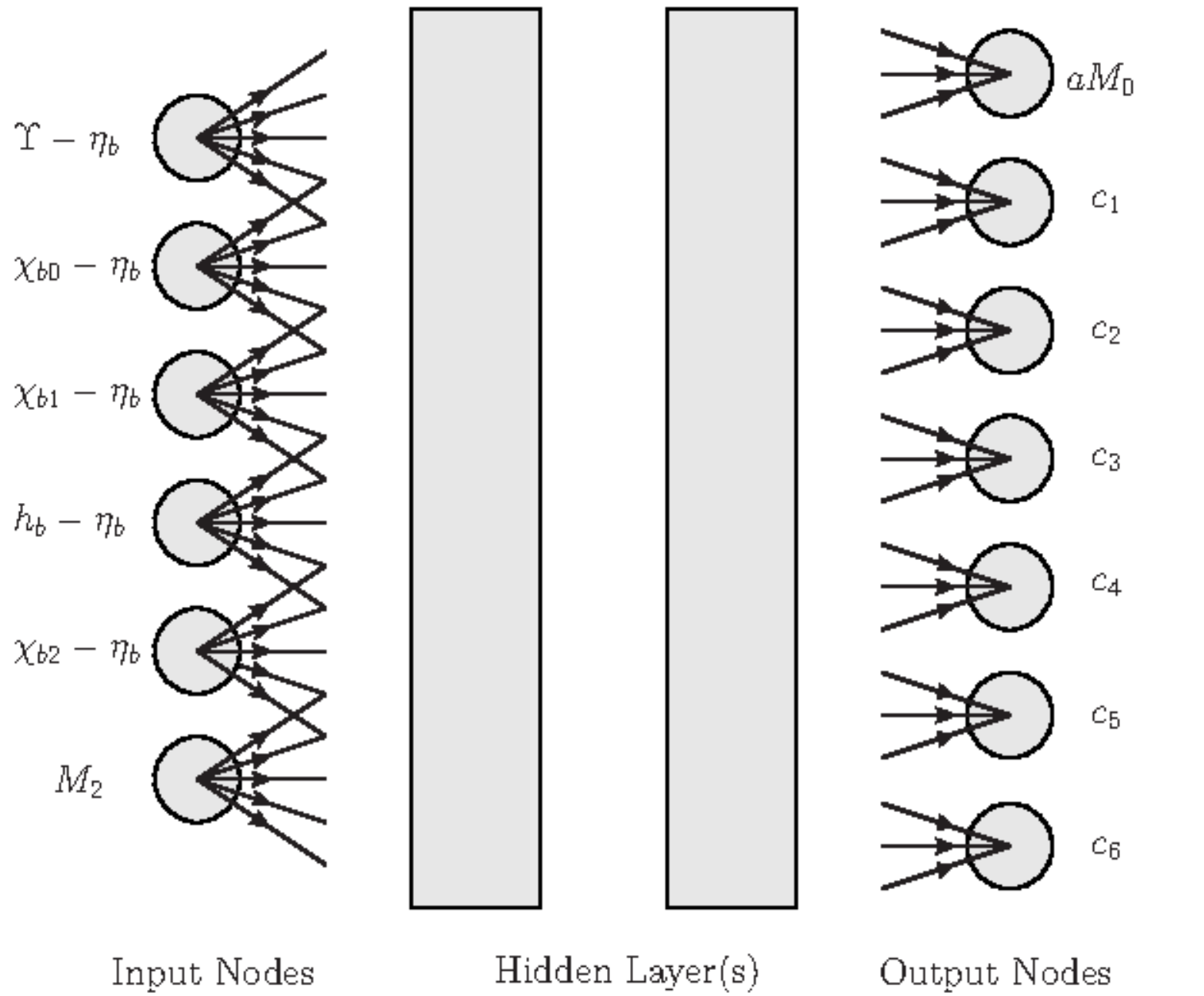}
  \caption{Schematic picture of our setup, lines indicate weights (not all are drawn as each node is connected to each within the hidden layer) and the arrows indicate the feed-forward nature of our network.}\label{fig:nnschem}
\end{figure}

We will follow our prior work of \cite{Hudspith:2021iqu} where we tuned a
relativistic charm-quark action nonperturbatively by using a neural network to
infer the dependence of states on the action parameters. This time we will
focus on the splittings of states, rather than explicit masses. A schematic of
our approach is shown in Fig.~\ref{fig:nnschem}, showing the inputs and
outputs we use to train the network. It should be noted that the parameters
are not universal: they will, in principle, be specific to our underlying
gauge action, fermionic action, our choice of tadpole improvement factor, our
choice of improvement terms in the NRQCD Hamiltonian, and even our renormalisation trajectory set by the observable(s) used to define it.

We will use a network with two hidden layers of just 12 nodes. The network uses the \verb|Adam| minimizer \cite{kingma2014method} with early stopping, mini-batches, and an adjusted learning rate. Approximately $20\%$ of our data is used for validation, and we generate 100 different runs of randomly-chosen Lattice NRQCD coefficients per ensemble.

One may wonder if our setup in Fig.~\ref{fig:nnschem} is underdetermined due to having more outputs than inputs, but we observe that the coefficients are all empirically well-determined. To further investigate this concern, we perform a test on ensemble A653 with fewer parameters (we fix $c_5=c_6=1$) to check the general features of our tuning in App.~\ref{sec:app_fixed}. We observe a slightly worse quality in the predicted parameters with this reduced parameter set but qualitatively similar results to our full tuning.

\section{Tuning results\label{results}}

\begin{table}[h]
\begin{tabular}{ccc|cc}
\toprule
Ensemble & T-boundary & a[fm] & $U_0$ & $\text{N}_{\text{Conf}}\times \text{N}_{\text{Prop}}$ \\ 
\hline
A653 & Periodic & 0.09929 & 0.85005 & $100\times12$ \\
A653$^*$ & Periodic & 0.09929 & 0.82918 & $100\times 12$ \\
\hline
U103 & Open & 0.08636 & 0.85248 & $500\times 2$ \\
B450 & Periodic & 0.07634 & 0.85812 & $400\times 8$ \\
H200 & Open & 0.06426 & 0.86153 & $500\times 2$ \\
\botrule
\end{tabular}
\caption{Flavor $\text{SU}(3)_f$-symmetric ensembles used in this work for the nonperturbative tuning. A653$^*$ indicates tadpole-improvement with the mean Landau link, all others use the fourth root of the plaquette determined from the entire lattice volume. $N_\text{Conf}$ indicates the number of independent gauge configurations used and $N_\text{Prop}$ indicates the number of NRQCD propagators used.}\label{tab:ensembles}
\end{table}

We will only consider gauge ensembles generated by the Coordinated Lattice Simulations (CLS) consortium at the $\text{SU}(3)_f$-symmetric point to determine our Lattice NRQCD parameters,
as dynamical light pions are not expected to significantly effect the
splittings considered \footnote{This was tested on our lightest pion-mass ensemble (C101) to be the case.}. These ensembles have a mixture of open- (U103 and H200) and periodic- (A653 and B450) temporal-boundaries. All of these ensembles are $n_f=2+1$ nonperturbatively-improved clover-Wilson, Symanzik gauge configurations and are listed in
Tab.~\ref{tab:ensembles}. Further details on their generation can be found
in \cite{Luscher:2012av} and \cite{Bruno:2014jqa}, with lattice spacings from
\cite{Bruno:2016plf} and an estimate for A653 from \cite{Chao:2021tvp}. For
the open-boundary configurations we sit in the middle of the lattice and
compute forward and backward Lattice NRQCD b-quark propagators away from this time-slice. For the
periodic-boundary configurations we only compute the forward-propagating state
to allow for a larger temporal range to fit to. For the open-boundary-condition ensembles we use the plaquette for the whole gauge field in our tadpole improvement, and not one measured in the bulk.

\begin{table}
  \begin{tabular}{c|ccccccc}
    \toprule
    Ensemble & $aM_0$ & $c_1$ & $c_2$ & $c_3$ & $c_4$ & $c_5$ & $c_6$ \\
    \hline
    A653 & 2.0585(25) & 0.611(18) & -1.255(44) & 1.121(11) & 0.996(10) & 0.688(17) & 0.701(17) \\
    A653$^*$ & 2.1088(55) & 0.790(14) & -1.014(73) & 1.062(15) & 0.931(18) & 0.700(13) & 0.695(13) \\
    \hline
    U103 & 1.8035(32) & 0.787(16) & -0.789(61) & 1.057(10) & 0.960(11) & 0.828(13) & 0.932(14) \\
    B450 & 1.5542(89) & 0.819(16) & -0.556(50) & 1.077(13) & 0.913(15) & 0.861(11) & 0.903(15) \\
    H200 & 1.3189(19) & 0.940(9) & -0.519(56) & 1.027(3) & 0.832(13) & 0.866(10) & 0.866(9) \\
    \botrule
  \end{tabular}
  \caption{Nonperturbative tuning parameters for the various $\text{SU}(3)_f$-symmetric ensembles considered in this work, bracketed values indicate the variations in the neural-network predicted parameters and not true errors. For each ensemble 100 tuning runs were performed.}\label{tab:nncoefs}
\end{table}

\begin{figure}[t]
\includegraphics[scale=0.27]{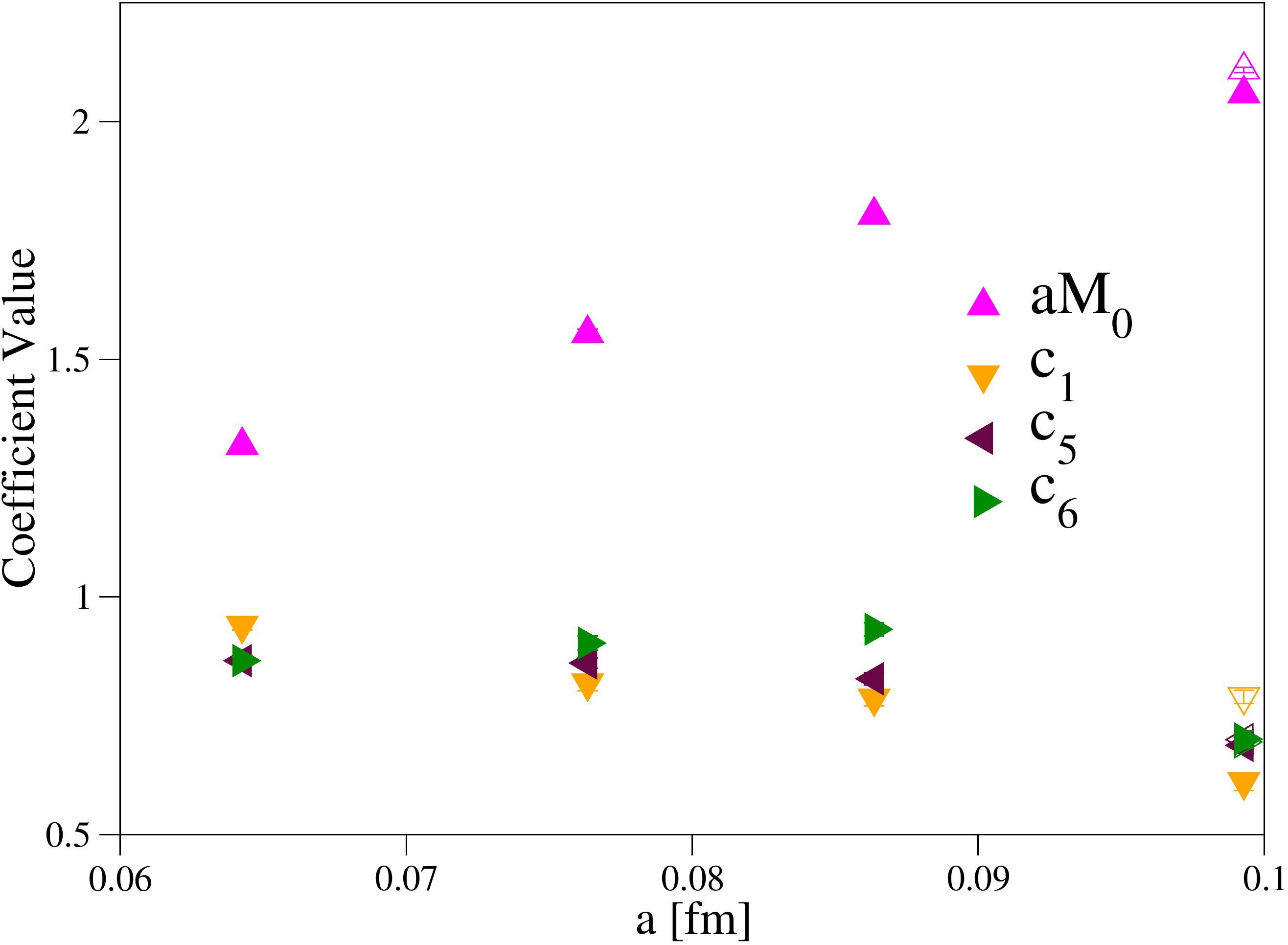}
\includegraphics[scale=0.27]{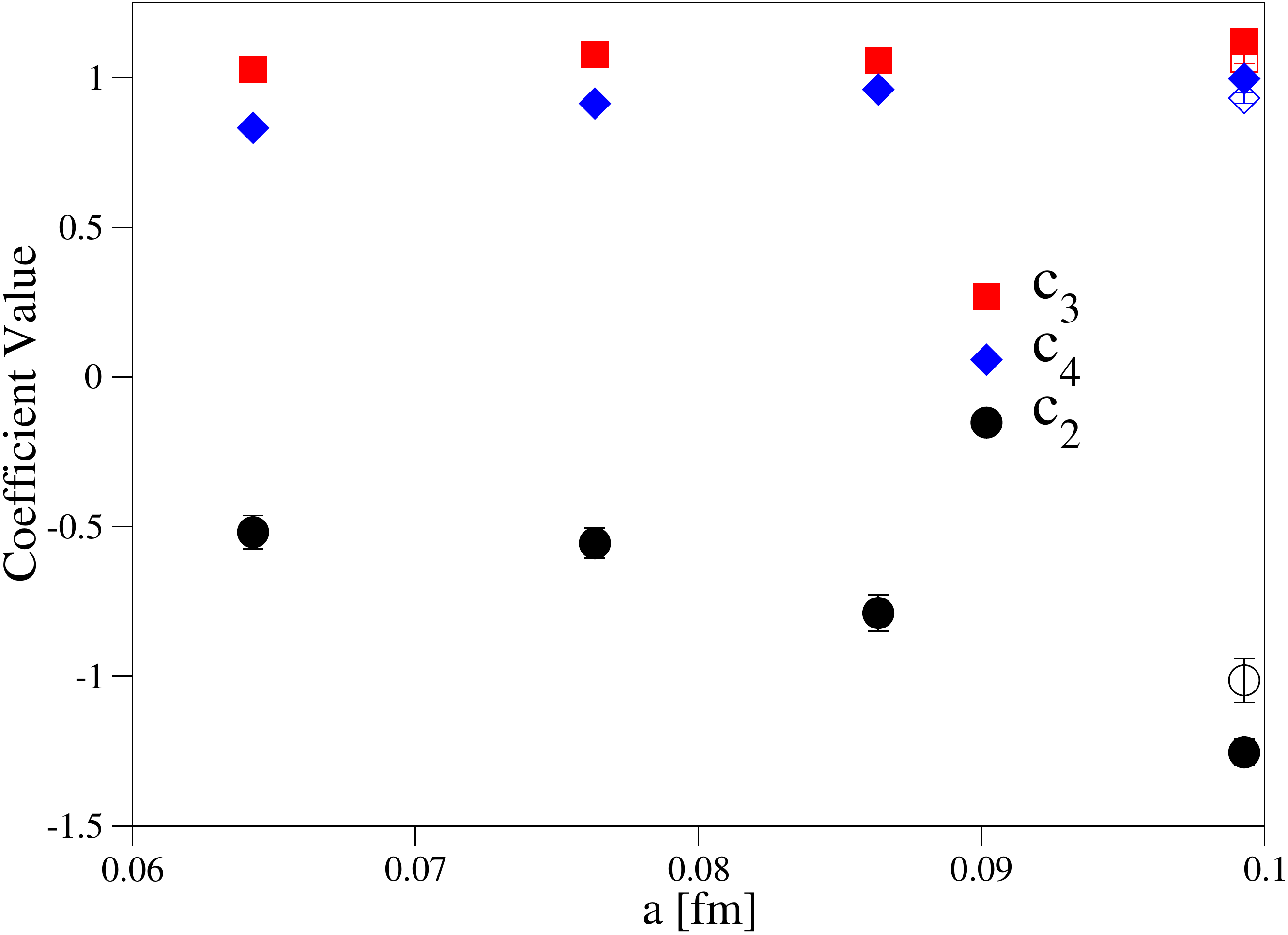}
\caption{Plots of the coefficient determinations (Tab.~\ref{tab:nncoefs}) from
  our neural network vs. the lattice spacing. Open symbols indicate the values
  from tuning with the $U_{0L}$ tadpole prescription.}\label{fig:coeffs}
\end{figure}

Tab.~\ref{tab:nncoefs} and Fig.~\ref{fig:coeffs} illustrate a curious
result: the neural network strongly prefers a negative value of the
coefficient $c_2$ with a significant dependence on both the lattice spacing
and the choice of tadpole factor
$U_0$. It is important to note that positive values of
$c_2$ exist in all of our training datasets, as do the tree-level
parameters, so this result is unlikely to be biased by our training data. The
parameters $c_1,\;c_5,$ and $c_6$ are the other spin-independent contributions to the
NRQCD Hamiltonian and are all smaller than 1 and seem to be inversely
proportional to $c_2$. This is further illustrated in our re-tuning with $c_5=c_6=1$ in
App.~\ref{sec:app_fixed} where we observe that $c_2$ decreases when $c_5$ and $c_6$ are larger. It is evident that the tuned $c_1$ acts similarly
as it multiplies the same operator as $c_6$, albeit with a different power of the bare mass and numerical prefactors.

A slightly unexpected behaviour is also
apparent for the determined coefficient $c_4$, as it is smaller than 1 for all
our ensembles, suggestive of interplay between the determinations of $c_2$ and
$c_4$. From our training runs we have seen that while
$c_2$ is sensitive to the S-wave -- P-wave splitting (with
positive $c_2$ reducing this and negative $c_2$ enhancing it), $c_4$ is mostly 
sensitive to the hyperfine $\Upsilon-\eta_b$ ($1S$) splitting with a larger value of $c_4$ increasing this splitting. This is further supported by our results in Sec.~\ref{sec:higher_order_syst} where tuning to the $B^*-B$ splitting increases the 1S-hyperfine and the parameter $c_4$ whilst also making $c_2$ positive.

Both $c_2$ and $c_4$ show strong dependence on the choice of tadpole factor,
which is unsurprising as they multiply the field strength tensor. Oddly, the
network gives the coefficient $c_1$ some visible tadpole dependence;
implying this coefficient is being used to compensate for changes in $c_2$
and $c_4$. In our bare-mass regime (and with our chosen stability parameter) $c_1$ is likely one of the largest spin-independent contributions and therefore a significant handle for the neural network.

The determined coefficient $c_3$ is always close to 1 with some tadpole-dependence and mild lattice-spacing variation. The parameters $c_5$ and $c_6$ show almost no tadpole-dependence but strong lattice-spacing dependence, which is to be expected as their job is to cancel higher-order lattice artifacts.

\begin{figure}[h!]
  \centering
  \includegraphics[scale=0.27]{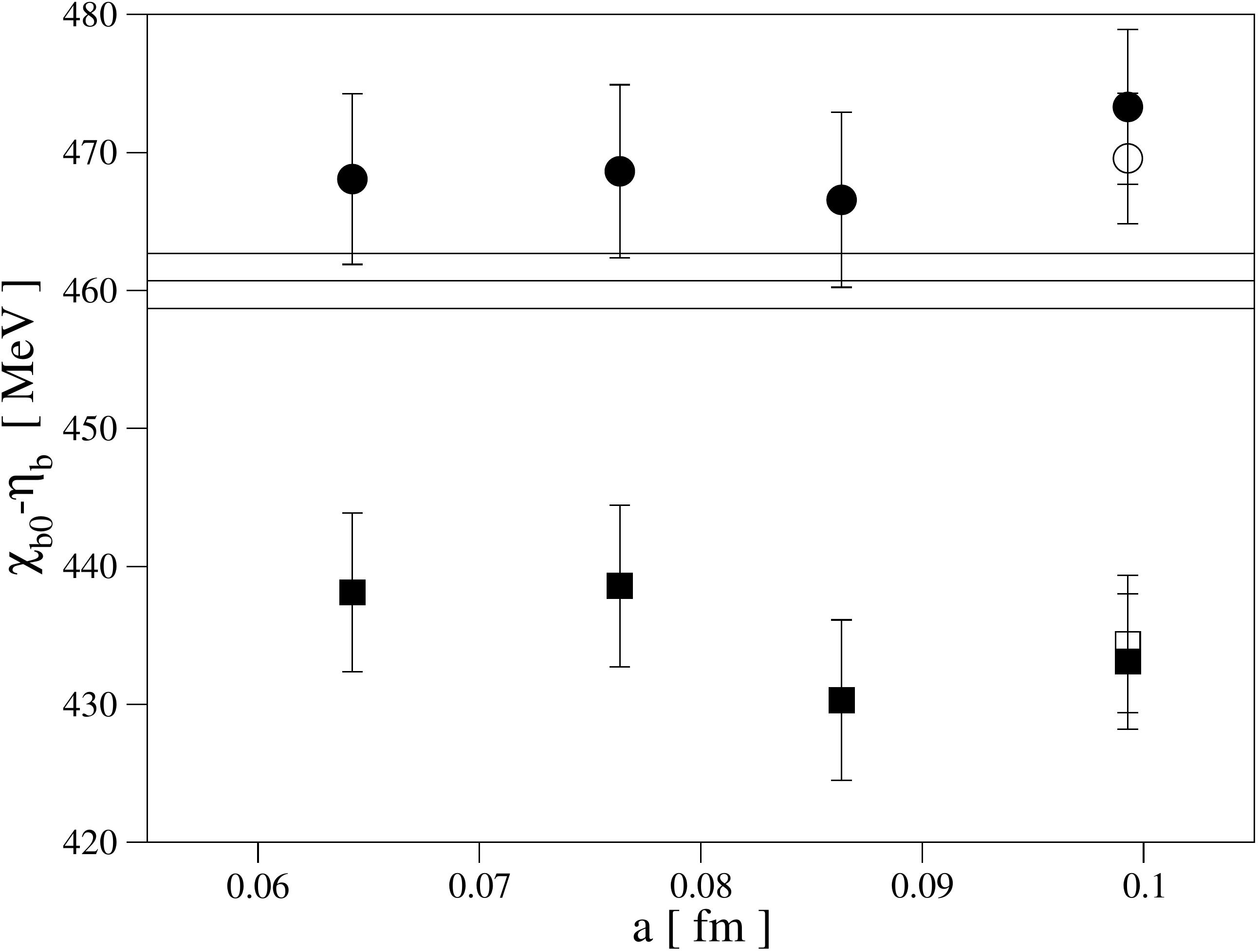}
  \includegraphics[scale=0.27]{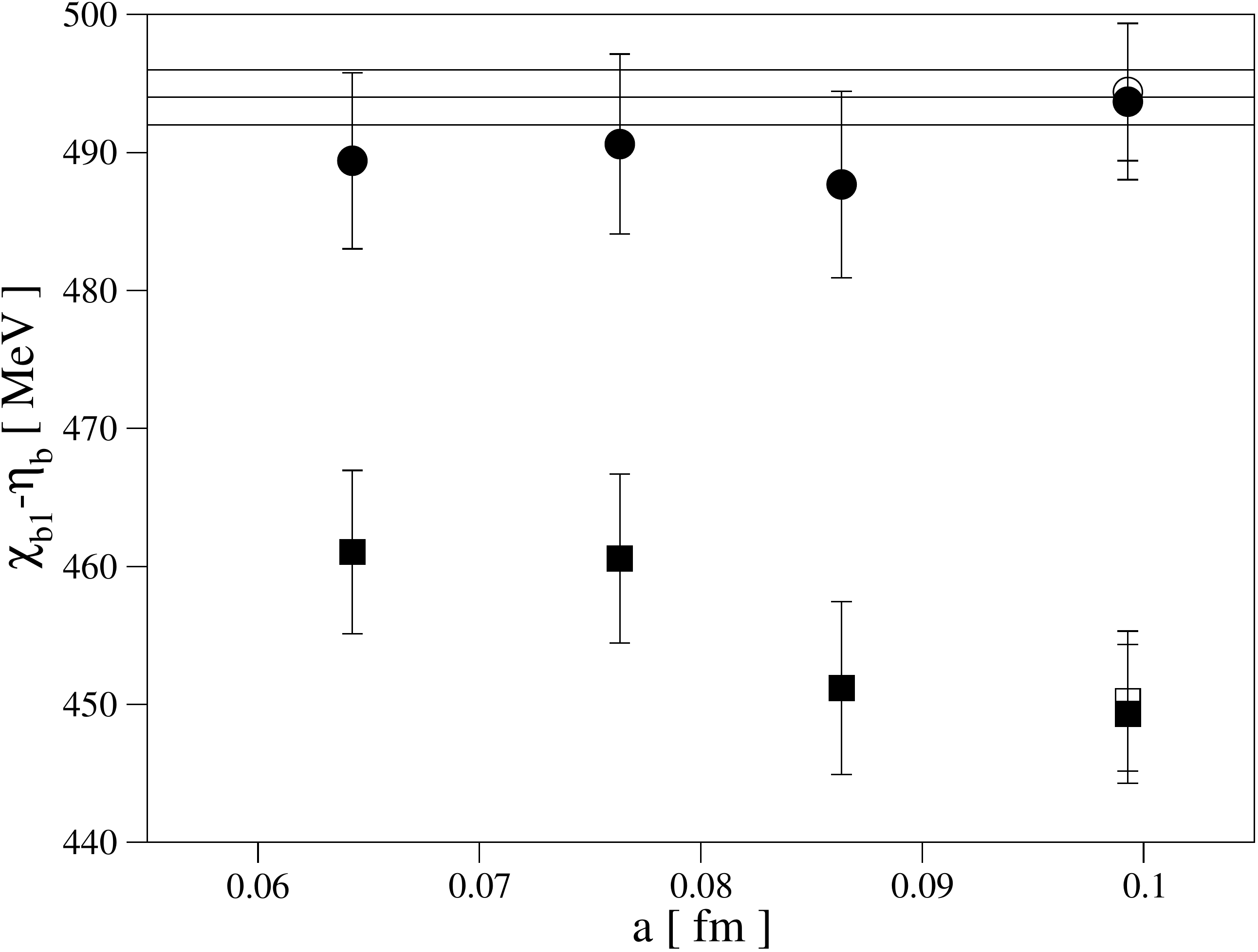}
  \includegraphics[scale=0.27]{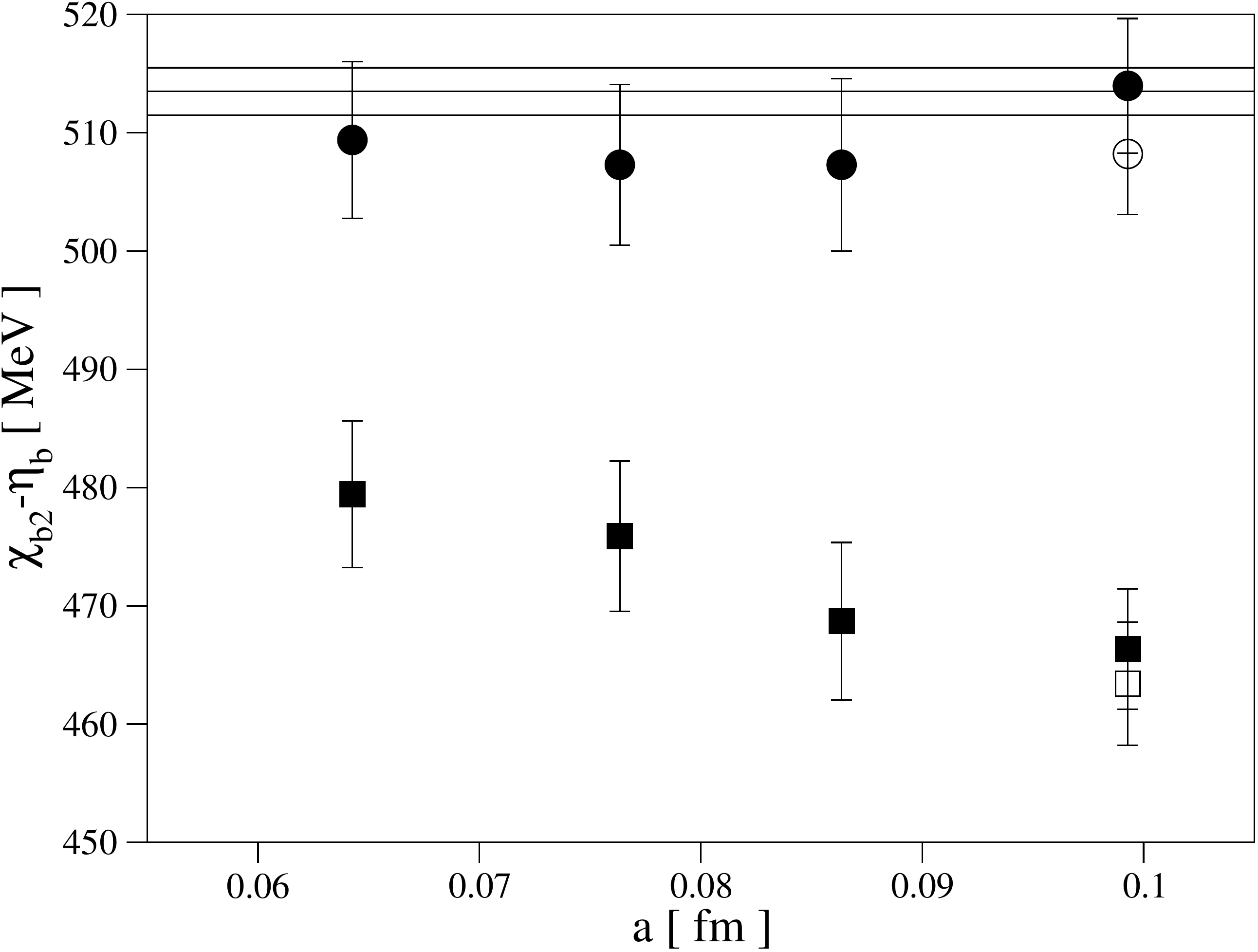}
  \includegraphics[scale=0.27]{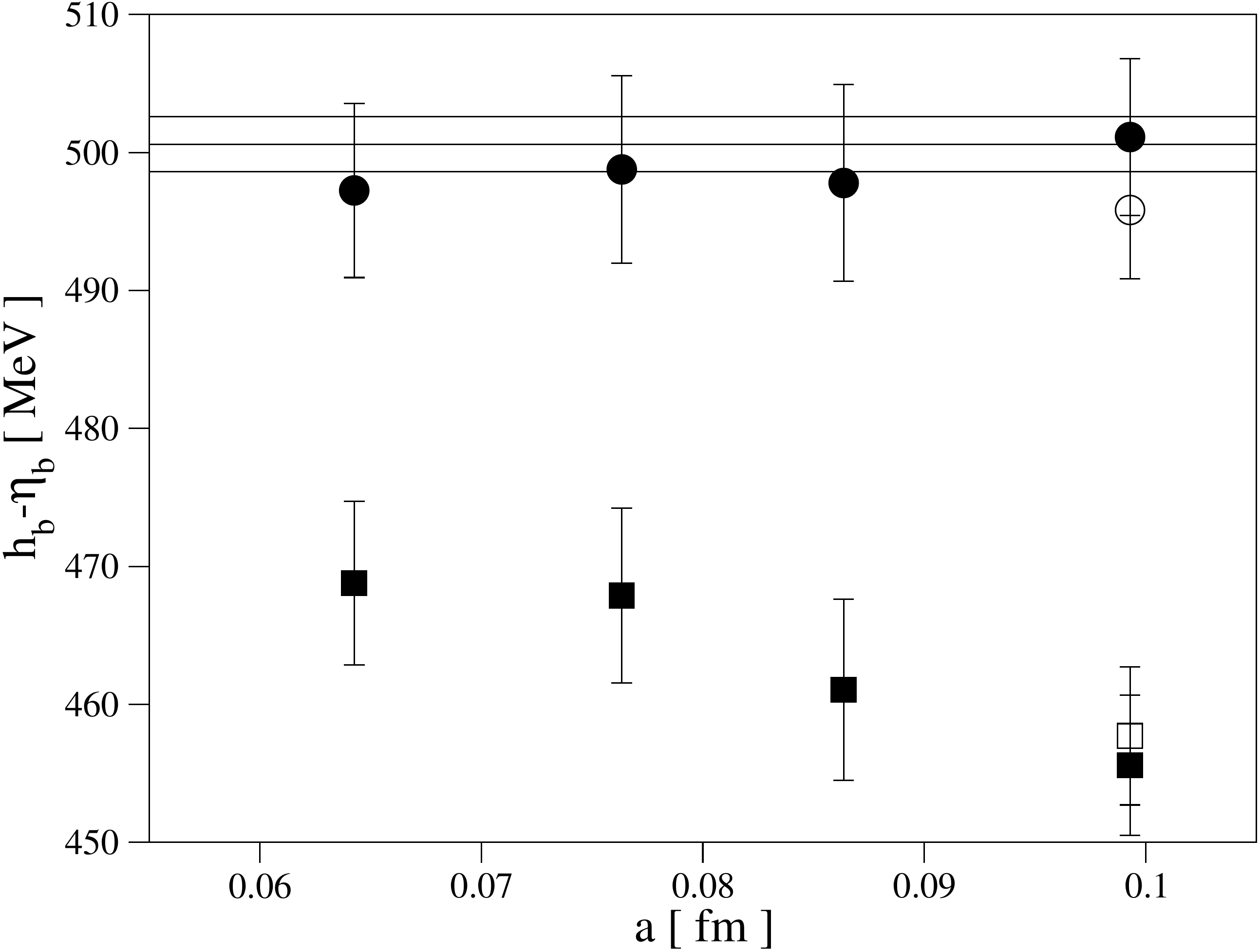}
  \includegraphics[scale=0.27]{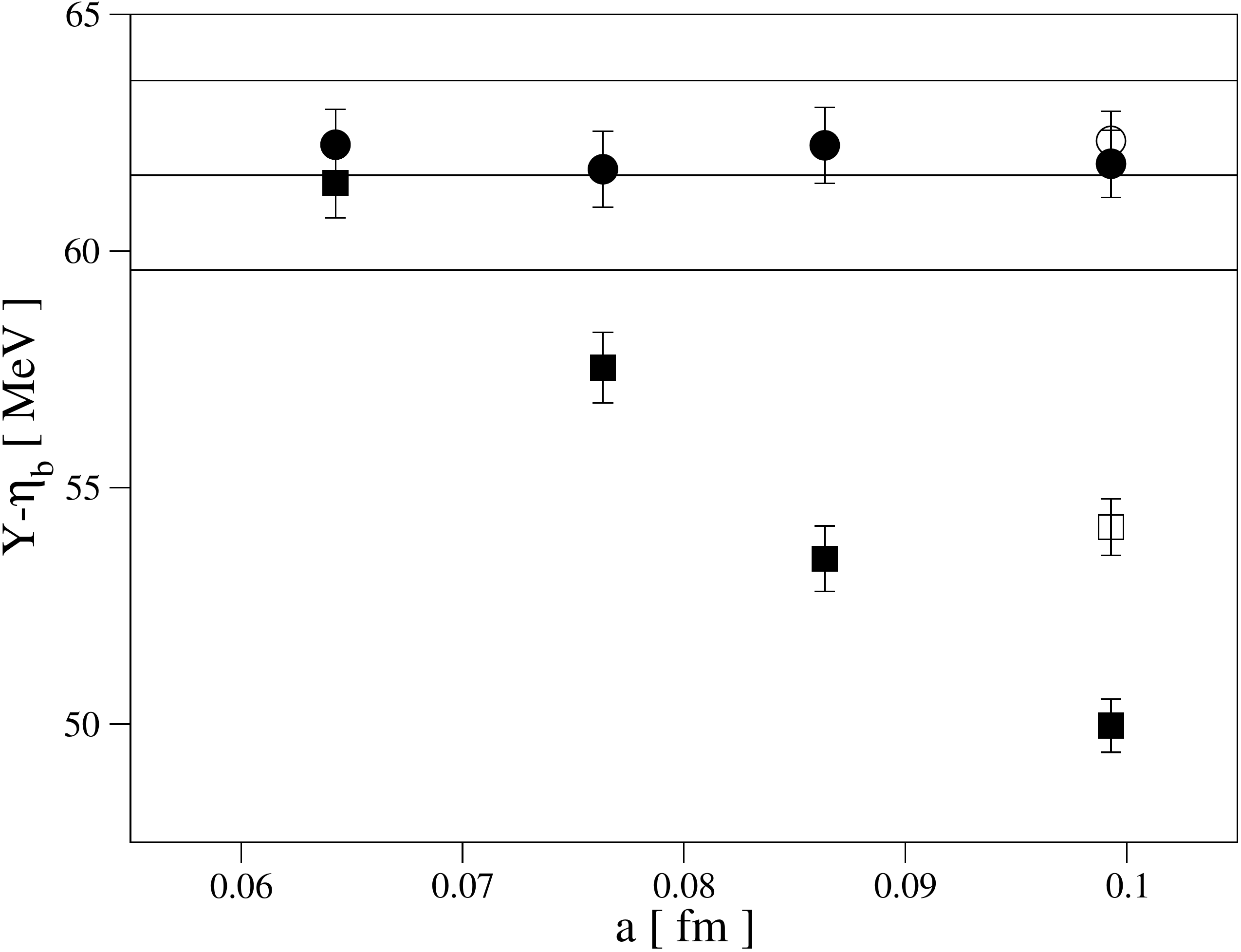}
  \caption{Measured splittings from our tuning (circles) vs. those obtained from using tree-level coefficients (squares). Open symbols illustrate the use of the mean Landau link tadpole improvement term, closed symbols from the fourth root of the plaquette. Horizontal lines indicate the PDG values for these splittings.}\label{fig:tunings}
\end{figure}

The bare-mass tuning appears to not strongly depend
on the other coefficients, only on the choice of tadpole improvement
factor. This can be inferred by comparing the tuned values of $aM_0$ for the
tree-level tuning (Tab.~\ref{tab:tree_tune}) with our neural network
parameters (Tab.~\ref{tab:nncoefs}), where $aM_0$ is more-or-less consistent
within errors. This suggests that one can likely tune $aM_0$ independently of
all the other parameters (with the interesting exception of A653 with the
$U_{0L}$ tadpole-term). For the nonperturbatively-tuned parameters
we see that for $U_{0P}$ or $U_{0L}$ the ratio of their $aM_0$s is consistent with the inverse of their ratios of tadpole improvement factors.

Fig.~\ref{fig:tunings} illustrates our ability to reproduce the experimental
low-lying S- and P-wave splittings for bottomonia in comparison to tree-level
NRQCD for different lattice spacings. Apart from our tuning being 
consistent with experiment, we note some interesting features: tree-level
NRQCD's approach to the continuum for the 1P-1S splittings appear to be linear with the lattice spacing,
indicating that on these lattices it will never be accurately reproduced within the
applicability window of NRQCD. For our neural network tuning $\chi_{b0}$ appears slightly too heavy, and $\chi_{b2}$ too
light, with $\chi_{b1}$ and $h_b$ matching experiment much better. This could be a
sign of missing higher-order terms in our NRQCD implementation as this feature
seems independent of the lattice spacing. Finally, we note that for the tree-level parameters' results with the tadpole factor $U_{0L}$ there is better agreement with the experimental hyperfine splitting than for those using $U_{0P}$. The two choices of tadpole-improvement factor for tree-level NRQCD are, however, equally-poor for the splittings between the P-wave states and the $\eta_b$. Whichever tadpole-improvement factor is used is completely irrelevant with the
neural network tuning as the choice gets absorbed into the nonperturbative coefficients.

\begin{table}
  \begin{tabular}{c|ccccc}
    \toprule
    Ensemble & A653 & A653$^*$ & U103 & B450 & H200 \\ 
    \hline
    $aM_0$ & 2.073(16) & 2.151(16) & 1.812(12) & 1.559(6) & 1.323(9)\\ 
    \botrule
  \end{tabular}
  \caption{Tree-level $aM_0$ tunings obtained from the spin-averaged kinetic mass by linear interpolation.}\label{tab:tree_tune}
\end{table}

We investigate our ability to reproduce the continuum spectrum via the simple absolute percentage deviation metric
\begin{equation}\label{eq:pcdev}
D = \frac{100}{N}\sum_i^{N} \bigg|\frac{\bar{S}^{\text{Latt.}}_i-\bar{S}^{\text{Cont.}}_i}{\bar{S}^{\text{Cont.}}_i}\bigg|,
\end{equation}
with $\bar{S}^{\text{Latt.}}$ being the average value of one of our splittings and $\bar{S}^{\text{Cont.}}$ being the expected central-value of its continuum counterpart. We present results comparing our states from our predicted parameters with those from the tree-level tuning in Tab.~\ref{tab:deviations}; the approach to the continuum for each of our splittings can be also be seen in Fig.~\ref{fig:tunings}.

Tab.~\ref{tab:deviations} illustrates quite clearly that the tree-level parameters do appear to approach the continuum experimental spectrum, albeit very slowly. Our nonperturbative tuning never deviates in central value by more than 1.1\% on average and within the experimental and lattice errors the agreement is near absolute. Of course, this is the intention of our tuning but it is worth noting that at the simple truncation of the NRQCD Hamiltonian such a reproduction of experiment is possible and a tuning with higher-order coefficients against this set of states will be unlikely to provide much improvement, and may not be well-constrained.

\begin{table}
  \begin{tabular}{c|ccccc}
    \toprule
    Ensemble & A653 & A653$^*$ & U103 & B450 & H200 \\ 
    \hline
    NN-tuned   &  0.7\% & 1.0\% & 1.1\% & 0.8\% & 1.0\% \\
    Tree-level & 10.4\% & 9.0\% & 9.0\% & 6.4\% & 5.0\% \\
    \botrule
  \end{tabular}
  \caption{Average absolute percentage deviation ``$D$'', Eq.~\ref{eq:pcdev}, from our target continuum splittings.}\label{tab:deviations}
\end{table}

\subsection{Excited states of bottomonium}\label{sec:excited_b}

As we have already seen in Fig.~\ref{fig:tunings}, tree-level Lattice NRQCD's ability to replicate even the basic 1P-1S splittings is poor and it seems unlikely that their excited couterparts' determination will be much better. In this section we consider the excited S-wave $\eta_b(3S/2S)$,  $\Upsilon(3S/2S)$ and P-wave $\chi_{b0}(3P/2P)$, $\chi_{b1}(3P/2P)$, $\chi_{b2}(3P/2P)$, and $h_b(3P/2P)$ states' splittings with respect to the $\eta_b(1S)$, and compare these to their PDG continuum counterparts where known.

\begin{table}[h]
\begin{tabular}{cc|cccc}
\toprule
Ensemble & $N_\text{Conf}\times N_\text{Prop}$ & 1 & 2 & 3 & 4 \\
\hline
A653 & $500\times 48$ & 2.5 & 5 & 10 & 20 \\
U103 & $500\times 16$ & 2.5 & 5 & 10 & 20 \\
B450 & $400\times 32$ & 4 & 8 & 16 & 32 \\ 
H200 & $500\times 8$  & 4 & 8 & 16 & 32 \\ 
\botrule
\end{tabular}
\caption{Gaussian smearing widths $\alpha$ used at source and sink to create the GEVP of Eq.~\ref{eq:smearmat} as well as the number of configurations and NRQCD bottom propagators used.}\label{tab:smalpha}
\end{table}

For the determination of our excited states we choose to create a symmetric Generalised Eigenvalue Problem (GEVP) \cite{Michael:1982gb,Luscher:1990ck,Blossier:2009kd} from a
matrix of correlation functions constructed with different Gaussian source (first S) and sink (second S) smearings:
\begin{equation}\label{eq:smearmat}
  \begin{pmatrix}
    S_1S_1 & S_1S_2 & S_1S_3 & S_1S_4 \\
    S_2S_1 & S_2S_2 & S_2S_3 & S_2S_4 \\
    S_3S_1 & S_3S_2 & S_3S_3 & S_3S_4 \\
    S_4S_1 & S_4S_2 & S_4S_3 & S_4S_4 
  \end{pmatrix}
  .
\end{equation}
As was the case in the tuning we use Coulomb gauge-fixed wall sources, where for the source
we simply apply an arbitrary function, and for the sink we apply the
convolution methodology outlined in the appendices of
\cite{Hudspith:2021iqu}. Each ``S'' denotes a different smearing radius
(squared), which we will call $\alpha$, as illustrated in
Tab.~\ref{tab:smalpha}. We are primarily
interested in the lowest 3 states, with the fourth typically lying above the
$B\bar{B}$-threshold, and expected to have contamination from higher states. We solve this GEVP and diagonalise the correlator matrix for a specific ``diagonalisation time" \cite{McNeile:2000hf} to obtain our principal correlators, which we then perform a correlated single-exponential fit to. Here, we needed increased statistics in comparison to our tuning runs (Tab.~\ref{tab:ensembles}) to be able to stably resolve the higher levels. For the P-wave states the calculation is quite complicated as the source has a derivative and must be matched similarly at the sink, and contracted with a propagator with smearing at the same source and sink going backwards, hence many Lattice NRQCD propagators are needed for these states.

\begin{figure}[h!]
\includegraphics[scale=0.45]{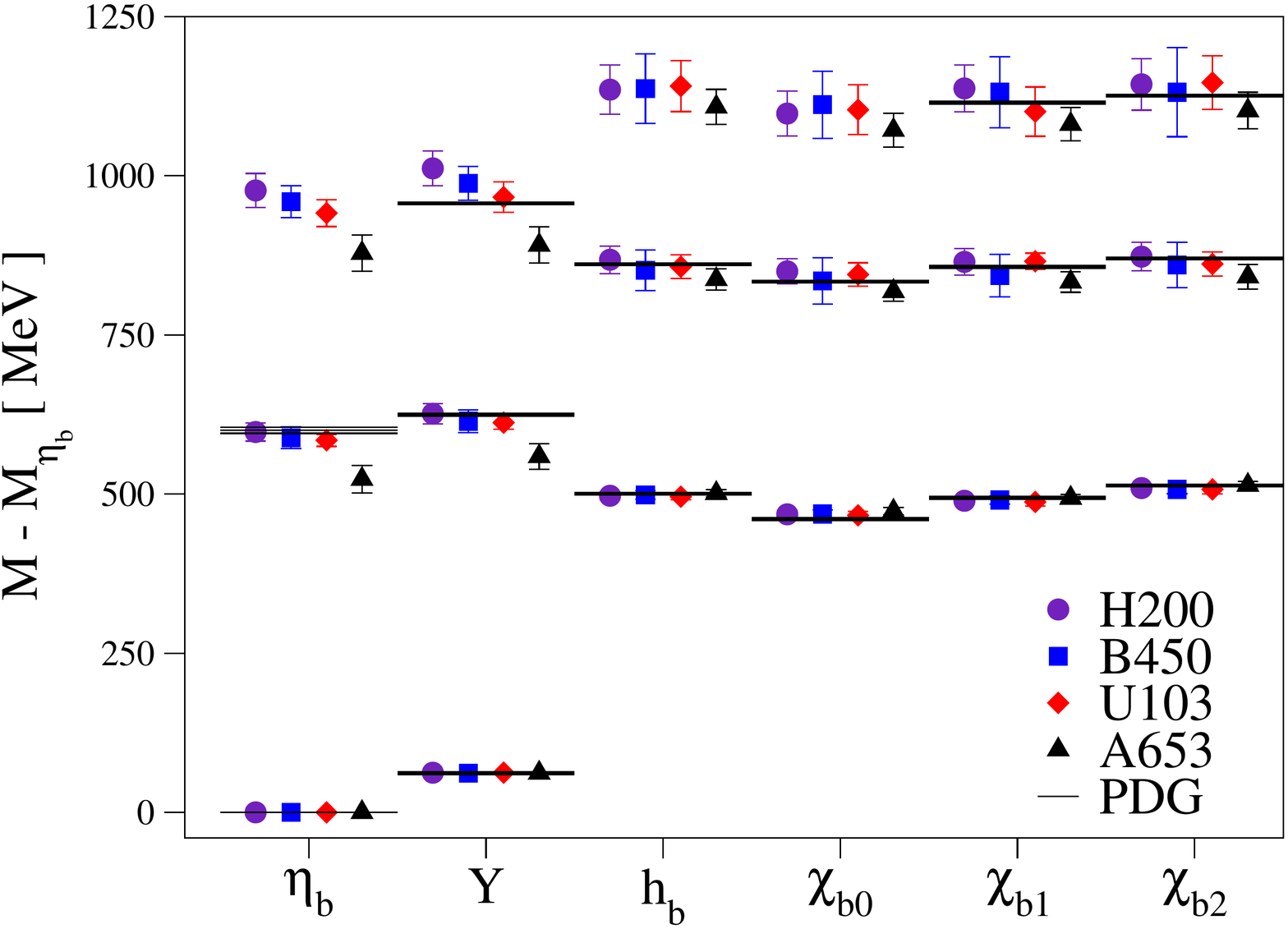}
\includegraphics[scale=0.45]{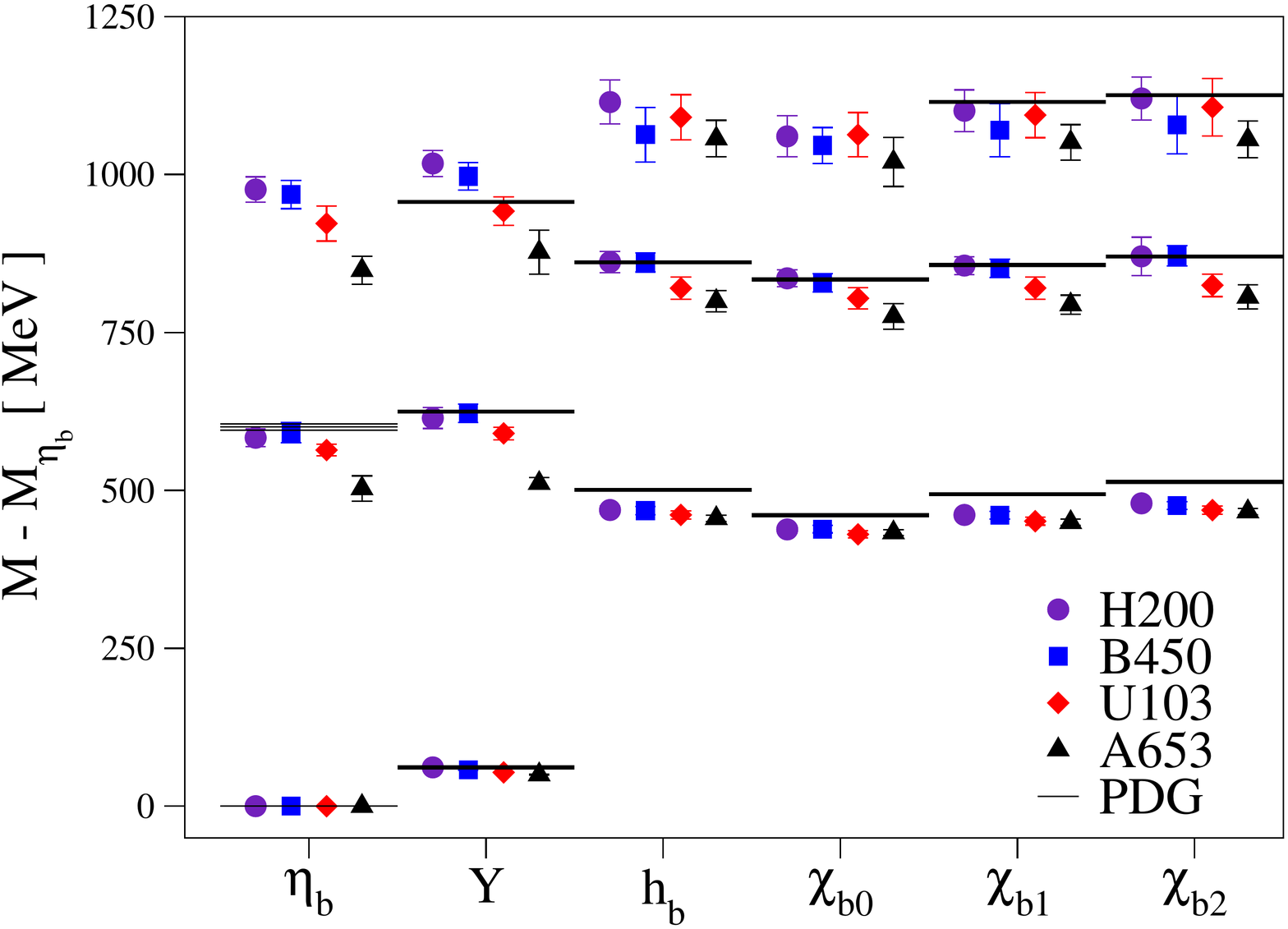}
\caption{S-wave and P-wave splittings of bottomonia with respect to the $\eta_b$ from the neural network tuning (top) and the tree-level coefficients (bottom). Shown are the 1S, 2S, 3S, and 1P, 2P, and 3P splittings as well as their values from the PDG.}\label{fig:highorder}
\end{figure}

Fig.\ref{fig:highorder} illustrates the excited S- and P-wave state splittings with respect to the $\eta_b(1S)$. As we have already seen earlier in this section, the tree-level coefficients under-estimate the 1P-1S
splitting. However, their 2P-1S splitting is closer to experiment and
the 3P-1S is again lower. This well-behaved 2P-1S splitting of tree-level Lattice NRQCD is likely a
coincidence as the 2P-1P splitting is significantly over-estimated as the
lattice-spacing decreases. In comparison, the neural network tuning for the 2P-1S splitting
is consistent with experiment and the spread of results over our lattice
spacings is much more narrow. For the 2S-1S and 3S-1S splittings the spread of
values is again smaller for the neural network, which illustrates that some
discretisation effect is being treated better in our tuning. The variation in errors is indicative of some of the ambiguities in finding a good fit range for these states, and the results presented in this section have errors which are statistical-only and intended to be indicative rather than quantitative.

\section{Exotic doubly-heavy tetraquark states}\label{sec:ram}

\begin{table}[h!]
\begin{tabular}{cc|cc}
\toprule
Ensemble & Mass trajectory & $L^3\times L_T$ & $N_\text{Conf}\times N_\text{Prop}$ \\
\hline
U103 & $\text{Tr}[M]=C$ & $24^3\times 128$ & $1000\times 23$ \\
H101 & $\text{Tr}[M]=C$ & $32^3\times 96$ & $500\times 12$ \\
\hline
U102 & $\text{Tr}[M]=C$ & $24^3\times 128$ & $732\times 18$ \\
H102 & $\text{Tr}[M]=C$ & $32^3\times 96$ & $500\times 16$ \\
\hline
U101 & $\text{Tr}[M]=C$ & $24^3\times 128$ & $600\times 18$ \\
H105 & $\text{Tr}[M]=C$ & $32^3\times 96$ & $500\times 16$ \\
N101 & $\text{Tr}[M]=C$ & $48^3\times 128$ & $537\times 18$ \\
\hline
C101 & $\text{Tr}[M]=C$ & $48^3\times 96$ & $400\times 16$ \\
\hline
\hline
H107 & $\widetilde{m_s}=\widetilde{m_s}^{\text{Phys.}}$ & $32^3\times 96$ & $500\times16$ \\
H106 & $\widetilde{m_s}=\widetilde{m_s}^{\text{Phys.}}$ & $32^3\times 96$ & $500\times16$ \\
\hline
\hline
H200 & $\text{Tr}[M]=C$ & $32^3\times 96$ & $500\times 28$ \\
\botrule
\end{tabular}
\caption{Ensemble details and gathered statistics used in our study of doubly-heavy tetraquarks and exotic $B_s$-mesons. $N_\text{Conf}$ indicates the number of independent gauge configurations used and $N_\text{Prop}$ indicates the number of propagators (light,strange, and $2\times$bottom) generated per configuration.}\label{tab:conftable}
\end{table}

Tab.~\ref{tab:conftable} gives the ensembles used for the analysis of
tetraquarks and $B_s$ mesons, and in terms of statistics at least
$6000$ light, strange, and bottom propagators were used for each
ensemble. In the following we will primarily consider only the coarse CLS ``1''
lattice spacing and use an ensemble at the finer ``2'' lattice spacing for
comparison of discretisation effects. We will consider two mass trajectories:
one where the sum of the quark masses $\frac{2}{\kappa_l}+\frac{1}{\kappa_s}$
is constant as proposed in \cite{Bietenholz:2010jr} and one where the
renormalised strange quark mass $\widetilde{m_s}\approx
\widetilde{m_s}^\text{Phys.}$ is kept approximately constant
\cite{Bali:2016umi}. Both of these quark-mass trajectories use the
same underlying gauge and quark action and should agree at the physical pion-mass point. With regard to the range of pion masses covered; we will use ensembles from the quark $\text{SU}(3)_f$-symmetric mass point $m_\pi \approx 420$ MeV down to $m_\pi \approx 220$ MeV. More details on the generation of these ensembles can be found in
\cite{Bruno:2014jqa} and \cite{RQCD:2022xux}.

From e.g. \cite{Francis:2016hui,Junnarkar:2018twb,Leskovec:2019ioa} it has
been observed that the dependence of the binding of the tetraquark states
discussed in the following sections is approximately linear in $m_\pi^2$, so the pion
mass-range encompassed by our study is suitable to obtain an extrapolated result at the physical pion mass. In this work we make use of the dimensionless quantity $\phi_2=8t_0 m_\pi^2$ (with $t_0$ measured on each ensemble) as it has less uncertainty than the dimensionful physical pion mass, due to the error on the lattice spacing from \cite{Bruno:2016plf}. The ensembles used in our study span a large range of $m_\pi,m_\pi L,$ and $m_K L$ which helps us control finite-volume and chiral-extrapolation systematics.

In the next sections we again use Coulomb gauge-fixed wall sources with Gaussian smeared sinks, and primarily focus on results from the ``1" lattice spacing ensembles. We will use a fixed physical sink-smearing $\alpha=22$ for the tetraquark correlators, and a slightly larger choice $\alpha=25$ for the $B$ and $B_s$ mesons. We use comparable physical choices of $\alpha$ for the cross-check at the finest lattice spacing on ensemble H200. 

\subsection{Determining the binding energy for a $ud\bar{b}\bar{b}$ tetraquark}\label{sec:udbb}

To arrive at precise and accurate results, systematic uncertainties on the
predictions of the proposed deeply-bound $I(J^P)=0(1^+)$, $ud\bar{b}\bar{b}$ tetraquark
are of considerable interest. Indeed, currently the main goal of the community is to
systematically improve calculations to the point where consensus can be
reached on the mass of this state. In this context, it is important to investigate 
the size of discretisation effects, and the impact of the heavy-quark formalism employed.

We start by calculating a $4\times 4$ matrix of correlators built from quasi-local operators to determine the lowest-lying tetraquark ground state, as in \cite{Hudspith:2020tdf}:
\begin{equation}
\begin{aligned}
D &= ({u_a}^T C\gamma_5 d_b)(\bar{b}_a C\gamma_i \bar{b}_b^T),\\
E &= ({u_a}^T C\gamma_t\gamma_5 d_b)(\bar{b}_a C\gamma_i\gamma_t \bar{b}_b^T),\\
M &= (\bar{b} \gamma_5 u )(\bar{b} \gamma_i d ) - [u\leftrightarrow d],\\
N &= (\bar{b} I u )(\bar{b} \gamma_5\gamma_i d ) - [u\leftrightarrow d].
\end{aligned}
\end{equation}
We will not use an operator that looks like $B^*B^*$, and subsequently this
expected state will not be seen in our GEVP. From a preliminary study it seems like this state overlaps more strongly with extended, derivative-type operators.

As we use a non-symmetric operator setup for the GEVP, we cannot guarantee that the resulting
eigenvalues are real (in turns out that they are, well into the region where
the signal degrades). We choose to approximately diagonalise the
correlator matrix using the left and right eigenvectors of the GEVP at
a specific ``diagonalisation time''
\cite{Francis:2018qch,Junnarkar:2018twb,Hudspith:2020tdf}. We
then determine the principal correlators from the diagonal of this matrix, performing a correlated fit to a single exponential Ansatz of the lowest-lying state at sufficiently-large separations to determine our tetraquark mass.

\begin{figure}[tb]
  \includegraphics[width=8cm]{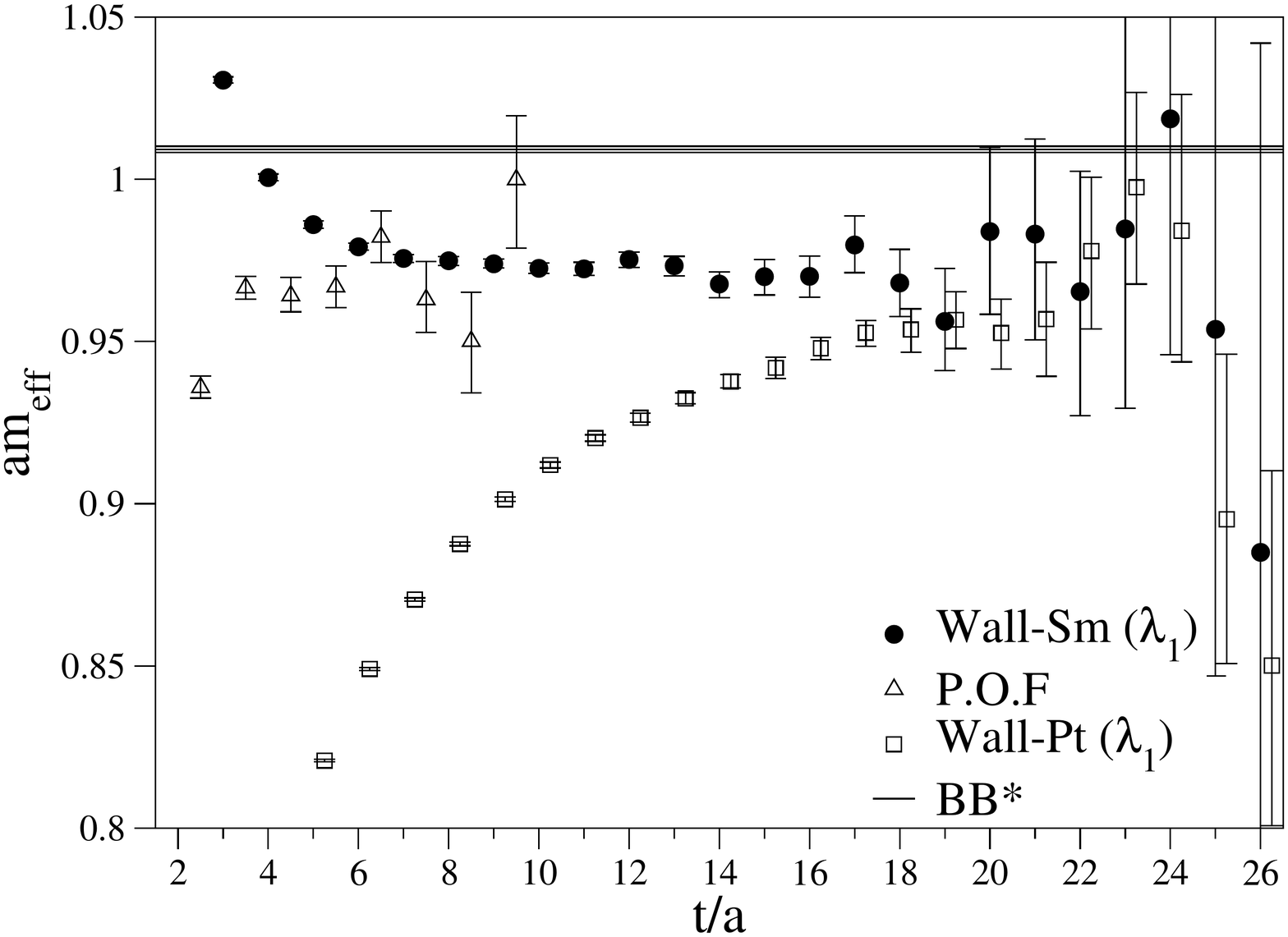}
  \includegraphics[width=8cm]{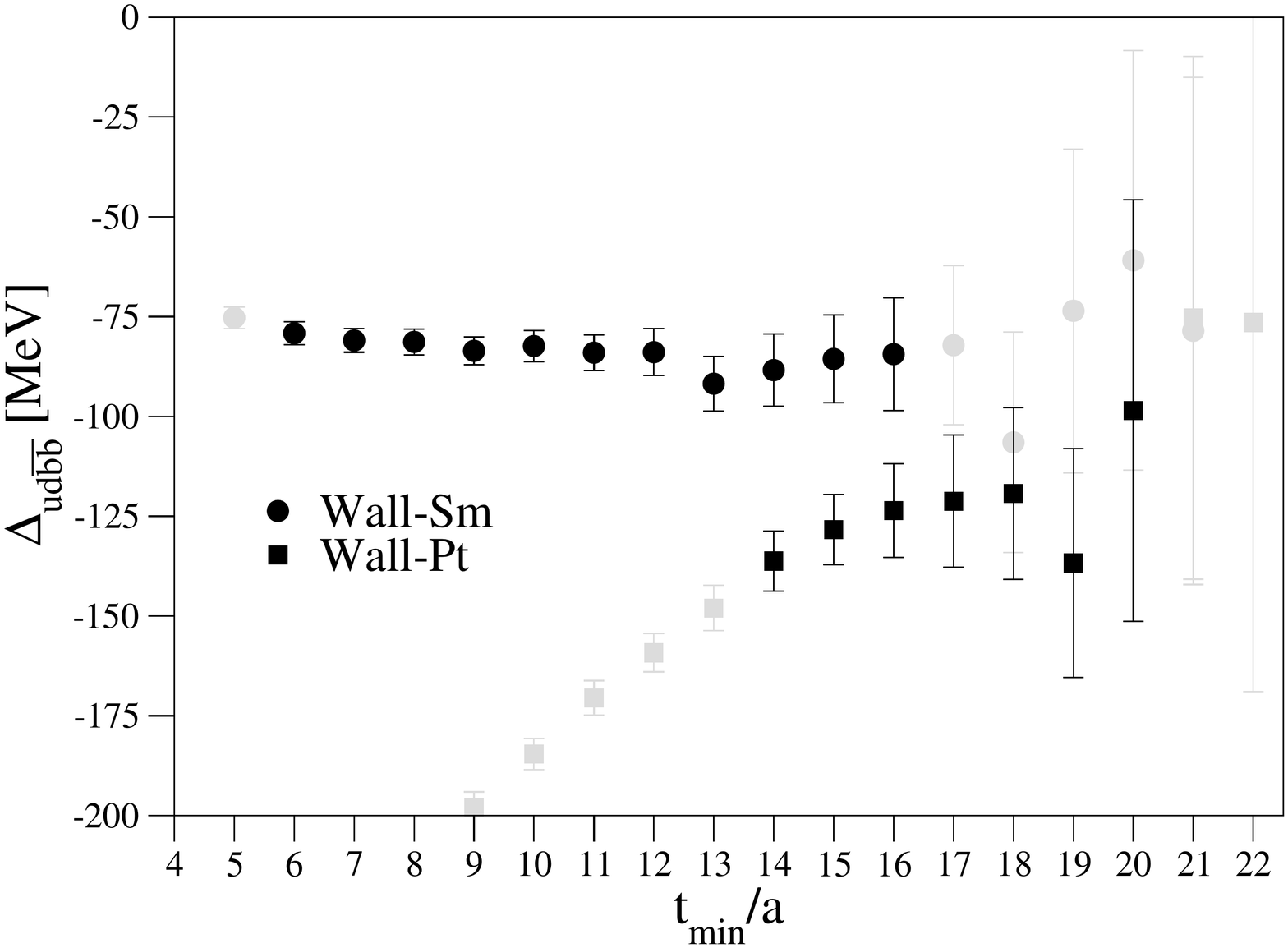}
  \caption{(Left) A comparison on ensemble U103 of the effective masses of the lowest-lying principal correlator for gauge-fixed wall source with point (Wall-Pt) or smeared (Wall-Sm) sink or from the $2\times 2$ pencil-of-functions (P.O.F) of a single Wall-Pt correlator. The data has been shifted for clarity. (Right) Dependence of the fit results on the lower fit bound (at fixed large upper fit bound) for a single exponential fit for both smeared-sink and point-sink; greyed-out points indicate poor fit quality.}\label{fig:effmass_U103}
\end{figure}

Fig.~\ref{fig:effmass_U103} illustrates the danger in using a single-exponential fit to gauge-fixed wall-source and point-sink data for this $ud\bar{b}\bar{b}$ tetraquark. We chose ensemble U103 for this comparison, as it has the highest statistics. Clearly, the effective mass for the Wall-Pt data is still trending upward as the signal is lost, whereas for the Wall-Sm determination good stability is seen over a large range in time. The right ``fit-stability'' plot illustrates that with noisy data it is quite easy to find a good-quality fit (in terms of $\chi^2/dof$ and p-value) over a small range and end up with a deeper value of the binding, and an unforeseen associated systematic. A two-exponential fit to the point-sink data is consistent with the smeared-sink determination, and the smeared-sink determination is also consistent with a $2\times 2$ generalized pencil-of-functions \cite{Aubin:2010jc,Green:2014xba} analysis of the point-sink diquark/anti-diquark correlator from operator D.

\begin{figure}[tb]
  \includegraphics[scale=0.32]{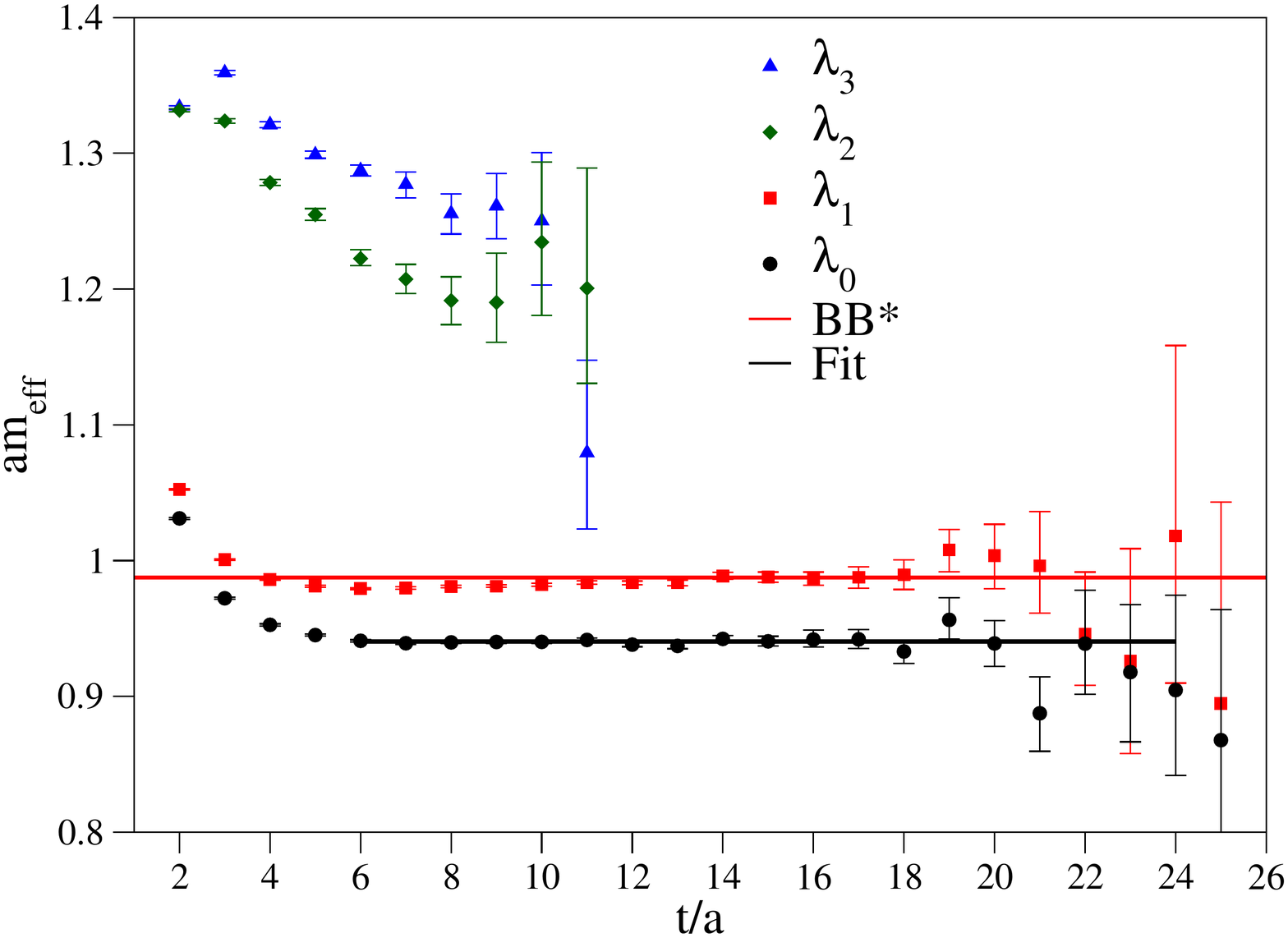}
  \caption{Eigenvalue effective masses for ensemble N101. In addition to the
    effective masses, the fit
    determination for the mass of the $ud\bar{b}\bar{b}$ tetraquark is shown
    as the black line and our determination of the first non-interacting
    $BB^*$ level is shown as a red line.}\label{fig:exemplary_udbb_N101}
\end{figure}

\begin{table}
\begin{tabular}{c|cc|cc}
\toprule
Ensemble &  $\phi_2$ & $m_\pi L$ & $B^*-B$ [MeV] & $\Delta_{ud\bar{b}\bar{b}}$ [MeV] \\
\hline
U103 & 0.7496 & 4.33 & 36.9(0.6) & -83.5(3.4) \\
H101 & 0.7561 & 5.83 & 35.3(1.0) & -87.9(2.3) \\
\hline
U102 & 0.5604 & 3.74 & 36.1(1.1) & -81.6(3.7) \\
H102 & 0.5469 & 4.93 & 34.8(1.0) & -95.0(3.2) \\
\hline
U101 & 0.3378 & 2.88 & 33.1(0.9) & -88.0(6.5) \\
H105 & 0.3451 & 3.91 & 36.3(0.8) & -104.9(3.5) \\
N101 & 0.3445 & 5.86 & 35.7(0.6) & -107.7(2.1) \\
\hline
C101 & 0.2195 & 4.66 & 33.9(0.7) & -113.2(2.7) \\
\hline
\hline
H107 & 0.5550 & 5.12 & 40.8(1.1) & -92.8(2.7) \\
H106 & 0.5550 & 3.88 & 39.1(1.0) & -107.1(3.5) \\
\hline
\hline
H200 & 0.7469 & 4.32 & 34.3(0.6) & -71.9(2.6)\\
\botrule
\end{tabular}
\caption{Measurements necessary for the $ud\bar{b}\bar{b}$ tetraquark candidate. Here, $m_\pi$ has been measured directly within this dataset.}\label{tab:udbb_table}
\end{table}

Fig.~\ref{fig:exemplary_udbb_N101} shows an exemplary effective-mass plot of
the four eigenvalues obtained from the GEVP for the ensemble N101, as well as the fit to
determine the $ud\bar{b}\bar{b}$ tetraquark mass and a line of the measured
lowest-lying non-interacting $BB^*$ threshold. In all cases our second
eigenvalue $\lambda_1$ is consistent with this threshold. The optimal smearing $\alpha$ for the $ud\bar{b}\bar{b}$ tetraquark is sub-optimal for the $BB^*$ threshold as there is still some visible excited-state contamination making this second level approach from below. It is this difference in excited-state contamination that makes a determination of the binding through a ratio of the lowest eigenvalue and the expected threshold dangerous for this data, therefore we do not perform such an analysis.

\begin{figure}[tb]
\includegraphics[scale=0.35]{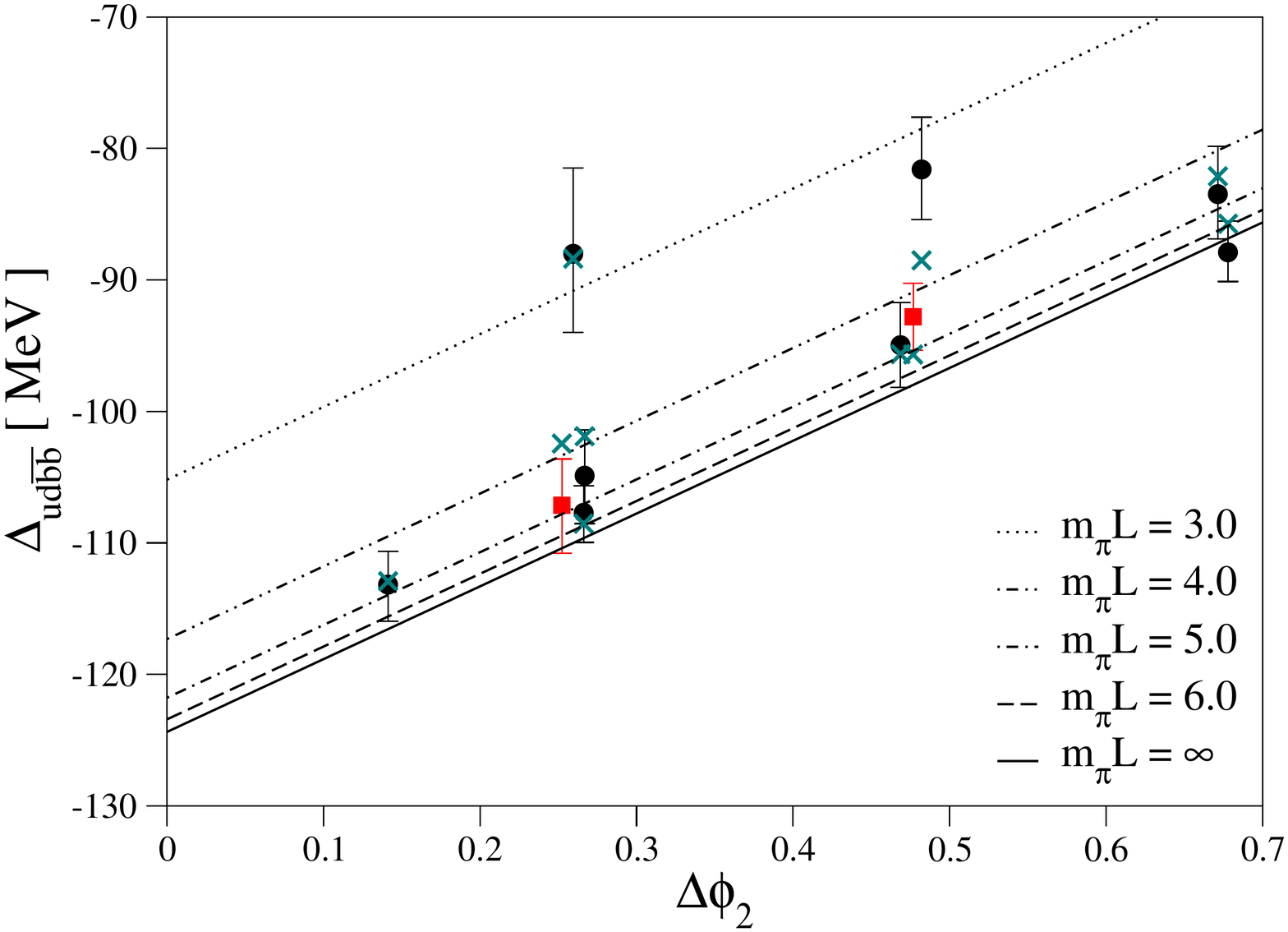}
\caption{Chiral and infinite-volume extrapolations of the $ud\bar{b}\bar{b}$ tetraquark, black points indicate the $\text{Tr}[M]=C$ data, red squares indicate the $\widetilde{m_s}=\widetilde{m_s}^{\text{Phys.}}$ trajectory. Lines of exemplary constant $m_\pi L$ are also plotted. Teal stars indicate the central value of the fit corresponding to the point at the same $\phi_2$.}\label{fig:infvol_tetra}
\end{figure}

Fig.~\ref{fig:infvol_tetra} shows the combined chiral and infinite-volume extrapolation of the mass of the ground state from the GEVP (the data of which is tabulated in Tab.~\ref{tab:udbb_table}), after subtracting the $BB^*$ mass (which cancels the additive mass renormalisation from NRQCD). We use the following Ansatz for a deeply-bound state:
\begin{equation}\label{eq:chiral_udbb}
\Delta_{ud\bar{b}\bar{b}}(\Delta \phi_2,m_\pi L,a) = \Delta_{ud\bar{b}\bar{b}}(0,\infty,a)(1+A \Delta \phi_2 + Be^{-m_\pi L}).
\end{equation}
With $\Delta_{ud\bar{b}\bar{b}}(0,\infty,a), A,$ and $B$ being shared fit
parameters between the two sets of mass-trajectories (as we expect minimal sea-strange contributions to this quantity), and $\Delta \phi_2 =
\phi_2^{\text{Lat}}-\phi_2^{\text{Phys}}$ being the difference between
the lattice-measured dimensionless $\phi_2$ and its physical value. For the
iso-symmetric ``physical'' pion mass we use the value of $134.8$ MeV
recommended in \cite{Aoki:2016frl} and for the physical $t_0$ we use the value
determined by the Flavor Lattice Averaging Group (FLAG)
\cite{FlavourLatticeAveragingGroupFLAG:2021npn}\footnote{We tested that using
  the CLS continuum value of $t_0$ instead produces a negligible shift in our
  final result}. We note that finite-volume effects here are significant and the deviation from our infinite-volume result in the chiral limit and that of $m_\pi L=4$ is still a $5.6\%$ correction.

From our extrapolation we obtain the binding energy at the ``1'' lattice-spacing,
\begin{equation}\label{eq:extrap1udbb}
\Delta_{ud\bar{b}\bar{b}}(0,\infty,a=0.08636\;\text{fm}) = -124.4(2.7) \text{ MeV}\;.
\end{equation}
The fit that produced this result has $\chi^2/dof=1.2$. We note that our extrapolation to physical pion mass could have higher-order contributions, so we fit Eq.~\ref{eq:chiral_udbb} with an extra term of $(\Delta \phi_2)^{3/2}$ or $(\Delta \phi_2)^2$, both of which can describe our data reasonably well and give values for $\Delta_{ud\bar{b}\bar{b}}(0,\infty,a=0.08636\;\text{fm})$ of $-133.4(6.7)$ and $-131.9(5.5)$ respectively. For our final result we use Eq.~\ref{eq:extrap1udbb} as the two higher-order fits do show some signs of over-fitting, and add half the difference of the larger result as a systematic, which we denote $\chi$. We do not have the precision or number of ensembles to fit a higher-order finite volume correction term in combination with the one we already have. Upon removing ensembles from our fit with $m_\pi L<4$ our result did not change within statistical errors.

Due to the reasonably large discrepancy between the results on ensembles H200 and U103 of
11.6 MeV, we decide to add half of this difference to the central value of Eq.~\ref{eq:extrap1udbb} and use all of this difference as an uncertainty estimate, such that our error encompasses the coarser and
finer lattice-spacing result. Later on in
Sec.~\ref{sec:higher_order_syst} we see a reasonably strong, positive, linear dependence of $\Delta_{ud\bar{b}\bar{b}}(0,\infty,a)$ on the 
$B^*-B$ splitting. As our tuning under-predicts this splitting (particularly for the $\text{Tr}[M]=C$ trajectory), we add 6.6 MeV (obtained from a linear interpolation of the data in Fig.~\ref{fig:hot_splits} to the physical $B^*-B$) to our determination (incorporating half of this correction as a systematic for our final error) to correct for this, giving our final result
\begin{equation}\label{eq:final_udbb}
\Delta_{ud\bar{b}\bar{b}}(0,\infty,0) = -112.0(2.7)_{\text{Stat.}}(4.5)_\chi(11.6)_{\text{a}}(3.3)_{B^*-B} \text{ MeV}.
\end{equation}
We observe that our data suggests the $a\rightarrow 0$ limit corresponds to a shallower
binding, although the resulting state is still deeply-bound and
strong-interaction stable. It is clear that our estimate is dominated by
systematics relating to discretisation effects, where the $a\rightarrow 0$
limit is not well-controlled in Lattice NRQCD.

\subsection{On the existence of an $\ell s\bar{b}\bar{b}$ tetraquark}\label{sec:lsbb}

We now briefly turn our attention to a possible $\ell s\bar{b}\bar{b}$
tetraquark ($T_{bbs}$), predicted from Lattice QCD in
\cite{Francis:2016hui,Junnarkar:2018twb}, and \cite{Meinel:2022lzo} to lie around 90 MeV
below the lowest-lying non-interacting two-meson threshold. This value from the literature seems already quite large as our $ud\bar{b}\bar{b}$
tetraquark is bound by roughly this magnitude and is expected to be more
deeply bound due to the good light-diquark $ud$ configuration. In addition to the $\ell s \bar{b}\bar{b}$ tetraquark state, there are now two
very close-by meson-meson states at the $B_sB^*$ and $B_s^* B$ thresholds, and
hence finite-volume effects may be non-trivial. As discussed earlier, most
phenomenological studies agree on the existence of a $ud\bar{b}\bar{b}$
tetraquark below threshold, while fewer have considered the $\ell s\bar{b}\bar{b}$.

For this calculation we will investigate results from a $5\times 5$ GEVP of the following quasi-local meson-meson operators with quantum numbers $I(J^P) = \frac{1}{2}(1^+)$:
\begin{equation}
  \begin{gathered}
M = (\bar{b} \gamma_5 u )(\bar{b} \gamma_i s ), \quad N = (\bar{b} I u )(\bar{b} \gamma_5\gamma_i s )\\
O = (\bar{b} \gamma_5 s )(\bar{b} \gamma_i u ), \quad P = (\bar{b} I s )(\bar{b} \gamma_5\gamma_i u )\\
Q = \epsilon_{ijk}(\bar{b} \gamma_j u )(\bar{b} \gamma_k s ).
  \end{gathered}
\end{equation}
Although we include an operator resembling $B^* B_s^*$ we will again
have trouble identifying this expected level in our GEVP. For all of our ensembles we find (as in
nature) that $B_sB^*$ is the lowest-lying non-interacting two-meson threshold,
and we determine the mass difference $\Delta_{\ell s\bar{b}\bar{b}}$ with regard to this threshold.

\begin{figure}[tb]
  \includegraphics[scale=0.32]{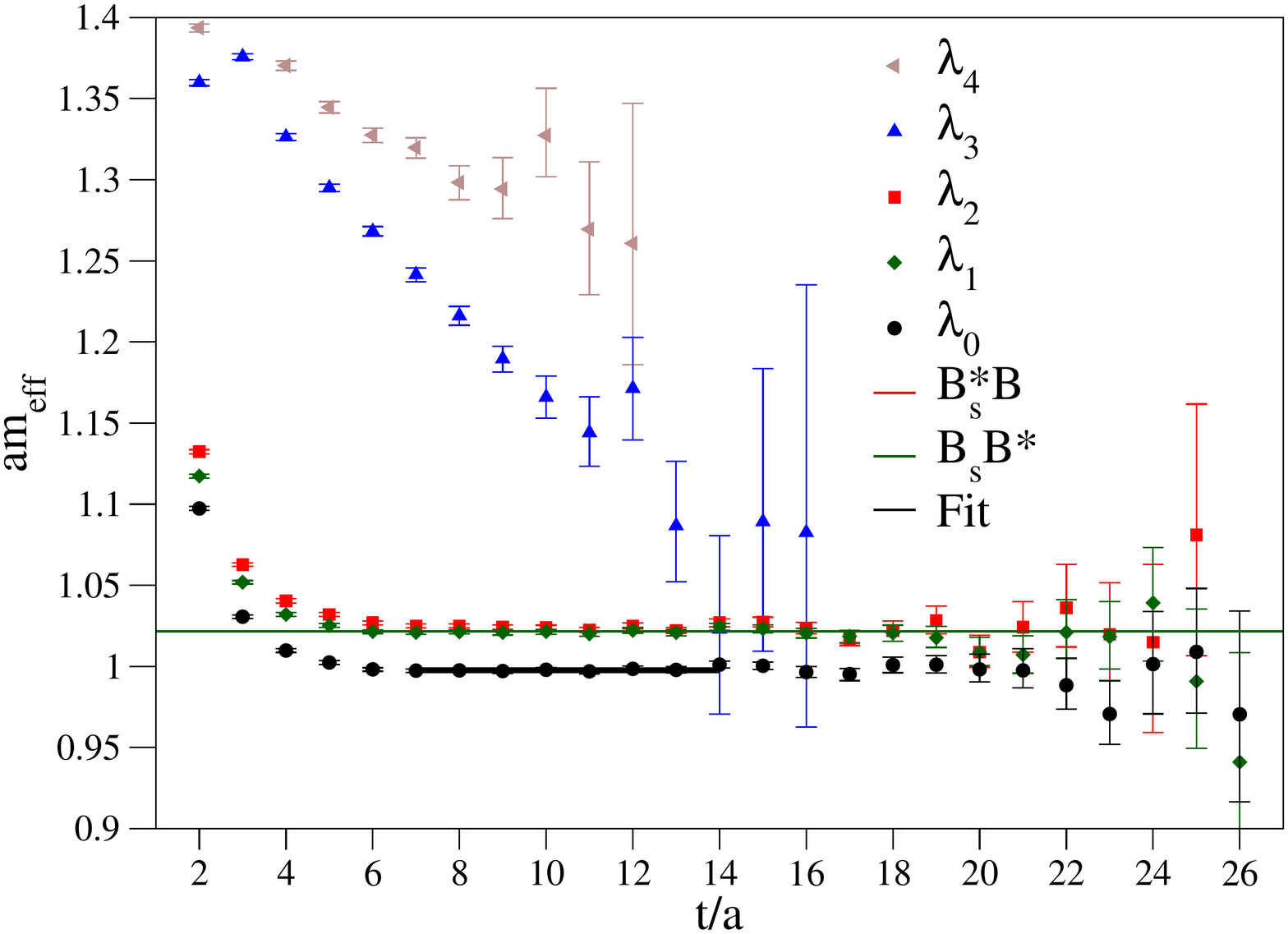}
  \caption{Eigenvalue effective masses for ensemble H106. The fit
    determination for the mass of the $\ell s\bar{b}\bar{b}$ tetraquark is
    shown as the black line and the determinations of the first two
    non-interacting levels $B_sB^*$ and $B_s^* B$ are shown as (almost identical)
    red and green lines respectively.}\label{fig:exemplary_lsbb_H106}
\end{figure}

Fig.~\ref{fig:exemplary_lsbb_H106} shows the effective masses of the eigenvalues for the $\widetilde{m_s}=\widetilde{m_s}^{\text{Phys.}}$ ensemble H106 as well as the two lowest-lying two-meson non-interacting thresholds $B_sB^*$ and $BB_s^*$, which are practically degenerate for this ensemble. As was the case for the $ud\bar{b}\bar{b}$, $\lambda_1$ is consistent with our lowest-lying non-interacting threshold state. We have two close-together levels above the ground state which appear to correspond to $B_sB^*$ and $BB_s^*$. Our fourth level, $\lambda_3$, is poorly-determined suggesting that our basis needs improvement to capture the expected $B^*B_s^*$ level. Either way, these levels are reasonably far from the ground state tetraquark candidate.

\begin{table}[tb]
\begin{tabular}{c|ccc}
  \toprule
  Ensemble & $m_K L$ & $B_s^*-B_s$ & $\Delta_{\ell s\bar{b}\bar{b}}$ [MeV] \\
  \hline
  U102 & 4.61 & 37.6(1.1) & -58.5(2.9) \\
  H102 & 6.10 & 35.8(1.8) & -67.1(3.1) \\
  \hline
  U101 & 4.82 & 37.1(1.1) & -57.8(4.2) \\
  H105 & 6.44 & 35.8(1.8) & -61.3(2.6) \\
  N101 & 9.69 & 38.4(0.6) & -59.7(1.3) \\
  \hline
  C101 & 9.87 & 36.9(0.8) & -60.8(1.5) \\
  \hline
  \hline
  H107 & 7.61 & 41.6(0.8) & -52.9(1.7) \\
  H106 & 7.21 & 38.8(0.8) & -55.0(2.4) \\
  \botrule
\end{tabular}
\caption{Results for the binding energy of the $\ell s\bar{b}\bar{b}$ tetraquark candidate.}\label{tab:lsbb_del}
\end{table}

\begin{figure}[tb]
\includegraphics[scale=0.35]{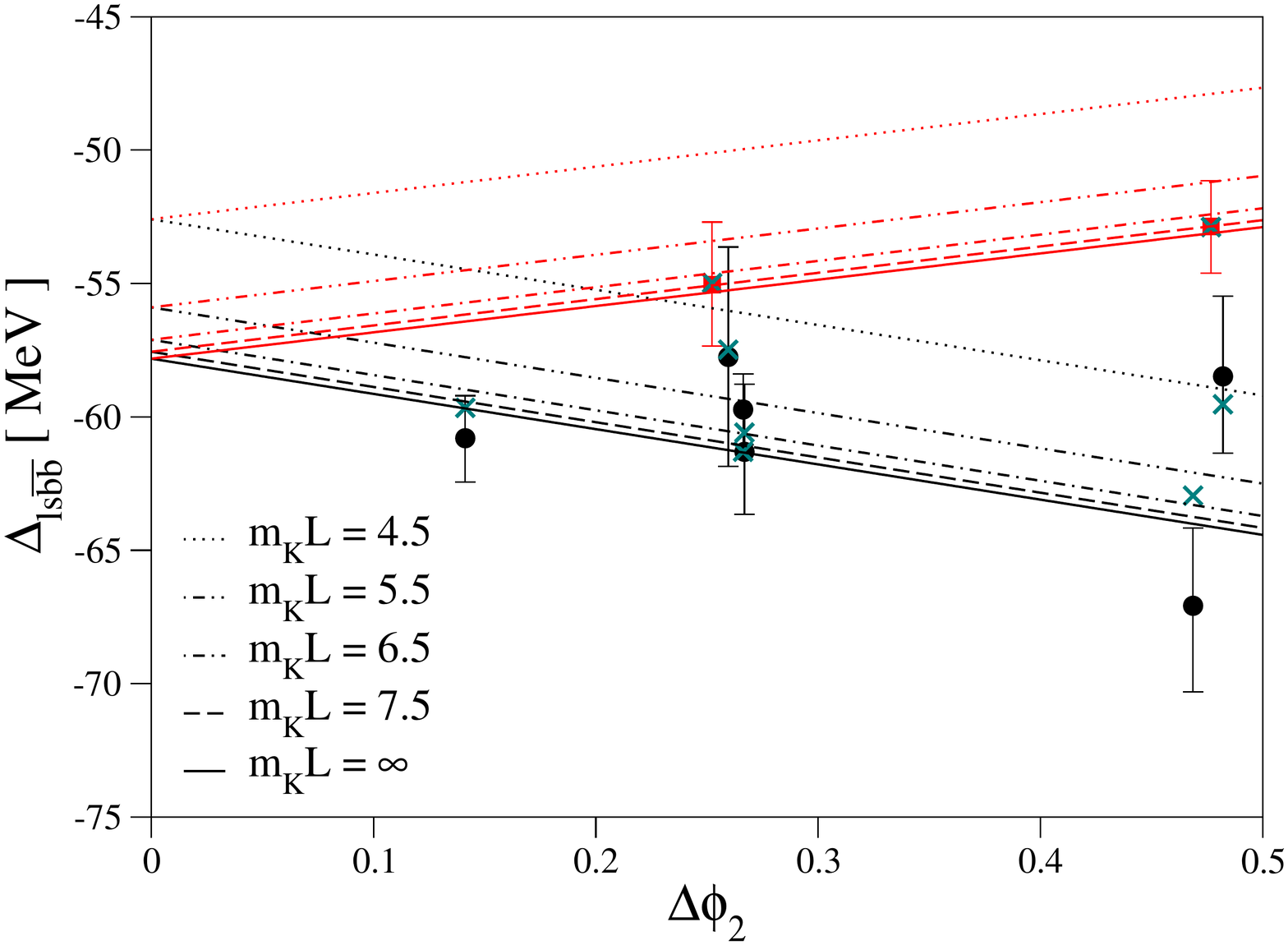}
\caption{Combined chiral and infinite-volume extrapolations of the $\ell s\bar{b}\bar{b}$ tetraquark. Red indicates the separate $\widetilde{m_s}=\widetilde{m_s}^{\text{Phys.}}$ mass-trajectory. Lines of exemplary constant $m_K L$ are also plotted. Teal stars indicate the central value of the fit corresponding to the point at the same $\phi_2$.}\label{fig:infvol_lsbb}
\end{figure}

Fig.~\ref{fig:infvol_lsbb} and Tab.~\ref{tab:lsbb_del} show the lattice data with broken
$\text{SU}(3)_f$, as these are the only ensembles we include in our fit. From this data we can see that as
the strange quark becomes heavier and the light quark lighter the
attractiveness of the good light-diquark diminishes. As the physical ($m_K L$) volume increases the state becomes more deeply-bound, as was the case for the $ud\bar{b}\bar{b}$.

We fit the data displayed in  Fig.~\ref{fig:infvol_lsbb} and tabulated in Tab.~\ref{tab:lsbb_del} to the combined chiral/infinite-volume Ansatz
\begin{equation}\label{eq:lsbbfvol}
\Delta_{\ell s\bar{b}\bar{b}}(\Delta \phi_2, m_K L,a) = \Delta_{\ell s\bar{b}\bar{b}}(0,\infty,a)\left( 1 + A \Delta \phi_2 + B e^{-m_K L}\right),
\end{equation}
where the mass coefficient $A$ is now different for the two mass-trajectories as sea and valence strange content is expected to be important. The parameters $\Delta_{\ell s\bar{b}\bar{b}}(0,\infty,a)$ and $B$ are again shared between the two trajectories. This fit has $\chi^2/dof\approx 1$. An extrapolation in $8\Delta t_0 m_K^2$ instead has effectively the same $\chi^2/dof$ and is entirely consistent with our final extrapolation result of $-57.8(2.4)$ . A fit with $e^{-m_\pi L}$ has a slightly worse $\chi^2/dof\approx 1.2$ and a larger central value ($-59.7(2.2)$), a fit without a finite volume term has $\chi^2/dof\approx 1.1$ with again a slightly larger central value $-59.1(2.1)$. As neither the fits with $e^{-m_K L}$ or $e^{-m_\pi L}$ have good significance for the parameter $B$, we choose to quote the average between their results and add half their difference as a finite-volume systematic. We note that given the quality of the data combined with the requirement of having more free fit parameters, it was not possible to fit higher-order terms in $\Delta\phi_2$. Our final result did not change within errors upon enforcing a cut of $m_KL>5$.

This fit at fixed ``1'' lattice-spacing gives the infinite-volume chiral-limit result:
\begin{equation}
\Delta_{\ell s\bar{b}\bar{b}}(0,\infty,a=0.08636\;\text{fm}) = -58.8(2.4)_\text{Stat.}(1.0)_\text{FV}\text{ MeV}.
\end{equation}
Again, taking the deviation between H200 and U103 as our lattice-spacing
systematic \footnote{neglecting the tiny differences in volume and pion mass between these two ensembles} and considering that our $B_s^*-B_s$ splitting is similarly as poor as our $B^*-B$ we perform the same two shifts as in the $ud\bar{b}\bar{b}$ case, yielding our final result:
\begin{equation}
\Delta_{ud\bar{b}\bar{b}}(0,\infty,0) = -46.4(2.4)_\text{Stat.}(1.0)_{\text{FV}}(11.6)_\text{a}(3.3)_{B_s^*-B_s}\text{ MeV},
\end{equation}
which is much less deeply-bound than the previous lattice determinations of
\cite{Francis:2016hui,Junnarkar:2018twb,Meinel:2022lzo}, but was already
hinted at in \cite{Hudspith:2020tdf} and is consistent with
the phenomenological predictions of
\cite{Semay:1994ht,Eichten:2017ffp}, and \cite{Braaten:2020nwp}. A shallow binding such as this poses a significant challenge for experimental detection. 

\section{The fate of the scalar and axial $B_s$-mesons}\label{sec:bs}

During our investigation of the $ud\bar{b}\bar{b}$ tetraquark it came to our attention that on the $\text{SU}(3)_f$-symmetric ensemble ``U103" the positive-parity B-mesons with the simple local operators,
\begin{equation}
B_{s0}^* = (\bar{b}Is),\quad B_{s1} = (\bar{b}\gamma_i\gamma_t s),
\end{equation}
lie below the expected $B\pi$ and $B^*\pi$ thresholds respectively. This was
not necessarily expected, as a previous finite-volume calculation of these states
with different techniques \cite{Lang:2015hza} required the inclusion of explicit meson-meson
interpolating fields to observe a state below threshold. In the following investigation of these excited $B_s$-mesons we simply use the same wall-source/smeared sink mesons that were computed in the previous sections' $ud\bar{b}\bar{b}$ and $\ell s \bar{b}\bar{b}$ determinations.

Tab.~\ref{tab:splittings_BS} gives the numerical results for the difference between the $B_{s0}^*$ and $BK$ threshold, denoted $\Delta_{B_{s0}^*}$, and the difference between the $B_{s1}$ and $B^*K$, denoted $\Delta_{B_{s1}}$. We also give our measured difference between the $B_{s0}^*$ and $B_{s1}$.

\begin{table}[h!]
\begin{tabular}{c|c|c|c}
\toprule
Ensemble & $\Delta_{B_{s0}^*}$ [MeV] & $\Delta_{B_{s1}}$ [MeV] & $B_{s0}^*-B_{s1}$ [MeV] \\
\hline
U103 & -78.5(5.2) & -85.2(5.4) & -30.1(1.5) \\
H101 & -49.8(6.0) & -55.7(6.1) & -29.4(2.5) \\
\hline
U102 & -91.6(7.3) & -90.9(7.7) & -36.9(3.5) \\
H102 & -61.3(6.0) & -62.1(6.9) & -33.9(2.7) \\
\hline
U101 & -78.5(5.4) & -84.2(5.1) & -27.4(2.0) \\
H105 & -59.2(5.2) & -70.1(7.6) & -39.8(6.7) \\
N101 & -57.9(4.1) & -63.2(4.5) & -30.4(2.7) \\
\hline
C101 & -63.7(2.9) & -65.8(3.8) & -34.4(2.0) \\
\hline
\hline
H107 & -101.2(6.5) & -110.4(7.1) & -31.7(2.5) \\
H106 & -90.4(5.1) & -96.6(5.1) & -32.9(2.2) \\
\hline
\hline
H200 & -92.2(7.0) & -98.6(6.8) & -27.9(2.2)\\
\botrule
\end{tabular}
\caption{$B_{s0}^*$ and $B_{s1}$ meson mass-splittings with regards to their respective measured, expected, non-interacting thresholds $BK$ and $B^*K$.}\label{tab:splittings_BS}
\end{table}

\begin{figure}
\includegraphics[scale=0.285]{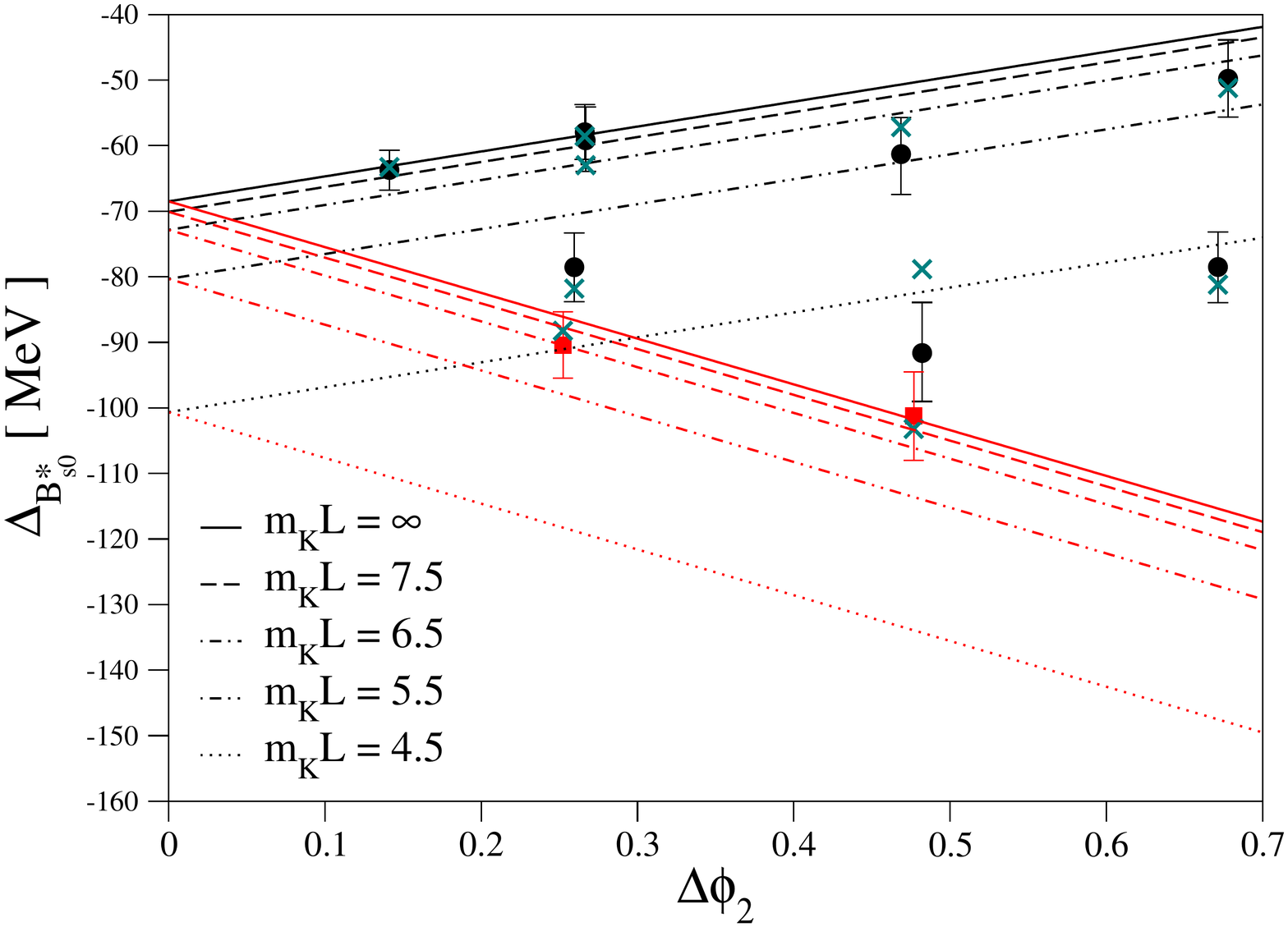}
\hspace{-24pt}
\includegraphics[scale=0.285]{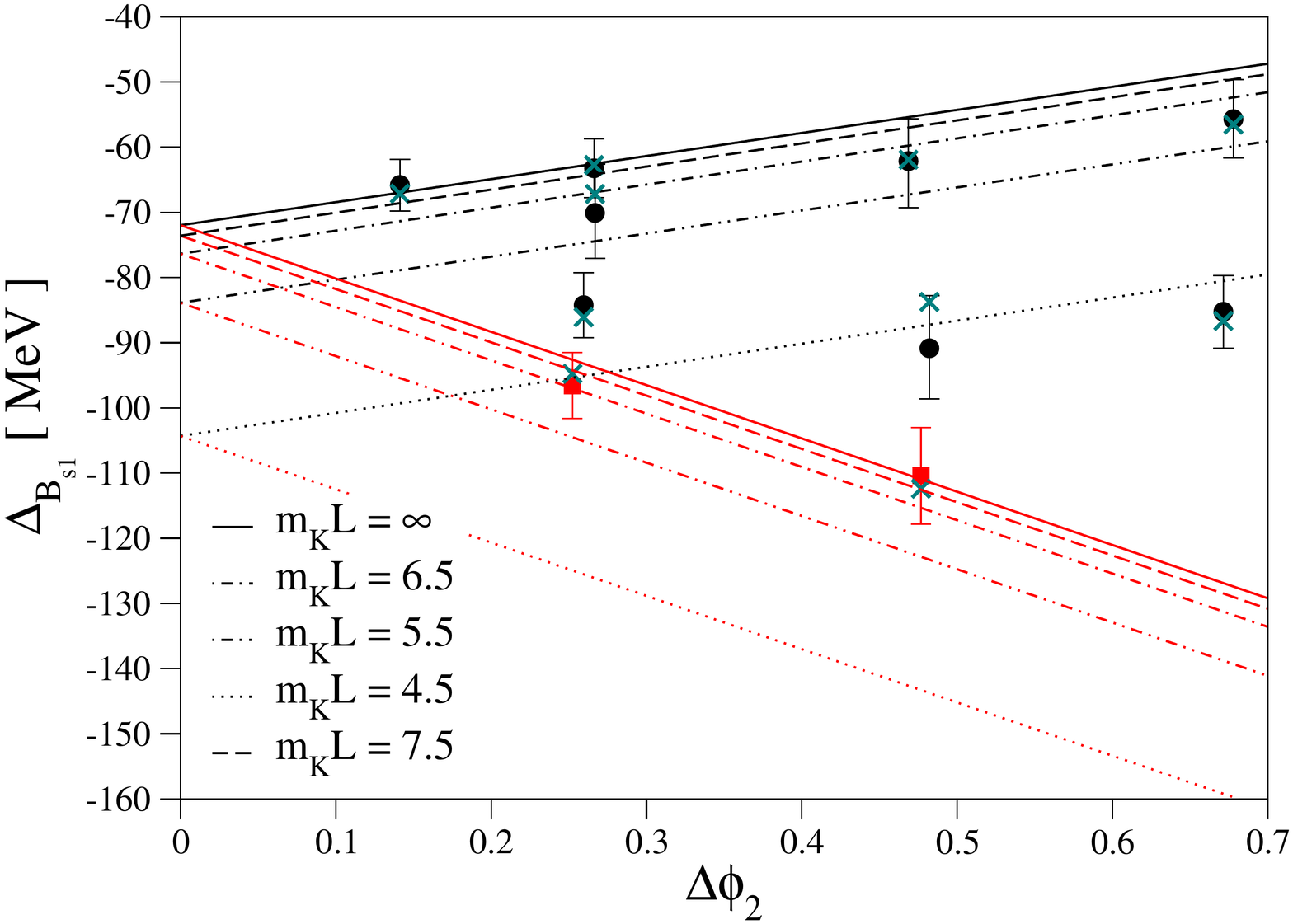}
\caption{(Left) scalar $B_{s0}^*$ combined chiral and infinite volume extrapolation. (Right) axial-vector $B_{s1}$ combined chiral and infinite volume extrapolation. Illustrative lines of constant $m_K L$ are displayed and red indicates the results from the $\widetilde{m_s}=\widetilde{m_s}^{\text{Phys.}}$ trajectory. Teal stars indicate the central value of the fit corresponding to the point at the same $\Delta \phi_2$.}\label{fig:scalar_extrap}
\end{figure}

In Fig.~\ref{fig:scalar_extrap} we show the combined chiral/infinite volume extrapolation of the binding energy for the scalar and axial $B_s$-mesons, from the neural network tuned b-quarks. Here we used the same simple fit Ansatz from the previous sections for the bound-state system of both splittings:
\begin{equation}\label{eq:bsfvolK}
\Delta_{B_{s0}^*/B_{s1}}(\Delta \phi_2, m_K L,a) = \Delta_{B_{s0}^*/B_{s1}}(0,\infty,a)\left( 1 + A \Delta \phi_2 + B e^{-m_K L}\right).
\end{equation}
As in the $\ell s\bar{b}\bar{b}$ case, we share the fit parameters $\Delta_{B_{s0}^*/B_{s1}}(0,\infty,a)$ and
$B$ between our mass-trajectories, and allow $A$ to be a free parameter for
each of them. We find the coefficient $B$ to be very large, suggesting significant finite-volume effects are present in this quantity. The plots (\ref{fig:effmass_Bs0} and \ref{fig:effmass_Bs1}) of the effective masses (in App.~\ref{app:effmassBS}) show, that even with our significant statistical resolution these quantities are noisy and often display large fluctuations in time.

The results for the two splittings at physical pion mass in the infinite-volume limit from our two different mass-trajectories are (see Tab.~\ref{tab:bsfvol} of App.~\ref{sec:app_bsfv} for more details):
\begin{equation}\label{eq:fvolBs}
\begin{aligned}
\Delta_{B_{s0}^*}(0,\infty,a=0.08636\;\text{fm}) &= -68.5(3.0) \text{ [MeV]},\\
\Delta_{B_{s1}}(0,\infty,a=0.08636\;\text{fm}) &= -72.0(3.7) \text{ [MeV]}.
\end{aligned}
\end{equation}
These fits have $\chi^2/dof = 0.8$ and $0.3$ respectively. We considered fits with another free parameter multiplying higher-order powers of $\phi_2$, but could not fit such expressions stably. We note that upon a cut of $m_K L>5$ our results do not change within error.

We again convince ourselves that our largest systematics come from lattice-spacing artifacts. To further quantify our systematics we first consider the difference between the extrapolated results using the tree-level NRQCD prescription or various
tunings to get the physical $B^*-B$ splitting right, but we find their effects
to be negligible (discussed in Sec.~\ref{sec:higher_order_syst}). We then
measure the same quantities using the neural network tuning on ensemble H200
as an indication of the light-quark cut-off effects. We observe that half the
difference between H200 and U103 is $6.85$ Mev for the $B_{s0}^*$ and $6.7$ MeV for the $B_{s1}$, so we add that to the
central values of Eq.~\ref{eq:fvolBs} and quote the full difference as the
systematic error. Unlike for the previous two tetraquark candidates, we will
show in the next section in Fig.~\ref{fig:hot_splits} that there is no
dependence of these quantities on the $B^*-B$ splitting at the $\text{SU(3)}_f$-symmetric point, so we assume we do not have this associated systematic. For our final result we quote:
\begin{equation}
\begin{aligned}
\Delta_{B_{s0}^*}(0,\infty,0) &= -75.4(3.0)_{\text{Stat.}}(13.7)_{\text{a}} \text{ [MeV]},\\
\Delta_{B_{s1}}(0,\infty,0) &= -78.7(3.7)_{\text{Stat.}}(13.4)_{\text{a}} \text{ [MeV]}.
\end{aligned}
\end{equation}
These determinations equate to masses of $B_{s0}^*=5698(14)$ and $B_{s1}=5741(14)$ MeV for these states, where the iso-symmetric kaon mass $m_K=494.2$ MeV was used. Again, our leading systematic is our conservative estimate emanating from discretisation effects.

\section{Cross-checks and NRQCD systematics}\label{sec:higher_order_syst}

At fixed bottomonium tuning, the measured $B^*-B$ splitting does not trend toward the physical result as the lattice spacing is decreased. As an alternative to our standard tuning, we are free in our philosophy to tune the NRQCD coefficients with the physical $B^*-B$ splitting as a parameter instead or in addition to the bottomonium hyperfine splitting. As a further possible choice, we also consider the tuning with the higher-order spin-dependent terms $c_7,c_8,$ and $c_9$ of Eq.~\ref{eq:nrqcd_hospin} set to their tree-level values. We will only perform this investigation on a single ensemble (U103) as this can be used to estimate a systematic for the $ud\bar{b}\bar{b}$ and $\ell s \bar{b}\bar{b}$ tetraquarks, or the $B_{s0}^*$ and $B_{s1}$ mesons.

\begin{table}
  \begin{tabular}{c|ccccccc|ccc}
    \toprule
    Tuning & $aM_0$ & $c_1$ & $c_2$ & $c_3$ & $c_4$ & $c_5$ & $c_6$ & $c_7$ & $c_8$ & $c_9$ \\
    \hline
    Bottom & 1.8035(32) & 0.787(16) & -0.789(61) & 1.057(10) & 0.960(11) & 0.828(13) & 0.932(14) & 0 & 0 & 0 \\
    Full & 1.8014(36) & 0.791(25) & 0.084(204) & 1.061(10) & 1.183(39) & 0.860(19) & 0.965(15) & 0 & 0 & 0 \\
    $B^*-B$ & 1.8020(11) & 0.838(7) & 0.371(38) & 1.073(5) & 1.384(13) & 0.831(9) & 0.944(9) & 0 & 0 & 0 \\
    Tree & 1.8120(120) & 1 & 1 & 1 & 1 & 1 & 1 & 0 & 0 & 0 \\
    \hline
    Bottom & 1.8643(20) & 0.861(14) & -0.548(58) & 1.134(18) & 1.092(16) & 0.810(19) & 0.980(14) & 1 & 1 & 1 \\
    Full & 1.8612(24) & 0.867(14) & -0.373(80) & 1.100(16) & 1.190(25) & 0.821(12) & 0.965(12) & 1 & 1 & 1 \\
    $B^*-B$ & 1.8613(14) & 0.895(10) & -0.167(40) & 1.095(12) & 1.312(9) & 0.794(11)& 0.946(9) & 1 & 1 & 1 \\
    Tree & 1.8825(110) & 1 & 1 & 1 & 1 & 1 & 1 & 1 & 1 & 1 \\
    \botrule
  \end{tabular}
  \caption{Investigation of different tuning-strategies and inclusion of higher-order spin-dependent terms of NRQCD for the ensemble U103. ``Bottom'' refers to our standard tuning while ``$B^*-B$'' refers to the tuning where the $B$-meson hyperfine splitting is used instead of the bottomonium hyperfine splitting, and ``Full'' refers to using both splittings.}\label{tab:higher_order_difftune}
\end{table}

\begin{figure}[h!]
  \includegraphics[scale=0.29]{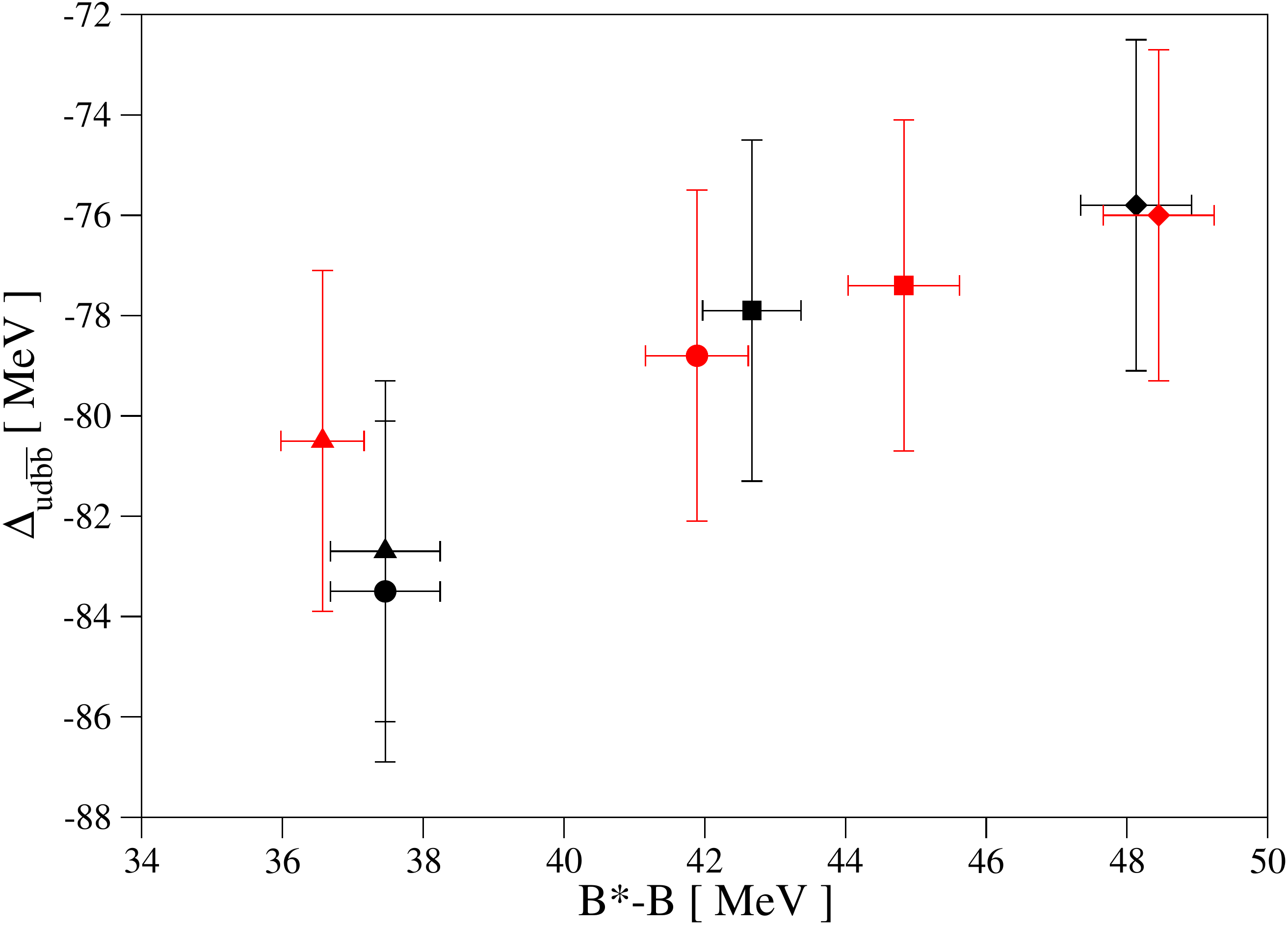}
  \includegraphics[scale=0.29]{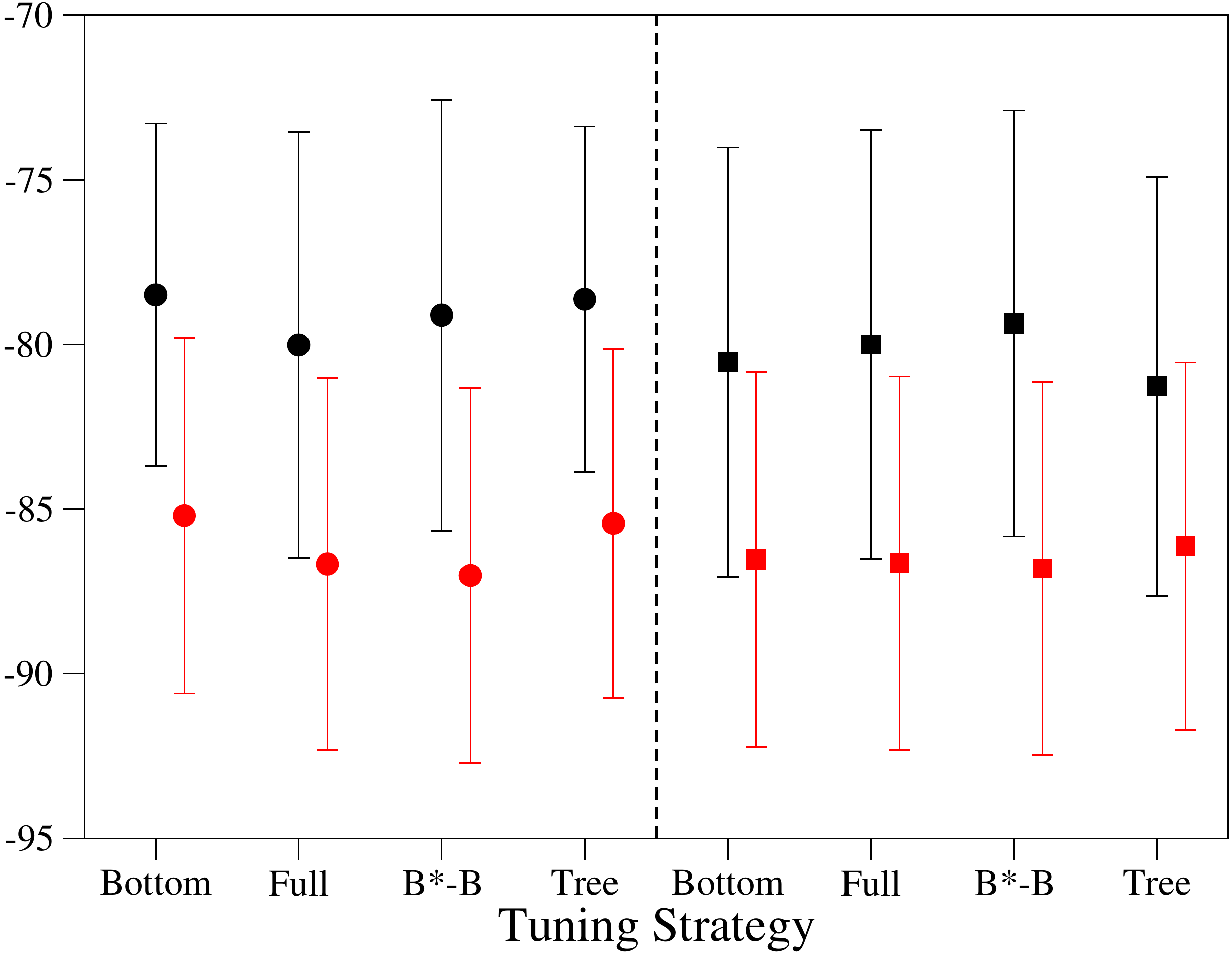}
  \caption{(Left) $ud\bar{b}\bar{b}$ tetraquark binding-energy dependence on the $B^*-B$ hyperfine splitting using the extended tunings of Tab.~\ref{tab:higher_order_difftune}. Red indicates that the tree-level spin-dependent $O(v^6)$ terms were used whereas black means $O(v^4)$, circles are our standard tuning, squares the "full''  tuning, diamonds the $B^*-B$, and triangles the tree-level coefficients. (Right) $B_{s0}^*$ (black) and $B_{s1}$ (red) binding-energy dependence on the tuning strategy, points on the right of the dashed line include spin-dependent $O(v^6)$ terms.}\label{fig:hot_splits}
\end{figure}

\begin{figure}[h!]
  \includegraphics[scale=0.35]{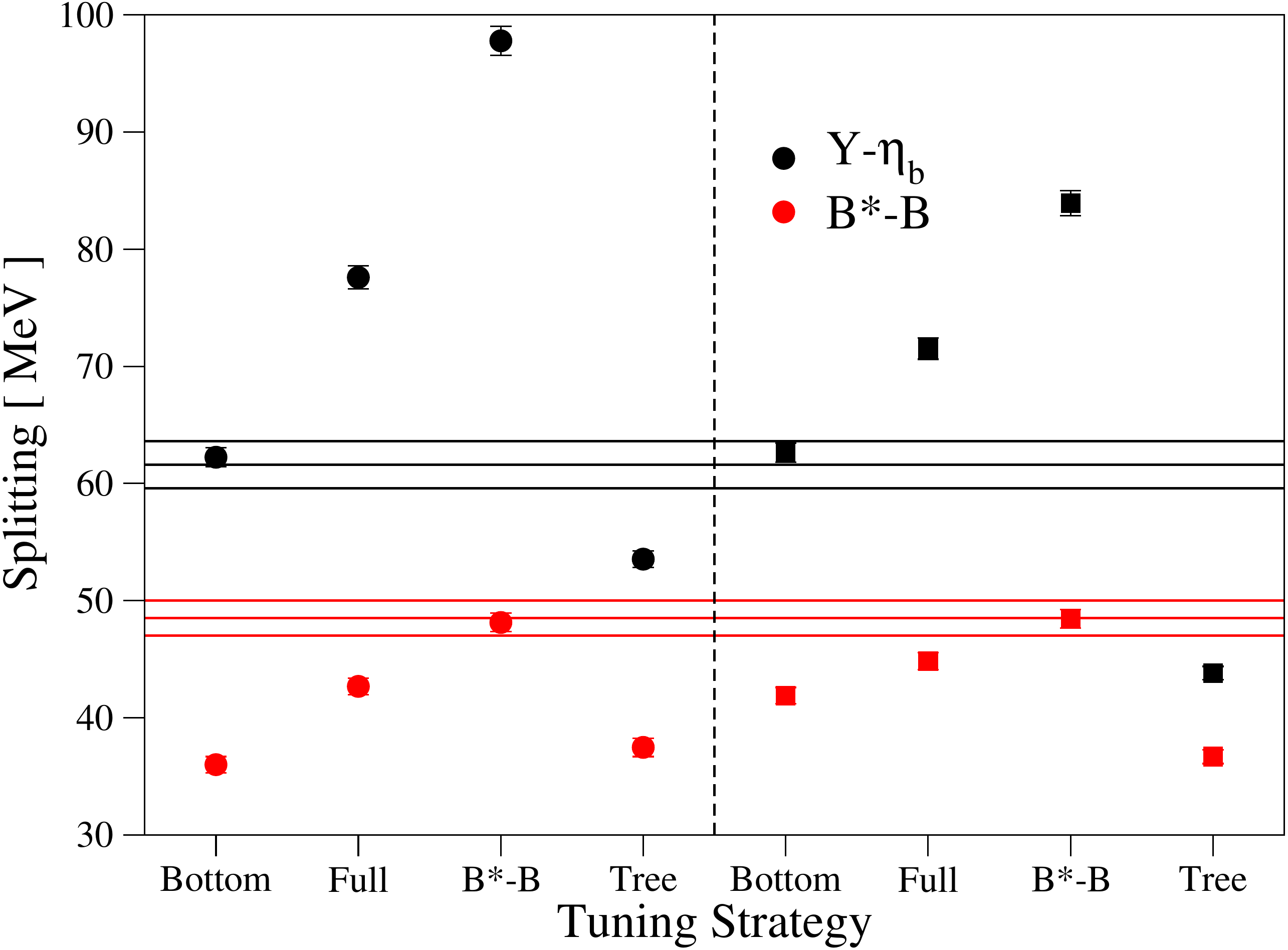}
  \caption{Hyperfine splittings for the various tuning strategies and parameters in Tab.~\ref{tab:higher_order_difftune}. The points left of the dashed line use the $O(v^4)$ NRQCD Hamiltonian, and those on the right include tree-level spin-dependent $O(v^6)$ terms. Horizontal lines give the PDG values for these splittings.}\label{fig:hyperfines_hot}
\end{figure}

In Tab.~\ref{tab:higher_order_difftune} we give the various nonperturbatively-tuned coefficients we obtain with the inclusion of $O(v^4)$ and partial-$O(v^6)$ terms in the NRQCD Hamiltonian. The parameter $c_7$ has a somewhat strong impact on the kinetic mass and a complete retuning of the bare mass $aM_0$ was needed for the inclusion of the tree-level higher-order terms. Already we can see some patterns in the coefficients: for purely tuning the $B^*-B$ splitting the parameter $c_4$ is the most important and is in strong conflict with the pure-bottomonium tuning. The larger the value of $c_4$ the less-negative the coefficient $c_2$ needs to become, suggesting there is a strong spin-independent contribution from $c_4$ that $c_2$ wants to counteract. At $O(v^4)$ $c_1$ and $c_3$ are positively-correlated with $c_4$, an increase in $c_4$ increases both. Upon inclusion of tree-level $O(v^6)$ spin-dependent terms $c_1$ still grows with $c_4$ but now $c_3$ decreases. 

The higher-order discretisation-effect correction terms $c_5$ and $c_6$ are quite consistent within the range of predictions provided by the network. The variation on the ``Full'' tuning's parameters is always larger, suggesting that adding an extra input that is in tension with the others produces worse results. Put another way: the network struggles to optimally determine the parameters that satisfy all the splittings used as inputs.

Fig.~\ref{fig:hot_splits} illustrates the dependence of the $ud\bar{b}\bar{b}$ tetraquark and the $B_{s0}^*$ and $B_{s1}$ meson binding energies on the $B^*$--$B$ hyperfine-splitting. We note that there is a somewhat strong dependence on this splitting for the tetraquark and none for the exotic B-mesons. For the tetraquarks in the previous section, we therefore shifted our final result and quantified the associated systematic uncertainty related to our $O(v^4)$-tuning. From Fig.~\ref{fig:hyperfines_hot} it is clear that we cannot simultaneously tune both the $\Upsilon-\eta_b$ and $B^*-B$ splittings to their physical value. There however seems to be convergence from tuning the pure bottomonia spectrum with tree-level higher-order terms. With the addition of more states to tune against, or with the fixing of other terms to 1, it should be possible to also allow $c_7$ to vary to see whether these hyperfine splittings can be improved further.

\section{Conclusions\label{conclusions}}

We have shown it is possible to reproduce the experimental splittings of
bottomonia with a simple Lattice NRQCD prescription that has nonperturbatively tuned parameters
from a neural network. We have investigated some of the S- and P-wave
excited states of bottomonia in order to test our tuning and we find good
consistency with experiment where available, and a more continuum-like
behaviour of the bottomonium spectrum at finite lattice spacing.  We have found that under our tuning the $B^*-B$ splitting is fairly far from the continuum result, with an indication that inclusion of higher-order spin-dependent terms improves the situation.

For all the heavy-light quantities we measured there was no discernible difference in the result from either using the tree-level NRQCD coefficients or those determined from a neural-network using pure-bottomonium ground states. It could be possible that significant differences from the NRQCD prescription are largely canceled by the threshold subtractions performed for the states we investigated. The $B^*-B$ splitting does however have a significant impact on the doubly-heavy tetraquark results.

Our physics objective was the calculation of the binding energy of the popular
$I(J^P)=0(1^+)$ $ud\bar{b}\bar{b}$, and $\frac{1}{2}(1^+)$ $\ell s\bar{b}\bar{b}$ tetraquark candidates. For the
$ud\bar{b}\bar{b}$ tetraquark we find a strong-interaction-stable bound state
$112.0(13.2)$ Mev below threshold, consistent with previous lattice studies
. Our determination comes from a combined
fit of 10 different lattice ensembles with a large variation in pion mass and $m_\pi L$. We
observe somewhat sizeable finite-volume effects, $B^*-B$ mis-tuning effects,
and most-importantly discretisation effects. The latter forms our
largest systematic uncertainty. A future calculation of this state with a relativistic heavy quark action to properly address this systematic is quite desirable.

\begin{figure}[tb]
\centering
\includegraphics[scale=0.285]{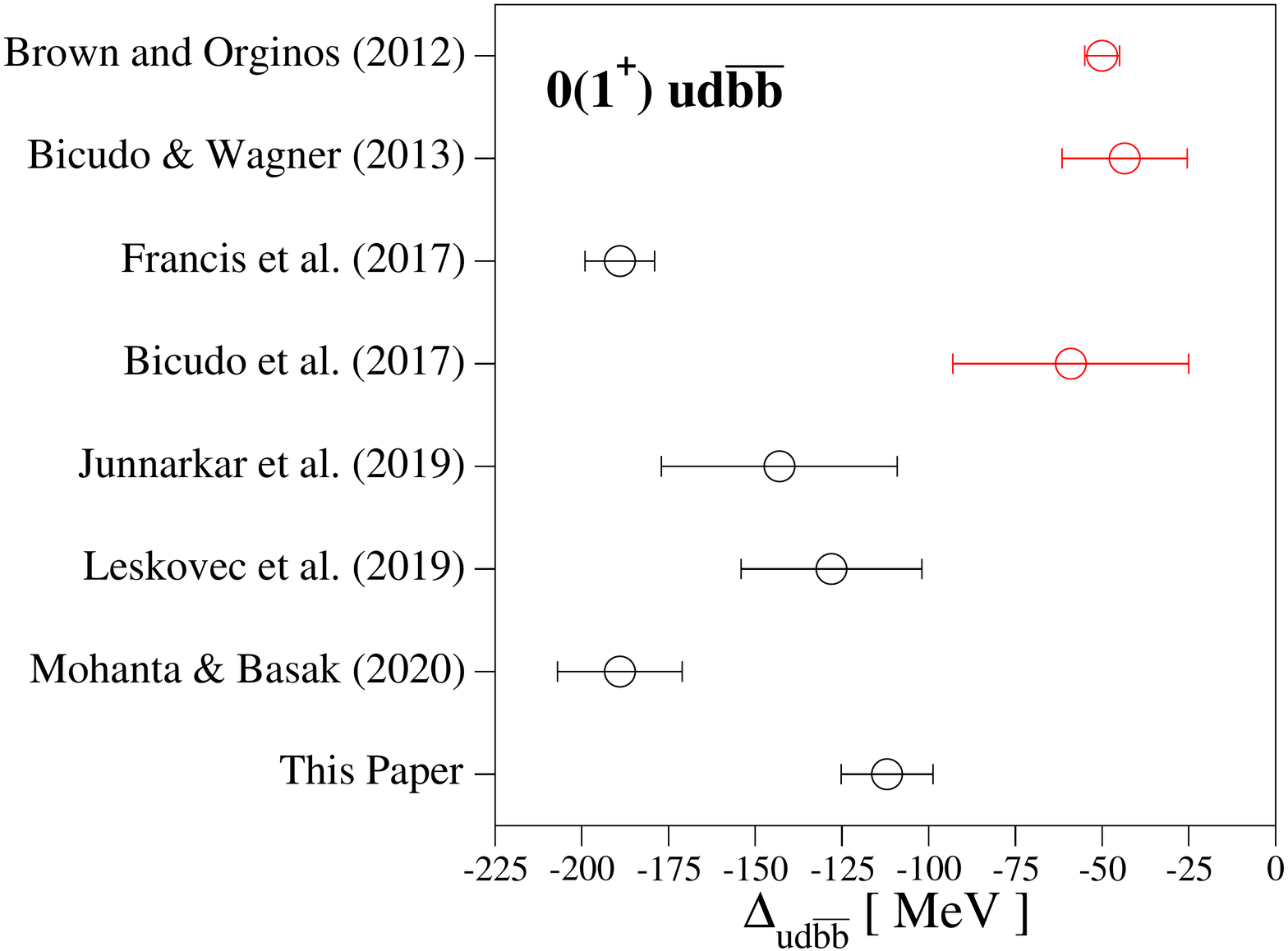}
\hspace{-24pt}
\includegraphics[scale=0.285]{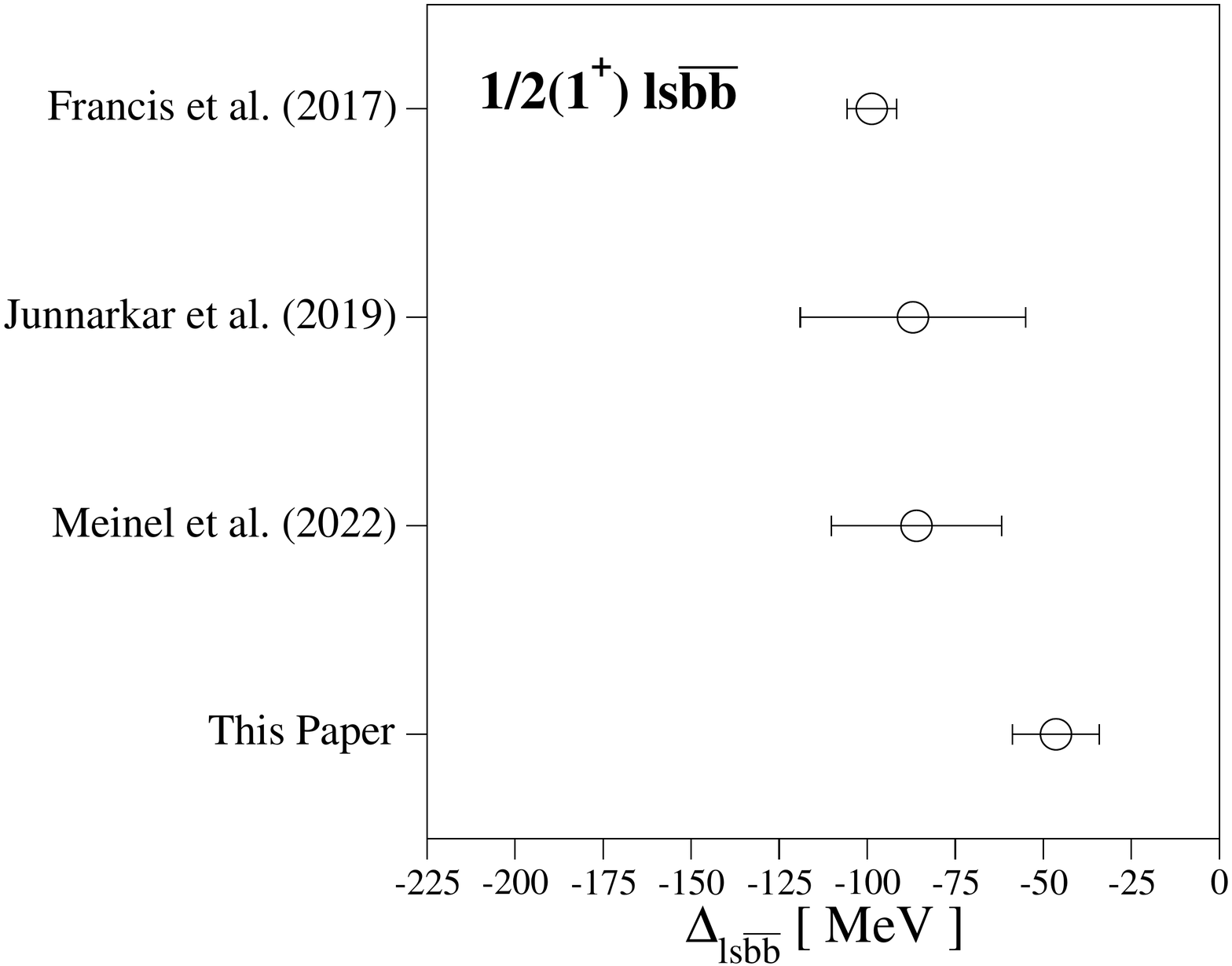}
\caption{Comparison plots of our result and other Lattice QCD determinations for the $ud\bar{b}\bar{b}$ (left) and $\ell s\bar{b}\bar{b}$ (right) tetraquarks. Black circles indicate calculations performed with Lattice NRQCD for the b-quarks and red circles indicate the use of static b-quarks. The years in parentheses are the publication dates of the papers. Statistical and systematic errors have been added in quadrature.}\label{fig:comparison_tetras}
\end{figure}

The related $\ell s \bar{b}\bar{b}$ tetraquark candidate
has typically been predicted by lattice studies to lie $\approx 90$ MeV below the lowest-lying
non-interacting two-meson threshold ($B_sB^*$); here we measure this state to be only $46.4(12.3)$ MeV below it. As was the case for the $ud\bar{b}\bar{b}$, our result is
dominated by our estimate of discretisation systematics. Finite-volume effects are small and an $e^{-m_K L}$ term is slightly preferred for our range of $m_KL$. A comparison of our doubly-heavy tetraquarks with other lattice determinations can be seen in Fig.~\ref{fig:comparison_tetras}, our measurements are at the bottom of the figures as they are all represented chronologically.

\begin{figure}[tb]
\centering
\includegraphics[scale=0.4]{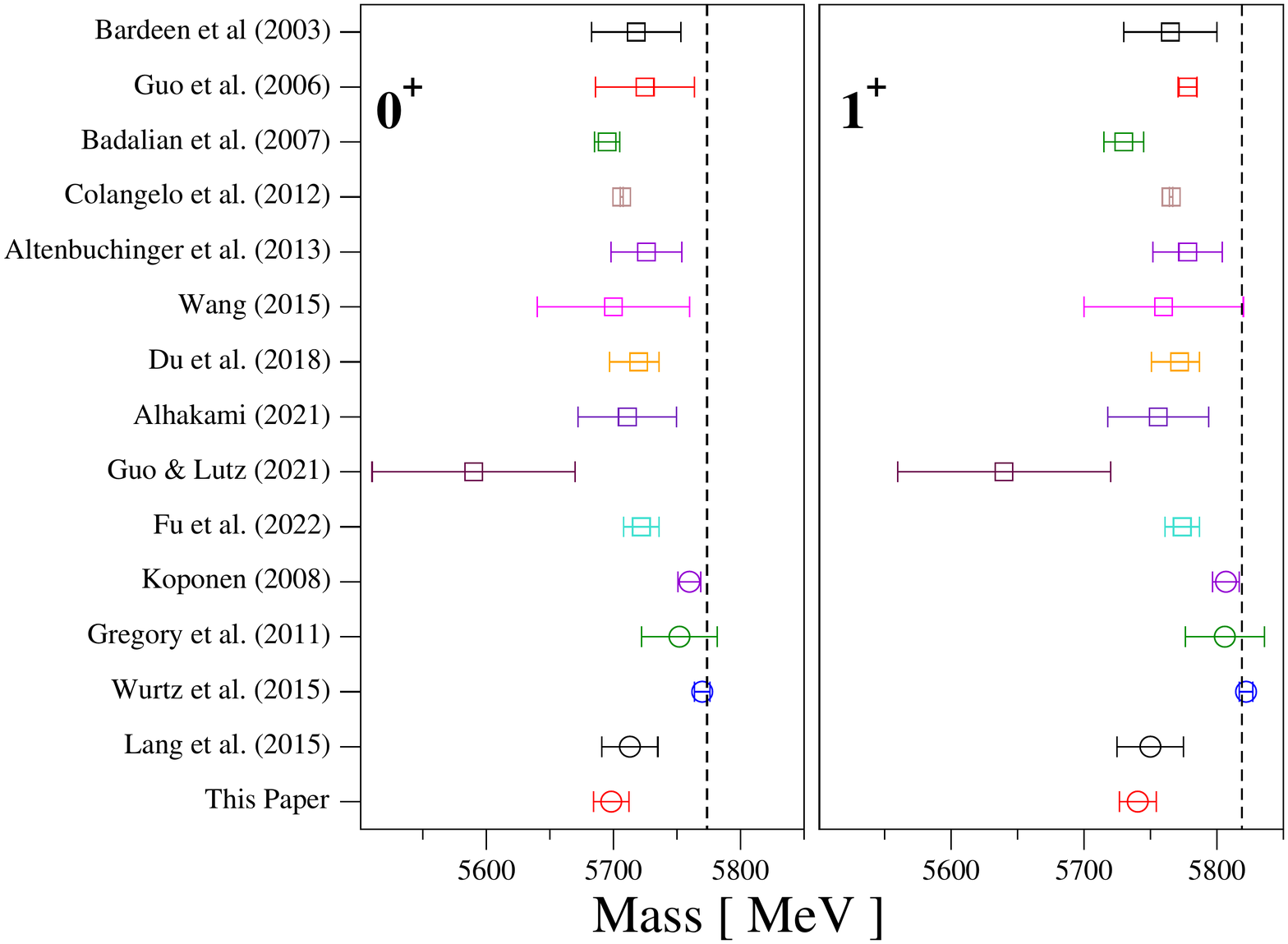}
\caption{Comparison of our results for the $B_{s0}^*$ (left pane) and $B_{s1}$
  (right pane) ground-state
  masses to various other results. Circles denote the results of lattice
  calculations, while squares denote the results from model/EFT
  calculations. The vertical line denotes the respective threshold. To
  translate our result for the binding energy to the result displayed in this
  plot an iso-symmetric kaon mass of $494.2$~MeV has been used.}\label{fig:comparison_bs}
\end{figure}

Finally, we investigated the $B_{s0}^*$ and $B_{s1}$ scalar and axial-vector
$B_s$-mesons in our setup and predict states $75.4(14.0)$ and
$78.7(13.9)$ MeV below the corresponding $BK$ and $B^*K$ thresholds respectively. Fig.~\ref{fig:comparison_bs} shows a comparison to various Lattice QCD and selected \footnote{We chose the references that attempt to quantify the
  uncertainty of the calculation. Reference \cite{Albaladejo:2016ztm} and \cite{Yang:2022vdb} use Lattice QCD input and
are not fully independent of previous lattice results and so we therefore mention
them here instead of in the overview plot. The results from \cite{Cleven:2010aw} are superseded by the results in \cite{Fu:2021wde} and we therefore omit the former from the overview plot.}
model/effective field theory calculations. Our results are fully consistent with
the previous pole-determination of \cite{Lang:2015hza} and with most of the calculations
based on effective field theory approaches and models based on chiral and
heavy-quark symmetry. Our results however show tension with some of the previous heavy-light
Lattice NRQCD calculations, and we observe pretty significant finite-volume and
pion-mass dependencies that must be taken into account in determining these
states on the lattice. Our study has a very different approach to the previous
works and different associated systematics. Hopefully further Lattice QCD studies of these states will be performed in the future to clarify their masses in anticipation of an experimental determination.

The use of NRQCD is still relevant in the field of Lattice QCD even as measurements with physical, dynamical b-quarks are beginning to be performed. Lattice NRQCD is useful as a very cheap and statistically-precise exploratory tool, but as it does not have a formal continuum limit one must ensure it is able to describe known states appropriately before performing physically-relevant predictions. We argue this can be precisely done with our tuning at little extra expense, and comes with the minor loss of not having simple bottomonia states as predictions.

\acknowledgments{The authors would like to acknowledge useful discussions
  with Matthias Lutz and Parikshit Junnarkar. We thank the CLS consortium for providing gauge configurations. D.M. acknowledges funding by the
  Heisenberg Programme of the Deutsche Forschungsgemeinschaft (DFG, German
  Research Foundation) – project number 454605793. Calculations for this
  project were partly performed on the HPC cluster “Mogon II” at JGU
  Mainz. This research was supported in part by the cluster computing resource
  provided by the IT Division at the GSI Helmholtzzentrum für
  Schwerionenforschung, Darmstadt, Germany (HPC cluster Virgo). For the neural network training and predictions we made use of the Keras API. For the light- and strange-quark propagator inversions we used the package OpenQCD \cite{Luscher:2012av}.}

\appendix

\section{Technical aspects of the NRQCD calculation}\label{app:NRQCD_tech}

In this appendix we describe some of the details of our NRQCD
implementation. Considering Eq.~\ref{eq:nrqcd_ac}, it should be clear that the
\textit{natural} operations will be on $N_C\times N_C$ color-matrices, so sink
and source color indices will be the fastest-moving in our propagator
solution. Outer (spin) indices are considered as a matrix of size $2\times 2$, i.e. the size of the Pauli matrices. We utilise the NRQCD implementation in \cite{Hudspith:2020tdf}, which is a pure thread-parallel (threaded using \verb|OpenMP| \cite{660313} 
pragmas) application of the evolution equation
(Eq.~\ref{eq:evolutionNRQCD}) in a time-slice by time-slice manner. Aside from
simple loop-unrolling and loop-fusion optimisation strategies, this
implementation applies all contributions (except $H_0$) of the Hamiltonian
accumulated on the time-slice into a buffer for better thread parallelism. As
such, an important optimisation is the pre-computation of the combinations of
links:
$U_\mu\left(x+\frac{a}{2}\hat{\mu}\right)U_\mu\left(x+3\frac{a}{2}\hat\mu\right)$,
and of course the improved field-strength tensors. Such an optimisation is
necessary to avoid implicit thread-barriers in \verb|OpenMP|'s parallel for
routines, as they cost many CPU-cycles. It turns out that these applications
of the higher-order contributions from the Hamiltonian ($\delta H$) are the most costly parts of performing the evolution.

The majority of the NRQCD evolution algorithm can be boiled-down to performing the product $A_{ab}S^{\alpha\beta}_{bc}$ i.e. a color-matrix multiplied by a propagator ($ab$ are inner color indices and $\alpha\beta$ the outer Dirac) and a specific, unrolled, advanced vector extensions/fused multiply-add (AVX/FMA) implementation is called that performs this operation reasonably optimally.

\section{Retuning with fixed parameters}\label{sec:app_fixed}

Here we detail an investigation into keeping the two higher-order lattice-spacing correction terms ($c_5$ and $c_6$) in the NRQCD action fixed to 1 and nonperturbatively tuning the others as one may be concerned with our setup having more outputs than inputs.

\begin{table}[h!]
  \begin{tabular}{ccccccc}
    \toprule
    $aM_0$ & $c_1$ & $c_2$ & $c_3$ & $c_4$ & $c_5$ & $c_6$ \\
    \hline
    2.0583(6) & 0.814(19) & -1.085(102) & 1.145(11) & 1.018(13) & 1 & 1 \\
    \botrule
  \end{tabular}
  \caption{Our nonperturbative tuning with $c_5$ and $c_6$ set to their tree-level values for the $U_{0P}$ tadpole factor using ensemble A653. To be compared with our results in Tab.~\ref{tab:nncoefs}.}\label{tab:fixed_test}
\end{table}

In Tab.~\ref{tab:fixed_test} 
we give the tuning parameters when $c_5=c_6=1$. It seems that $aM_0$ is not greatly affected by this change, $c_3$ appears affected by a few \%. We do see $c_4$ grow similarly, and $c_1$ tends closer to 1. $c_2$ remains strongly negative as all of the tuning runs have illustrated. $c_1$ and $c_2$ vary significantly compared to our full tuning. By not tuning $c_5$ and $c_6$ we obtain a worse percentage deviation (Eq.~\ref{eq:pcdev}) of $1.8\%$ for our predicted values compared to the value of $0.7\%$ in Tab.~\ref{tab:deviations}. This mostly manifests in smaller splittings between the P-wave states. It is quite interesting that the range of neural-network predictions for $c_2$ is so much larger than with the full 7-parameter tuning. This could be that the fit is trying to suppress higher-order spin-independent effects and tadpole factors directly with $c_1$ and $c_2$ as there is no longer freedom to do so with $c_5$ and $c_6$.

\section{On the finite-volume effects of the $B_{s0}^*$ and $B_{s1}$ mesons}\label{sec:app_bsfv}

\begin{table}[h!]
  \begin{tabular}{c|cc|c}
    \toprule
    Fit & $\Delta_{B_{s0}^*}$ [MeV] & $\Delta_{B_{s1}}$ [MeV] & $\chi^2/\text{dof}$ \\
    \hline
    Combined $m_K L$ & $-67.5(3.4)$ & $-71.9(3.8)$ & $0.5$ \\
    Combined $m_\pi L$ & $-54.7(4.1)$ & $-58.4(4.6)$ & $2.9$ \\
    Combined $m_\pi L+m_K L$ & $-68.8(5.5)$ & $-73.4(6.1)$ & $0.5$ \\
    \hline
    Individual $m_K L$ & $-68.0(3.3)$ & $-71.4(3.9)$ & $0.9,\; 0.3$ \\
    Individual $m_\pi L$ & $-55.3(3.9)$ & $-57.6(4.8)$ & $4.2,\; 2.8$ \\
    Individual $m_\pi L+m_K L$ & $-70.1(5.4)$ & $-72.0(6.2)$ & $1.1,\; 0.3$\\
    \botrule
  \end{tabular}
  \caption{Investigation of the combined chiral/finite-volume fits to the $B_{s0}^*$ and $B_{s1}$ mass differences along only the $\text{Tr}[M]=C$ mass trajectory. ``Combined'' indicates that the results are from a simultaneous fit to both $B_{s0}^*$ and $B_{s1}$ mass-differences, whereas ``Individual'' indicates fits individually to the data. ``$m_K L$'' indicates a fit to the form of Eq.~\ref{eq:bsfvolK_trM}, ``$m_\pi L$'' to Eq.~\ref{eq:bsfv_mpil}, and ``$m_\pi L+m_K L$'' to Eq.~\ref{eq:bsfv_mkmpil}.}\label{tab:bsfvol}
\end{table}

Here we investigate further variations of the fits with regard to the
finite-volume effects for the threshold-subtracted $B_{s0}^*$ and $B_{s1}$
mesons along the $\text{Tr}[M]=C$ mass-trajectory. Our preferred form is given by
\begin{equation}\label{eq:bsfvolK_trM}
\Delta_{B_{s0}^*/B_{s1}}(\phi_2, m_K L,a) = \Delta_{B_{s0}^*/B_{s1}}(0,\infty,a)\left( 1 + A \phi_2 + C e^{-m_K L}\right),
\end{equation}
but it could be possible that this is sub-leading to a finite-volume term such as
\begin{equation}\label{eq:bsfv_mpil}
\Delta_{B_{s0}^*/B_{s1}}(\phi_2, m_\pi L,a) = \Delta_{B_{s0}^*/B_{s1}}(0,\infty,a)\left( 1 + A \phi_2 + C e^{-m_\pi L}\right),
\end{equation}
or even a combination of the two:
\begin{equation}\label{eq:bsfv_mkmpil}
\Delta_{B_{s0}^*/B_{s1}}(\phi_2, m_K L, m_\pi L, a) = \Delta_{B_{s0}^*/B_{s1}}(0,\infty,\infty,a)\left( 1 + A \phi_2 + B e^{-m_K L} + C e^{-m_\pi L}\right).
\end{equation}
We also note that in our individual fits the coefficients A and B are consistent with one-another for both the $B_{s0}^*$ and $B_{s1}$, so we consider a simultaneous fit with these ($A$, $B$, and $C$) parameters shared and the $\Delta$s free. We tabulate these results in Tab.~\ref{tab:bsfvol}.

Tab.~\ref{tab:bsfvol} illustrates that our data cannot be well described by a single $e^{-m_\pi L}$ finite-volume term, whereas any fit (combined or not) with an $m_K L$ term describes the data extremely well. We can include both $m_\pi L$ and $m_K L$ terms in our combined chiral/finite-volume fit but the result remains consistent with just the $m_K L$ form. In fact, the coefficient ``C'' is consistent with zero for these fits and typically the statistical error of the determination increases with no real improvement in fit quality. This suggests that $m_\pi L$ finite-volume effect terms serve only as nuisance parameters within the precision of our data, and we can safely conclude that they are sub-leading in both our $B_{s0}^*$ and $B_{s1}$ determinations.

\section{Effective masses - exotic $B_s$ mesons}\label{app:effmassBS}

In this appendix we show the effective masses of our $B_{s0}^*$ and $B_{s1}$ data for all ensembles that enter the analysis.

\newpage

\begin{figure}[h!]
  \centering
  \includegraphics[scale=0.16]{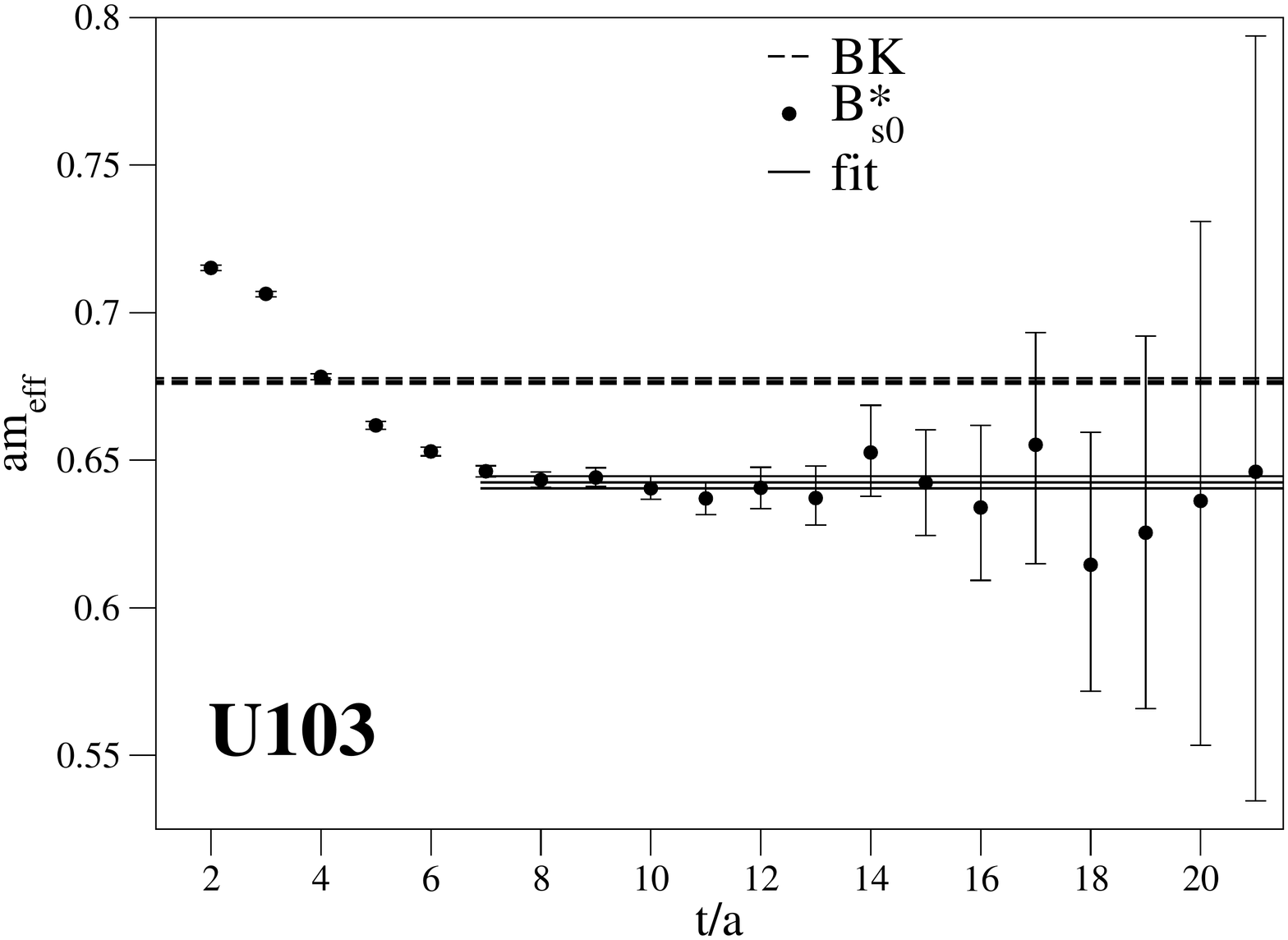}
  \includegraphics[scale=0.16]{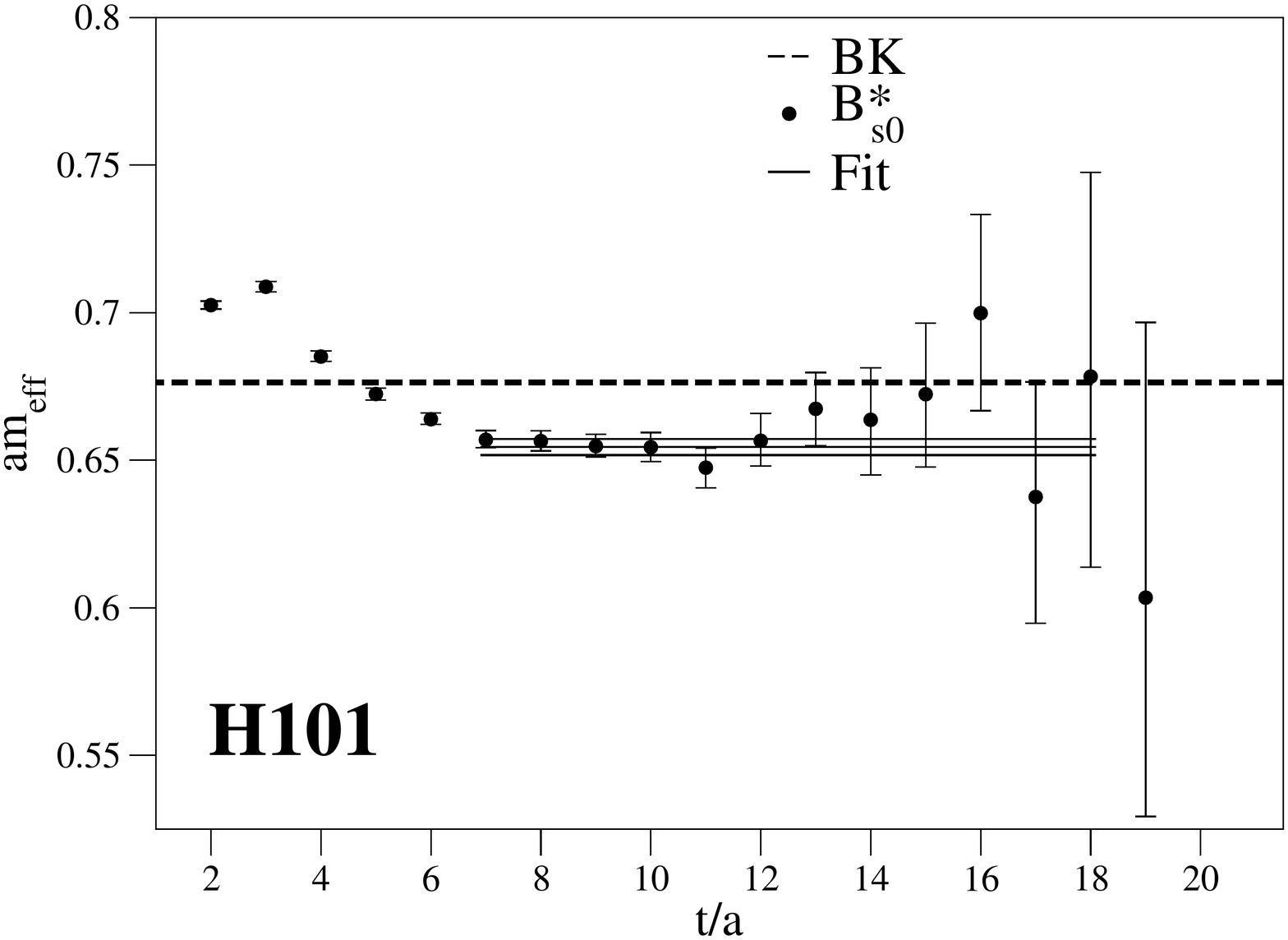}
  \includegraphics[scale=0.16]{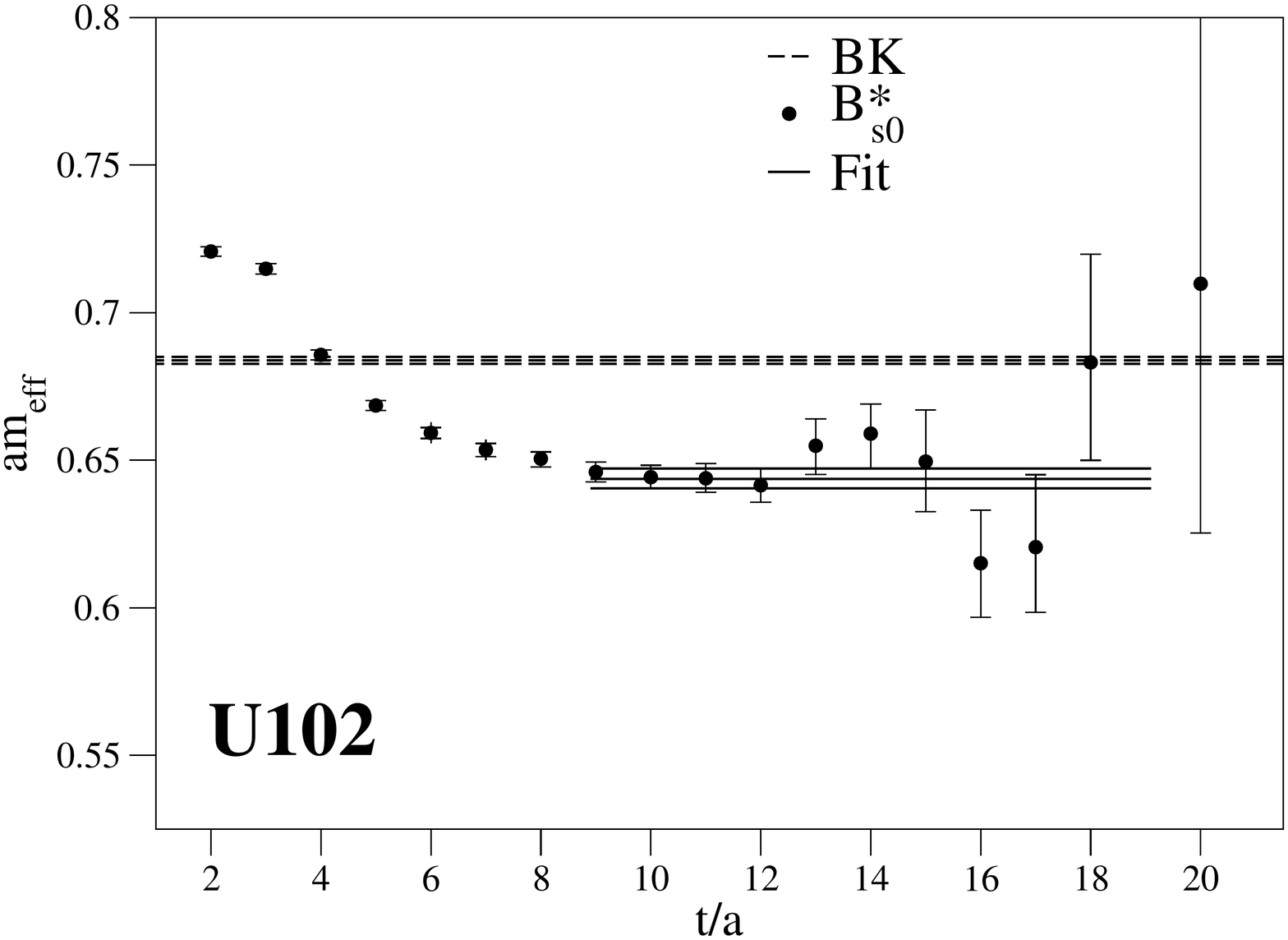}
  \newline
  \includegraphics[scale=0.16]{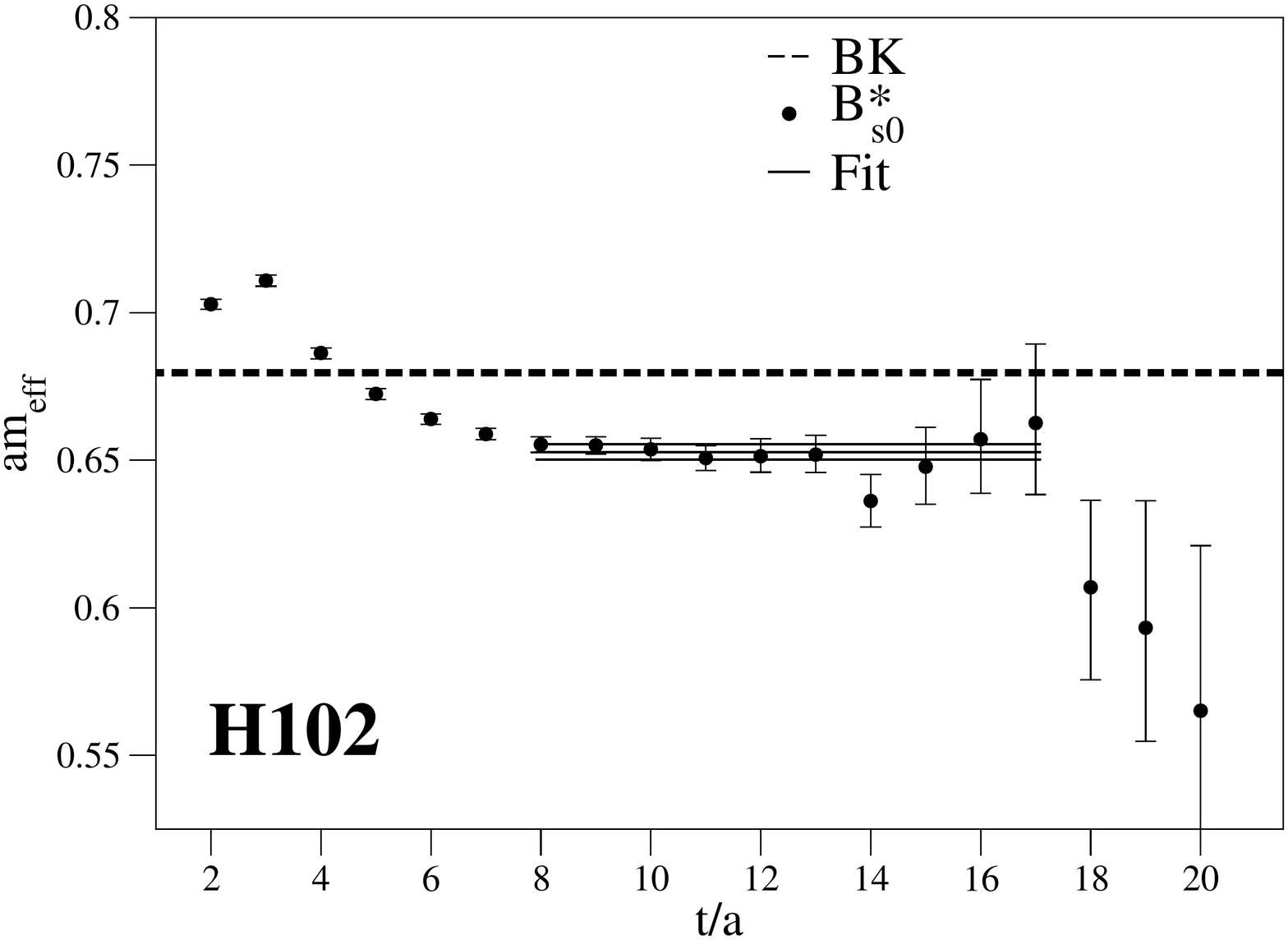}
  \includegraphics[scale=0.16]{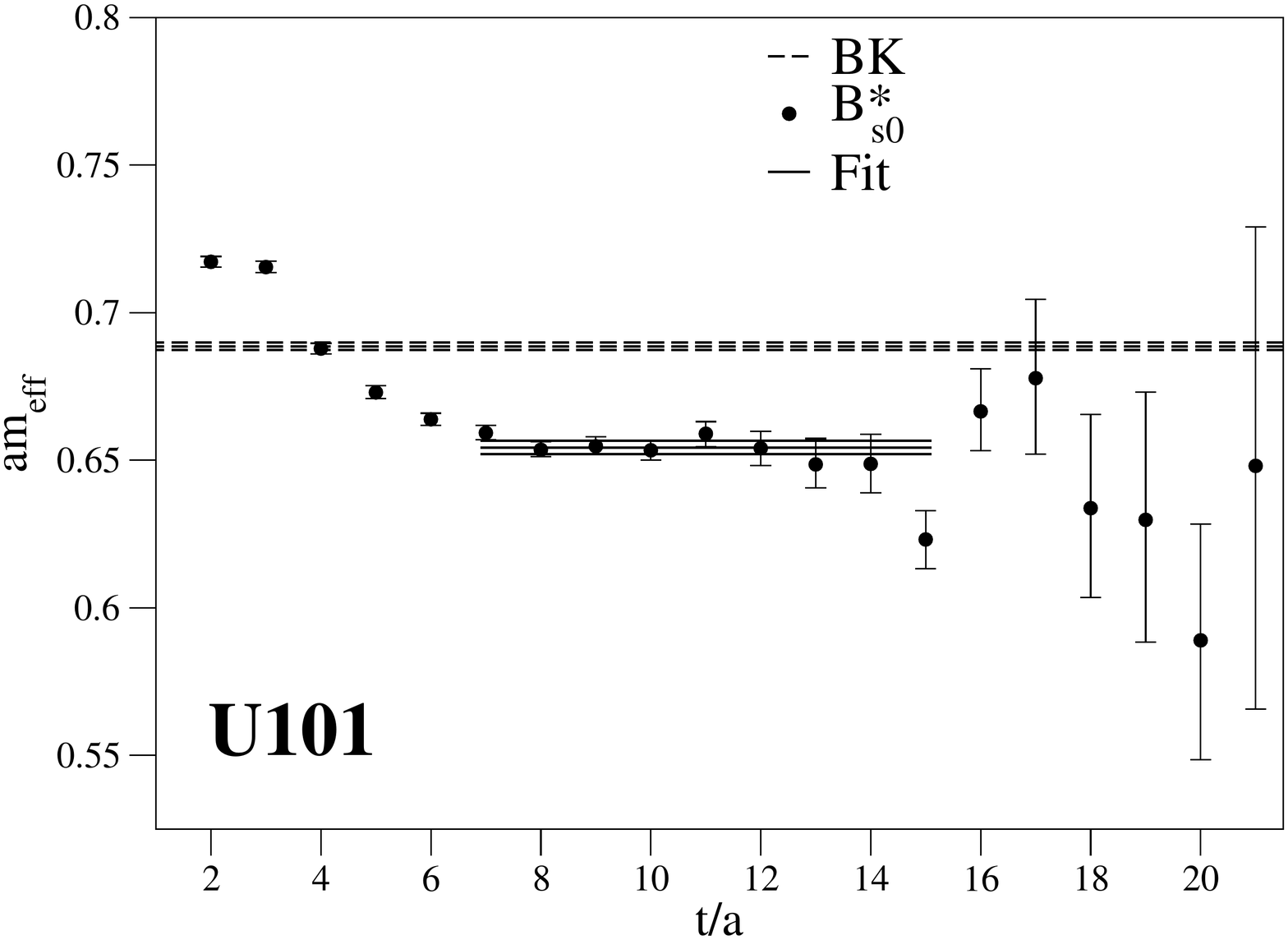}
  \includegraphics[scale=0.16]{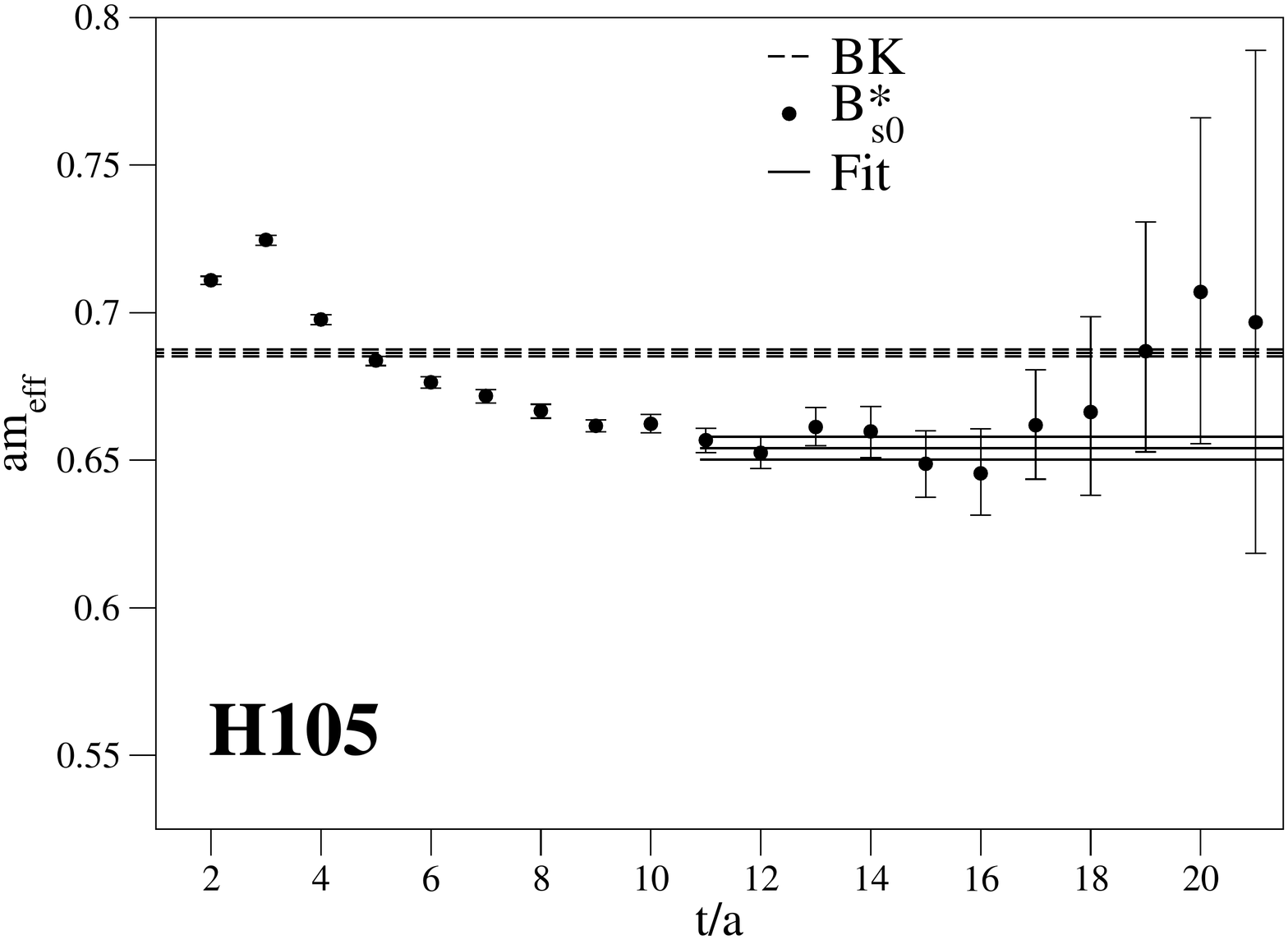}
  \newline	
  \includegraphics[scale=0.16]{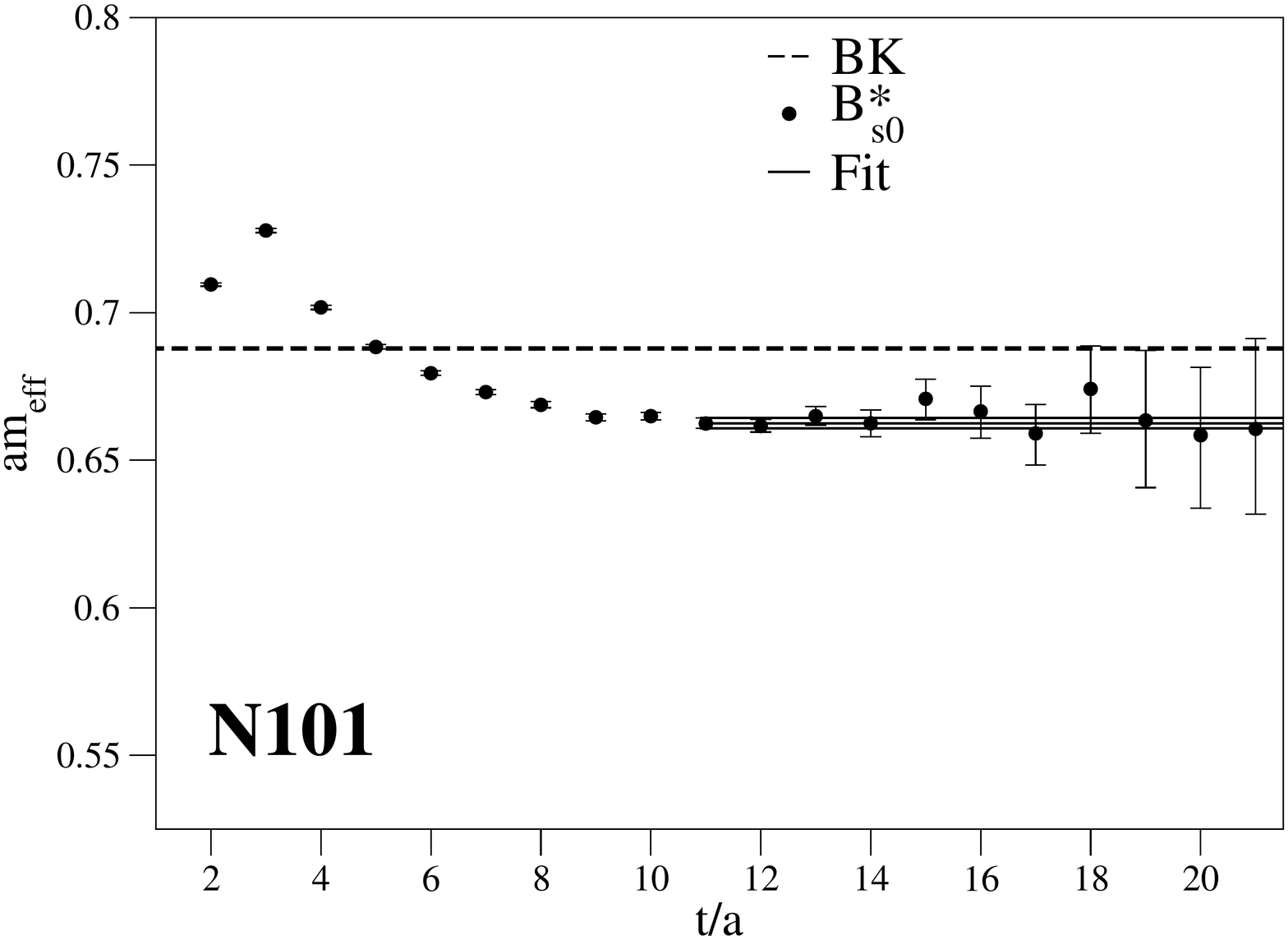}
  \includegraphics[scale=0.16]{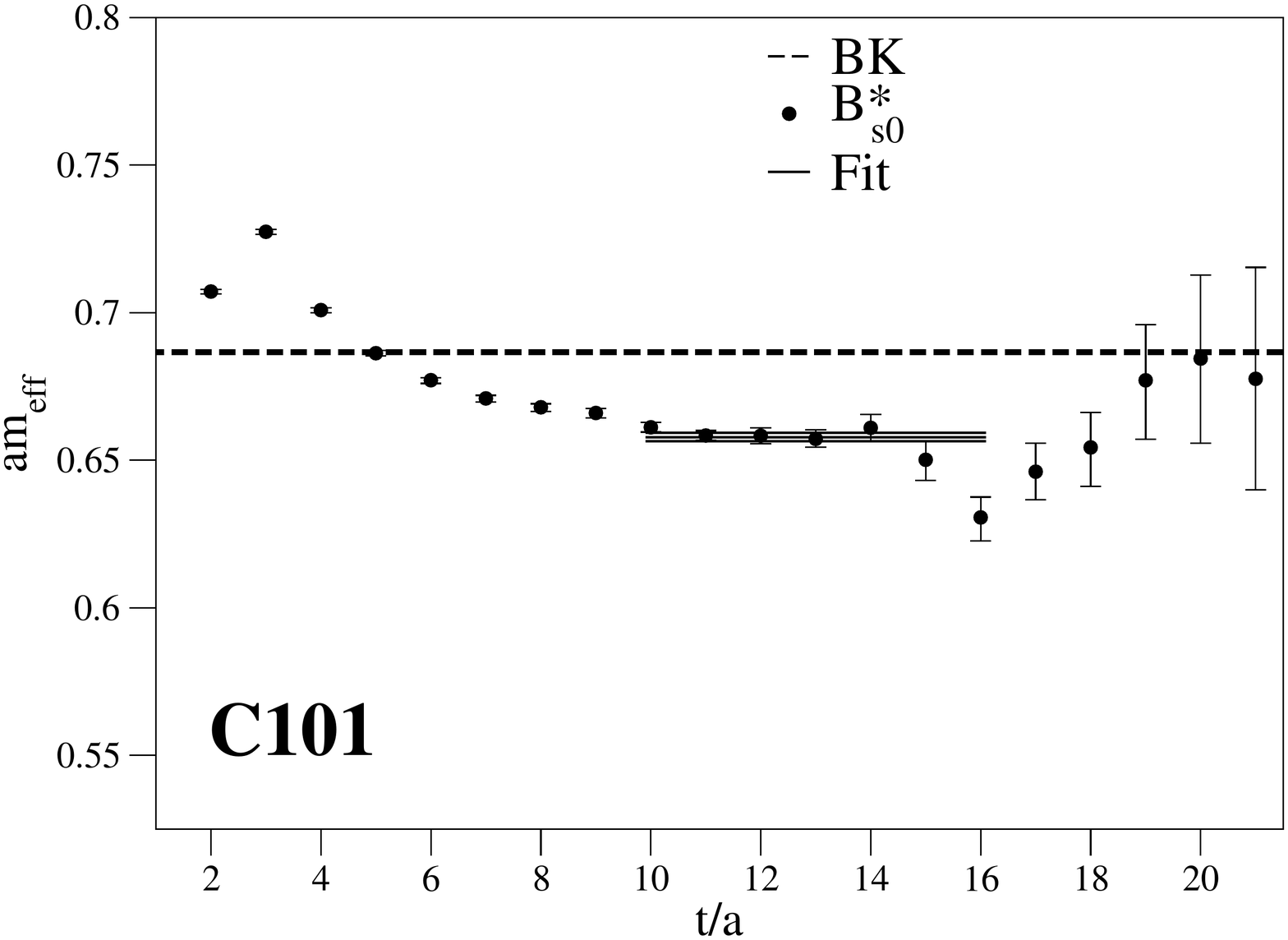}
  \includegraphics[scale=0.16]{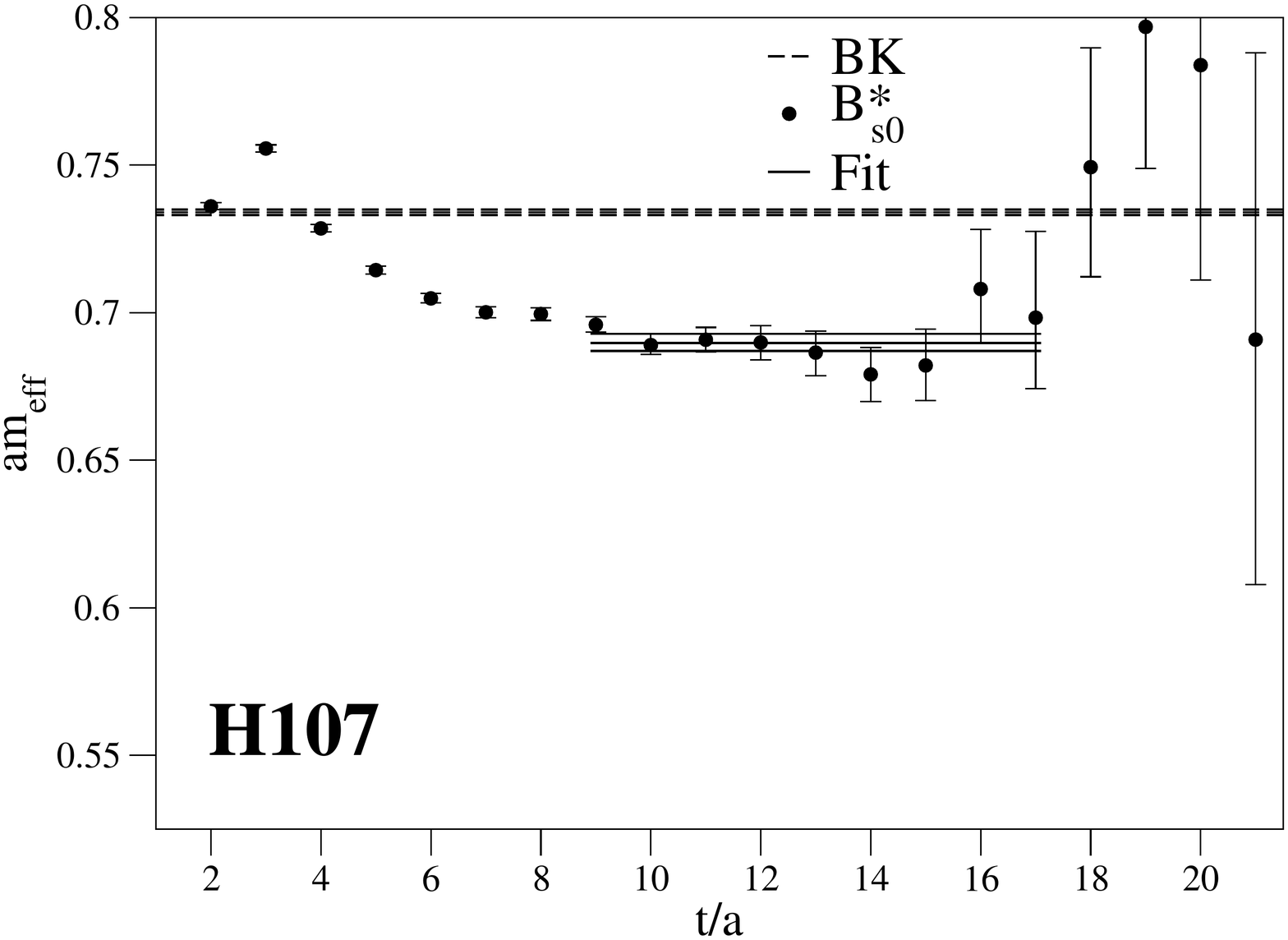}
  \newline
  \includegraphics[scale=0.16]{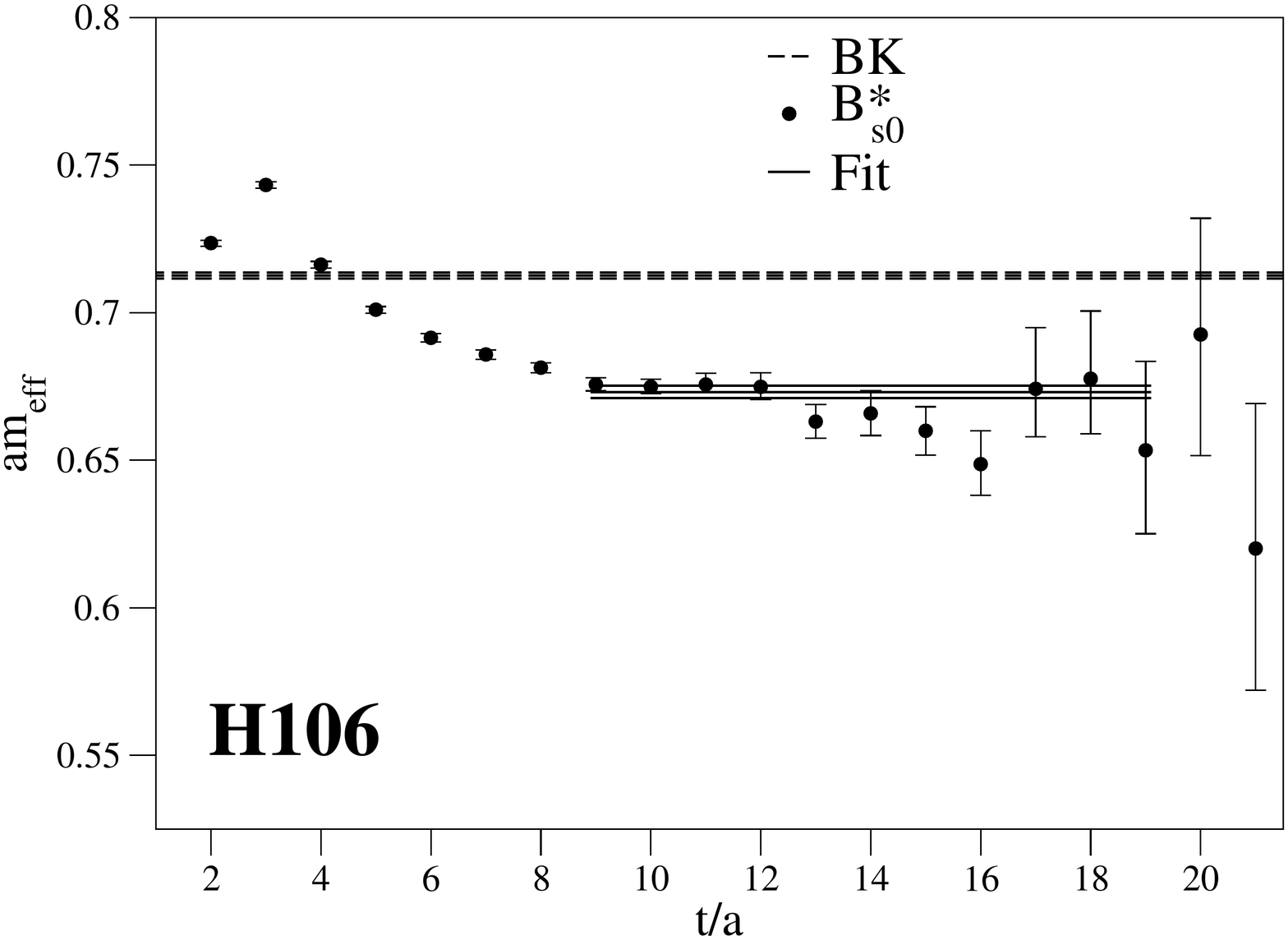}
  \includegraphics[scale=0.16]{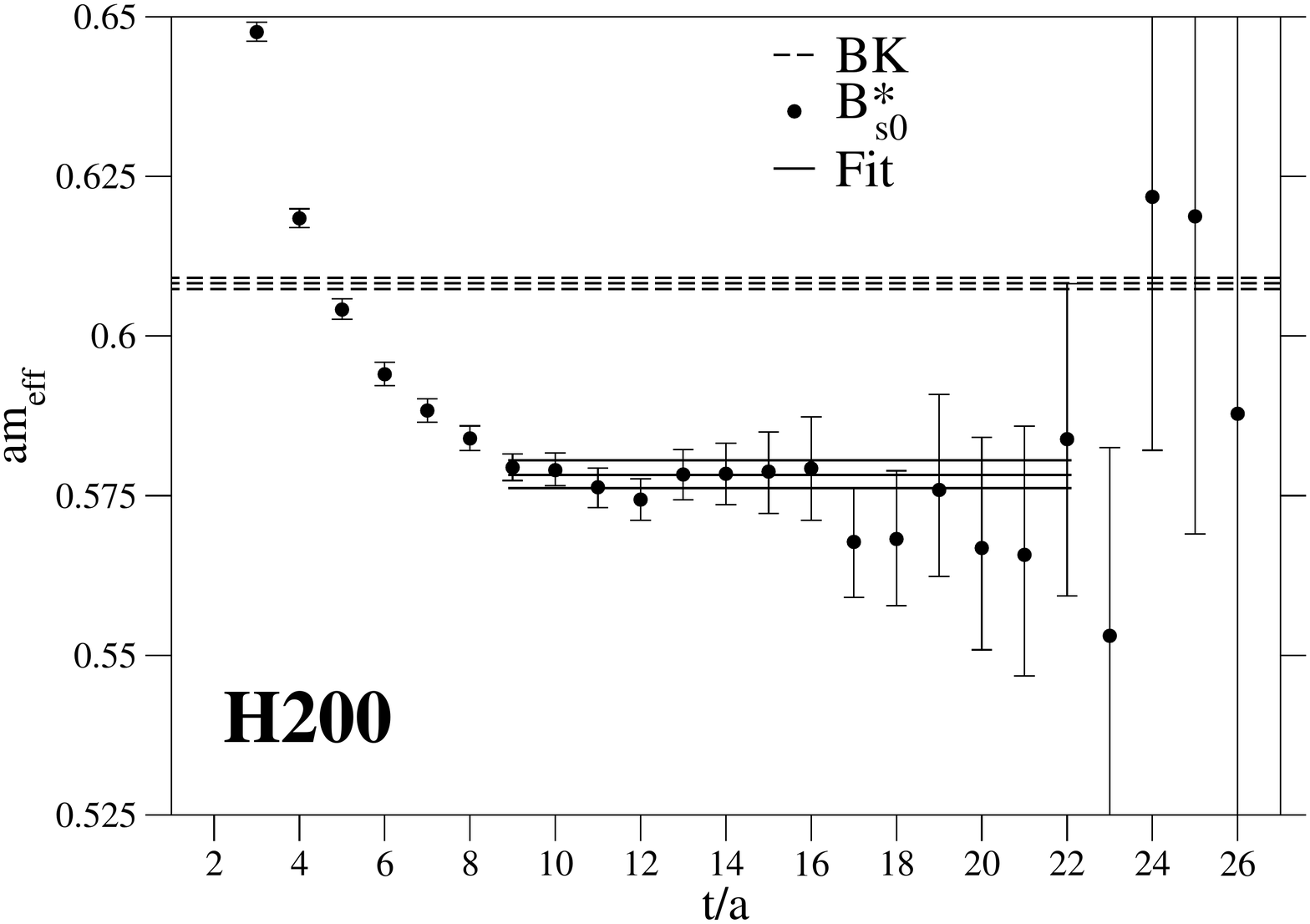}
\caption{Effective mass plots of the $B_{s0}^*$ meson with the measured ($BK$) threshold plotted. From left to right and top to bottom we show results from: U103, H101, U102, H102, U101, H105, N101, C101, H107, H106, H200.}\label{fig:effmass_Bs0}
\end{figure}

\newpage

\begin{figure}[h!]
  \centering
  \includegraphics[scale=0.16]{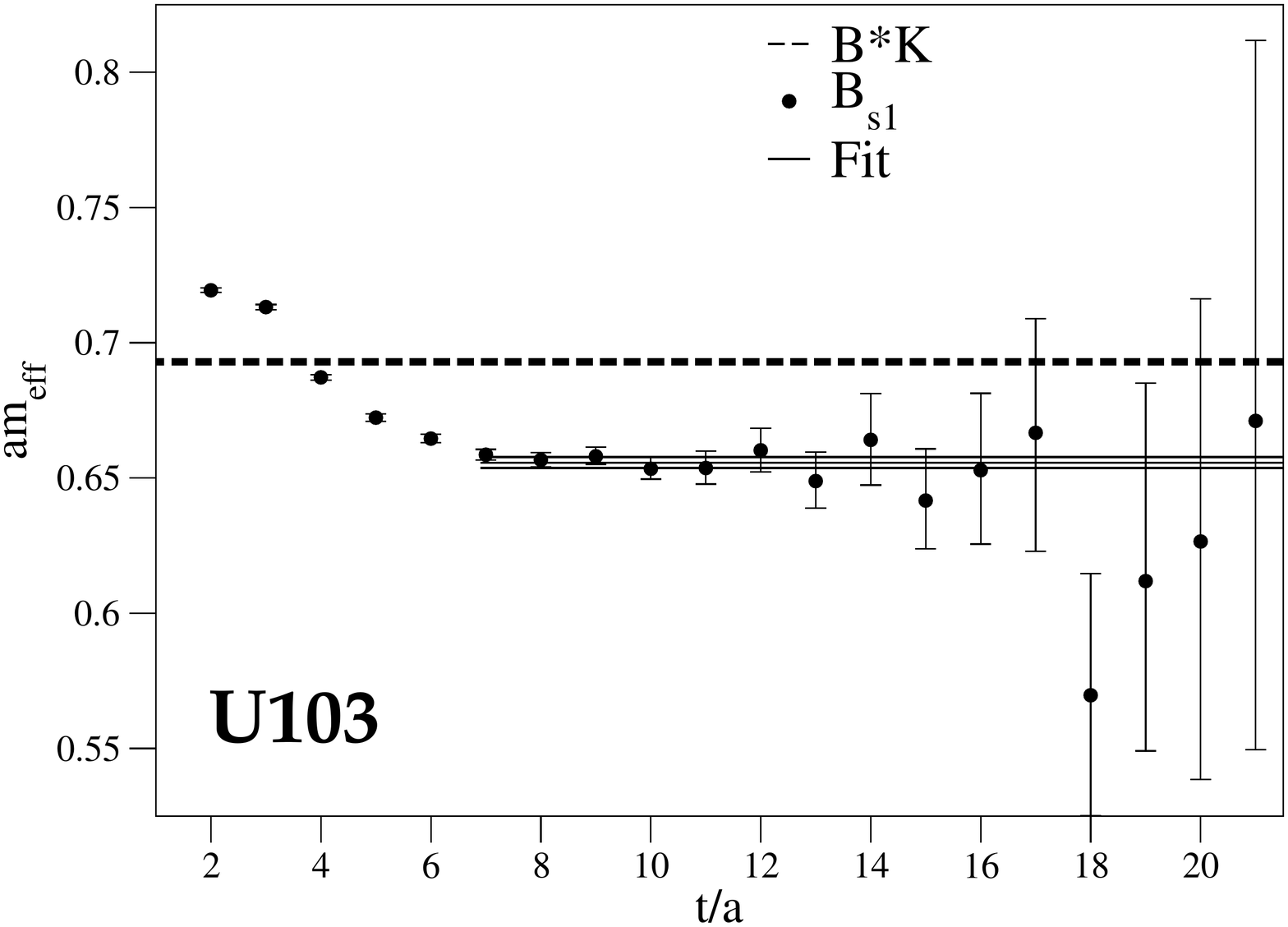}
  \includegraphics[scale=0.16]{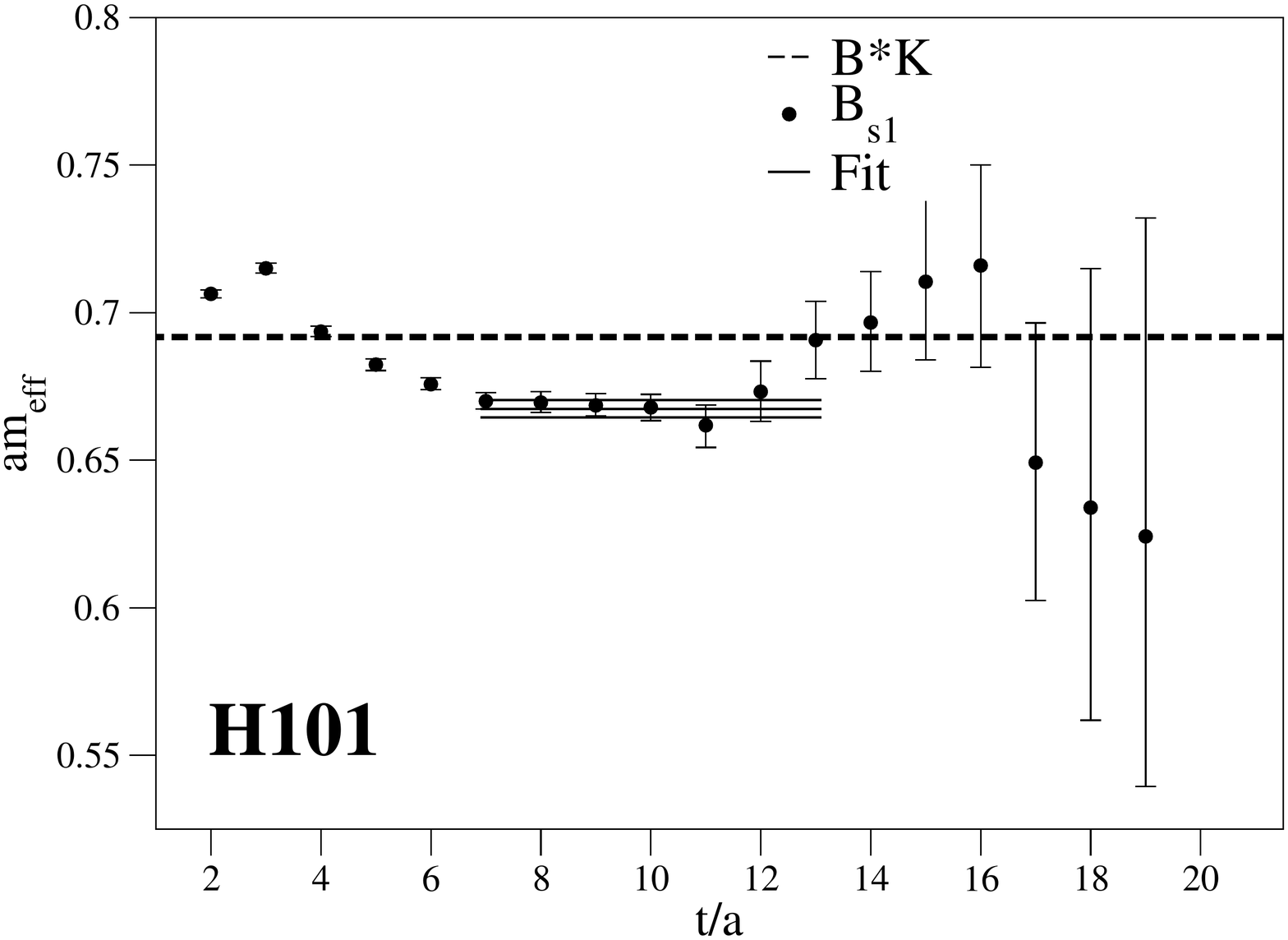}
  \includegraphics[scale=0.16]{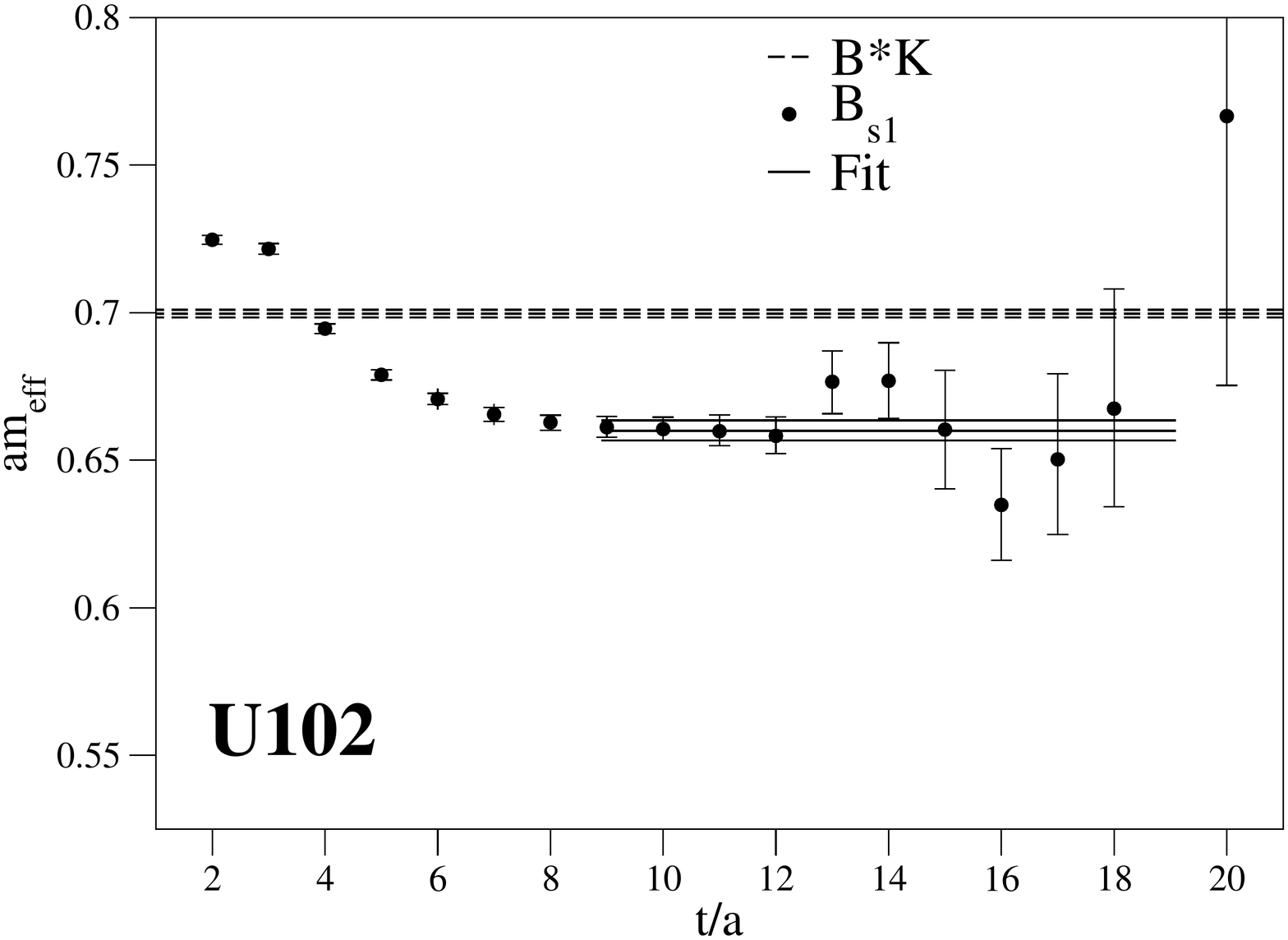}
  \newline
  \includegraphics[scale=0.16]{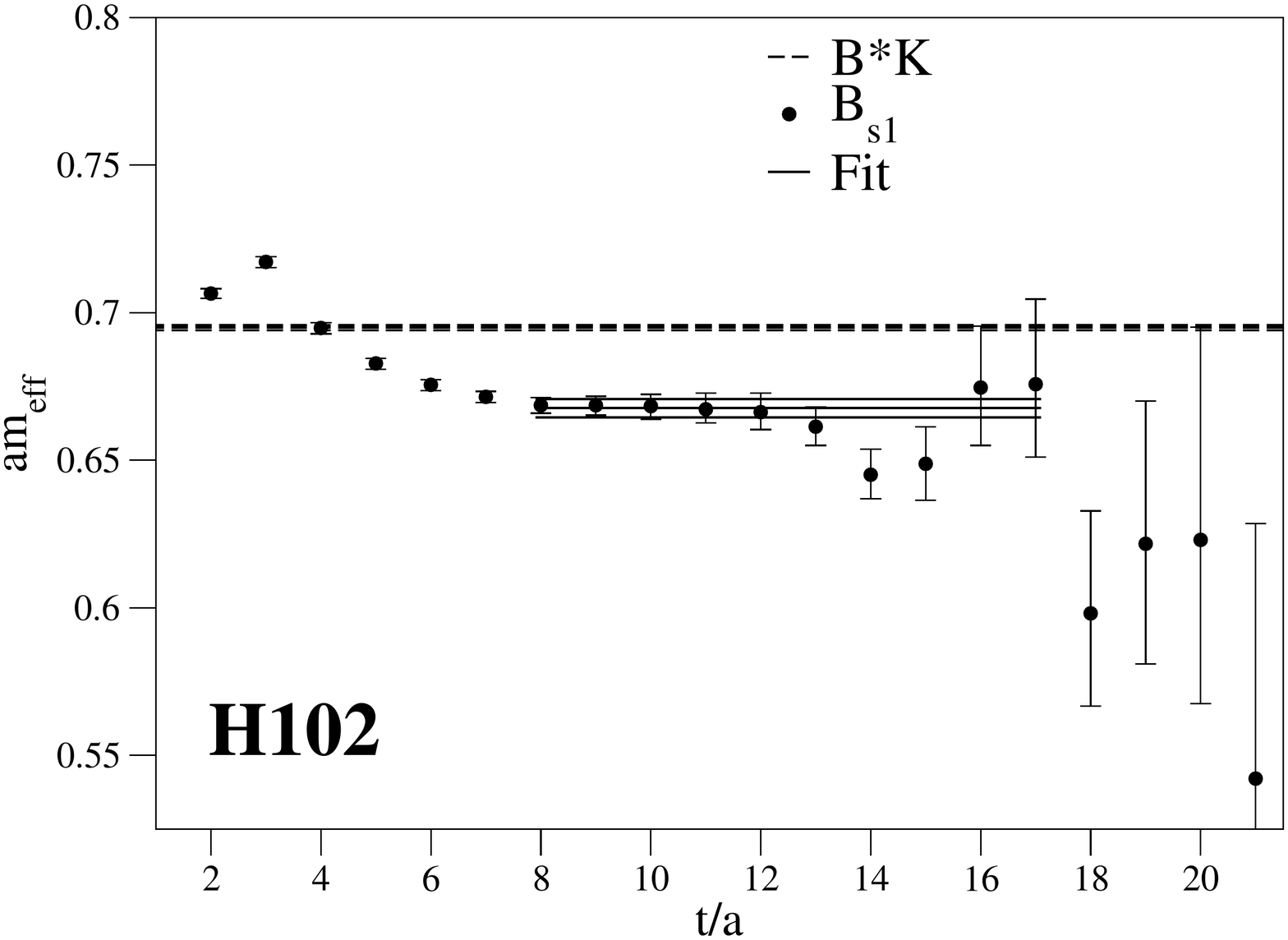}
  \includegraphics[scale=0.16]{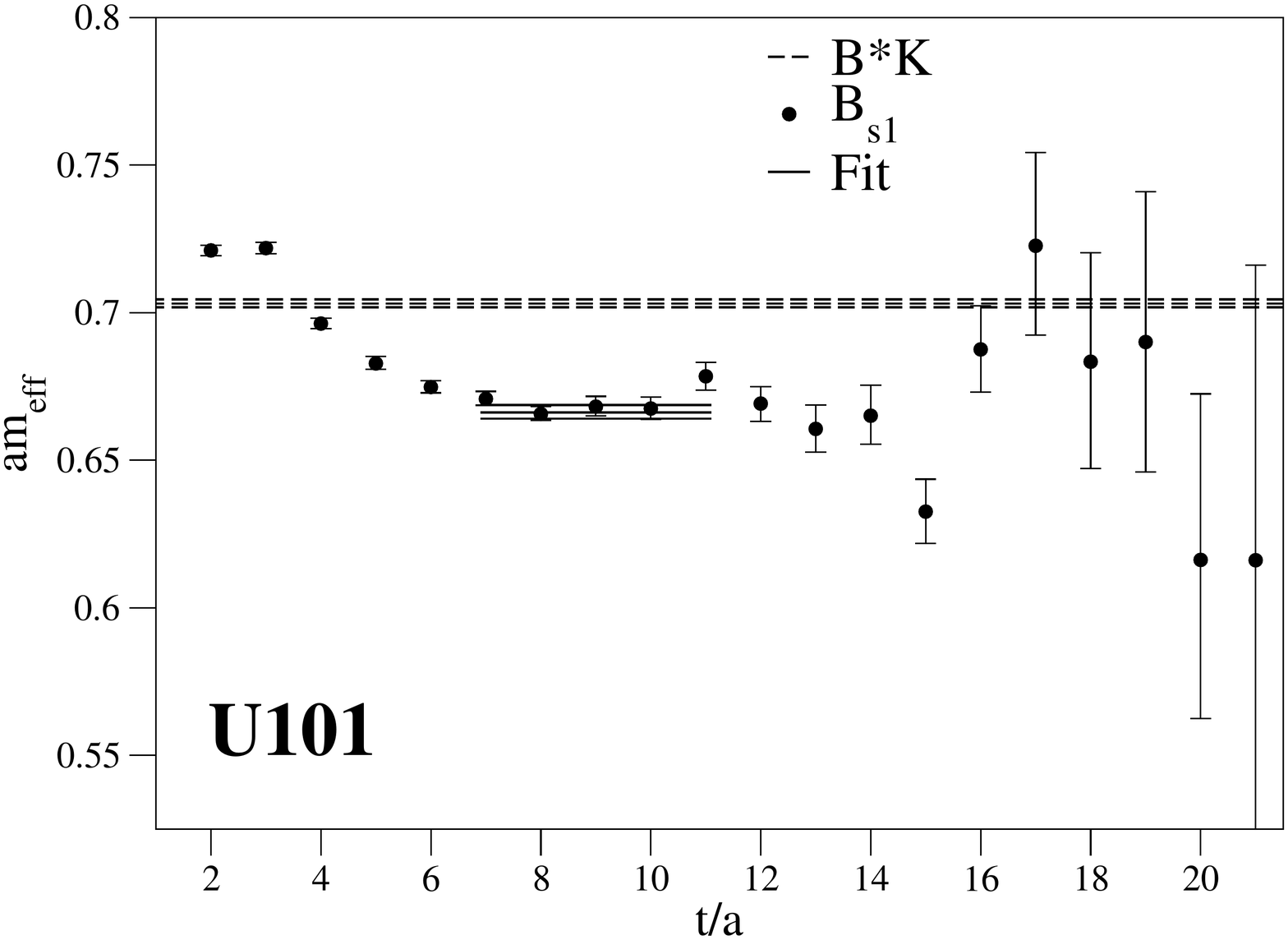}
  \includegraphics[scale=0.16]{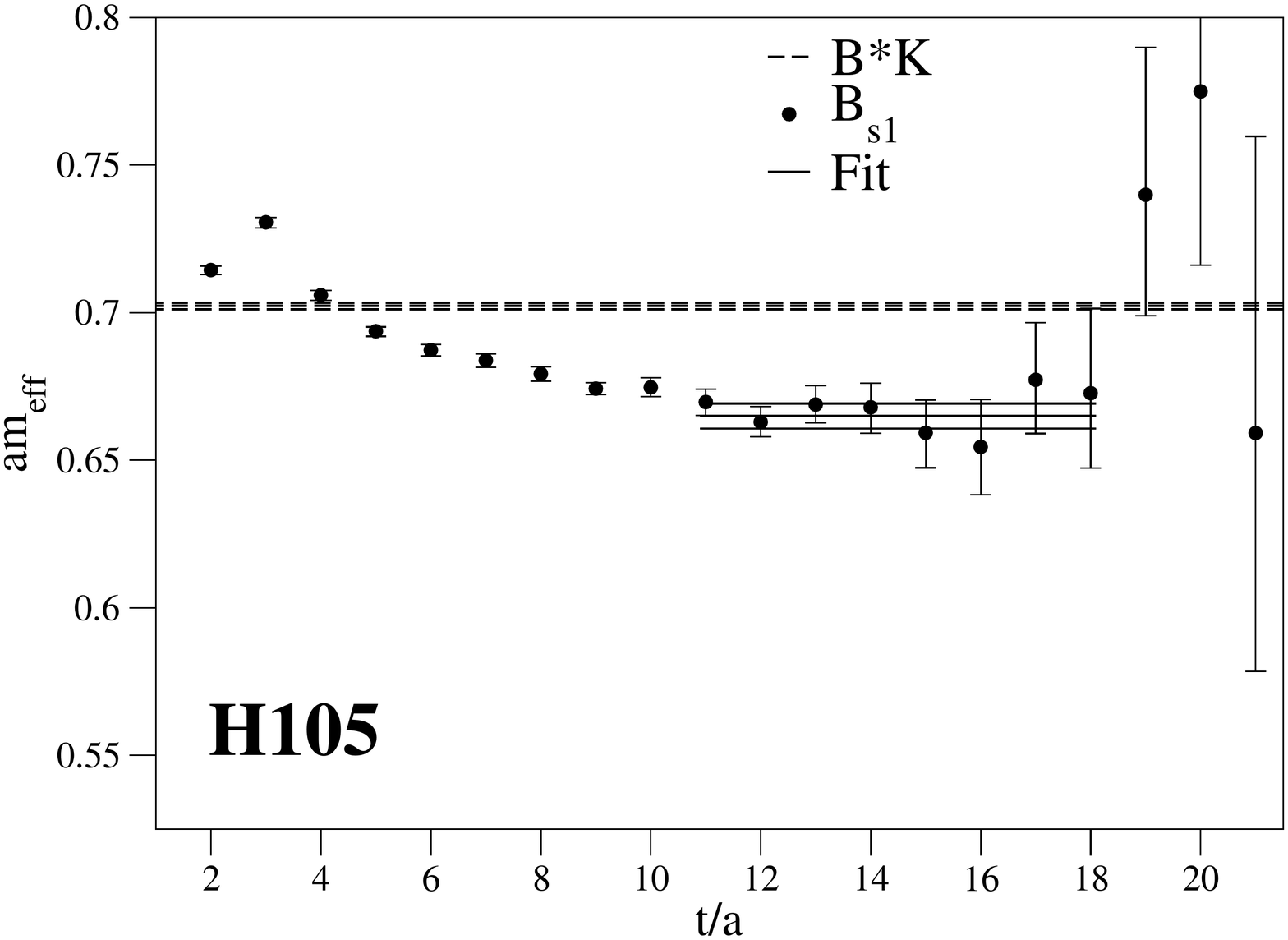}
  \newline
  \includegraphics[scale=0.16]{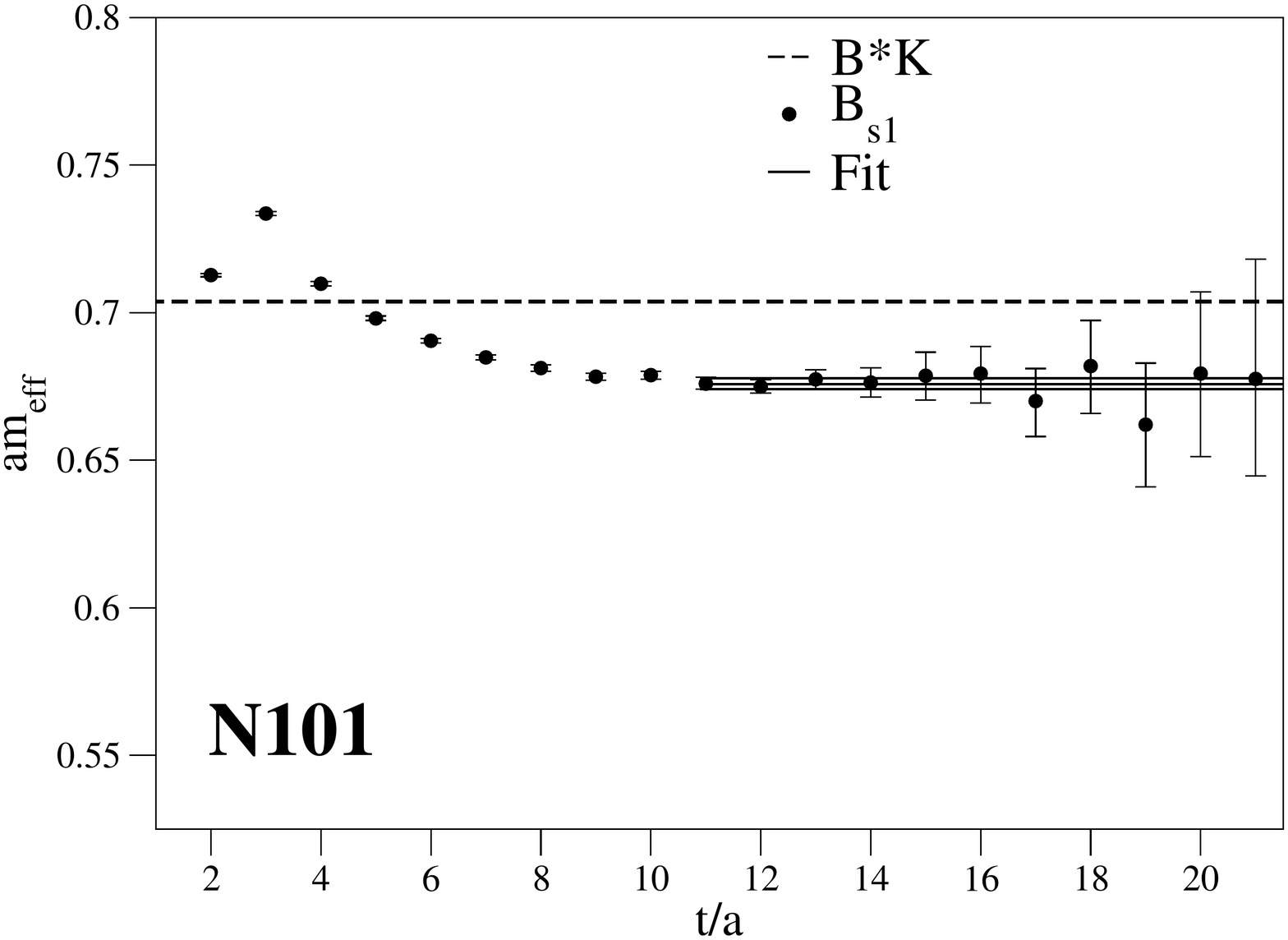}
  \includegraphics[scale=0.16]{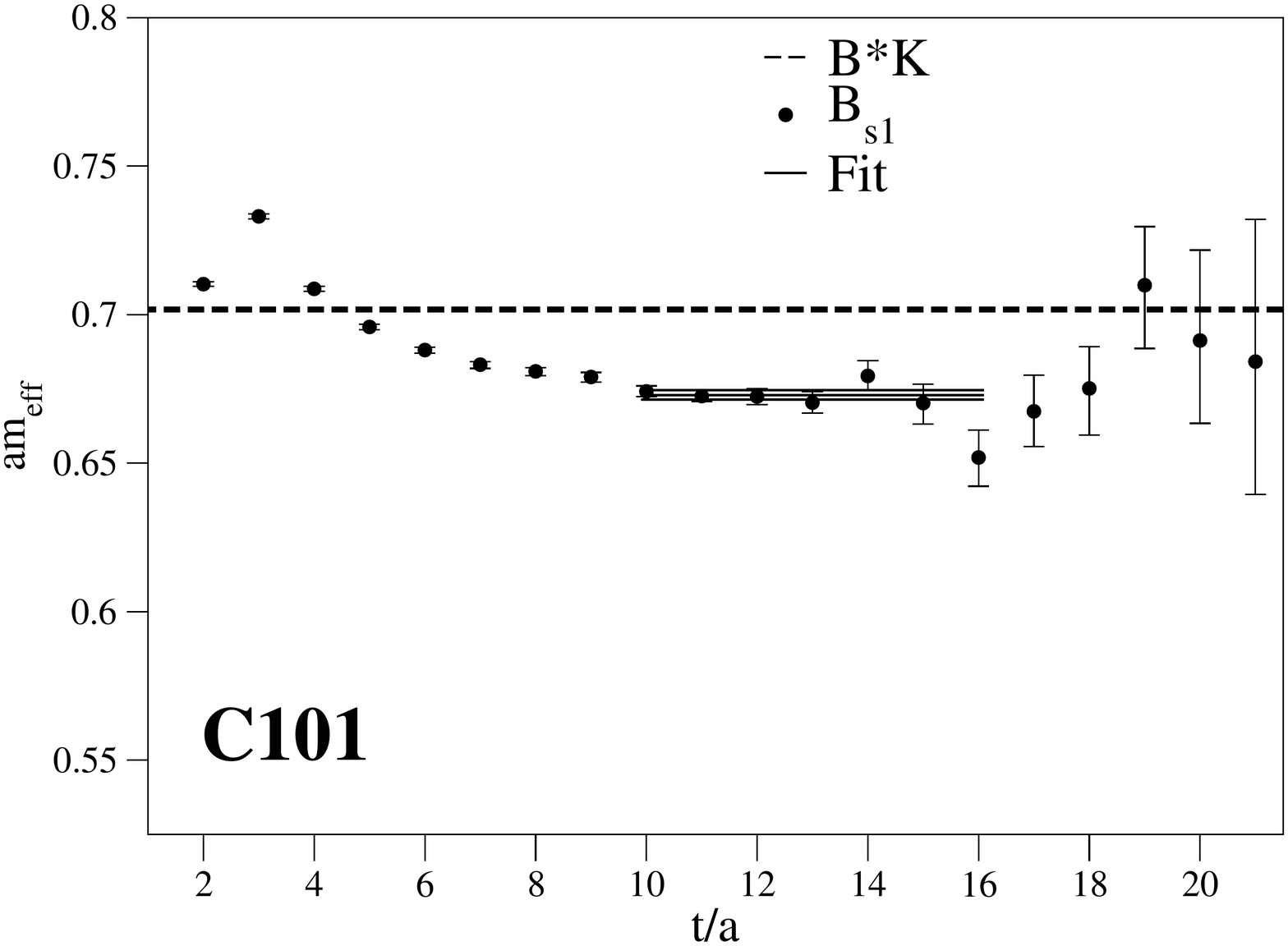}
  \includegraphics[scale=0.16]{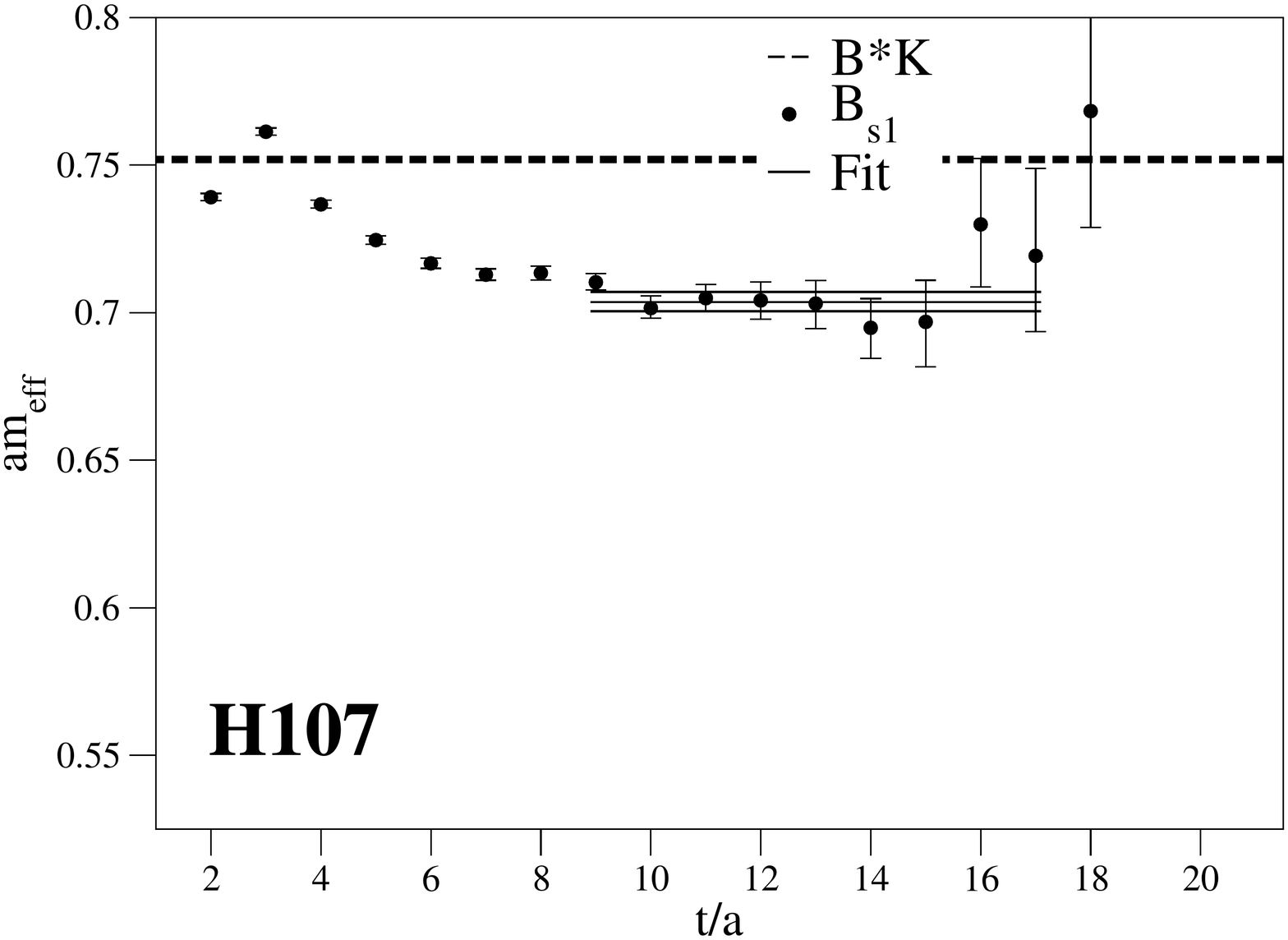}
  \newline
  \includegraphics[scale=0.16]{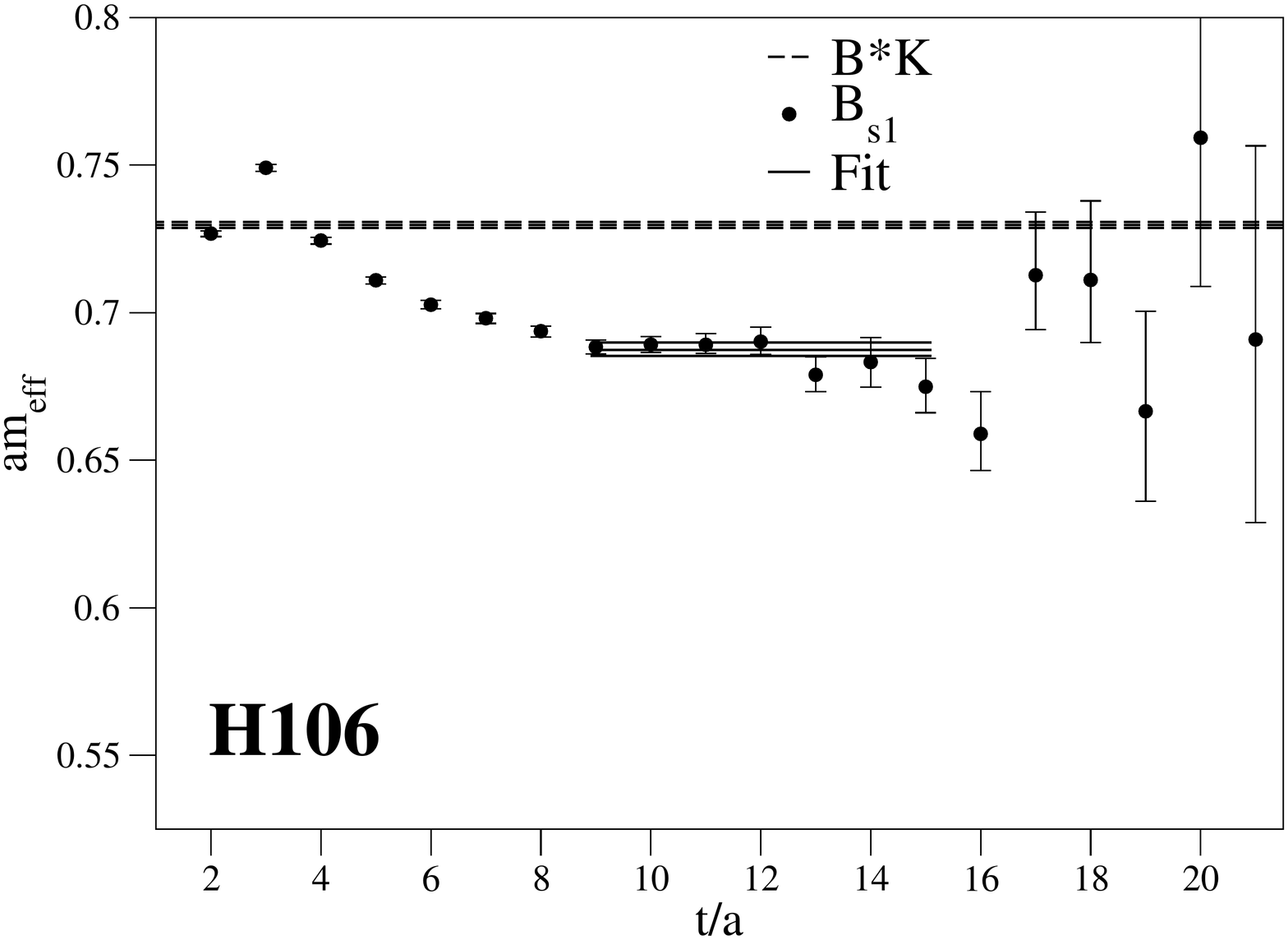}
  \includegraphics[scale=0.16]{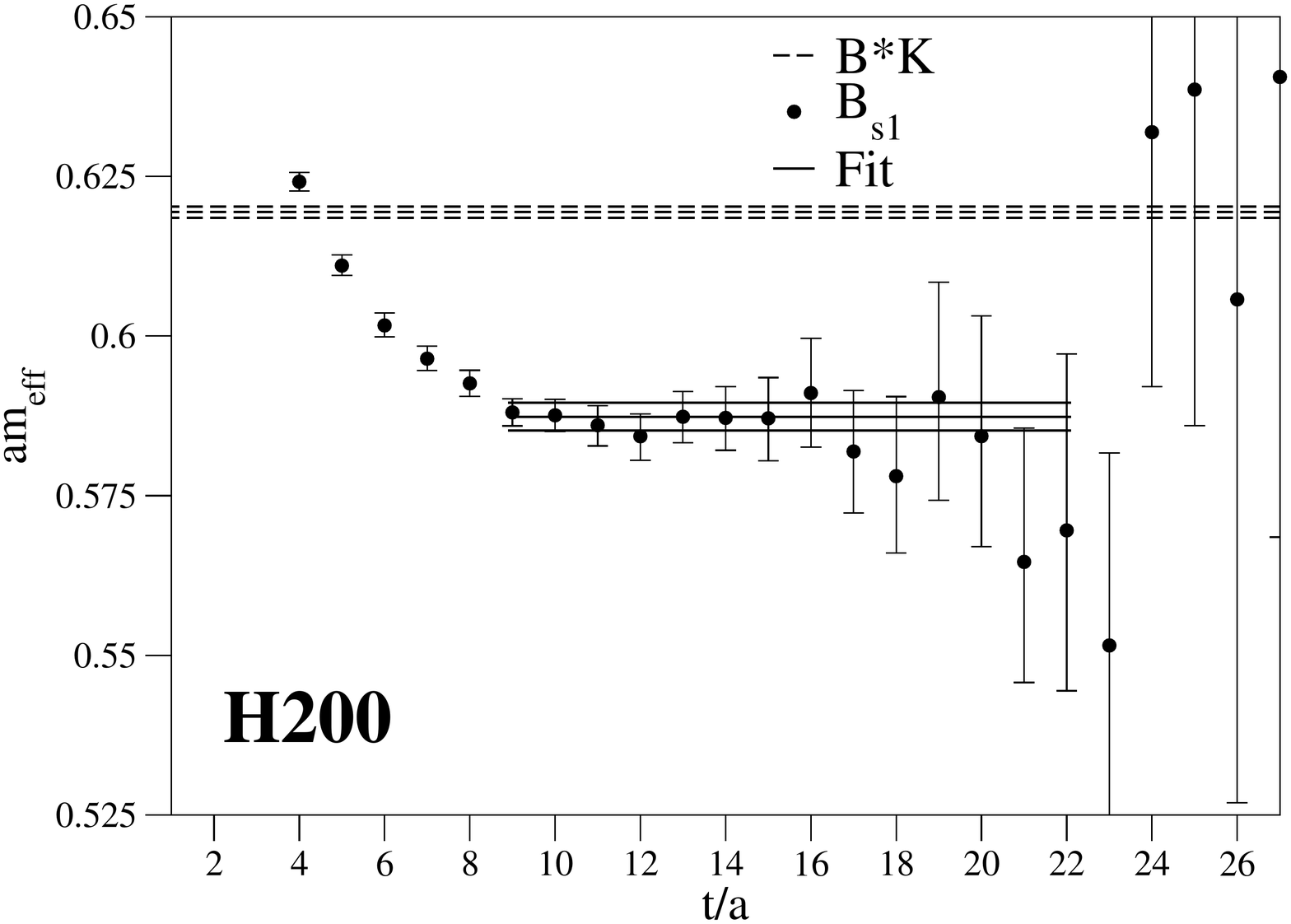}
\caption{Similarly to Fig.~\ref{fig:effmass_Bs0} but of the $B_{s1}$ meson with the measured expected threshold ($B^*K$) plotted.}\label{fig:effmass_Bs1}
\end{figure}

\bibliography{NN}

\begin{thebibliography}{142}%
\makeatletter
\providecommand \@ifxundefined [1]{%
 \@ifx{#1\undefined}
}%
\providecommand \@ifnum [1]{%
 \ifnum #1\expandafter \@firstoftwo
 \else \expandafter \@secondoftwo
 \fi
}%
\providecommand \@ifx [1]{%
 \ifx #1\expandafter \@firstoftwo
 \else \expandafter \@secondoftwo
 \fi
}%
\providecommand \natexlab [1]{#1}%
\providecommand \enquote  [1]{``#1''}%
\providecommand \bibnamefont  [1]{#1}%
\providecommand \bibfnamefont [1]{#1}%
\providecommand \citenamefont [1]{#1}%
\providecommand \href@noop [0]{\@secondoftwo}%
\providecommand \href [0]{\begingroup \@sanitize@url \@href}%
\providecommand \@href[1]{\@@startlink{#1}\@@href}%
\providecommand \@@href[1]{\endgroup#1\@@endlink}%
\providecommand \@sanitize@url [0]{\catcode `\\12\catcode `\$12\catcode
  `\&12\catcode `\#12\catcode `\^12\catcode `\_12\catcode `\%12\relax}%
\providecommand \@@startlink[1]{}%
\providecommand \@@endlink[0]{}%
\providecommand \url  [0]{\begingroup\@sanitize@url \@url }%
\providecommand \@url [1]{\endgroup\@href {#1}{\urlprefix }}%
\providecommand \urlprefix  [0]{URL }%
\providecommand \Eprint [0]{\href }%
\providecommand \doibase [0]{https://doi.org/}%
\providecommand \selectlanguage [0]{\@gobble}%
\providecommand \bibinfo  [0]{\@secondoftwo}%
\providecommand \bibfield  [0]{\@secondoftwo}%
\providecommand \translation [1]{[#1]}%
\providecommand \BibitemOpen [0]{}%
\providecommand \bibitemStop [0]{}%
\providecommand \bibitemNoStop [0]{.\EOS\space}%
\providecommand \EOS [0]{\spacefactor3000\relax}%
\providecommand \BibitemShut  [1]{\csname bibitem#1\endcsname}%
\let\auto@bib@innerbib\@empty
\bibitem [{\citenamefont {Bicudo}\ \emph {et~al.}(2015)\citenamefont {Bicudo},
  \citenamefont {Cichy}, \citenamefont {Peters}, \citenamefont {Wagenbach},\
  and\ \citenamefont {Wagner}}]{Bicudo:2015vta}%
  \BibitemOpen
  \bibfield  {author} {\bibinfo {author} {\bibfnamefont {P.}~\bibnamefont
  {Bicudo}}, \bibinfo {author} {\bibfnamefont {K.}~\bibnamefont {Cichy}},
  \bibinfo {author} {\bibfnamefont {A.}~\bibnamefont {Peters}}, \bibinfo
  {author} {\bibfnamefont {B.}~\bibnamefont {Wagenbach}},\ and\ \bibinfo
  {author} {\bibfnamefont {M.}~\bibnamefont {Wagner}},\ }\bibfield  {title}
  {\bibinfo {title} {{Evidence for the existence of $u d \bar{b} \bar{b}$ and
  the non-existence of $s s \bar{b} \bar{b}$ and $c c \bar{b} \bar{b}$
  tetraquarks from lattice QCD}},\ }\href
  {https://doi.org/10.1103/PhysRevD.92.014507} {\bibfield  {journal} {\bibinfo
  {journal} {Phys. Rev. D}\ }\textbf {\bibinfo {volume} {92}},\ \bibinfo
  {pages} {014507} (\bibinfo {year} {2015})},\ \Eprint
  {https://arxiv.org/abs/1505.00613} {arXiv:1505.00613 [hep-lat]} \BibitemShut
  {NoStop}%
\bibitem [{\citenamefont {Francis}\ \emph {et~al.}(2017)\citenamefont
  {Francis}, \citenamefont {Hudspith}, \citenamefont {Lewis},\ and\
  \citenamefont {Maltman}}]{Francis:2016hui}%
  \BibitemOpen
  \bibfield  {author} {\bibinfo {author} {\bibfnamefont {A.}~\bibnamefont
  {Francis}}, \bibinfo {author} {\bibfnamefont {R.~J.}\ \bibnamefont
  {Hudspith}}, \bibinfo {author} {\bibfnamefont {R.}~\bibnamefont {Lewis}},\
  and\ \bibinfo {author} {\bibfnamefont {K.}~\bibnamefont {Maltman}},\
  }\bibfield  {title} {\bibinfo {title} {{Lattice Prediction for Deeply Bound
  Doubly Heavy Tetraquarks}},\ }\href
  {https://doi.org/10.1103/PhysRevLett.118.142001} {\bibfield  {journal}
  {\bibinfo  {journal} {Phys. Rev. Lett.}\ }\textbf {\bibinfo {volume} {118}},\
  \bibinfo {pages} {142001} (\bibinfo {year} {2017})},\ \Eprint
  {https://arxiv.org/abs/1607.05214} {arXiv:1607.05214 [hep-lat]} \BibitemShut
  {NoStop}%
\bibitem [{\citenamefont {Richards}\ \emph {et~al.}(1990)\citenamefont
  {Richards}, \citenamefont {Sinclair},\ and\ \citenamefont
  {Sivers}}]{Richards:1990xf}%
  \BibitemOpen
  \bibfield  {author} {\bibinfo {author} {\bibfnamefont {D.~G.}\ \bibnamefont
  {Richards}}, \bibinfo {author} {\bibfnamefont {D.~K.}\ \bibnamefont
  {Sinclair}},\ and\ \bibinfo {author} {\bibfnamefont {D.~W.}\ \bibnamefont
  {Sivers}},\ }\bibfield  {title} {\bibinfo {title} {{Lattice QCD simulation of
  meson exchange forces}},\ }\href {https://doi.org/10.1103/PhysRevD.42.3191}
  {\bibfield  {journal} {\bibinfo  {journal} {Phys. Rev. D}\ }\textbf {\bibinfo
  {volume} {42}},\ \bibinfo {pages} {3191} (\bibinfo {year}
  {1990})}\BibitemShut {NoStop}%
\bibitem [{\citenamefont {Mihaly}\ \emph {et~al.}(1997)\citenamefont {Mihaly},
  \citenamefont {Fiebig}, \citenamefont {Markum},\ and\ \citenamefont
  {Rabitsch}}]{Mihaly:1996ue}%
  \BibitemOpen
  \bibfield  {author} {\bibinfo {author} {\bibfnamefont {A.}~\bibnamefont
  {Mihaly}}, \bibinfo {author} {\bibfnamefont {H.~R.}\ \bibnamefont {Fiebig}},
  \bibinfo {author} {\bibfnamefont {H.}~\bibnamefont {Markum}},\ and\ \bibinfo
  {author} {\bibfnamefont {K.}~\bibnamefont {Rabitsch}},\ }\bibfield  {title}
  {\bibinfo {title} {{Interactions between heavy - light mesons in lattice
  QCD}},\ }\href {https://doi.org/10.1103/PhysRevD.55.3077} {\bibfield
  {journal} {\bibinfo  {journal} {Phys. Rev. D}\ }\textbf {\bibinfo {volume}
  {55}},\ \bibinfo {pages} {3077} (\bibinfo {year} {1997})}\BibitemShut
  {NoStop}%
\bibitem [{\citenamefont {Green}\ and\ \citenamefont
  {Pennanen}(1998)}]{Green:1998nt}%
  \BibitemOpen
  \bibfield  {author} {\bibinfo {author} {\bibfnamefont {A.~M.}\ \bibnamefont
  {Green}}\ and\ \bibinfo {author} {\bibfnamefont {P.}~\bibnamefont
  {Pennanen}},\ }\bibfield  {title} {\bibinfo {title} {{An Interquark potential
  model for multiquark systems}},\ }\href
  {https://doi.org/10.1103/PhysRevC.57.3384} {\bibfield  {journal} {\bibinfo
  {journal} {Phys. Rev. C}\ }\textbf {\bibinfo {volume} {57}},\ \bibinfo
  {pages} {3384} (\bibinfo {year} {1998})},\ \Eprint
  {https://arxiv.org/abs/hep-lat/9804003} {arXiv:hep-lat/9804003} \BibitemShut
  {NoStop}%
\bibitem [{\citenamefont {Stewart}\ and\ \citenamefont
  {Koniuk}(1998)}]{Stewart:1998hk}%
  \BibitemOpen
  \bibfield  {author} {\bibinfo {author} {\bibfnamefont {C.}~\bibnamefont
  {Stewart}}\ and\ \bibinfo {author} {\bibfnamefont {R.}~\bibnamefont
  {Koniuk}},\ }\bibfield  {title} {\bibinfo {title} {{Hadronic molecules in
  lattice QCD}},\ }\href {https://doi.org/10.1103/PhysRevD.57.5581} {\bibfield
  {journal} {\bibinfo  {journal} {Phys. Rev. D}\ }\textbf {\bibinfo {volume}
  {57}},\ \bibinfo {pages} {5581} (\bibinfo {year} {1998})},\ \Eprint
  {https://arxiv.org/abs/hep-lat/9803003} {arXiv:hep-lat/9803003} \BibitemShut
  {NoStop}%
\bibitem [{\citenamefont {Pennanen}\ \emph {et~al.}(2000)\citenamefont
  {Pennanen}, \citenamefont {Michael},\ and\ \citenamefont
  {Green}}]{Pennanen:1999xi}%
  \BibitemOpen
  \bibfield  {author} {\bibinfo {author} {\bibfnamefont {P.}~\bibnamefont
  {Pennanen}}, \bibinfo {author} {\bibfnamefont {C.}~\bibnamefont {Michael}},\
  and\ \bibinfo {author} {\bibfnamefont {A.~M.}\ \bibnamefont {Green}}
  (\bibinfo {collaboration} {UKQCD}),\ }\bibfield  {title} {\bibinfo {title}
  {{Interactions of heavy light mesons}},\ }\href
  {https://doi.org/10.1016/S0920-5632(00)91622-0} {\bibfield  {journal}
  {\bibinfo  {journal} {Nucl. Phys. B Proc. Suppl.}\ }\textbf {\bibinfo
  {volume} {83}},\ \bibinfo {pages} {200} (\bibinfo {year} {2000})},\ \Eprint
  {https://arxiv.org/abs/hep-lat/9908032} {arXiv:hep-lat/9908032} \BibitemShut
  {NoStop}%
\bibitem [{\citenamefont {Detmold}\ \emph {et~al.}(2007)\citenamefont
  {Detmold}, \citenamefont {Orginos},\ and\ \citenamefont
  {Savage}}]{Detmold:2007wk}%
  \BibitemOpen
  \bibfield  {author} {\bibinfo {author} {\bibfnamefont {W.}~\bibnamefont
  {Detmold}}, \bibinfo {author} {\bibfnamefont {K.}~\bibnamefont {Orginos}},\
  and\ \bibinfo {author} {\bibfnamefont {M.~J.}\ \bibnamefont {Savage}},\
  }\bibfield  {title} {\bibinfo {title} {{BB Potentials in Quenched Lattice
  QCD}},\ }\href {https://doi.org/10.1103/PhysRevD.76.114503} {\bibfield
  {journal} {\bibinfo  {journal} {Phys. Rev. D}\ }\textbf {\bibinfo {volume}
  {76}},\ \bibinfo {pages} {114503} (\bibinfo {year} {2007})},\ \Eprint
  {https://arxiv.org/abs/hep-lat/0703009} {arXiv:hep-lat/0703009} \BibitemShut
  {NoStop}%
\bibitem [{\citenamefont {Brown}\ and\ \citenamefont
  {Orginos}(2012)}]{Brown:2012tm}%
  \BibitemOpen
  \bibfield  {author} {\bibinfo {author} {\bibfnamefont {Z.~S.}\ \bibnamefont
  {Brown}}\ and\ \bibinfo {author} {\bibfnamefont {K.}~\bibnamefont
  {Orginos}},\ }\bibfield  {title} {\bibinfo {title} {{Tetraquark bound states
  in the heavy-light heavy-light system}},\ }\href
  {https://doi.org/10.1103/PhysRevD.86.114506} {\bibfield  {journal} {\bibinfo
  {journal} {Phys. Rev. D}\ }\textbf {\bibinfo {volume} {86}},\ \bibinfo
  {pages} {114506} (\bibinfo {year} {2012})},\ \Eprint
  {https://arxiv.org/abs/1210.1953} {arXiv:1210.1953 [hep-lat]} \BibitemShut
  {NoStop}%
\bibitem [{\citenamefont {Wagner}(2011)}]{Wagner:2011ev}%
  \BibitemOpen
  \bibfield  {author} {\bibinfo {author} {\bibfnamefont {M.}~\bibnamefont
  {Wagner}} (\bibinfo {collaboration} {ETM}),\ }\bibfield  {title} {\bibinfo
  {title} {{Static-static-light-light tetraquarks in lattice QCD}},\ }\href
  {https://doi.org/10.5506/APhysPolBSupp.4.747} {\bibfield  {journal} {\bibinfo
   {journal} {Acta Phys. Polon. Supp.}\ }\textbf {\bibinfo {volume} {4}},\
  \bibinfo {pages} {747} (\bibinfo {year} {2011})},\ \Eprint
  {https://arxiv.org/abs/1103.5147} {arXiv:1103.5147 [hep-lat]} \BibitemShut
  {NoStop}%
\bibitem [{\citenamefont {Bicudo}\ and\ \citenamefont
  {Wagner}(2013)}]{Bicudo:2012qt}%
  \BibitemOpen
  \bibfield  {author} {\bibinfo {author} {\bibfnamefont {P.}~\bibnamefont
  {Bicudo}}\ and\ \bibinfo {author} {\bibfnamefont {M.}~\bibnamefont {Wagner}}
  (\bibinfo {collaboration} {European Twisted Mass}),\ }\bibfield  {title}
  {\bibinfo {title} {{Lattice QCD signal for a bottom-bottom tetraquark}},\
  }\href {https://doi.org/10.1103/PhysRevD.87.114511} {\bibfield  {journal}
  {\bibinfo  {journal} {Phys. Rev. D}\ }\textbf {\bibinfo {volume} {87}},\
  \bibinfo {pages} {114511} (\bibinfo {year} {2013})},\ \Eprint
  {https://arxiv.org/abs/1209.6274} {arXiv:1209.6274 [hep-ph]} \BibitemShut
  {NoStop}%
\bibitem [{\citenamefont {Bicudo}\ \emph {et~al.}(2016)\citenamefont {Bicudo},
  \citenamefont {Cichy}, \citenamefont {Peters},\ and\ \citenamefont
  {Wagner}}]{Bicudo:2015kna}%
  \BibitemOpen
  \bibfield  {author} {\bibinfo {author} {\bibfnamefont {P.}~\bibnamefont
  {Bicudo}}, \bibinfo {author} {\bibfnamefont {K.}~\bibnamefont {Cichy}},
  \bibinfo {author} {\bibfnamefont {A.}~\bibnamefont {Peters}},\ and\ \bibinfo
  {author} {\bibfnamefont {M.}~\bibnamefont {Wagner}},\ }\bibfield  {title}
  {\bibinfo {title} {{BB interactions with static bottom quarks from Lattice
  QCD}},\ }\href {https://doi.org/10.1103/PhysRevD.93.034501} {\bibfield
  {journal} {\bibinfo  {journal} {Phys. Rev. D}\ }\textbf {\bibinfo {volume}
  {93}},\ \bibinfo {pages} {034501} (\bibinfo {year} {2016})},\ \Eprint
  {https://arxiv.org/abs/1510.03441} {arXiv:1510.03441 [hep-lat]} \BibitemShut
  {NoStop}%
\bibitem [{\citenamefont {Bicudo}\ \emph {et~al.}(2017)\citenamefont {Bicudo},
  \citenamefont {Scheunert},\ and\ \citenamefont {Wagner}}]{Bicudo:2016ooe}%
  \BibitemOpen
  \bibfield  {author} {\bibinfo {author} {\bibfnamefont {P.}~\bibnamefont
  {Bicudo}}, \bibinfo {author} {\bibfnamefont {J.}~\bibnamefont {Scheunert}},\
  and\ \bibinfo {author} {\bibfnamefont {M.}~\bibnamefont {Wagner}},\
  }\bibfield  {title} {\bibinfo {title} {{Including heavy spin effects in the
  prediction of a $\bar{b} \bar{b} u d$ tetraquark with lattice QCD
  potentials}},\ }\href {https://doi.org/10.1103/PhysRevD.95.034502} {\bibfield
   {journal} {\bibinfo  {journal} {Phys. Rev. D}\ }\textbf {\bibinfo {volume}
  {95}},\ \bibinfo {pages} {034502} (\bibinfo {year} {2017})},\ \Eprint
  {https://arxiv.org/abs/1612.02758} {arXiv:1612.02758 [hep-lat]} \BibitemShut
  {NoStop}%
\bibitem [{\citenamefont {Junnarkar}\ \emph {et~al.}(2019)\citenamefont
  {Junnarkar}, \citenamefont {Mathur},\ and\ \citenamefont
  {Padmanath}}]{Junnarkar:2018twb}%
  \BibitemOpen
  \bibfield  {author} {\bibinfo {author} {\bibfnamefont {P.}~\bibnamefont
  {Junnarkar}}, \bibinfo {author} {\bibfnamefont {N.}~\bibnamefont {Mathur}},\
  and\ \bibinfo {author} {\bibfnamefont {M.}~\bibnamefont {Padmanath}},\
  }\bibfield  {title} {\bibinfo {title} {{Study of doubly heavy tetraquarks in
  Lattice QCD}},\ }\href {https://doi.org/10.1103/PhysRevD.99.034507}
  {\bibfield  {journal} {\bibinfo  {journal} {Phys. Rev. D}\ }\textbf {\bibinfo
  {volume} {99}},\ \bibinfo {pages} {034507} (\bibinfo {year} {2019})},\
  \Eprint {https://arxiv.org/abs/1810.12285} {arXiv:1810.12285 [hep-lat]}
  \BibitemShut {NoStop}%
\bibitem [{\citenamefont {Leskovec}\ \emph {et~al.}(2019)\citenamefont
  {Leskovec}, \citenamefont {Meinel}, \citenamefont {Pflaumer},\ and\
  \citenamefont {Wagner}}]{Leskovec:2019ioa}%
  \BibitemOpen
  \bibfield  {author} {\bibinfo {author} {\bibfnamefont {L.}~\bibnamefont
  {Leskovec}}, \bibinfo {author} {\bibfnamefont {S.}~\bibnamefont {Meinel}},
  \bibinfo {author} {\bibfnamefont {M.}~\bibnamefont {Pflaumer}},\ and\
  \bibinfo {author} {\bibfnamefont {M.}~\bibnamefont {Wagner}},\ }\bibfield
  {title} {\bibinfo {title} {{Lattice QCD investigation of a doubly-bottom
  $\bar{b} \bar{b} u d$ tetraquark with quantum numbers $I(J^P) = 0(1^+)$}},\
  }\href {https://doi.org/10.1103/PhysRevD.100.014503} {\bibfield  {journal}
  {\bibinfo  {journal} {Phys. Rev. D}\ }\textbf {\bibinfo {volume} {100}},\
  \bibinfo {pages} {014503} (\bibinfo {year} {2019})},\ \Eprint
  {https://arxiv.org/abs/1904.04197} {arXiv:1904.04197 [hep-lat]} \BibitemShut
  {NoStop}%
\bibitem [{\citenamefont {Mohanta}\ and\ \citenamefont
  {Basak}(2020)}]{Mohanta:2020eed}%
  \BibitemOpen
  \bibfield  {author} {\bibinfo {author} {\bibfnamefont {P.}~\bibnamefont
  {Mohanta}}\ and\ \bibinfo {author} {\bibfnamefont {S.}~\bibnamefont
  {Basak}},\ }\bibfield  {title} {\bibinfo {title} {{Construction of
  $bb\bar{u}\bar{d}$ tetraquark states on lattice with NRQCD bottom and HISQ up
  and down quarks}},\ }\href {https://doi.org/10.1103/PhysRevD.102.094516}
  {\bibfield  {journal} {\bibinfo  {journal} {Phys. Rev. D}\ }\textbf {\bibinfo
  {volume} {102}},\ \bibinfo {pages} {094516} (\bibinfo {year} {2020})},\
  \Eprint {https://arxiv.org/abs/2008.11146} {arXiv:2008.11146 [hep-lat]}
  \BibitemShut {NoStop}%
\bibitem [{\citenamefont {Zouzou}\ \emph {et~al.}(1986)\citenamefont {Zouzou},
  \citenamefont {Silvestre-Brac}, \citenamefont {Gignoux},\ and\ \citenamefont
  {Richard}}]{Zouzou:1986qh}%
  \BibitemOpen
  \bibfield  {author} {\bibinfo {author} {\bibfnamefont {S.}~\bibnamefont
  {Zouzou}}, \bibinfo {author} {\bibfnamefont {B.}~\bibnamefont
  {Silvestre-Brac}}, \bibinfo {author} {\bibfnamefont {C.}~\bibnamefont
  {Gignoux}},\ and\ \bibinfo {author} {\bibfnamefont {J.~M.}\ \bibnamefont
  {Richard}},\ }\bibfield  {title} {\bibinfo {title} {{FOUR QUARK BOUND
  STATES}},\ }\href {https://doi.org/10.1007/BF01557611} {\bibfield  {journal}
  {\bibinfo  {journal} {Z. Phys. C}\ }\textbf {\bibinfo {volume} {30}},\
  \bibinfo {pages} {457} (\bibinfo {year} {1986})}\BibitemShut {NoStop}%
\bibitem [{\citenamefont {Carlson}\ \emph {et~al.}(1988)\citenamefont
  {Carlson}, \citenamefont {Heller},\ and\ \citenamefont
  {Tjon}}]{Carlson:1987hh}%
  \BibitemOpen
  \bibfield  {author} {\bibinfo {author} {\bibfnamefont {J.}~\bibnamefont
  {Carlson}}, \bibinfo {author} {\bibfnamefont {L.}~\bibnamefont {Heller}},\
  and\ \bibinfo {author} {\bibfnamefont {J.~A.}\ \bibnamefont {Tjon}},\
  }\bibfield  {title} {\bibinfo {title} {{Stability of Dimesons}},\ }\href
  {https://doi.org/10.1103/PhysRevD.37.744} {\bibfield  {journal} {\bibinfo
  {journal} {Phys. Rev. D}\ }\textbf {\bibinfo {volume} {37}},\ \bibinfo
  {pages} {744} (\bibinfo {year} {1988})}\BibitemShut {NoStop}%
\bibitem [{\citenamefont {Semay}\ and\ \citenamefont
  {Silvestre-Brac}(1994)}]{Semay:1994ht}%
  \BibitemOpen
  \bibfield  {author} {\bibinfo {author} {\bibfnamefont {C.}~\bibnamefont
  {Semay}}\ and\ \bibinfo {author} {\bibfnamefont {B.}~\bibnamefont
  {Silvestre-Brac}},\ }\bibfield  {title} {\bibinfo {title} {{Diquonia and
  potential models}},\ }\href {https://doi.org/10.1007/BF01413104} {\bibfield
  {journal} {\bibinfo  {journal} {Z. Phys. C}\ }\textbf {\bibinfo {volume}
  {61}},\ \bibinfo {pages} {271} (\bibinfo {year} {1994})}\BibitemShut
  {NoStop}%
\bibitem [{\citenamefont {Pepin}\ \emph {et~al.}(1997)\citenamefont {Pepin},
  \citenamefont {Stancu}, \citenamefont {Genovese},\ and\ \citenamefont
  {Richard}}]{Pepin:1996id}%
  \BibitemOpen
  \bibfield  {author} {\bibinfo {author} {\bibfnamefont {S.}~\bibnamefont
  {Pepin}}, \bibinfo {author} {\bibfnamefont {F.}~\bibnamefont {Stancu}},
  \bibinfo {author} {\bibfnamefont {M.}~\bibnamefont {Genovese}},\ and\
  \bibinfo {author} {\bibfnamefont {J.~M.}\ \bibnamefont {Richard}},\
  }\bibfield  {title} {\bibinfo {title} {{Tetraquarks with color blind forces
  in chiral quark models}},\ }\href
  {https://doi.org/10.1016/S0370-2693(96)01597-3} {\bibfield  {journal}
  {\bibinfo  {journal} {Phys. Lett. B}\ }\textbf {\bibinfo {volume} {393}},\
  \bibinfo {pages} {119} (\bibinfo {year} {1997})},\ \Eprint
  {https://arxiv.org/abs/hep-ph/9609348} {arXiv:hep-ph/9609348} \BibitemShut
  {NoStop}%
\bibitem [{\citenamefont {Brink}\ and\ \citenamefont
  {Stancu}(1998)}]{Brink:1998as}%
  \BibitemOpen
  \bibfield  {author} {\bibinfo {author} {\bibfnamefont {D.~M.}\ \bibnamefont
  {Brink}}\ and\ \bibinfo {author} {\bibfnamefont {F.}~\bibnamefont {Stancu}},\
  }\bibfield  {title} {\bibinfo {title} {{Tetraquarks with heavy flavors}},\
  }\href {https://doi.org/10.1103/PhysRevD.57.6778} {\bibfield  {journal}
  {\bibinfo  {journal} {Phys. Rev. D}\ }\textbf {\bibinfo {volume} {57}},\
  \bibinfo {pages} {6778} (\bibinfo {year} {1998})}\BibitemShut {NoStop}%
\bibitem [{\citenamefont {Vijande}\ \emph {et~al.}(2004)\citenamefont
  {Vijande}, \citenamefont {Fernandez}, \citenamefont {Valcarce},\ and\
  \citenamefont {Silvestre-Brac}}]{Vijande:2003ki}%
  \BibitemOpen
  \bibfield  {author} {\bibinfo {author} {\bibfnamefont {J.}~\bibnamefont
  {Vijande}}, \bibinfo {author} {\bibfnamefont {F.}~\bibnamefont {Fernandez}},
  \bibinfo {author} {\bibfnamefont {A.}~\bibnamefont {Valcarce}},\ and\
  \bibinfo {author} {\bibfnamefont {B.}~\bibnamefont {Silvestre-Brac}},\
  }\bibfield  {title} {\bibinfo {title} {{Tetraquarks in a chiral constituent
  quark model}},\ }\href {https://doi.org/10.1140/epja/i2003-10128-9}
  {\bibfield  {journal} {\bibinfo  {journal} {Eur. Phys. J. A}\ }\textbf
  {\bibinfo {volume} {19}},\ \bibinfo {pages} {383} (\bibinfo {year} {2004})},\
  \Eprint {https://arxiv.org/abs/hep-ph/0310007} {arXiv:hep-ph/0310007}
  \BibitemShut {NoStop}%
\bibitem [{\citenamefont {Ebert}\ \emph {et~al.}(2007)\citenamefont {Ebert},
  \citenamefont {Faustov}, \citenamefont {Galkin},\ and\ \citenamefont
  {Lucha}}]{Ebert:2007rn}%
  \BibitemOpen
  \bibfield  {author} {\bibinfo {author} {\bibfnamefont {D.}~\bibnamefont
  {Ebert}}, \bibinfo {author} {\bibfnamefont {R.~N.}\ \bibnamefont {Faustov}},
  \bibinfo {author} {\bibfnamefont {V.~O.}\ \bibnamefont {Galkin}},\ and\
  \bibinfo {author} {\bibfnamefont {W.}~\bibnamefont {Lucha}},\ }\bibfield
  {title} {\bibinfo {title} {{Masses of tetraquarks with two heavy quarks in
  the relativistic quark model}},\ }\href
  {https://doi.org/10.1103/PhysRevD.76.114015} {\bibfield  {journal} {\bibinfo
  {journal} {Phys. Rev. D}\ }\textbf {\bibinfo {volume} {76}},\ \bibinfo
  {pages} {114015} (\bibinfo {year} {2007})},\ \Eprint
  {https://arxiv.org/abs/0706.3853} {arXiv:0706.3853 [hep-ph]} \BibitemShut
  {NoStop}%
\bibitem [{\citenamefont {Vijande}\ \emph {et~al.}(2007)\citenamefont
  {Vijande}, \citenamefont {Weissman}, \citenamefont {Valcarce},\ and\
  \citenamefont {Barnea}}]{Vijande:2007rf}%
  \BibitemOpen
  \bibfield  {author} {\bibinfo {author} {\bibfnamefont {J.}~\bibnamefont
  {Vijande}}, \bibinfo {author} {\bibfnamefont {E.}~\bibnamefont {Weissman}},
  \bibinfo {author} {\bibfnamefont {A.}~\bibnamefont {Valcarce}},\ and\
  \bibinfo {author} {\bibfnamefont {N.}~\bibnamefont {Barnea}},\ }\bibfield
  {title} {\bibinfo {title} {{Are there compact heavy four-quark bound
  states?}},\ }\href {https://doi.org/10.1103/PhysRevD.76.094027} {\bibfield
  {journal} {\bibinfo  {journal} {Phys. Rev. D}\ }\textbf {\bibinfo {volume}
  {76}},\ \bibinfo {pages} {094027} (\bibinfo {year} {2007})},\ \Eprint
  {https://arxiv.org/abs/0710.2516} {arXiv:0710.2516 [hep-ph]} \BibitemShut
  {NoStop}%
\bibitem [{\citenamefont {Zhang}\ \emph {et~al.}(2008)\citenamefont {Zhang},
  \citenamefont {Zhang},\ and\ \citenamefont {Zhang}}]{Zhang:2007mu}%
  \BibitemOpen
  \bibfield  {author} {\bibinfo {author} {\bibfnamefont {M.}~\bibnamefont
  {Zhang}}, \bibinfo {author} {\bibfnamefont {H.~X.}\ \bibnamefont {Zhang}},\
  and\ \bibinfo {author} {\bibfnamefont {Z.~Y.}\ \bibnamefont {Zhang}},\
  }\bibfield  {title} {\bibinfo {title} {{QQ anti-q anti-q four-quark bound
  states in chiral SU(3) quark model}},\ }\href
  {https://doi.org/10.1088/0253-6102/50/2/31} {\bibfield  {journal} {\bibinfo
  {journal} {Commun. Theor. Phys.}\ }\textbf {\bibinfo {volume} {50}},\
  \bibinfo {pages} {437} (\bibinfo {year} {2008})},\ \Eprint
  {https://arxiv.org/abs/0711.1029} {arXiv:0711.1029 [nucl-th]} \BibitemShut
  {NoStop}%
\bibitem [{\citenamefont {Lee}\ and\ \citenamefont {Yasui}(2009)}]{Lee:2009rt}%
  \BibitemOpen
  \bibfield  {author} {\bibinfo {author} {\bibfnamefont {S.~H.}\ \bibnamefont
  {Lee}}\ and\ \bibinfo {author} {\bibfnamefont {S.}~\bibnamefont {Yasui}},\
  }\bibfield  {title} {\bibinfo {title} {{Stable multiquark states with heavy
  quarks in a diquark model}},\ }\href
  {https://doi.org/10.1140/epjc/s10052-009-1140-x} {\bibfield  {journal}
  {\bibinfo  {journal} {Eur. Phys. J. C}\ }\textbf {\bibinfo {volume} {64}},\
  \bibinfo {pages} {283} (\bibinfo {year} {2009})},\ \Eprint
  {https://arxiv.org/abs/0901.2977} {arXiv:0901.2977 [hep-ph]} \BibitemShut
  {NoStop}%
\bibitem [{\citenamefont {Yang}\ \emph {et~al.}(2009)\citenamefont {Yang},
  \citenamefont {Deng}, \citenamefont {Ping},\ and\ \citenamefont
  {Goldman}}]{Yang:2009zzp}%
  \BibitemOpen
  \bibfield  {author} {\bibinfo {author} {\bibfnamefont {Y.}~\bibnamefont
  {Yang}}, \bibinfo {author} {\bibfnamefont {C.}~\bibnamefont {Deng}}, \bibinfo
  {author} {\bibfnamefont {J.}~\bibnamefont {Ping}},\ and\ \bibinfo {author}
  {\bibfnamefont {T.}~\bibnamefont {Goldman}},\ }\bibfield  {title} {\bibinfo
  {title} {{S-wave Q Q anti-q anti-q state in the constituent quark model}},\
  }\href {https://doi.org/10.1103/PhysRevD.80.114023} {\bibfield  {journal}
  {\bibinfo  {journal} {Phys. Rev. D}\ }\textbf {\bibinfo {volume} {80}},\
  \bibinfo {pages} {114023} (\bibinfo {year} {2009})}\BibitemShut {NoStop}%
\bibitem [{\citenamefont {Wang}(2018)}]{Wang:2017uld}%
  \BibitemOpen
  \bibfield  {author} {\bibinfo {author} {\bibfnamefont {Z.-G.}\ \bibnamefont
  {Wang}},\ }\bibfield  {title} {\bibinfo {title} {{Analysis of the axialvector
  doubly heavy tetraquark states with QCD sum rules}},\ }\href
  {https://doi.org/10.5506/APhysPolB.49.1781} {\bibfield  {journal} {\bibinfo
  {journal} {Acta Phys. Polon. B}\ }\textbf {\bibinfo {volume} {49}},\ \bibinfo
  {pages} {1781} (\bibinfo {year} {2018})},\ \Eprint
  {https://arxiv.org/abs/1708.04545} {arXiv:1708.04545 [hep-ph]} \BibitemShut
  {NoStop}%
\bibitem [{\citenamefont {Karliner}\ and\ \citenamefont
  {Rosner}(2017)}]{Karliner:2017qjm}%
  \BibitemOpen
  \bibfield  {author} {\bibinfo {author} {\bibfnamefont {M.}~\bibnamefont
  {Karliner}}\ and\ \bibinfo {author} {\bibfnamefont {J.~L.}\ \bibnamefont
  {Rosner}},\ }\bibfield  {title} {\bibinfo {title} {{Discovery of
  doubly-charmed $\Xi_{cc}$ baryon implies a stable ($b b \bar{u} \bar{d}$)
  tetraquark}},\ }\href {https://doi.org/10.1103/PhysRevLett.119.202001}
  {\bibfield  {journal} {\bibinfo  {journal} {Phys. Rev. Lett.}\ }\textbf
  {\bibinfo {volume} {119}},\ \bibinfo {pages} {202001} (\bibinfo {year}
  {2017})},\ \Eprint {https://arxiv.org/abs/1707.07666} {arXiv:1707.07666
  [hep-ph]} \BibitemShut {NoStop}%
\bibitem [{\citenamefont {Eichten}\ and\ \citenamefont
  {Quigg}(2017)}]{Eichten:2017ffp}%
  \BibitemOpen
  \bibfield  {author} {\bibinfo {author} {\bibfnamefont {E.~J.}\ \bibnamefont
  {Eichten}}\ and\ \bibinfo {author} {\bibfnamefont {C.}~\bibnamefont
  {Quigg}},\ }\bibfield  {title} {\bibinfo {title} {{Heavy-quark symmetry
  implies stable heavy tetraquark mesons $Q_iQ_j \bar q_k \bar q_l$}},\ }\href
  {https://doi.org/10.1103/PhysRevLett.119.202002} {\bibfield  {journal}
  {\bibinfo  {journal} {Phys. Rev. Lett.}\ }\textbf {\bibinfo {volume} {119}},\
  \bibinfo {pages} {202002} (\bibinfo {year} {2017})},\ \Eprint
  {https://arxiv.org/abs/1707.09575} {arXiv:1707.09575 [hep-ph]} \BibitemShut
  {NoStop}%
\bibitem [{\citenamefont {Czarnecki}\ \emph {et~al.}(2018)\citenamefont
  {Czarnecki}, \citenamefont {Leng},\ and\ \citenamefont
  {Voloshin}}]{Czarnecki:2017vco}%
  \BibitemOpen
  \bibfield  {author} {\bibinfo {author} {\bibfnamefont {A.}~\bibnamefont
  {Czarnecki}}, \bibinfo {author} {\bibfnamefont {B.}~\bibnamefont {Leng}},\
  and\ \bibinfo {author} {\bibfnamefont {M.~B.}\ \bibnamefont {Voloshin}},\
  }\bibfield  {title} {\bibinfo {title} {{Stability of tetrons}},\ }\href
  {https://doi.org/10.1016/j.physletb.2018.01.034} {\bibfield  {journal}
  {\bibinfo  {journal} {Phys. Lett. B}\ }\textbf {\bibinfo {volume} {778}},\
  \bibinfo {pages} {233} (\bibinfo {year} {2018})},\ \Eprint
  {https://arxiv.org/abs/1708.04594} {arXiv:1708.04594 [hep-ph]} \BibitemShut
  {NoStop}%
\bibitem [{\citenamefont {Mehen}(2017)}]{Mehen:2017nrh}%
  \BibitemOpen
  \bibfield  {author} {\bibinfo {author} {\bibfnamefont {T.}~\bibnamefont
  {Mehen}},\ }\bibfield  {title} {\bibinfo {title} {{Implications of Heavy
  Quark-Diquark Symmetry for Excited Doubly Heavy Baryons and Tetraquarks}},\
  }\href {https://doi.org/10.1103/PhysRevD.96.094028} {\bibfield  {journal}
  {\bibinfo  {journal} {Phys. Rev. D}\ }\textbf {\bibinfo {volume} {96}},\
  \bibinfo {pages} {094028} (\bibinfo {year} {2017})},\ \Eprint
  {https://arxiv.org/abs/1708.05020} {arXiv:1708.05020 [hep-ph]} \BibitemShut
  {NoStop}%
\bibitem [{\citenamefont {Maiani}\ \emph {et~al.}(2019)\citenamefont {Maiani},
  \citenamefont {Polosa},\ and\ \citenamefont {Riquer}}]{Maiani:2019lpu}%
  \BibitemOpen
  \bibfield  {author} {\bibinfo {author} {\bibfnamefont {L.}~\bibnamefont
  {Maiani}}, \bibinfo {author} {\bibfnamefont {A.~D.}\ \bibnamefont {Polosa}},\
  and\ \bibinfo {author} {\bibfnamefont {V.}~\bibnamefont {Riquer}},\
  }\bibfield  {title} {\bibinfo {title} {{Hydrogen bond of QCD in doubly heavy
  baryons and tetraquarks}},\ }\href
  {https://doi.org/10.1103/PhysRevD.100.074002} {\bibfield  {journal} {\bibinfo
   {journal} {Phys. Rev. D}\ }\textbf {\bibinfo {volume} {100}},\ \bibinfo
  {pages} {074002} (\bibinfo {year} {2019})},\ \Eprint
  {https://arxiv.org/abs/1908.03244} {arXiv:1908.03244 [hep-ph]} \BibitemShut
  {NoStop}%
\bibitem [{\citenamefont {Hern\'andez}\ \emph {et~al.}(2020)\citenamefont
  {Hern\'andez}, \citenamefont {Vijande}, \citenamefont {Valcarce},\ and\
  \citenamefont {Richard}}]{Hernandez:2019eox}%
  \BibitemOpen
  \bibfield  {author} {\bibinfo {author} {\bibfnamefont {E.}~\bibnamefont
  {Hern\'andez}}, \bibinfo {author} {\bibfnamefont {J.}~\bibnamefont
  {Vijande}}, \bibinfo {author} {\bibfnamefont {A.}~\bibnamefont {Valcarce}},\
  and\ \bibinfo {author} {\bibfnamefont {J.-M.}\ \bibnamefont {Richard}},\
  }\bibfield  {title} {\bibinfo {title} {{Spectroscopy, lifetime and decay
  modes of the $T^-_{bb}$ tetraquark}},\ }\href
  {https://doi.org/10.1016/j.physletb.2019.135073} {\bibfield  {journal}
  {\bibinfo  {journal} {Phys. Lett. B}\ }\textbf {\bibinfo {volume} {800}},\
  \bibinfo {pages} {135073} (\bibinfo {year} {2020})},\ \Eprint
  {https://arxiv.org/abs/1910.13394} {arXiv:1910.13394 [hep-ph]} \BibitemShut
  {NoStop}%
\bibitem [{\citenamefont {L\"u}\ \emph {et~al.}(2020)\citenamefont {L\"u},
  \citenamefont {Chen},\ and\ \citenamefont {Dong}}]{Lu:2020rog}%
  \BibitemOpen
  \bibfield  {author} {\bibinfo {author} {\bibfnamefont {Q.-F.}\ \bibnamefont
  {L\"u}}, \bibinfo {author} {\bibfnamefont {D.-Y.}\ \bibnamefont {Chen}},\
  and\ \bibinfo {author} {\bibfnamefont {Y.-B.}\ \bibnamefont {Dong}},\
  }\bibfield  {title} {\bibinfo {title} {{Masses of doubly heavy tetraquarks
  $T_{QQ^\prime}$ in a relativized quark model}},\ }\href
  {https://doi.org/10.1103/PhysRevD.102.034012} {\bibfield  {journal} {\bibinfo
   {journal} {Phys. Rev. D}\ }\textbf {\bibinfo {volume} {102}},\ \bibinfo
  {pages} {034012} (\bibinfo {year} {2020})},\ \Eprint
  {https://arxiv.org/abs/2006.08087} {arXiv:2006.08087 [hep-ph]} \BibitemShut
  {NoStop}%
\bibitem [{\citenamefont {Braaten}\ \emph {et~al.}(2021)\citenamefont
  {Braaten}, \citenamefont {He},\ and\ \citenamefont
  {Mohapatra}}]{Braaten:2020nwp}%
  \BibitemOpen
  \bibfield  {author} {\bibinfo {author} {\bibfnamefont {E.}~\bibnamefont
  {Braaten}}, \bibinfo {author} {\bibfnamefont {L.-P.}\ \bibnamefont {He}},\
  and\ \bibinfo {author} {\bibfnamefont {A.}~\bibnamefont {Mohapatra}},\
  }\bibfield  {title} {\bibinfo {title} {{Masses of doubly heavy tetraquarks
  with error bars}},\ }\href {https://doi.org/10.1103/PhysRevD.103.016001}
  {\bibfield  {journal} {\bibinfo  {journal} {Phys. Rev. D}\ }\textbf {\bibinfo
  {volume} {103}},\ \bibinfo {pages} {016001} (\bibinfo {year} {2021})},\
  \Eprint {https://arxiv.org/abs/2006.08650} {arXiv:2006.08650 [hep-ph]}
  \BibitemShut {NoStop}%
\bibitem [{\citenamefont {Cheng}\ \emph {et~al.}(2021)\citenamefont {Cheng},
  \citenamefont {Li}, \citenamefont {Liu}, \citenamefont {Si},\ and\
  \citenamefont {Yao}}]{Cheng:2020wxa}%
  \BibitemOpen
  \bibfield  {author} {\bibinfo {author} {\bibfnamefont {J.-B.}\ \bibnamefont
  {Cheng}}, \bibinfo {author} {\bibfnamefont {S.-Y.}\ \bibnamefont {Li}},
  \bibinfo {author} {\bibfnamefont {Y.-R.}\ \bibnamefont {Liu}}, \bibinfo
  {author} {\bibfnamefont {Z.-G.}\ \bibnamefont {Si}},\ and\ \bibinfo {author}
  {\bibfnamefont {T.}~\bibnamefont {Yao}},\ }\bibfield  {title} {\bibinfo
  {title} {{Double-heavy tetraquark states with heavy diquark-antiquark
  symmetry}},\ }\href {https://doi.org/10.1088/1674-1137/abde2f} {\bibfield
  {journal} {\bibinfo  {journal} {Chin. Phys. C}\ }\textbf {\bibinfo {volume}
  {45}},\ \bibinfo {pages} {043102} (\bibinfo {year} {2021})},\ \Eprint
  {https://arxiv.org/abs/2008.00737} {arXiv:2008.00737 [hep-ph]} \BibitemShut
  {NoStop}%
\bibitem [{\citenamefont {Meng}\ \emph {et~al.}(2021)\citenamefont {Meng},
  \citenamefont {Hiyama}, \citenamefont {Hosaka}, \citenamefont {Oka},
  \citenamefont {Gubler}, \citenamefont {Can}, \citenamefont {Takahashi},\ and\
  \citenamefont {Zong}}]{Meng:2020knc}%
  \BibitemOpen
  \bibfield  {author} {\bibinfo {author} {\bibfnamefont {Q.}~\bibnamefont
  {Meng}}, \bibinfo {author} {\bibfnamefont {E.}~\bibnamefont {Hiyama}},
  \bibinfo {author} {\bibfnamefont {A.}~\bibnamefont {Hosaka}}, \bibinfo
  {author} {\bibfnamefont {M.}~\bibnamefont {Oka}}, \bibinfo {author}
  {\bibfnamefont {P.}~\bibnamefont {Gubler}}, \bibinfo {author} {\bibfnamefont
  {K.~U.}\ \bibnamefont {Can}}, \bibinfo {author} {\bibfnamefont {T.~T.}\
  \bibnamefont {Takahashi}},\ and\ \bibinfo {author} {\bibfnamefont {H.~S.}\
  \bibnamefont {Zong}},\ }\bibfield  {title} {\bibinfo {title} {{Stable
  double-heavy tetraquarks: spectrum and structure}},\ }\href
  {https://doi.org/10.1016/j.physletb.2021.136095} {\bibfield  {journal}
  {\bibinfo  {journal} {Phys. Lett. B}\ }\textbf {\bibinfo {volume} {814}},\
  \bibinfo {pages} {136095} (\bibinfo {year} {2021})},\ \Eprint
  {https://arxiv.org/abs/2009.14493} {arXiv:2009.14493 [nucl-th]} \BibitemShut
  {NoStop}%
\bibitem [{\citenamefont {Weng}\ \emph {et~al.}(2022)\citenamefont {Weng},
  \citenamefont {Deng},\ and\ \citenamefont {Zhu}}]{Weng:2021hje}%
  \BibitemOpen
  \bibfield  {author} {\bibinfo {author} {\bibfnamefont {X.-Z.}\ \bibnamefont
  {Weng}}, \bibinfo {author} {\bibfnamefont {W.-Z.}\ \bibnamefont {Deng}},\
  and\ \bibinfo {author} {\bibfnamefont {S.-L.}\ \bibnamefont {Zhu}},\
  }\bibfield  {title} {\bibinfo {title} {{Doubly heavy tetraquarks in an
  extended chromomagnetic model *}},\ }\href
  {https://doi.org/10.1088/1674-1137/ac2ed0} {\bibfield  {journal} {\bibinfo
  {journal} {Chin. Phys. C}\ }\textbf {\bibinfo {volume} {46}},\ \bibinfo
  {pages} {013102} (\bibinfo {year} {2022})},\ \Eprint
  {https://arxiv.org/abs/2108.07242} {arXiv:2108.07242 [hep-ph]} \BibitemShut
  {NoStop}%
\bibitem [{\citenamefont {Faustov}\ \emph {et~al.}(2021)\citenamefont
  {Faustov}, \citenamefont {Galkin},\ and\ \citenamefont
  {Savchenko}}]{Faustov:2021hjs}%
  \BibitemOpen
  \bibfield  {author} {\bibinfo {author} {\bibfnamefont {R.~N.}\ \bibnamefont
  {Faustov}}, \bibinfo {author} {\bibfnamefont {V.~O.}\ \bibnamefont
  {Galkin}},\ and\ \bibinfo {author} {\bibfnamefont {E.~M.}\ \bibnamefont
  {Savchenko}},\ }\bibfield  {title} {\bibinfo {title} {{Heavy tetraquarks in
  the relativistic quark model}},\ }\href
  {https://doi.org/10.3390/universe7040094} {\bibfield  {journal} {\bibinfo
  {journal} {Universe}\ }\textbf {\bibinfo {volume} {7}},\ \bibinfo {pages}
  {94} (\bibinfo {year} {2021})},\ \Eprint {https://arxiv.org/abs/2103.01763}
  {arXiv:2103.01763 [hep-ph]} \BibitemShut {NoStop}%
\bibitem [{\citenamefont {Deng}\ and\ \citenamefont
  {Zhu}(2022)}]{Deng:2021gnb}%
  \BibitemOpen
  \bibfield  {author} {\bibinfo {author} {\bibfnamefont {C.}~\bibnamefont
  {Deng}}\ and\ \bibinfo {author} {\bibfnamefont {S.-L.}\ \bibnamefont {Zhu}},\
  }\bibfield  {title} {\bibinfo {title} {{Tcc+ and its partners}},\ }\href
  {https://doi.org/10.1103/PhysRevD.105.054015} {\bibfield  {journal} {\bibinfo
   {journal} {Phys. Rev. D}\ }\textbf {\bibinfo {volume} {105}},\ \bibinfo
  {pages} {054015} (\bibinfo {year} {2022})},\ \Eprint
  {https://arxiv.org/abs/2112.12472} {arXiv:2112.12472 [hep-ph]} \BibitemShut
  {NoStop}%
\bibitem [{\citenamefont {Kim}\ \emph {et~al.}(2022)\citenamefont {Kim},
  \citenamefont {Oka},\ and\ \citenamefont {Suzuki}}]{Kim:2022mpa}%
  \BibitemOpen
  \bibfield  {author} {\bibinfo {author} {\bibfnamefont {Y.}~\bibnamefont
  {Kim}}, \bibinfo {author} {\bibfnamefont {M.}~\bibnamefont {Oka}},\ and\
  \bibinfo {author} {\bibfnamefont {K.}~\bibnamefont {Suzuki}},\ }\bibfield
  {title} {\bibinfo {title} {{Doubly heavy tetraquarks in a chiral-diquark
  picture}},\ }\href {https://doi.org/10.1103/PhysRevD.105.074021} {\bibfield
  {journal} {\bibinfo  {journal} {Phys. Rev. D}\ }\textbf {\bibinfo {volume}
  {105}},\ \bibinfo {pages} {074021} (\bibinfo {year} {2022})},\ \Eprint
  {https://arxiv.org/abs/2202.06520} {arXiv:2202.06520 [hep-ph]} \BibitemShut
  {NoStop}%
\bibitem [{\citenamefont {Meinel}\ \emph {et~al.}(2022)\citenamefont {Meinel},
  \citenamefont {Pflaumer},\ and\ \citenamefont {Wagner}}]{Meinel:2022lzo}%
  \BibitemOpen
  \bibfield  {author} {\bibinfo {author} {\bibfnamefont {S.}~\bibnamefont
  {Meinel}}, \bibinfo {author} {\bibfnamefont {M.}~\bibnamefont {Pflaumer}},\
  and\ \bibinfo {author} {\bibfnamefont {M.}~\bibnamefont {Wagner}},\
  }\bibfield  {title} {\bibinfo {title} {{Search for
  b\textasciimacron{}b\textasciimacron{}us and
  b\textasciimacron{}c\textasciimacron{}ud tetraquark bound states using
  lattice QCD}},\ }\href {https://doi.org/10.1103/PhysRevD.106.034507}
  {\bibfield  {journal} {\bibinfo  {journal} {Phys. Rev. D}\ }\textbf {\bibinfo
  {volume} {106}},\ \bibinfo {pages} {034507} (\bibinfo {year} {2022})},\
  \Eprint {https://arxiv.org/abs/2205.13982} {arXiv:2205.13982 [hep-lat]}
  \BibitemShut {NoStop}%
\bibitem [{\citenamefont {Silvestre-Brac}\ and\ \citenamefont
  {Semay}(1993)}]{Silvestre-Brac:1993zem}%
  \BibitemOpen
  \bibfield  {author} {\bibinfo {author} {\bibfnamefont {B.}~\bibnamefont
  {Silvestre-Brac}}\ and\ \bibinfo {author} {\bibfnamefont {C.}~\bibnamefont
  {Semay}},\ }\bibfield  {title} {\bibinfo {title} {{Systematics of L = 0 q-2
  anti-q-2 systems}},\ }\href {https://doi.org/10.1007/BF01565058} {\bibfield
  {journal} {\bibinfo  {journal} {Z. Phys. C}\ }\textbf {\bibinfo {volume}
  {57}},\ \bibinfo {pages} {273} (\bibinfo {year} {1993})}\BibitemShut
  {NoStop}%
\bibitem [{\citenamefont {Du}\ \emph {et~al.}(2013)\citenamefont {Du},
  \citenamefont {Chen}, \citenamefont {Chen},\ and\ \citenamefont
  {Zhu}}]{Du:2012wp}%
  \BibitemOpen
  \bibfield  {author} {\bibinfo {author} {\bibfnamefont {M.-L.}\ \bibnamefont
  {Du}}, \bibinfo {author} {\bibfnamefont {W.}~\bibnamefont {Chen}}, \bibinfo
  {author} {\bibfnamefont {X.-L.}\ \bibnamefont {Chen}},\ and\ \bibinfo
  {author} {\bibfnamefont {S.-L.}\ \bibnamefont {Zhu}},\ }\bibfield  {title}
  {\bibinfo {title} {{Exotic $QQ\bar{q}\bar{q}$, $QQ\bar{q}\bar{s}$ and
  $QQ\bar{s}\bar{s}$ states}},\ }\href
  {https://doi.org/10.1103/PhysRevD.87.014003} {\bibfield  {journal} {\bibinfo
  {journal} {Phys. Rev. D}\ }\textbf {\bibinfo {volume} {87}},\ \bibinfo
  {pages} {014003} (\bibinfo {year} {2013})},\ \Eprint
  {https://arxiv.org/abs/1209.5134} {arXiv:1209.5134 [hep-ph]} \BibitemShut
  {NoStop}%
\bibitem [{\citenamefont {Park}\ \emph {et~al.}(2019)\citenamefont {Park},
  \citenamefont {Noh},\ and\ \citenamefont {Lee}}]{Park:2018wjk}%
  \BibitemOpen
  \bibfield  {author} {\bibinfo {author} {\bibfnamefont {W.}~\bibnamefont
  {Park}}, \bibinfo {author} {\bibfnamefont {S.}~\bibnamefont {Noh}},\ and\
  \bibinfo {author} {\bibfnamefont {S.~H.}\ \bibnamefont {Lee}},\ }\bibfield
  {title} {\bibinfo {title} {{Masses of the doubly heavy tetraquarks in a
  constituent quark model}},\ }\href
  {https://doi.org/10.1016/j.nuclphysa.2018.12.019} {\bibfield  {journal}
  {\bibinfo  {journal} {Nucl. Phys. A}\ }\textbf {\bibinfo {volume} {983}},\
  \bibinfo {pages} {1} (\bibinfo {year} {2019})},\ \Eprint
  {https://arxiv.org/abs/1809.05257} {arXiv:1809.05257 [nucl-th]} \BibitemShut
  {NoStop}%
\bibitem [{\citenamefont {Deng}\ \emph {et~al.}(2020)\citenamefont {Deng},
  \citenamefont {Chen},\ and\ \citenamefont {Ping}}]{deng:2018kly}%
  \BibitemOpen
  \bibfield  {author} {\bibinfo {author} {\bibfnamefont {C.}~\bibnamefont
  {Deng}}, \bibinfo {author} {\bibfnamefont {H.}~\bibnamefont {Chen}},\ and\
  \bibinfo {author} {\bibfnamefont {J.}~\bibnamefont {Ping}},\ }\bibfield
  {title} {\bibinfo {title} {{Systematical investigation on the stability of
  doubly heavy tetraquark states}},\ }\href
  {https://doi.org/10.1140/epja/s10050-019-00012-y} {\bibfield  {journal}
  {\bibinfo  {journal} {Eur. Phys. J. A}\ }\textbf {\bibinfo {volume} {56}},\
  \bibinfo {pages} {9} (\bibinfo {year} {2020})},\ \Eprint
  {https://arxiv.org/abs/1811.06462} {arXiv:1811.06462 [hep-ph]} \BibitemShut
  {NoStop}%
\bibitem [{\citenamefont {Dai}\ \emph {et~al.}(2022)\citenamefont {Dai},
  \citenamefont {Oset}, \citenamefont {Feijoo}, \citenamefont {Molina},
  \citenamefont {Roca}, \citenamefont {Torres},\ and\ \citenamefont
  {Khemchandani}}]{Dai:2022ulk}%
  \BibitemOpen
  \bibfield  {author} {\bibinfo {author} {\bibfnamefont {L.~R.}\ \bibnamefont
  {Dai}}, \bibinfo {author} {\bibfnamefont {E.}~\bibnamefont {Oset}}, \bibinfo
  {author} {\bibfnamefont {A.}~\bibnamefont {Feijoo}}, \bibinfo {author}
  {\bibfnamefont {R.}~\bibnamefont {Molina}}, \bibinfo {author} {\bibfnamefont
  {L.}~\bibnamefont {Roca}}, \bibinfo {author} {\bibfnamefont {A.~M.}\
  \bibnamefont {Torres}},\ and\ \bibinfo {author} {\bibfnamefont {K.~P.}\
  \bibnamefont {Khemchandani}},\ }\bibfield  {title} {\bibinfo {title} {{Masses
  and widths of the exotic molecular B(s)(*)B(s)(*) states}},\ }\href
  {https://doi.org/10.1103/PhysRevD.105.074017} {\bibfield  {journal} {\bibinfo
   {journal} {Phys. Rev. D}\ }\textbf {\bibinfo {volume} {105}},\ \bibinfo
  {pages} {074017} (\bibinfo {year} {2022})},\ \bibinfo {note} {[Erratum:
  Phys.Rev.D 106, 099904 (2022)]},\ \Eprint {https://arxiv.org/abs/2201.04840}
  {arXiv:2201.04840 [hep-ph]} \BibitemShut {NoStop}%
\bibitem [{\citenamefont {Godfrey}\ and\ \citenamefont
  {Isgur}(1985)}]{Godfrey:1985xj}%
  \BibitemOpen
  \bibfield  {author} {\bibinfo {author} {\bibfnamefont {S.}~\bibnamefont
  {Godfrey}}\ and\ \bibinfo {author} {\bibfnamefont {N.}~\bibnamefont
  {Isgur}},\ }\bibfield  {title} {\bibinfo {title} {{Mesons in a Relativized
  Quark Model with Chromodynamics}},\ }\href
  {https://doi.org/10.1103/PhysRevD.32.189} {\bibfield  {journal} {\bibinfo
  {journal} {Phys. Rev. D}\ }\textbf {\bibinfo {volume} {32}},\ \bibinfo
  {pages} {189} (\bibinfo {year} {1985})}\BibitemShut {NoStop}%
\bibitem [{\citenamefont {Aubert}\ \emph {et~al.}(2003)\citenamefont {Aubert}
  \emph {et~al.}}]{BaBar:2003oey}%
  \BibitemOpen
  \bibfield  {author} {\bibinfo {author} {\bibfnamefont {B.}~\bibnamefont
  {Aubert}} \emph {et~al.} (\bibinfo {collaboration} {BaBar}),\ }\bibfield
  {title} {\bibinfo {title} {{Observation of a narrow meson decaying to $D_s^+
  \pi^0$ at a mass of 2.32-GeV/c$^2$}},\ }\href
  {https://doi.org/10.1103/PhysRevLett.90.242001} {\bibfield  {journal}
  {\bibinfo  {journal} {Phys. Rev. Lett.}\ }\textbf {\bibinfo {volume} {90}},\
  \bibinfo {pages} {242001} (\bibinfo {year} {2003})},\ \Eprint
  {https://arxiv.org/abs/hep-ex/0304021} {arXiv:hep-ex/0304021} \BibitemShut
  {NoStop}%
\bibitem [{\citenamefont {Besson}\ \emph {et~al.}(2003)\citenamefont {Besson}
  \emph {et~al.}}]{CLEO:2003ggt}%
  \BibitemOpen
  \bibfield  {author} {\bibinfo {author} {\bibfnamefont {D.}~\bibnamefont
  {Besson}} \emph {et~al.} (\bibinfo {collaboration} {CLEO}),\ }\bibfield
  {title} {\bibinfo {title} {{Observation of a narrow resonance of mass
  2.46-GeV/c**2 decaying to D*+(s) pi0 and confirmation of the D*(sJ)(2317)
  state}},\ }\href {https://doi.org/10.1103/PhysRevD.68.032002} {\bibfield
  {journal} {\bibinfo  {journal} {Phys. Rev. D}\ }\textbf {\bibinfo {volume}
  {68}},\ \bibinfo {pages} {032002} (\bibinfo {year} {2003})},\ \bibinfo {note}
  {[Erratum: Phys.Rev.D 75, 119908 (2007)]},\ \Eprint
  {https://arxiv.org/abs/hep-ex/0305100} {arXiv:hep-ex/0305100} \BibitemShut
  {NoStop}%
\bibitem [{\citenamefont {Lang}\ \emph {et~al.}(2015)\citenamefont {Lang},
  \citenamefont {Mohler}, \citenamefont {Prelovsek},\ and\ \citenamefont
  {Woloshyn}}]{Lang:2015hza}%
  \BibitemOpen
  \bibfield  {author} {\bibinfo {author} {\bibfnamefont {C.~B.}\ \bibnamefont
  {Lang}}, \bibinfo {author} {\bibfnamefont {D.}~\bibnamefont {Mohler}},
  \bibinfo {author} {\bibfnamefont {S.}~\bibnamefont {Prelovsek}},\ and\
  \bibinfo {author} {\bibfnamefont {R.~M.}\ \bibnamefont {Woloshyn}},\
  }\bibfield  {title} {\bibinfo {title} {{Predicting positive parity B$_s$
  mesons from lattice QCD}},\ }\href
  {https://doi.org/10.1016/j.physletb.2015.08.038} {\bibfield  {journal}
  {\bibinfo  {journal} {Phys. Lett. B}\ }\textbf {\bibinfo {volume} {750}},\
  \bibinfo {pages} {17} (\bibinfo {year} {2015})},\ \Eprint
  {https://arxiv.org/abs/1501.01646} {arXiv:1501.01646 [hep-lat]} \BibitemShut
  {NoStop}%
\bibitem [{\citenamefont {Gregory}\ \emph {et~al.}(2011)\citenamefont {Gregory}
  \emph {et~al.}}]{Gregory:2010gm}%
  \BibitemOpen
  \bibfield  {author} {\bibinfo {author} {\bibfnamefont {E.~B.}\ \bibnamefont
  {Gregory}} \emph {et~al.},\ }\bibfield  {title} {\bibinfo {title} {{Precise
  $B, B_s$ and $B_c$ meson spectroscopy from full lattice QCD}},\ }\href
  {https://doi.org/10.1103/PhysRevD.83.014506} {\bibfield  {journal} {\bibinfo
  {journal} {Phys. Rev. D}\ }\textbf {\bibinfo {volume} {83}},\ \bibinfo
  {pages} {014506} (\bibinfo {year} {2011})},\ \Eprint
  {https://arxiv.org/abs/1010.3848} {arXiv:1010.3848 [hep-lat]} \BibitemShut
  {NoStop}%
\bibitem [{\citenamefont {Wurtz}\ \emph {et~al.}(2015)\citenamefont {Wurtz},
  \citenamefont {Lewis},\ and\ \citenamefont {Woloshyn}}]{Wurtz:2015mqa}%
  \BibitemOpen
  \bibfield  {author} {\bibinfo {author} {\bibfnamefont {M.}~\bibnamefont
  {Wurtz}}, \bibinfo {author} {\bibfnamefont {R.}~\bibnamefont {Lewis}},\ and\
  \bibinfo {author} {\bibfnamefont {R.~M.}\ \bibnamefont {Woloshyn}},\
  }\bibfield  {title} {\bibinfo {title} {{Free-form smearing for bottomonium
  and B meson spectroscopy}},\ }\href
  {https://doi.org/10.1103/PhysRevD.92.054504} {\bibfield  {journal} {\bibinfo
  {journal} {Phys. Rev. D}\ }\textbf {\bibinfo {volume} {92}},\ \bibinfo
  {pages} {054504} (\bibinfo {year} {2015})},\ \Eprint
  {https://arxiv.org/abs/1505.04410} {arXiv:1505.04410 [hep-lat]} \BibitemShut
  {NoStop}%
\bibitem [{\citenamefont {Koponen}(2008)}]{Koponen:2007nr}%
  \BibitemOpen
  \bibfield  {author} {\bibinfo {author} {\bibfnamefont {J.}~\bibnamefont
  {Koponen}} (\bibinfo {collaboration} {UKQCD}),\ }\bibfield  {title} {\bibinfo
  {title} {{Energies of $B_s$ meson excited states: A Lattice study}},\ }\href
  {https://doi.org/10.1103/PhysRevD.78.074509} {\bibfield  {journal} {\bibinfo
  {journal} {Phys. Rev. D}\ }\textbf {\bibinfo {volume} {78}},\ \bibinfo
  {pages} {074509} (\bibinfo {year} {2008})},\ \Eprint
  {https://arxiv.org/abs/0708.2807} {arXiv:0708.2807 [hep-lat]} \BibitemShut
  {NoStop}%
\bibitem [{\citenamefont {Orsland}\ and\ \citenamefont
  {Hogaasen}(1999)}]{Orsland:1998de}%
  \BibitemOpen
  \bibfield  {author} {\bibinfo {author} {\bibfnamefont {A.~H.}\ \bibnamefont
  {Orsland}}\ and\ \bibinfo {author} {\bibfnamefont {H.}~\bibnamefont
  {Hogaasen}},\ }\bibfield  {title} {\bibinfo {title} {{Strong and
  electromagnetic decays for excited heavy mesons}},\ }\href
  {https://doi.org/10.1007/s100529900042} {\bibfield  {journal} {\bibinfo
  {journal} {Eur. Phys. J. C}\ }\textbf {\bibinfo {volume} {9}},\ \bibinfo
  {pages} {503} (\bibinfo {year} {1999})},\ \Eprint
  {https://arxiv.org/abs/hep-ph/9812347} {arXiv:hep-ph/9812347} \BibitemShut
  {NoStop}%
\bibitem [{\citenamefont {Bardeen}\ \emph {et~al.}(2003)\citenamefont
  {Bardeen}, \citenamefont {Eichten},\ and\ \citenamefont
  {Hill}}]{Bardeen:2003kt}%
  \BibitemOpen
  \bibfield  {author} {\bibinfo {author} {\bibfnamefont {W.~A.}\ \bibnamefont
  {Bardeen}}, \bibinfo {author} {\bibfnamefont {E.~J.}\ \bibnamefont
  {Eichten}},\ and\ \bibinfo {author} {\bibfnamefont {C.~T.}\ \bibnamefont
  {Hill}},\ }\bibfield  {title} {\bibinfo {title} {{Chiral multiplets of heavy
  - light mesons}},\ }\href {https://doi.org/10.1103/PhysRevD.68.054024}
  {\bibfield  {journal} {\bibinfo  {journal} {Phys. Rev. D}\ }\textbf {\bibinfo
  {volume} {68}},\ \bibinfo {pages} {054024} (\bibinfo {year} {2003})},\
  \Eprint {https://arxiv.org/abs/hep-ph/0305049} {arXiv:hep-ph/0305049}
  \BibitemShut {NoStop}%
\bibitem [{\citenamefont {Kolomeitsev}\ and\ \citenamefont
  {Lutz}(2004)}]{Kolomeitsev:2003ac}%
  \BibitemOpen
  \bibfield  {author} {\bibinfo {author} {\bibfnamefont {E.~E.}\ \bibnamefont
  {Kolomeitsev}}\ and\ \bibinfo {author} {\bibfnamefont {M.~F.~M.}\
  \bibnamefont {Lutz}},\ }\bibfield  {title} {\bibinfo {title} {{On Heavy light
  meson resonances and chiral symmetry}},\ }\href
  {https://doi.org/10.1016/j.physletb.2003.10.118} {\bibfield  {journal}
  {\bibinfo  {journal} {Phys. Lett. B}\ }\textbf {\bibinfo {volume} {582}},\
  \bibinfo {pages} {39} (\bibinfo {year} {2004})},\ \Eprint
  {https://arxiv.org/abs/hep-ph/0307133} {arXiv:hep-ph/0307133} \BibitemShut
  {NoStop}%
\bibitem [{\citenamefont {Matsuki}\ \emph {et~al.}(2005)\citenamefont
  {Matsuki}, \citenamefont {Mawatari}, \citenamefont {Morii},\ and\
  \citenamefont {Sudoh}}]{Matsuki:2004db}%
  \BibitemOpen
  \bibfield  {author} {\bibinfo {author} {\bibfnamefont {T.}~\bibnamefont
  {Matsuki}}, \bibinfo {author} {\bibfnamefont {K.}~\bibnamefont {Mawatari}},
  \bibinfo {author} {\bibfnamefont {T.}~\bibnamefont {Morii}},\ and\ \bibinfo
  {author} {\bibfnamefont {K.}~\bibnamefont {Sudoh}},\ }\bibfield  {title}
  {\bibinfo {title} {{$0^+$ and $1^+$ states of $B$ and $B_s$ mesons}},\ }\href
  {https://doi.org/10.1016/j.physletb.2004.12.012} {\bibfield  {journal}
  {\bibinfo  {journal} {Phys. Lett. B}\ }\textbf {\bibinfo {volume} {606}},\
  \bibinfo {pages} {329} (\bibinfo {year} {2005})},\ \Eprint
  {https://arxiv.org/abs/hep-ph/0411034} {arXiv:hep-ph/0411034} \BibitemShut
  {NoStop}%
\bibitem [{\citenamefont {Guo}\ \emph {et~al.}(2006)\citenamefont {Guo},
  \citenamefont {Shen}, \citenamefont {Chiang}, \citenamefont {Ping},\ and\
  \citenamefont {Zou}}]{Guo:2006fu}%
  \BibitemOpen
  \bibfield  {author} {\bibinfo {author} {\bibfnamefont {F.-K.}\ \bibnamefont
  {Guo}}, \bibinfo {author} {\bibfnamefont {P.-N.}\ \bibnamefont {Shen}},
  \bibinfo {author} {\bibfnamefont {H.-C.}\ \bibnamefont {Chiang}}, \bibinfo
  {author} {\bibfnamefont {R.-G.}\ \bibnamefont {Ping}},\ and\ \bibinfo
  {author} {\bibfnamefont {B.-S.}\ \bibnamefont {Zou}},\ }\bibfield  {title}
  {\bibinfo {title} {{Dynamically generated 0+ heavy mesons in a heavy chiral
  unitary approach}},\ }\href {https://doi.org/10.1016/j.physletb.2006.08.064}
  {\bibfield  {journal} {\bibinfo  {journal} {Phys. Lett. B}\ }\textbf
  {\bibinfo {volume} {641}},\ \bibinfo {pages} {278} (\bibinfo {year}
  {2006})},\ \Eprint {https://arxiv.org/abs/hep-ph/0603072}
  {arXiv:hep-ph/0603072} \BibitemShut {NoStop}%
\bibitem [{\citenamefont {Guo}\ \emph {et~al.}(2007)\citenamefont {Guo},
  \citenamefont {Shen},\ and\ \citenamefont {Chiang}}]{Guo:2006rp}%
  \BibitemOpen
  \bibfield  {author} {\bibinfo {author} {\bibfnamefont {F.-K.}\ \bibnamefont
  {Guo}}, \bibinfo {author} {\bibfnamefont {P.-N.}\ \bibnamefont {Shen}},\ and\
  \bibinfo {author} {\bibfnamefont {H.-C.}\ \bibnamefont {Chiang}},\ }\bibfield
   {title} {\bibinfo {title} {{Dynamically generated 1+ heavy mesons}},\ }\href
  {https://doi.org/10.1016/j.physletb.2007.01.050} {\bibfield  {journal}
  {\bibinfo  {journal} {Phys. Lett. B}\ }\textbf {\bibinfo {volume} {647}},\
  \bibinfo {pages} {133} (\bibinfo {year} {2007})},\ \Eprint
  {https://arxiv.org/abs/hep-ph/0610008} {arXiv:hep-ph/0610008} \BibitemShut
  {NoStop}%
\bibitem [{\citenamefont {Badalian}\ \emph {et~al.}(2008)\citenamefont
  {Badalian}, \citenamefont {Simonov},\ and\ \citenamefont
  {Trusov}}]{Badalian:2007yr}%
  \BibitemOpen
  \bibfield  {author} {\bibinfo {author} {\bibfnamefont {A.~M.}\ \bibnamefont
  {Badalian}}, \bibinfo {author} {\bibfnamefont {Y.~A.}\ \bibnamefont
  {Simonov}},\ and\ \bibinfo {author} {\bibfnamefont {M.~A.}\ \bibnamefont
  {Trusov}},\ }\bibfield  {title} {\bibinfo {title} {{The Chiral transitions in
  heavy-light mesons}},\ }\href {https://doi.org/10.1103/PhysRevD.77.074017}
  {\bibfield  {journal} {\bibinfo  {journal} {Phys. Rev. D}\ }\textbf {\bibinfo
  {volume} {77}},\ \bibinfo {pages} {074017} (\bibinfo {year} {2008})},\
  \Eprint {https://arxiv.org/abs/0712.3943} {arXiv:0712.3943 [hep-ph]}
  \BibitemShut {NoStop}%
\bibitem [{\citenamefont {Vijande}\ \emph {et~al.}(2008)\citenamefont
  {Vijande}, \citenamefont {Valcarce},\ and\ \citenamefont
  {Fernandez}}]{Vijande:2007ke}%
  \BibitemOpen
  \bibfield  {author} {\bibinfo {author} {\bibfnamefont {J.}~\bibnamefont
  {Vijande}}, \bibinfo {author} {\bibfnamefont {A.}~\bibnamefont {Valcarce}},\
  and\ \bibinfo {author} {\bibfnamefont {F.}~\bibnamefont {Fernandez}},\
  }\bibfield  {title} {\bibinfo {title} {{B meson spectroscopy}},\ }\href
  {https://doi.org/10.1103/PhysRevD.77.017501} {\bibfield  {journal} {\bibinfo
  {journal} {Phys. Rev. D}\ }\textbf {\bibinfo {volume} {77}},\ \bibinfo
  {pages} {017501} (\bibinfo {year} {2008})},\ \Eprint
  {https://arxiv.org/abs/0711.2359} {arXiv:0711.2359 [hep-ph]} \BibitemShut
  {NoStop}%
\bibitem [{\citenamefont {Cleven}\ \emph {et~al.}(2011)\citenamefont {Cleven},
  \citenamefont {Guo}, \citenamefont {Hanhart},\ and\ \citenamefont
  {Meissner}}]{Cleven:2010aw}%
  \BibitemOpen
  \bibfield  {author} {\bibinfo {author} {\bibfnamefont {M.}~\bibnamefont
  {Cleven}}, \bibinfo {author} {\bibfnamefont {F.-K.}\ \bibnamefont {Guo}},
  \bibinfo {author} {\bibfnamefont {C.}~\bibnamefont {Hanhart}},\ and\ \bibinfo
  {author} {\bibfnamefont {U.-G.}\ \bibnamefont {Meissner}},\ }\bibfield
  {title} {\bibinfo {title} {{Light meson mass dependence of the positive
  parity heavy-strange mesons}},\ }\href
  {https://doi.org/10.1140/epja/i2011-11019-2} {\bibfield  {journal} {\bibinfo
  {journal} {Eur. Phys. J. A}\ }\textbf {\bibinfo {volume} {47}},\ \bibinfo
  {pages} {19} (\bibinfo {year} {2011})},\ \Eprint
  {https://arxiv.org/abs/1009.3804} {arXiv:1009.3804 [hep-ph]} \BibitemShut
  {NoStop}%
\bibitem [{\citenamefont {Dmitrasinovic}(2012)}]{Dmitrasinovic:2012zz}%
  \BibitemOpen
  \bibfield  {author} {\bibinfo {author} {\bibfnamefont {V.}~\bibnamefont
  {Dmitrasinovic}},\ }\bibfield  {title} {\bibinfo {title} {{Chiral symmetry of
  heavy-light scalar mesons with $U_A(1)$ symmetry breaking}},\ }\href
  {https://doi.org/10.1103/PhysRevD.86.016006} {\bibfield  {journal} {\bibinfo
  {journal} {Phys. Rev. D}\ }\textbf {\bibinfo {volume} {86}},\ \bibinfo
  {pages} {016006} (\bibinfo {year} {2012})}\BibitemShut {NoStop}%
\bibitem [{\citenamefont {Colangelo}\ \emph {et~al.}(2012)\citenamefont
  {Colangelo}, \citenamefont {De~Fazio}, \citenamefont {Giannuzzi},\ and\
  \citenamefont {Nicotri}}]{Colangelo:2012xi}%
  \BibitemOpen
  \bibfield  {author} {\bibinfo {author} {\bibfnamefont {P.}~\bibnamefont
  {Colangelo}}, \bibinfo {author} {\bibfnamefont {F.}~\bibnamefont {De~Fazio}},
  \bibinfo {author} {\bibfnamefont {F.}~\bibnamefont {Giannuzzi}},\ and\
  \bibinfo {author} {\bibfnamefont {S.}~\bibnamefont {Nicotri}},\ }\bibfield
  {title} {\bibinfo {title} {{New meson spectroscopy with open charm and
  beauty}},\ }\href {https://doi.org/10.1103/PhysRevD.86.054024} {\bibfield
  {journal} {\bibinfo  {journal} {Phys. Rev. D}\ }\textbf {\bibinfo {volume}
  {86}},\ \bibinfo {pages} {054024} (\bibinfo {year} {2012})},\ \Eprint
  {https://arxiv.org/abs/1207.6940} {arXiv:1207.6940 [hep-ph]} \BibitemShut
  {NoStop}%
\bibitem [{\citenamefont {Altenbuchinger}\ \emph {et~al.}(2014)\citenamefont
  {Altenbuchinger}, \citenamefont {Geng},\ and\ \citenamefont
  {Weise}}]{Altenbuchinger:2013vwa}%
  \BibitemOpen
  \bibfield  {author} {\bibinfo {author} {\bibfnamefont {M.}~\bibnamefont
  {Altenbuchinger}}, \bibinfo {author} {\bibfnamefont {L.~S.}\ \bibnamefont
  {Geng}},\ and\ \bibinfo {author} {\bibfnamefont {W.}~\bibnamefont {Weise}},\
  }\bibfield  {title} {\bibinfo {title} {{Scattering lengths of Nambu-Goldstone
  bosons off $D$ mesons and dynamically generated heavy-light mesons}},\ }\href
  {https://doi.org/10.1103/PhysRevD.89.014026} {\bibfield  {journal} {\bibinfo
  {journal} {Phys. Rev. D}\ }\textbf {\bibinfo {volume} {89}},\ \bibinfo
  {pages} {014026} (\bibinfo {year} {2014})},\ \Eprint
  {https://arxiv.org/abs/1309.4743} {arXiv:1309.4743 [hep-ph]} \BibitemShut
  {NoStop}%
\bibitem [{\citenamefont {Torres-Rincon}\ \emph {et~al.}(2014)\citenamefont
  {Torres-Rincon}, \citenamefont {Tolos},\ and\ \citenamefont
  {Romanets}}]{Torres-Rincon:2014ffa}%
  \BibitemOpen
  \bibfield  {author} {\bibinfo {author} {\bibfnamefont {J.~M.}\ \bibnamefont
  {Torres-Rincon}}, \bibinfo {author} {\bibfnamefont {L.}~\bibnamefont
  {Tolos}},\ and\ \bibinfo {author} {\bibfnamefont {O.}~\bibnamefont
  {Romanets}},\ }\bibfield  {title} {\bibinfo {title} {{Open bottom states and
  the $\bar B$-meson propagation in hadronic matter}},\ }\href
  {https://doi.org/10.1103/PhysRevD.89.074042} {\bibfield  {journal} {\bibinfo
  {journal} {Phys. Rev. D}\ }\textbf {\bibinfo {volume} {89}},\ \bibinfo
  {pages} {074042} (\bibinfo {year} {2014})},\ \Eprint
  {https://arxiv.org/abs/1403.1371} {arXiv:1403.1371 [hep-ph]} \BibitemShut
  {NoStop}%
\bibitem [{\citenamefont {Wang}(2015)}]{Wang:2015mxa}%
  \BibitemOpen
  \bibfield  {author} {\bibinfo {author} {\bibfnamefont {Z.-G.}\ \bibnamefont
  {Wang}},\ }\bibfield  {title} {\bibinfo {title} {{Analysis of the masses and
  decay constants of the heavy-light mesons with QCD sum rules}},\ }\href
  {https://doi.org/10.1140/epjc/s10052-015-3653-9} {\bibfield  {journal}
  {\bibinfo  {journal} {Eur. Phys. J. C}\ }\textbf {\bibinfo {volume} {75}},\
  \bibinfo {pages} {427} (\bibinfo {year} {2015})},\ \Eprint
  {https://arxiv.org/abs/1506.01993} {arXiv:1506.01993 [hep-ph]} \BibitemShut
  {NoStop}%
\bibitem [{\citenamefont {Ortega}\ \emph {et~al.}(2017)\citenamefont {Ortega},
  \citenamefont {Segovia}, \citenamefont {Entem},\ and\ \citenamefont
  {Fern\'andez}}]{Ortega:2016pgg}%
  \BibitemOpen
  \bibfield  {author} {\bibinfo {author} {\bibfnamefont {P.~G.}\ \bibnamefont
  {Ortega}}, \bibinfo {author} {\bibfnamefont {J.}~\bibnamefont {Segovia}},
  \bibinfo {author} {\bibfnamefont {D.~R.}\ \bibnamefont {Entem}},\ and\
  \bibinfo {author} {\bibfnamefont {F.}~\bibnamefont {Fern\'andez}},\
  }\bibfield  {title} {\bibinfo {title} {{Threshold effects in P-wave
  bottom-strange mesons}},\ }\href {https://doi.org/10.1103/PhysRevD.95.034010}
  {\bibfield  {journal} {\bibinfo  {journal} {Phys. Rev. D}\ }\textbf {\bibinfo
  {volume} {95}},\ \bibinfo {pages} {034010} (\bibinfo {year} {2017})},\
  \Eprint {https://arxiv.org/abs/1612.04826} {arXiv:1612.04826 [hep-ph]}
  \BibitemShut {NoStop}%
\bibitem [{\citenamefont {Albaladejo}\ \emph {et~al.}(2017)\citenamefont
  {Albaladejo}, \citenamefont {Fernandez-Soler}, \citenamefont {Nieves},\ and\
  \citenamefont {Ortega}}]{Albaladejo:2016ztm}%
  \BibitemOpen
  \bibfield  {author} {\bibinfo {author} {\bibfnamefont {M.}~\bibnamefont
  {Albaladejo}}, \bibinfo {author} {\bibfnamefont {P.}~\bibnamefont
  {Fernandez-Soler}}, \bibinfo {author} {\bibfnamefont {J.}~\bibnamefont
  {Nieves}},\ and\ \bibinfo {author} {\bibfnamefont {P.~G.}\ \bibnamefont
  {Ortega}},\ }\bibfield  {title} {\bibinfo {title} {{Lowest-lying even-parity
  ${\bar{B}}_s$ mesons: heavy-quark spin-flavor symmetry, chiral dynamics, and
  constituent quark-model bare masses}},\ }\href
  {https://doi.org/10.1140/epjc/s10052-017-4735-7} {\bibfield  {journal}
  {\bibinfo  {journal} {Eur. Phys. J. C}\ }\textbf {\bibinfo {volume} {77}},\
  \bibinfo {pages} {170} (\bibinfo {year} {2017})},\ \Eprint
  {https://arxiv.org/abs/1612.07782} {arXiv:1612.07782 [hep-ph]} \BibitemShut
  {NoStop}%
\bibitem [{\citenamefont {Cheng}\ and\ \citenamefont
  {Yu}(2017)}]{Cheng:2017oqh}%
  \BibitemOpen
  \bibfield  {author} {\bibinfo {author} {\bibfnamefont {H.-Y.}\ \bibnamefont
  {Cheng}}\ and\ \bibinfo {author} {\bibfnamefont {F.-S.}\ \bibnamefont {Yu}},\
  }\bibfield  {title} {\bibinfo {title} {{Masses of Scalar and Axial-Vector B
  Mesons Revisited}},\ }\href {https://doi.org/10.1140/epjc/s10052-017-5252-4}
  {\bibfield  {journal} {\bibinfo  {journal} {Eur. Phys. J. C}\ }\textbf
  {\bibinfo {volume} {77}},\ \bibinfo {pages} {668} (\bibinfo {year} {2017})},\
  \Eprint {https://arxiv.org/abs/1704.01208} {arXiv:1704.01208 [hep-ph]}
  \BibitemShut {NoStop}%
\bibitem [{\citenamefont {Du}\ \emph {et~al.}(2018)\citenamefont {Du},
  \citenamefont {Albaladejo}, \citenamefont {Fern\'andez-Soler}, \citenamefont
  {Guo}, \citenamefont {Hanhart}, \citenamefont {Mei\ss{}ner}, \citenamefont
  {Nieves},\ and\ \citenamefont {Yao}}]{Du:2017zvv}%
  \BibitemOpen
  \bibfield  {author} {\bibinfo {author} {\bibfnamefont {M.-L.}\ \bibnamefont
  {Du}}, \bibinfo {author} {\bibfnamefont {M.}~\bibnamefont {Albaladejo}},
  \bibinfo {author} {\bibfnamefont {P.}~\bibnamefont {Fern\'andez-Soler}},
  \bibinfo {author} {\bibfnamefont {F.-K.}\ \bibnamefont {Guo}}, \bibinfo
  {author} {\bibfnamefont {C.}~\bibnamefont {Hanhart}}, \bibinfo {author}
  {\bibfnamefont {U.-G.}\ \bibnamefont {Mei\ss{}ner}}, \bibinfo {author}
  {\bibfnamefont {J.}~\bibnamefont {Nieves}},\ and\ \bibinfo {author}
  {\bibfnamefont {D.-L.}\ \bibnamefont {Yao}},\ }\bibfield  {title} {\bibinfo
  {title} {{Towards a new paradigm for heavy-light meson spectroscopy}},\
  }\href {https://doi.org/10.1103/PhysRevD.98.094018} {\bibfield  {journal}
  {\bibinfo  {journal} {Phys. Rev. D}\ }\textbf {\bibinfo {volume} {98}},\
  \bibinfo {pages} {094018} (\bibinfo {year} {2018})},\ \Eprint
  {https://arxiv.org/abs/1712.07957} {arXiv:1712.07957 [hep-ph]} \BibitemShut
  {NoStop}%
\bibitem [{\citenamefont {Alhakami}(2021)}]{Alhakami:2020vil}%
  \BibitemOpen
  \bibfield  {author} {\bibinfo {author} {\bibfnamefont {M.~H.}\ \bibnamefont
  {Alhakami}},\ }\bibfield  {title} {\bibinfo {title} {{Predictions for the
  beauty meson spectrum}},\ }\href
  {https://doi.org/10.1103/PhysRevD.103.034009} {\bibfield  {journal} {\bibinfo
   {journal} {Phys. Rev. D}\ }\textbf {\bibinfo {volume} {103}},\ \bibinfo
  {pages} {034009} (\bibinfo {year} {2021})},\ \Eprint
  {https://arxiv.org/abs/2006.16878} {arXiv:2006.16878 [hep-ph]} \BibitemShut
  {NoStop}%
\bibitem [{\citenamefont {Fu}\ \emph {et~al.}(2022)\citenamefont {Fu},
  \citenamefont {Grie\ss{}hammer}, \citenamefont {Guo}, \citenamefont
  {Hanhart},\ and\ \citenamefont {Mei\ss{}ner}}]{Fu:2021wde}%
  \BibitemOpen
  \bibfield  {author} {\bibinfo {author} {\bibfnamefont {H.-L.}\ \bibnamefont
  {Fu}}, \bibinfo {author} {\bibfnamefont {H.~W.}\ \bibnamefont
  {Grie\ss{}hammer}}, \bibinfo {author} {\bibfnamefont {F.-K.}\ \bibnamefont
  {Guo}}, \bibinfo {author} {\bibfnamefont {C.}~\bibnamefont {Hanhart}},\ and\
  \bibinfo {author} {\bibfnamefont {U.-G.}\ \bibnamefont {Mei\ss{}ner}},\
  }\bibfield  {title} {\bibinfo {title} {{Update on strong and radiative decays
  of the $D_{s0}^*(2317)$ and $D_{s1}(2460)$ and their bottom cousins}},\
  }\href {https://doi.org/10.1140/epja/s10050-022-00724-8} {\bibfield
  {journal} {\bibinfo  {journal} {Eur. Phys. J. A}\ }\textbf {\bibinfo {volume}
  {58}},\ \bibinfo {pages} {70} (\bibinfo {year} {2022})},\ \Eprint
  {https://arxiv.org/abs/2111.09481} {arXiv:2111.09481 [hep-ph]} \BibitemShut
  {NoStop}%
\bibitem [{\citenamefont {Yang}\ \emph {et~al.}(2022)\citenamefont {Yang},
  \citenamefont {Wang}, \citenamefont {Wu}, \citenamefont {Oka},\ and\
  \citenamefont {Zhu}}]{Yang:2022vdb}%
  \BibitemOpen
  \bibfield  {author} {\bibinfo {author} {\bibfnamefont {Z.}~\bibnamefont
  {Yang}}, \bibinfo {author} {\bibfnamefont {G.-J.}\ \bibnamefont {Wang}},
  \bibinfo {author} {\bibfnamefont {J.-J.}\ \bibnamefont {Wu}}, \bibinfo
  {author} {\bibfnamefont {M.}~\bibnamefont {Oka}},\ and\ \bibinfo {author}
  {\bibfnamefont {S.-L.}\ \bibnamefont {Zhu}},\ }\bibfield  {title} {\bibinfo
  {title} {{The investigations of the $P$-wave $B_s$ states combining quark
  model and lattice QCD in the coupled channel framework}},\ }\href@noop {} {\
  (\bibinfo {year} {2022})},\ \Eprint {https://arxiv.org/abs/2207.07320}
  {arXiv:2207.07320 [hep-lat]} \BibitemShut {NoStop}%
\bibitem [{\citenamefont {Godfrey}\ and\ \citenamefont
  {Kokoski}(1991)}]{Godfrey:1986wj}%
  \BibitemOpen
  \bibfield  {author} {\bibinfo {author} {\bibfnamefont {S.}~\bibnamefont
  {Godfrey}}\ and\ \bibinfo {author} {\bibfnamefont {R.}~\bibnamefont
  {Kokoski}},\ }\bibfield  {title} {\bibinfo {title} {{The Properties of p Wave
  Mesons with One Heavy Quark}},\ }\href
  {https://doi.org/10.1103/PhysRevD.43.1679} {\bibfield  {journal} {\bibinfo
  {journal} {Phys. Rev. D}\ }\textbf {\bibinfo {volume} {43}},\ \bibinfo
  {pages} {1679} (\bibinfo {year} {1991})}\BibitemShut {NoStop}%
\bibitem [{\citenamefont {Lahde}\ \emph {et~al.}(2000)\citenamefont {Lahde},
  \citenamefont {Nyfalt},\ and\ \citenamefont {Riska}}]{Lahde:1999ih}%
  \BibitemOpen
  \bibfield  {author} {\bibinfo {author} {\bibfnamefont {T.~A.}\ \bibnamefont
  {Lahde}}, \bibinfo {author} {\bibfnamefont {C.~J.}\ \bibnamefont {Nyfalt}},\
  and\ \bibinfo {author} {\bibfnamefont {D.~O.}\ \bibnamefont {Riska}},\
  }\bibfield  {title} {\bibinfo {title} {{Spectra and M1 decay widths of heavy
  light mesons}},\ }\href {https://doi.org/10.1016/S0375-9474(00)00154-8}
  {\bibfield  {journal} {\bibinfo  {journal} {Nucl. Phys. A}\ }\textbf
  {\bibinfo {volume} {674}},\ \bibinfo {pages} {141} (\bibinfo {year}
  {2000})},\ \Eprint {https://arxiv.org/abs/hep-ph/9908485}
  {arXiv:hep-ph/9908485} \BibitemShut {NoStop}%
\bibitem [{\citenamefont {Di~Pierro}\ and\ \citenamefont
  {Eichten}(2001)}]{DiPierro:2001dwf}%
  \BibitemOpen
  \bibfield  {author} {\bibinfo {author} {\bibfnamefont {M.}~\bibnamefont
  {Di~Pierro}}\ and\ \bibinfo {author} {\bibfnamefont {E.}~\bibnamefont
  {Eichten}},\ }\bibfield  {title} {\bibinfo {title} {{Excited Heavy - Light
  Systems and Hadronic Transitions}},\ }\href
  {https://doi.org/10.1103/PhysRevD.64.114004} {\bibfield  {journal} {\bibinfo
  {journal} {Phys. Rev. D}\ }\textbf {\bibinfo {volume} {64}},\ \bibinfo
  {pages} {114004} (\bibinfo {year} {2001})},\ \Eprint
  {https://arxiv.org/abs/hep-ph/0104208} {arXiv:hep-ph/0104208} \BibitemShut
  {NoStop}%
\bibitem [{\citenamefont {Lakhina}\ and\ \citenamefont
  {Swanson}(2007)}]{Lakhina:2006fy}%
  \BibitemOpen
  \bibfield  {author} {\bibinfo {author} {\bibfnamefont {O.}~\bibnamefont
  {Lakhina}}\ and\ \bibinfo {author} {\bibfnamefont {E.~S.}\ \bibnamefont
  {Swanson}},\ }\bibfield  {title} {\bibinfo {title} {{A Canonical
  Ds(2317)?}},\ }\href {https://doi.org/10.1016/j.physletb.2007.01.075}
  {\bibfield  {journal} {\bibinfo  {journal} {Phys. Lett. B}\ }\textbf
  {\bibinfo {volume} {650}},\ \bibinfo {pages} {159} (\bibinfo {year}
  {2007})},\ \Eprint {https://arxiv.org/abs/hep-ph/0608011}
  {arXiv:hep-ph/0608011} \BibitemShut {NoStop}%
\bibitem [{\citenamefont {Ebert}\ \emph {et~al.}(2010)\citenamefont {Ebert},
  \citenamefont {Faustov},\ and\ \citenamefont {Galkin}}]{Ebert:2009ua}%
  \BibitemOpen
  \bibfield  {author} {\bibinfo {author} {\bibfnamefont {D.}~\bibnamefont
  {Ebert}}, \bibinfo {author} {\bibfnamefont {R.~N.}\ \bibnamefont {Faustov}},\
  and\ \bibinfo {author} {\bibfnamefont {V.~O.}\ \bibnamefont {Galkin}},\
  }\bibfield  {title} {\bibinfo {title} {{Heavy-light meson spectroscopy and
  Regge trajectories in the relativistic quark model}},\ }\href
  {https://doi.org/10.1140/epjc/s10052-010-1233-6} {\bibfield  {journal}
  {\bibinfo  {journal} {Eur. Phys. J. C}\ }\textbf {\bibinfo {volume} {66}},\
  \bibinfo {pages} {197} (\bibinfo {year} {2010})},\ \Eprint
  {https://arxiv.org/abs/0910.5612} {arXiv:0910.5612 [hep-ph]} \BibitemShut
  {NoStop}%
\bibitem [{\citenamefont {Sun}\ \emph {et~al.}(2014)\citenamefont {Sun},
  \citenamefont {Song}, \citenamefont {Chen}, \citenamefont {Liu},\ and\
  \citenamefont {Zhu}}]{Sun:2014wea}%
  \BibitemOpen
  \bibfield  {author} {\bibinfo {author} {\bibfnamefont {Y.}~\bibnamefont
  {Sun}}, \bibinfo {author} {\bibfnamefont {Q.-T.}\ \bibnamefont {Song}},
  \bibinfo {author} {\bibfnamefont {D.-Y.}\ \bibnamefont {Chen}}, \bibinfo
  {author} {\bibfnamefont {X.}~\bibnamefont {Liu}},\ and\ \bibinfo {author}
  {\bibfnamefont {S.-L.}\ \bibnamefont {Zhu}},\ }\bibfield  {title} {\bibinfo
  {title} {{Higher bottom and bottom-strange mesons}},\ }\href
  {https://doi.org/10.1103/PhysRevD.89.054026} {\bibfield  {journal} {\bibinfo
  {journal} {Phys. Rev. D}\ }\textbf {\bibinfo {volume} {89}},\ \bibinfo
  {pages} {054026} (\bibinfo {year} {2014})},\ \Eprint
  {https://arxiv.org/abs/1401.1595} {arXiv:1401.1595 [hep-ph]} \BibitemShut
  {NoStop}%
\bibitem [{\citenamefont {Godfrey}\ \emph {et~al.}(2016)\citenamefont
  {Godfrey}, \citenamefont {Moats},\ and\ \citenamefont
  {Swanson}}]{Godfrey:2016nwn}%
  \BibitemOpen
  \bibfield  {author} {\bibinfo {author} {\bibfnamefont {S.}~\bibnamefont
  {Godfrey}}, \bibinfo {author} {\bibfnamefont {K.}~\bibnamefont {Moats}},\
  and\ \bibinfo {author} {\bibfnamefont {E.~S.}\ \bibnamefont {Swanson}},\
  }\bibfield  {title} {\bibinfo {title} {{$B$ and $B_s$ Meson Spectroscopy}},\
  }\href {https://doi.org/10.1103/PhysRevD.94.054025} {\bibfield  {journal}
  {\bibinfo  {journal} {Phys. Rev. D}\ }\textbf {\bibinfo {volume} {94}},\
  \bibinfo {pages} {054025} (\bibinfo {year} {2016})},\ \Eprint
  {https://arxiv.org/abs/1607.02169} {arXiv:1607.02169 [hep-ph]} \BibitemShut
  {NoStop}%
\bibitem [{\citenamefont {McNeile}\ \emph {et~al.}(2010)\citenamefont
  {McNeile}, \citenamefont {Davies}, \citenamefont {Follana}, \citenamefont
  {Hornbostel},\ and\ \citenamefont {Lepage}}]{McNeile:2010ji}%
  \BibitemOpen
  \bibfield  {author} {\bibinfo {author} {\bibfnamefont {C.}~\bibnamefont
  {McNeile}}, \bibinfo {author} {\bibfnamefont {C.~T.~H.}\ \bibnamefont
  {Davies}}, \bibinfo {author} {\bibfnamefont {E.}~\bibnamefont {Follana}},
  \bibinfo {author} {\bibfnamefont {K.}~\bibnamefont {Hornbostel}},\ and\
  \bibinfo {author} {\bibfnamefont {G.~P.}\ \bibnamefont {Lepage}},\ }\bibfield
   {title} {\bibinfo {title} {{High-Precision c and b Masses, and QCD Coupling
  from Current-Current Correlators in Lattice and Continuum QCD}},\ }\href
  {https://doi.org/10.1103/PhysRevD.82.034512} {\bibfield  {journal} {\bibinfo
  {journal} {Phys. Rev. D}\ }\textbf {\bibinfo {volume} {82}},\ \bibinfo
  {pages} {034512} (\bibinfo {year} {2010})},\ \Eprint
  {https://arxiv.org/abs/1004.4285} {arXiv:1004.4285 [hep-lat]} \BibitemShut
  {NoStop}%
\bibitem [{\citenamefont {Davies}\ \emph
  {et~al.}(1994{\natexlab{a}})\citenamefont {Davies}, \citenamefont
  {Hornbostel}, \citenamefont {Langnau}, \citenamefont {Lepage}, \citenamefont
  {Lidsey}, \citenamefont {Morningstar}, \citenamefont {Shigemitsu},\ and\
  \citenamefont {Sloan}}]{Davies:1994pz}%
  \BibitemOpen
  \bibfield  {author} {\bibinfo {author} {\bibfnamefont {C.~T.~H.}\
  \bibnamefont {Davies}}, \bibinfo {author} {\bibfnamefont {K.}~\bibnamefont
  {Hornbostel}}, \bibinfo {author} {\bibfnamefont {A.}~\bibnamefont {Langnau}},
  \bibinfo {author} {\bibfnamefont {G.~P.}\ \bibnamefont {Lepage}}, \bibinfo
  {author} {\bibfnamefont {A.}~\bibnamefont {Lidsey}}, \bibinfo {author}
  {\bibfnamefont {C.~J.}\ \bibnamefont {Morningstar}}, \bibinfo {author}
  {\bibfnamefont {J.}~\bibnamefont {Shigemitsu}},\ and\ \bibinfo {author}
  {\bibfnamefont {J.~H.}\ \bibnamefont {Sloan}},\ }\bibfield  {title} {\bibinfo
  {title} {{A New determination of M(b) using lattice QCD}},\ }\href
  {https://doi.org/10.1103/PhysRevLett.73.2654} {\bibfield  {journal} {\bibinfo
   {journal} {Phys. Rev. Lett.}\ }\textbf {\bibinfo {volume} {73}},\ \bibinfo
  {pages} {2654} (\bibinfo {year} {1994}{\natexlab{a}})},\ \Eprint
  {https://arxiv.org/abs/hep-lat/9404012} {arXiv:hep-lat/9404012} \BibitemShut
  {NoStop}%
\bibitem [{\citenamefont {Gray}\ \emph {et~al.}(2005)\citenamefont {Gray},
  \citenamefont {Allison}, \citenamefont {Davies}, \citenamefont {Dalgic},
  \citenamefont {Lepage}, \citenamefont {Shigemitsu},\ and\ \citenamefont
  {Wingate}}]{Gray:2005ur}%
  \BibitemOpen
  \bibfield  {author} {\bibinfo {author} {\bibfnamefont {A.}~\bibnamefont
  {Gray}}, \bibinfo {author} {\bibfnamefont {I.}~\bibnamefont {Allison}},
  \bibinfo {author} {\bibfnamefont {C.~T.~H.}\ \bibnamefont {Davies}}, \bibinfo
  {author} {\bibfnamefont {E.}~\bibnamefont {Dalgic}}, \bibinfo {author}
  {\bibfnamefont {G.~P.}\ \bibnamefont {Lepage}}, \bibinfo {author}
  {\bibfnamefont {J.}~\bibnamefont {Shigemitsu}},\ and\ \bibinfo {author}
  {\bibfnamefont {M.}~\bibnamefont {Wingate}},\ }\bibfield  {title} {\bibinfo
  {title} {{The Upsilon spectrum and m(b) from full lattice QCD}},\ }\href
  {https://doi.org/10.1103/PhysRevD.72.094507} {\bibfield  {journal} {\bibinfo
  {journal} {Phys. Rev. D}\ }\textbf {\bibinfo {volume} {72}},\ \bibinfo
  {pages} {094507} (\bibinfo {year} {2005})},\ \Eprint
  {https://arxiv.org/abs/hep-lat/0507013} {arXiv:hep-lat/0507013} \BibitemShut
  {NoStop}%
\bibitem [{\citenamefont {Na}\ \emph {et~al.}(2015)\citenamefont {Na},
  \citenamefont {Bouchard}, \citenamefont {Lepage}, \citenamefont {Monahan},\
  and\ \citenamefont {Shigemitsu}}]{Na:2015kha}%
  \BibitemOpen
  \bibfield  {author} {\bibinfo {author} {\bibfnamefont {H.}~\bibnamefont
  {Na}}, \bibinfo {author} {\bibfnamefont {C.~M.}\ \bibnamefont {Bouchard}},
  \bibinfo {author} {\bibfnamefont {G.~P.}\ \bibnamefont {Lepage}}, \bibinfo
  {author} {\bibfnamefont {C.}~\bibnamefont {Monahan}},\ and\ \bibinfo {author}
  {\bibfnamefont {J.}~\bibnamefont {Shigemitsu}} (\bibinfo {collaboration}
  {HPQCD}),\ }\bibfield  {title} {\bibinfo {title} {{$B \rightarrow D l \nu$
  form factors at nonzero recoil and extraction of $|V_{cb}|$}},\ }\href
  {https://doi.org/10.1103/PhysRevD.93.119906} {\bibfield  {journal} {\bibinfo
  {journal} {Phys. Rev. D}\ }\textbf {\bibinfo {volume} {92}},\ \bibinfo
  {pages} {054510} (\bibinfo {year} {2015})},\ \bibinfo {note} {[Erratum:
  Phys.Rev.D 93, 119906 (2016)]},\ \Eprint {https://arxiv.org/abs/1505.03925}
  {arXiv:1505.03925 [hep-lat]} \BibitemShut {NoStop}%
\bibitem [{\citenamefont {El-Khadra}\ \emph {et~al.}(1997)\citenamefont
  {El-Khadra}, \citenamefont {Kronfeld},\ and\ \citenamefont
  {Mackenzie}}]{El-Khadra:1996wdx}%
  \BibitemOpen
  \bibfield  {author} {\bibinfo {author} {\bibfnamefont {A.~X.}\ \bibnamefont
  {El-Khadra}}, \bibinfo {author} {\bibfnamefont {A.~S.}\ \bibnamefont
  {Kronfeld}},\ and\ \bibinfo {author} {\bibfnamefont {P.~B.}\ \bibnamefont
  {Mackenzie}},\ }\bibfield  {title} {\bibinfo {title} {{Massive fermions in
  lattice gauge theory}},\ }\href {https://doi.org/10.1103/PhysRevD.55.3933}
  {\bibfield  {journal} {\bibinfo  {journal} {Phys. Rev. D}\ }\textbf {\bibinfo
  {volume} {55}},\ \bibinfo {pages} {3933} (\bibinfo {year} {1997})},\ \Eprint
  {https://arxiv.org/abs/hep-lat/9604004} {arXiv:hep-lat/9604004} \BibitemShut
  {NoStop}%
\bibitem [{\citenamefont {Christ}\ \emph {et~al.}(2007)\citenamefont {Christ},
  \citenamefont {Li},\ and\ \citenamefont {Lin}}]{Christ:2006us}%
  \BibitemOpen
  \bibfield  {author} {\bibinfo {author} {\bibfnamefont {N.~H.}\ \bibnamefont
  {Christ}}, \bibinfo {author} {\bibfnamefont {M.}~\bibnamefont {Li}},\ and\
  \bibinfo {author} {\bibfnamefont {H.-W.}\ \bibnamefont {Lin}},\ }\bibfield
  {title} {\bibinfo {title} {{Relativistic Heavy Quark Effective Action}},\
  }\href {https://doi.org/10.1103/PhysRevD.76.074505} {\bibfield  {journal}
  {\bibinfo  {journal} {Phys. Rev. D}\ }\textbf {\bibinfo {volume} {76}},\
  \bibinfo {pages} {074505} (\bibinfo {year} {2007})},\ \Eprint
  {https://arxiv.org/abs/hep-lat/0608006} {arXiv:hep-lat/0608006} \BibitemShut
  {NoStop}%
\bibitem [{\citenamefont {Oktay}\ and\ \citenamefont
  {Kronfeld}(2008)}]{Oktay:2008ex}%
  \BibitemOpen
  \bibfield  {author} {\bibinfo {author} {\bibfnamefont {M.~B.}\ \bibnamefont
  {Oktay}}\ and\ \bibinfo {author} {\bibfnamefont {A.~S.}\ \bibnamefont
  {Kronfeld}},\ }\bibfield  {title} {\bibinfo {title} {{New lattice action for
  heavy quarks}},\ }\href {https://doi.org/10.1103/PhysRevD.78.014504}
  {\bibfield  {journal} {\bibinfo  {journal} {Phys. Rev. D}\ }\textbf {\bibinfo
  {volume} {78}},\ \bibinfo {pages} {014504} (\bibinfo {year} {2008})},\
  \Eprint {https://arxiv.org/abs/0803.0523} {arXiv:0803.0523 [hep-lat]}
  \BibitemShut {NoStop}%
\bibitem [{\citenamefont {Heitger}\ and\ \citenamefont
  {Sommer}(2004)}]{Heitger:2003nj}%
  \BibitemOpen
  \bibfield  {author} {\bibinfo {author} {\bibfnamefont {J.}~\bibnamefont
  {Heitger}}\ and\ \bibinfo {author} {\bibfnamefont {R.}~\bibnamefont {Sommer}}
  (\bibinfo {collaboration} {ALPHA}),\ }\bibfield  {title} {\bibinfo {title}
  {{Nonperturbative heavy quark effective theory}},\ }\href
  {https://doi.org/10.1088/1126-6708/2004/02/022} {\bibfield  {journal}
  {\bibinfo  {journal} {JHEP}\ }\textbf {\bibinfo {volume} {02}},\ \bibinfo
  {pages} {022}},\ \Eprint {https://arxiv.org/abs/hep-lat/0310035}
  {arXiv:hep-lat/0310035} \BibitemShut {NoStop}%
\bibitem [{\citenamefont {Blossier}\ \emph {et~al.}(2010)\citenamefont
  {Blossier} \emph {et~al.}}]{ETM:2009sed}%
  \BibitemOpen
  \bibfield  {author} {\bibinfo {author} {\bibfnamefont {B.}~\bibnamefont
  {Blossier}} \emph {et~al.} (\bibinfo {collaboration} {ETM}),\ }\bibfield
  {title} {\bibinfo {title} {{A Proposal for B-physics on current lattices}},\
  }\href {https://doi.org/10.1007/JHEP04(2010)049} {\bibfield  {journal}
  {\bibinfo  {journal} {JHEP}\ }\textbf {\bibinfo {volume} {04}},\ \bibinfo
  {pages} {049}},\ \Eprint {https://arxiv.org/abs/0909.3187} {arXiv:0909.3187
  [hep-lat]} \BibitemShut {NoStop}%
\bibitem [{\citenamefont {Parrott}\ \emph {et~al.}(2021)\citenamefont
  {Parrott}, \citenamefont {Bouchard}, \citenamefont {Davies},\ and\
  \citenamefont {Hatton}}]{Parrott:2020vbe}%
  \BibitemOpen
  \bibfield  {author} {\bibinfo {author} {\bibfnamefont {W.~G.}\ \bibnamefont
  {Parrott}}, \bibinfo {author} {\bibfnamefont {C.}~\bibnamefont {Bouchard}},
  \bibinfo {author} {\bibfnamefont {C.~T.~H.}\ \bibnamefont {Davies}},\ and\
  \bibinfo {author} {\bibfnamefont {D.}~\bibnamefont {Hatton}},\ }\bibfield
  {title} {\bibinfo {title} {{Toward accurate form factors for $B$-to-light
  meson decay from lattice QCD}},\ }\href
  {https://doi.org/10.1103/PhysRevD.103.094506} {\bibfield  {journal} {\bibinfo
   {journal} {Phys. Rev. D}\ }\textbf {\bibinfo {volume} {103}},\ \bibinfo
  {pages} {094506} (\bibinfo {year} {2021})},\ \Eprint
  {https://arxiv.org/abs/2010.07980} {arXiv:2010.07980 [hep-lat]} \BibitemShut
  {NoStop}%
\bibitem [{\citenamefont {Colquhoun}\ \emph {et~al.}(2022)\citenamefont
  {Colquhoun}, \citenamefont {Hashimoto}, \citenamefont {Kaneko},\ and\
  \citenamefont {Koponen}}]{Colquhoun:2022atw}%
  \BibitemOpen
  \bibfield  {author} {\bibinfo {author} {\bibfnamefont {B.}~\bibnamefont
  {Colquhoun}}, \bibinfo {author} {\bibfnamefont {S.}~\bibnamefont
  {Hashimoto}}, \bibinfo {author} {\bibfnamefont {T.}~\bibnamefont {Kaneko}},\
  and\ \bibinfo {author} {\bibfnamefont {J.}~\bibnamefont {Koponen}} (\bibinfo
  {collaboration} {JLQCD}),\ }\bibfield  {title} {\bibinfo {title} {{Form
  factors of
  B\textrightarrow{}\ensuremath{\pi}\ensuremath{\ell}\ensuremath{\nu} and a
  determination of |Vub| with M\"obius domain-wall fermions}},\ }\href
  {https://doi.org/10.1103/PhysRevD.106.054502} {\bibfield  {journal} {\bibinfo
   {journal} {Phys. Rev. D}\ }\textbf {\bibinfo {volume} {106}},\ \bibinfo
  {pages} {054502} (\bibinfo {year} {2022})},\ \Eprint
  {https://arxiv.org/abs/2203.04938} {arXiv:2203.04938 [hep-lat]} \BibitemShut
  {NoStop}%
\bibitem [{\citenamefont {Lepage}\ \emph {et~al.}(1992)\citenamefont {Lepage},
  \citenamefont {Magnea}, \citenamefont {Nakhleh}, \citenamefont {Magnea},\
  and\ \citenamefont {Hornbostel}}]{Lepage:1992tx}%
  \BibitemOpen
  \bibfield  {author} {\bibinfo {author} {\bibfnamefont {G.~P.}\ \bibnamefont
  {Lepage}}, \bibinfo {author} {\bibfnamefont {L.}~\bibnamefont {Magnea}},
  \bibinfo {author} {\bibfnamefont {C.}~\bibnamefont {Nakhleh}}, \bibinfo
  {author} {\bibfnamefont {U.}~\bibnamefont {Magnea}},\ and\ \bibinfo {author}
  {\bibfnamefont {K.}~\bibnamefont {Hornbostel}},\ }\bibfield  {title}
  {\bibinfo {title} {{Improved nonrelativistic QCD for heavy quark physics}},\
  }\href {https://doi.org/10.1103/PhysRevD.46.4052} {\bibfield  {journal}
  {\bibinfo  {journal} {Phys. Rev. D}\ }\textbf {\bibinfo {volume} {46}},\
  \bibinfo {pages} {4052} (\bibinfo {year} {1992})},\ \Eprint
  {https://arxiv.org/abs/hep-lat/9205007} {arXiv:hep-lat/9205007} \BibitemShut
  {NoStop}%
\bibitem [{\citenamefont {Meinel}(2010)}]{Meinel:2010pv}%
  \BibitemOpen
  \bibfield  {author} {\bibinfo {author} {\bibfnamefont {S.}~\bibnamefont
  {Meinel}},\ }\bibfield  {title} {\bibinfo {title} {{Bottomonium spectrum at
  order $v^6$ from domain-wall lattice QCD: Precise results for hyperfine
  splittings}},\ }\href {https://doi.org/10.1103/PhysRevD.82.114502} {\bibfield
   {journal} {\bibinfo  {journal} {Phys. Rev. D}\ }\textbf {\bibinfo {volume}
  {82}},\ \bibinfo {pages} {114502} (\bibinfo {year} {2010})},\ \Eprint
  {https://arxiv.org/abs/1007.3966} {arXiv:1007.3966 [hep-lat]} \BibitemShut
  {NoStop}%
\bibitem [{\citenamefont {Davies}\ \emph {et~al.}(2019)\citenamefont {Davies},
  \citenamefont {Harrison}, \citenamefont {Hughes}, \citenamefont {Horgan},
  \citenamefont {von Hippel},\ and\ \citenamefont {Wingate}}]{Davies:2018fwg}%
  \BibitemOpen
  \bibfield  {author} {\bibinfo {author} {\bibfnamefont {C.~T.~H.}\
  \bibnamefont {Davies}}, \bibinfo {author} {\bibfnamefont {J.}~\bibnamefont
  {Harrison}}, \bibinfo {author} {\bibfnamefont {C.}~\bibnamefont {Hughes}},
  \bibinfo {author} {\bibfnamefont {R.~R.}\ \bibnamefont {Horgan}}, \bibinfo
  {author} {\bibfnamefont {G.~M.}\ \bibnamefont {von Hippel}},\ and\ \bibinfo
  {author} {\bibfnamefont {M.}~\bibnamefont {Wingate}},\ }\bibfield  {title}
  {\bibinfo {title} {{Improving the kinetic couplings in lattice
  nonrelativistic QCD}},\ }\href {https://doi.org/10.1103/PhysRevD.99.054502}
  {\bibfield  {journal} {\bibinfo  {journal} {Phys. Rev. D}\ }\textbf {\bibinfo
  {volume} {99}},\ \bibinfo {pages} {054502} (\bibinfo {year} {2019})},\
  \Eprint {https://arxiv.org/abs/1812.11639} {arXiv:1812.11639 [hep-lat]}
  \BibitemShut {NoStop}%
\bibitem [{\citenamefont {Lepage}\ and\ \citenamefont
  {Mackenzie}(1993)}]{Lepage:1992xa}%
  \BibitemOpen
  \bibfield  {author} {\bibinfo {author} {\bibfnamefont {G.~P.}\ \bibnamefont
  {Lepage}}\ and\ \bibinfo {author} {\bibfnamefont {P.~B.}\ \bibnamefont
  {Mackenzie}},\ }\bibfield  {title} {\bibinfo {title} {{On the viability of
  lattice perturbation theory}},\ }\href
  {https://doi.org/10.1103/PhysRevD.48.2250} {\bibfield  {journal} {\bibinfo
  {journal} {Phys. Rev. D}\ }\textbf {\bibinfo {volume} {48}},\ \bibinfo
  {pages} {2250} (\bibinfo {year} {1993})},\ \Eprint
  {https://arxiv.org/abs/hep-lat/9209022} {arXiv:hep-lat/9209022} \BibitemShut
  {NoStop}%
\bibitem [{\citenamefont {Lepage}(1998)}]{Lepage:1997id}%
  \BibitemOpen
  \bibfield  {author} {\bibinfo {author} {\bibfnamefont {P.}~\bibnamefont
  {Lepage}},\ }\bibfield  {title} {\bibinfo {title} {{Perturbative improvement
  for lattice QCD: An Update}},\ }\href
  {https://doi.org/10.1016/S0920-5632(97)00489-1} {\bibfield  {journal}
  {\bibinfo  {journal} {Nucl. Phys. B Proc. Suppl.}\ }\textbf {\bibinfo
  {volume} {60}},\ \bibinfo {pages} {267} (\bibinfo {year} {1998})},\ \Eprint
  {https://arxiv.org/abs/hep-lat/9707026} {arXiv:hep-lat/9707026} \BibitemShut
  {NoStop}%
\bibitem [{\citenamefont {Morningstar}(1994)}]{Morningstar:1994qe}%
  \BibitemOpen
  \bibfield  {author} {\bibinfo {author} {\bibfnamefont {C.~J.}\ \bibnamefont
  {Morningstar}},\ }\bibfield  {title} {\bibinfo {title} {{Radiative
  corrections to the kinetic couplings in nonrelativistic lattice QCD}},\
  }\href {https://doi.org/10.1103/PhysRevD.50.5902} {\bibfield  {journal}
  {\bibinfo  {journal} {Phys. Rev. D}\ }\textbf {\bibinfo {volume} {50}},\
  \bibinfo {pages} {5902} (\bibinfo {year} {1994})},\ \Eprint
  {https://arxiv.org/abs/hep-lat/9406002} {arXiv:hep-lat/9406002} \BibitemShut
  {NoStop}%
\bibitem [{\citenamefont {Morningstar}(1993)}]{Morningstar:1993de}%
  \BibitemOpen
  \bibfield  {author} {\bibinfo {author} {\bibfnamefont {C.~J.}\ \bibnamefont
  {Morningstar}},\ }\bibfield  {title} {\bibinfo {title} {{The Heavy quark
  selfenergy in nonrelativistic lattice QCD}},\ }\href
  {https://doi.org/10.1103/PhysRevD.48.2265} {\bibfield  {journal} {\bibinfo
  {journal} {Phys. Rev. D}\ }\textbf {\bibinfo {volume} {48}},\ \bibinfo
  {pages} {2265} (\bibinfo {year} {1993})},\ \Eprint
  {https://arxiv.org/abs/hep-lat/9301005} {arXiv:hep-lat/9301005} \BibitemShut
  {NoStop}%
\bibitem [{\citenamefont {Mathur}\ \emph {et~al.}(2002)\citenamefont {Mathur},
  \citenamefont {Lewis},\ and\ \citenamefont {Woloshyn}}]{Mathur:2002ce}%
  \BibitemOpen
  \bibfield  {author} {\bibinfo {author} {\bibfnamefont {N.}~\bibnamefont
  {Mathur}}, \bibinfo {author} {\bibfnamefont {R.}~\bibnamefont {Lewis}},\ and\
  \bibinfo {author} {\bibfnamefont {R.~M.}\ \bibnamefont {Woloshyn}},\
  }\bibfield  {title} {\bibinfo {title} {{Charmed and bottom baryons from
  lattice NRQCD}},\ }\href {https://doi.org/10.1103/PhysRevD.66.014502}
  {\bibfield  {journal} {\bibinfo  {journal} {Phys. Rev. D}\ }\textbf {\bibinfo
  {volume} {66}},\ \bibinfo {pages} {014502} (\bibinfo {year} {2002})},\
  \Eprint {https://arxiv.org/abs/hep-ph/0203253} {arXiv:hep-ph/0203253}
  \BibitemShut {NoStop}%
\bibitem [{Note1()}]{Note1}%
  \BibitemOpen
  \bibinfo {note} {As we will use a mix of periodic and open boundary condition
  ensembles in time, we will look at a single, coarse, periodic box to do this
  comparison.}\BibitemShut {Stop}%
\bibitem [{\citenamefont {Dowdall}\ \emph {et~al.}(2012)\citenamefont
  {Dowdall}, \citenamefont {Davies}, \citenamefont {Hammant},\ and\
  \citenamefont {Horgan}}]{Dowdall:2012ab}%
  \BibitemOpen
  \bibfield  {author} {\bibinfo {author} {\bibfnamefont {R.~J.}\ \bibnamefont
  {Dowdall}}, \bibinfo {author} {\bibfnamefont {C.~T.~H.}\ \bibnamefont
  {Davies}}, \bibinfo {author} {\bibfnamefont {T.~C.}\ \bibnamefont
  {Hammant}},\ and\ \bibinfo {author} {\bibfnamefont {R.~R.}\ \bibnamefont
  {Horgan}},\ }\bibfield  {title} {\bibinfo {title} {{Precise heavy-light meson
  masses and hyperfine splittings from lattice QCD including charm quarks in
  the sea}},\ }\href {https://doi.org/10.1103/PhysRevD.86.094510} {\bibfield
  {journal} {\bibinfo  {journal} {Phys. Rev. D}\ }\textbf {\bibinfo {volume}
  {86}},\ \bibinfo {pages} {094510} (\bibinfo {year} {2012})},\ \Eprint
  {https://arxiv.org/abs/1207.5149} {arXiv:1207.5149 [hep-lat]} \BibitemShut
  {NoStop}%
\bibitem [{\citenamefont {Manohar}(1997)}]{Manohar:1997qy}%
  \BibitemOpen
  \bibfield  {author} {\bibinfo {author} {\bibfnamefont {A.~V.}\ \bibnamefont
  {Manohar}},\ }\bibfield  {title} {\bibinfo {title} {{The HQET / NRQCD
  Lagrangian to order alpha / m-3}},\ }\href
  {https://doi.org/10.1103/PhysRevD.56.230} {\bibfield  {journal} {\bibinfo
  {journal} {Phys. Rev. D}\ }\textbf {\bibinfo {volume} {56}},\ \bibinfo
  {pages} {230} (\bibinfo {year} {1997})},\ \Eprint
  {https://arxiv.org/abs/hep-ph/9701294} {arXiv:hep-ph/9701294} \BibitemShut
  {NoStop}%
\bibitem [{\citenamefont {Lewis}\ and\ \citenamefont
  {Woloshyn}(1998)}]{Lewis:1998ka}%
  \BibitemOpen
  \bibfield  {author} {\bibinfo {author} {\bibfnamefont {R.}~\bibnamefont
  {Lewis}}\ and\ \bibinfo {author} {\bibfnamefont {R.~M.}\ \bibnamefont
  {Woloshyn}},\ }\bibfield  {title} {\bibinfo {title} {{O(1 / M**3) effects for
  heavy - light mesons in lattice NRQCD}},\ }\href
  {https://doi.org/10.1103/PhysRevD.58.074506} {\bibfield  {journal} {\bibinfo
  {journal} {Phys. Rev. D}\ }\textbf {\bibinfo {volume} {58}},\ \bibinfo
  {pages} {074506} (\bibinfo {year} {1998})},\ \Eprint
  {https://arxiv.org/abs/hep-lat/9803004} {arXiv:hep-lat/9803004} \BibitemShut
  {NoStop}%
\bibitem [{\citenamefont {Hughes}\ \emph {et~al.}(2015)\citenamefont {Hughes},
  \citenamefont {Dowdall}, \citenamefont {Davies}, \citenamefont {Horgan},
  \citenamefont {von Hippel},\ and\ \citenamefont {Wingate}}]{Hughes:2015dba}%
  \BibitemOpen
  \bibfield  {author} {\bibinfo {author} {\bibfnamefont {C.}~\bibnamefont
  {Hughes}}, \bibinfo {author} {\bibfnamefont {R.~J.}\ \bibnamefont {Dowdall}},
  \bibinfo {author} {\bibfnamefont {C.~T.~H.}\ \bibnamefont {Davies}}, \bibinfo
  {author} {\bibfnamefont {R.~R.}\ \bibnamefont {Horgan}}, \bibinfo {author}
  {\bibfnamefont {G.}~\bibnamefont {von Hippel}},\ and\ \bibinfo {author}
  {\bibfnamefont {M.}~\bibnamefont {Wingate}},\ }\bibfield  {title} {\bibinfo
  {title} {{Hindered M1 Radiative Decay of $\Upsilon(2S)$ from Lattice
  NRQCD}},\ }\href {https://doi.org/10.1103/PhysRevD.92.094501} {\bibfield
  {journal} {\bibinfo  {journal} {Phys. Rev. D}\ }\textbf {\bibinfo {volume}
  {92}},\ \bibinfo {pages} {094501} (\bibinfo {year} {2015})},\ \Eprint
  {https://arxiv.org/abs/1508.01694} {arXiv:1508.01694 [hep-lat]} \BibitemShut
  {NoStop}%
\bibitem [{\citenamefont {Meinel}(2012)}]{Meinel:2012qz}%
  \BibitemOpen
  \bibfield  {author} {\bibinfo {author} {\bibfnamefont {S.}~\bibnamefont
  {Meinel}},\ }\bibfield  {title} {\bibinfo {title} {{Excited-state
  spectroscopy of triply-bottom baryons from lattice QCD}},\ }\href
  {https://doi.org/10.1103/PhysRevD.85.114510} {\bibfield  {journal} {\bibinfo
  {journal} {Phys. Rev. D}\ }\textbf {\bibinfo {volume} {85}},\ \bibinfo
  {pages} {114510} (\bibinfo {year} {2012})},\ \Eprint
  {https://arxiv.org/abs/1202.1312} {arXiv:1202.1312 [hep-lat]} \BibitemShut
  {NoStop}%
\bibitem [{\citenamefont {Hammant}\ \emph {et~al.}(2011)\citenamefont
  {Hammant}, \citenamefont {Hart}, \citenamefont {von Hippel}, \citenamefont
  {Horgan},\ and\ \citenamefont {Monahan}}]{Hammant:2011bt}%
  \BibitemOpen
  \bibfield  {author} {\bibinfo {author} {\bibfnamefont {T.~C.}\ \bibnamefont
  {Hammant}}, \bibinfo {author} {\bibfnamefont {A.~G.}\ \bibnamefont {Hart}},
  \bibinfo {author} {\bibfnamefont {G.~M.}\ \bibnamefont {von Hippel}},
  \bibinfo {author} {\bibfnamefont {R.~R.}\ \bibnamefont {Horgan}},\ and\
  \bibinfo {author} {\bibfnamefont {C.~J.}\ \bibnamefont {Monahan}},\
  }\bibfield  {title} {\bibinfo {title} {{Radiative improvement of the lattice
  NRQCD action using the background field method and application to the
  hyperfine splitting of quarkonium states}},\ }\href
  {https://doi.org/10.1103/PhysRevLett.107.112002} {\bibfield  {journal}
  {\bibinfo  {journal} {Phys. Rev. Lett.}\ }\textbf {\bibinfo {volume} {107}},\
  \bibinfo {pages} {112002} (\bibinfo {year} {2011})},\ \bibinfo {note}
  {[Erratum: Phys.Rev.Lett. 115, 039901 (2015)]},\ \Eprint
  {https://arxiv.org/abs/1105.5309} {arXiv:1105.5309 [hep-lat]} \BibitemShut
  {NoStop}%
\bibitem [{\citenamefont {Hammant}\ \emph {et~al.}(2013)\citenamefont
  {Hammant}, \citenamefont {Hart}, \citenamefont {von Hippel}, \citenamefont
  {Horgan},\ and\ \citenamefont {Monahan}}]{Hammant:2013sca}%
  \BibitemOpen
  \bibfield  {author} {\bibinfo {author} {\bibfnamefont {T.~C.}\ \bibnamefont
  {Hammant}}, \bibinfo {author} {\bibfnamefont {A.~G.}\ \bibnamefont {Hart}},
  \bibinfo {author} {\bibfnamefont {G.~M.}\ \bibnamefont {von Hippel}},
  \bibinfo {author} {\bibfnamefont {R.~R.}\ \bibnamefont {Horgan}},\ and\
  \bibinfo {author} {\bibfnamefont {C.~J.}\ \bibnamefont {Monahan}},\
  }\bibfield  {title} {\bibinfo {title} {{Radiative improvement of the lattice
  nonrelativistic QCD action using the background field method with
  applications to quarkonium spectroscopy}},\ }\href
  {https://doi.org/10.1103/PhysRevD.88.014505} {\bibfield  {journal} {\bibinfo
  {journal} {Phys. Rev. D}\ }\textbf {\bibinfo {volume} {88}},\ \bibinfo
  {pages} {014505} (\bibinfo {year} {2013})},\ \bibinfo {note} {[Erratum:
  Phys.Rev.D 92, 119904 (2015)]},\ \Eprint {https://arxiv.org/abs/1303.3234}
  {arXiv:1303.3234 [hep-lat]} \BibitemShut {NoStop}%
\bibitem [{\citenamefont {Bedaque}(2004)}]{Bedaque:2004kc}%
  \BibitemOpen
  \bibfield  {author} {\bibinfo {author} {\bibfnamefont {P.~F.}\ \bibnamefont
  {Bedaque}},\ }\bibfield  {title} {\bibinfo {title} {{Aharonov-Bohm effect and
  nucleon nucleon phase shifts on the lattice}},\ }\href
  {https://doi.org/10.1016/j.physletb.2004.04.045} {\bibfield  {journal}
  {\bibinfo  {journal} {Phys. Lett. B}\ }\textbf {\bibinfo {volume} {593}},\
  \bibinfo {pages} {82} (\bibinfo {year} {2004})},\ \Eprint
  {https://arxiv.org/abs/nucl-th/0402051} {arXiv:nucl-th/0402051} \BibitemShut
  {NoStop}%
\bibitem [{\citenamefont {Sachrajda}\ and\ \citenamefont
  {Villadoro}(2005)}]{Sachrajda:2004mi}%
  \BibitemOpen
  \bibfield  {author} {\bibinfo {author} {\bibfnamefont {C.~T.}\ \bibnamefont
  {Sachrajda}}\ and\ \bibinfo {author} {\bibfnamefont {G.}~\bibnamefont
  {Villadoro}},\ }\bibfield  {title} {\bibinfo {title} {{Twisted boundary
  conditions in lattice simulations}},\ }\href
  {https://doi.org/10.1016/j.physletb.2005.01.033} {\bibfield  {journal}
  {\bibinfo  {journal} {Phys. Lett. B}\ }\textbf {\bibinfo {volume} {609}},\
  \bibinfo {pages} {73} (\bibinfo {year} {2005})},\ \Eprint
  {https://arxiv.org/abs/hep-lat/0411033} {arXiv:hep-lat/0411033} \BibitemShut
  {NoStop}%
\bibitem [{\citenamefont {Davies}\ \emph
  {et~al.}(1994{\natexlab{b}})\citenamefont {Davies}, \citenamefont
  {Hornbostel}, \citenamefont {Langnau}, \citenamefont {Lepage}, \citenamefont
  {Lidsey}, \citenamefont {Shigemitsu},\ and\ \citenamefont
  {Sloan}}]{Davies:1994mp}%
  \BibitemOpen
  \bibfield  {author} {\bibinfo {author} {\bibfnamefont {C.~T.~H.}\
  \bibnamefont {Davies}}, \bibinfo {author} {\bibfnamefont {K.}~\bibnamefont
  {Hornbostel}}, \bibinfo {author} {\bibfnamefont {A.}~\bibnamefont {Langnau}},
  \bibinfo {author} {\bibfnamefont {G.~P.}\ \bibnamefont {Lepage}}, \bibinfo
  {author} {\bibfnamefont {A.}~\bibnamefont {Lidsey}}, \bibinfo {author}
  {\bibfnamefont {J.}~\bibnamefont {Shigemitsu}},\ and\ \bibinfo {author}
  {\bibfnamefont {J.~H.}\ \bibnamefont {Sloan}},\ }\bibfield  {title} {\bibinfo
  {title} {{Precision Upsilon spectroscopy from nonrelativistic lattice QCD}},\
  }\href {https://doi.org/10.1103/PhysRevD.50.6963} {\bibfield  {journal}
  {\bibinfo  {journal} {Phys. Rev. D}\ }\textbf {\bibinfo {volume} {50}},\
  \bibinfo {pages} {6963} (\bibinfo {year} {1994}{\natexlab{b}})},\ \Eprint
  {https://arxiv.org/abs/hep-lat/9406017} {arXiv:hep-lat/9406017} \BibitemShut
  {NoStop}%
\bibitem [{Note2()}]{Note2}%
  \BibitemOpen
  \bibinfo {note} {Fixed to a precision of $10^{-14}$ using the FACG algorithm
  \cite {Hudspith:2014oja}.}\BibitemShut {Stop}%
\bibitem [{\citenamefont {Zyla}\ \emph {et~al.}(2020)\citenamefont {Zyla} \emph
  {et~al.}}]{Zyla:2020zbs}%
  \BibitemOpen
  \bibfield  {author} {\bibinfo {author} {\bibfnamefont {P.}~\bibnamefont
  {Zyla}} \emph {et~al.} (\bibinfo {collaboration} {Particle Data Group}),\
  }\bibfield  {title} {\bibinfo {title} {{Review of Particle Physics}},\ }\href
  {https://doi.org/10.1093/ptep/ptaa104} {\bibfield  {journal} {\bibinfo
  {journal} {PTEP}\ }\textbf {\bibinfo {volume} {2020}},\ \bibinfo {pages}
  {083C01} (\bibinfo {year} {2020})}\BibitemShut {NoStop}%
\bibitem [{\citenamefont {Davies}\ \emph {et~al.}(2004)\citenamefont {Davies}
  \emph {et~al.}}]{HPQCD:2003rsu}%
  \BibitemOpen
  \bibfield  {author} {\bibinfo {author} {\bibfnamefont {C.~T.~H.}\
  \bibnamefont {Davies}} \emph {et~al.} (\bibinfo {collaboration} {HPQCD,
  UKQCD, MILC, Fermilab Lattice}),\ }\bibfield  {title} {\bibinfo {title}
  {{High precision lattice QCD confronts experiment}},\ }\href
  {https://doi.org/10.1103/PhysRevLett.92.022001} {\bibfield  {journal}
  {\bibinfo  {journal} {Phys. Rev. Lett.}\ }\textbf {\bibinfo {volume} {92}},\
  \bibinfo {pages} {022001} (\bibinfo {year} {2004})},\ \Eprint
  {https://arxiv.org/abs/hep-lat/0304004} {arXiv:hep-lat/0304004} \BibitemShut
  {NoStop}%
\bibitem [{\citenamefont {Meinel}(2009)}]{Meinel:2009rd}%
  \BibitemOpen
  \bibfield  {author} {\bibinfo {author} {\bibfnamefont {S.}~\bibnamefont
  {Meinel}},\ }\bibfield  {title} {\bibinfo {title} {{The Bottomonium spectrum
  from lattice QCD with 2+1 flavors of domain wall fermions}},\ }\href
  {https://doi.org/10.1103/PhysRevD.79.094501} {\bibfield  {journal} {\bibinfo
  {journal} {Phys. Rev. D}\ }\textbf {\bibinfo {volume} {79}},\ \bibinfo
  {pages} {094501} (\bibinfo {year} {2009})},\ \Eprint
  {https://arxiv.org/abs/0903.3224} {arXiv:0903.3224 [hep-lat]} \BibitemShut
  {NoStop}%
\bibitem [{\citenamefont {Hudspith}\ and\ \citenamefont
  {Mohler}(2021)}]{Hudspith:2021iqu}%
  \BibitemOpen
  \bibfield  {author} {\bibinfo {author} {\bibfnamefont {R.~J.}\ \bibnamefont
  {Hudspith}}\ and\ \bibinfo {author} {\bibfnamefont {D.}~\bibnamefont
  {Mohler}},\ }\bibfield  {title} {\bibinfo {title} {{A fully non-perturbative
  charm-quark tuning using machine learning}},\ }\href@noop {} {\bibfield
  {journal} {\bibinfo  {journal} {Arxiv}\ } (\bibinfo {year} {2021})},\ \Eprint
  {https://arxiv.org/abs/2112.01997} {arXiv:2112.01997 [hep-lat]} \BibitemShut
  {NoStop}%
\bibitem [{\citenamefont {Kingma}\ and\ \citenamefont
  {Ba}(2014)}]{kingma2014method}%
  \BibitemOpen
  \bibfield  {author} {\bibinfo {author} {\bibfnamefont {D.~P.}\ \bibnamefont
  {Kingma}}\ and\ \bibinfo {author} {\bibfnamefont {J.}~\bibnamefont {Ba}},\
  }\href {http://arxiv.org/abs/1412.6980} {\bibinfo {title} {Adam: A method for
  stochastic optimization}} (\bibinfo {year} {2014}),\ \bibinfo {note} {cite
  arxiv:1412.6980, Published as a conference paper at the 3rd International
  Conference for Learning Representations, San Diego, 2015}\BibitemShut
  {NoStop}%
\bibitem [{Note3()}]{Note3}%
  \BibitemOpen
  \bibinfo {note} {This was tested on our lightest pion-mass ensemble (C101) to
  be the case.}\BibitemShut {Stop}%
\bibitem [{\citenamefont {Luscher}\ and\ \citenamefont
  {Schaefer}(2013)}]{Luscher:2012av}%
  \BibitemOpen
  \bibfield  {author} {\bibinfo {author} {\bibfnamefont {M.}~\bibnamefont
  {Luscher}}\ and\ \bibinfo {author} {\bibfnamefont {S.}~\bibnamefont
  {Schaefer}},\ }\bibfield  {title} {\bibinfo {title} {{Lattice QCD with open
  boundary conditions and twisted-mass reweighting}},\ }\href
  {https://doi.org/10.1016/j.cpc.2012.10.003} {\bibfield  {journal} {\bibinfo
  {journal} {Comput. Phys. Commun.}\ }\textbf {\bibinfo {volume} {184}},\
  \bibinfo {pages} {519} (\bibinfo {year} {2013})},\ \Eprint
  {https://arxiv.org/abs/1206.2809} {arXiv:1206.2809 [hep-lat]} \BibitemShut
  {NoStop}%
\bibitem [{\citenamefont {Bruno}\ \emph {et~al.}(2015)\citenamefont {Bruno}
  \emph {et~al.}}]{Bruno:2014jqa}%
  \BibitemOpen
  \bibfield  {author} {\bibinfo {author} {\bibfnamefont {M.}~\bibnamefont
  {Bruno}} \emph {et~al.},\ }\bibfield  {title} {\bibinfo {title} {{Simulation
  of QCD with N$_{f} =$ 2 $+$ 1 flavors of non-perturbatively improved Wilson
  fermions}},\ }\href {https://doi.org/10.1007/JHEP02(2015)043} {\bibfield
  {journal} {\bibinfo  {journal} {JHEP}\ }\textbf {\bibinfo {volume} {02}},\
  \bibinfo {pages} {043}},\ \Eprint {https://arxiv.org/abs/1411.3982}
  {arXiv:1411.3982 [hep-lat]} \BibitemShut {NoStop}%
\bibitem [{\citenamefont {Bruno}\ \emph {et~al.}(2017)\citenamefont {Bruno},
  \citenamefont {Korzec},\ and\ \citenamefont {Schaefer}}]{Bruno:2016plf}%
  \BibitemOpen
  \bibfield  {author} {\bibinfo {author} {\bibfnamefont {M.}~\bibnamefont
  {Bruno}}, \bibinfo {author} {\bibfnamefont {T.}~\bibnamefont {Korzec}},\ and\
  \bibinfo {author} {\bibfnamefont {S.}~\bibnamefont {Schaefer}},\ }\bibfield
  {title} {\bibinfo {title} {{Setting the scale for the CLS $2 + 1$ flavor
  ensembles}},\ }\href {https://doi.org/10.1103/PhysRevD.95.074504} {\bibfield
  {journal} {\bibinfo  {journal} {Phys. Rev. D}\ }\textbf {\bibinfo {volume}
  {95}},\ \bibinfo {pages} {074504} (\bibinfo {year} {2017})},\ \Eprint
  {https://arxiv.org/abs/1608.08900} {arXiv:1608.08900 [hep-lat]} \BibitemShut
  {NoStop}%
\bibitem [{\citenamefont {Chao}\ \emph {et~al.}(2021)\citenamefont {Chao},
  \citenamefont {Hudspith}, \citenamefont {G\'erardin}, \citenamefont {Green},
  \citenamefont {Meyer},\ and\ \citenamefont {Ottnad}}]{Chao:2021tvp}%
  \BibitemOpen
  \bibfield  {author} {\bibinfo {author} {\bibfnamefont {E.-H.}\ \bibnamefont
  {Chao}}, \bibinfo {author} {\bibfnamefont {R.~J.}\ \bibnamefont {Hudspith}},
  \bibinfo {author} {\bibfnamefont {A.}~\bibnamefont {G\'erardin}}, \bibinfo
  {author} {\bibfnamefont {J.~R.}\ \bibnamefont {Green}}, \bibinfo {author}
  {\bibfnamefont {H.~B.}\ \bibnamefont {Meyer}},\ and\ \bibinfo {author}
  {\bibfnamefont {K.}~\bibnamefont {Ottnad}},\ }\bibfield  {title} {\bibinfo
  {title} {{Hadronic light-by-light contribution to $(g-2)_\mu $ from lattice
  QCD: a complete calculation}},\ }\href
  {https://doi.org/10.1140/epjc/s10052-021-09455-4} {\bibfield  {journal}
  {\bibinfo  {journal} {Eur. Phys. J. C}\ }\textbf {\bibinfo {volume} {81}},\
  \bibinfo {pages} {651} (\bibinfo {year} {2021})},\ \Eprint
  {https://arxiv.org/abs/2104.02632} {arXiv:2104.02632 [hep-lat]} \BibitemShut
  {NoStop}%
\bibitem [{\citenamefont {Michael}\ and\ \citenamefont
  {Teasdale}(1983)}]{Michael:1982gb}%
  \BibitemOpen
  \bibfield  {author} {\bibinfo {author} {\bibfnamefont {C.}~\bibnamefont
  {Michael}}\ and\ \bibinfo {author} {\bibfnamefont {I.}~\bibnamefont
  {Teasdale}},\ }\bibfield  {title} {\bibinfo {title} {{Extracting Glueball
  Masses From Lattice {QCD}}},\ }\href
  {https://doi.org/10.1016/0550-3213(83)90674-0} {\bibfield  {journal}
  {\bibinfo  {journal} {Nucl. Phys. B}\ }\textbf {\bibinfo {volume} {215}},\
  \bibinfo {pages} {433} (\bibinfo {year} {1983})}\BibitemShut {NoStop}%
\bibitem [{\citenamefont {Luscher}\ and\ \citenamefont
  {Wolff}(1990)}]{Luscher:1990ck}%
  \BibitemOpen
  \bibfield  {author} {\bibinfo {author} {\bibfnamefont {M.}~\bibnamefont
  {Luscher}}\ and\ \bibinfo {author} {\bibfnamefont {U.}~\bibnamefont
  {Wolff}},\ }\bibfield  {title} {\bibinfo {title} {{How to Calculate the
  Elastic Scattering Matrix in Two-dimensional Quantum Field Theories by
  Numerical Simulation}},\ }\href
  {https://doi.org/10.1016/0550-3213(90)90540-T} {\bibfield  {journal}
  {\bibinfo  {journal} {Nucl. Phys. B}\ }\textbf {\bibinfo {volume} {339}},\
  \bibinfo {pages} {222} (\bibinfo {year} {1990})}\BibitemShut {NoStop}%
\bibitem [{\citenamefont {Blossier}\ \emph {et~al.}(2009)\citenamefont
  {Blossier}, \citenamefont {Della~Morte}, \citenamefont {von Hippel},
  \citenamefont {Mendes},\ and\ \citenamefont {Sommer}}]{Blossier:2009kd}%
  \BibitemOpen
  \bibfield  {author} {\bibinfo {author} {\bibfnamefont {B.}~\bibnamefont
  {Blossier}}, \bibinfo {author} {\bibfnamefont {M.}~\bibnamefont
  {Della~Morte}}, \bibinfo {author} {\bibfnamefont {G.}~\bibnamefont {von
  Hippel}}, \bibinfo {author} {\bibfnamefont {T.}~\bibnamefont {Mendes}},\ and\
  \bibinfo {author} {\bibfnamefont {R.}~\bibnamefont {Sommer}},\ }\bibfield
  {title} {\bibinfo {title} {{On the generalized eigenvalue method for energies
  and matrix elements in lattice field theory}},\ }\href
  {https://doi.org/10.1088/1126-6708/2009/04/094} {\bibfield  {journal}
  {\bibinfo  {journal} {JHEP}\ }\textbf {\bibinfo {volume} {04}},\ \bibinfo
  {pages} {094}},\ \Eprint {https://arxiv.org/abs/0902.1265} {arXiv:0902.1265
  [hep-lat]} \BibitemShut {NoStop}%
\bibitem [{\citenamefont {McNeile}\ and\ \citenamefont
  {Michael}(2000)}]{McNeile:2000hf}%
  \BibitemOpen
  \bibfield  {author} {\bibinfo {author} {\bibfnamefont {C.}~\bibnamefont
  {McNeile}}\ and\ \bibinfo {author} {\bibfnamefont {C.}~\bibnamefont
  {Michael}} (\bibinfo {collaboration} {UKQCD}),\ }\bibfield  {title} {\bibinfo
  {title} {{The $\eta$ and $\eta$' mesons in QCD}},\ }\href
  {https://doi.org/10.1016/S0370-2693(00)01010-8} {\bibfield  {journal}
  {\bibinfo  {journal} {Phys. Lett. B}\ }\textbf {\bibinfo {volume} {491}},\
  \bibinfo {pages} {123} (\bibinfo {year} {2000})},\ \bibinfo {note} {[Erratum:
  Phys.Lett.B 551, 391--391 (2003)]},\ \Eprint
  {https://arxiv.org/abs/hep-lat/0006020} {arXiv:hep-lat/0006020} \BibitemShut
  {NoStop}%
\bibitem [{\citenamefont {Bietenholz}\ \emph {et~al.}(2010)\citenamefont
  {Bietenholz} \emph {et~al.}}]{Bietenholz:2010jr}%
  \BibitemOpen
  \bibfield  {author} {\bibinfo {author} {\bibfnamefont {W.}~\bibnamefont
  {Bietenholz}} \emph {et~al.},\ }\bibfield  {title} {\bibinfo {title} {{Tuning
  the strange quark mass in lattice simulations}},\ }\href
  {https://doi.org/10.1016/j.physletb.2010.05.067} {\bibfield  {journal}
  {\bibinfo  {journal} {Phys. Lett. B}\ }\textbf {\bibinfo {volume} {690}},\
  \bibinfo {pages} {436} (\bibinfo {year} {2010})},\ \Eprint
  {https://arxiv.org/abs/1003.1114} {arXiv:1003.1114 [hep-lat]} \BibitemShut
  {NoStop}%
\bibitem [{\citenamefont {Bali}\ \emph {et~al.}(2016)\citenamefont {Bali},
  \citenamefont {Scholz}, \citenamefont {Simeth},\ and\ \citenamefont
  {S\"oldner}}]{Bali:2016umi}%
  \BibitemOpen
  \bibfield  {author} {\bibinfo {author} {\bibfnamefont {G.~S.}\ \bibnamefont
  {Bali}}, \bibinfo {author} {\bibfnamefont {E.~E.}\ \bibnamefont {Scholz}},
  \bibinfo {author} {\bibfnamefont {J.}~\bibnamefont {Simeth}},\ and\ \bibinfo
  {author} {\bibfnamefont {W.}~\bibnamefont {S\"oldner}} (\bibinfo
  {collaboration} {RQCD}),\ }\bibfield  {title} {\bibinfo {title} {{Lattice
  simulations with $N_f=2+1$ improved Wilson fermions at a fixed strange quark
  mass}},\ }\href {https://doi.org/10.1103/PhysRevD.94.074501} {\bibfield
  {journal} {\bibinfo  {journal} {Phys. Rev. D}\ }\textbf {\bibinfo {volume}
  {94}},\ \bibinfo {pages} {074501} (\bibinfo {year} {2016})},\ \Eprint
  {https://arxiv.org/abs/1606.09039} {arXiv:1606.09039 [hep-lat]} \BibitemShut
  {NoStop}%
\bibitem [{\citenamefont {Bali}\ \emph {et~al.}(2022)\citenamefont {Bali},
  \citenamefont {Collins}, \citenamefont {Georg}, \citenamefont {Jenkins},
  \citenamefont {Korcyl}, \citenamefont {Sch\"afer}, \citenamefont {Scholz},
  \citenamefont {Simeth}, \citenamefont {S\"oldner},\ and\ \citenamefont
  {Weish\"aupl}}]{RQCD:2022xux}%
  \BibitemOpen
  \bibfield  {author} {\bibinfo {author} {\bibfnamefont {G.~S.}\ \bibnamefont
  {Bali}}, \bibinfo {author} {\bibfnamefont {S.}~\bibnamefont {Collins}},
  \bibinfo {author} {\bibfnamefont {P.}~\bibnamefont {Georg}}, \bibinfo
  {author} {\bibfnamefont {D.}~\bibnamefont {Jenkins}}, \bibinfo {author}
  {\bibfnamefont {P.}~\bibnamefont {Korcyl}}, \bibinfo {author} {\bibfnamefont
  {A.}~\bibnamefont {Sch\"afer}}, \bibinfo {author} {\bibfnamefont {E.~E.}\
  \bibnamefont {Scholz}}, \bibinfo {author} {\bibfnamefont {J.}~\bibnamefont
  {Simeth}}, \bibinfo {author} {\bibfnamefont {W.}~\bibnamefont {S\"oldner}},\
  and\ \bibinfo {author} {\bibfnamefont {S.}~\bibnamefont {Weish\"aupl}}
  (\bibinfo {collaboration} {RQCD}),\ }\bibfield  {title} {\bibinfo {title}
  {{Scale setting and the light baryon spectrum in $N_f=2+1$ QCD with Wilson
  fermions}},\ }\href@noop {} {\  (\bibinfo {year} {2022})},\ \Eprint
  {https://arxiv.org/abs/2211.03744} {arXiv:2211.03744 [hep-lat]} \BibitemShut
  {NoStop}%
\bibitem [{\citenamefont {Hudspith}\ \emph {et~al.}(2020)\citenamefont
  {Hudspith}, \citenamefont {Colquhoun}, \citenamefont {Francis}, \citenamefont
  {Lewis},\ and\ \citenamefont {Maltman}}]{Hudspith:2020tdf}%
  \BibitemOpen
  \bibfield  {author} {\bibinfo {author} {\bibfnamefont {R.~J.}\ \bibnamefont
  {Hudspith}}, \bibinfo {author} {\bibfnamefont {B.}~\bibnamefont {Colquhoun}},
  \bibinfo {author} {\bibfnamefont {A.}~\bibnamefont {Francis}}, \bibinfo
  {author} {\bibfnamefont {R.}~\bibnamefont {Lewis}},\ and\ \bibinfo {author}
  {\bibfnamefont {K.}~\bibnamefont {Maltman}},\ }\bibfield  {title} {\bibinfo
  {title} {{A lattice investigation of exotic tetraquark channels}},\ }\href
  {https://doi.org/10.1103/PhysRevD.102.114506} {\bibfield  {journal} {\bibinfo
   {journal} {Phys. Rev. D}\ }\textbf {\bibinfo {volume} {102}},\ \bibinfo
  {pages} {114506} (\bibinfo {year} {2020})},\ \Eprint
  {https://arxiv.org/abs/2006.14294} {arXiv:2006.14294 [hep-lat]} \BibitemShut
  {NoStop}%
\bibitem [{\citenamefont {Francis}\ \emph {et~al.}(2019)\citenamefont
  {Francis}, \citenamefont {Green}, \citenamefont {Junnarkar}, \citenamefont
  {Miao}, \citenamefont {Rae},\ and\ \citenamefont {Wittig}}]{Francis:2018qch}%
  \BibitemOpen
  \bibfield  {author} {\bibinfo {author} {\bibfnamefont {A.}~\bibnamefont
  {Francis}}, \bibinfo {author} {\bibfnamefont {J.~R.}\ \bibnamefont {Green}},
  \bibinfo {author} {\bibfnamefont {P.~M.}\ \bibnamefont {Junnarkar}}, \bibinfo
  {author} {\bibfnamefont {C.}~\bibnamefont {Miao}}, \bibinfo {author}
  {\bibfnamefont {T.~D.}\ \bibnamefont {Rae}},\ and\ \bibinfo {author}
  {\bibfnamefont {H.}~\bibnamefont {Wittig}},\ }\bibfield  {title} {\bibinfo
  {title} {{Lattice QCD study of the $H$ dibaryon using hexaquark and
  two-baryon interpolators}},\ }\href
  {https://doi.org/10.1103/PhysRevD.99.074505} {\bibfield  {journal} {\bibinfo
  {journal} {Phys. Rev. D}\ }\textbf {\bibinfo {volume} {99}},\ \bibinfo
  {pages} {074505} (\bibinfo {year} {2019})},\ \Eprint
  {https://arxiv.org/abs/1805.03966} {arXiv:1805.03966 [hep-lat]} \BibitemShut
  {NoStop}%
\bibitem [{\citenamefont {Aubin}\ and\ \citenamefont
  {Orginos}(2011)}]{Aubin:2010jc}%
  \BibitemOpen
  \bibfield  {author} {\bibinfo {author} {\bibfnamefont {C.}~\bibnamefont
  {Aubin}}\ and\ \bibinfo {author} {\bibfnamefont {K.}~\bibnamefont
  {Orginos}},\ }\bibfield  {title} {\bibinfo {title} {{A new approach for Delta
  form factors}},\ }\href {https://doi.org/10.1063/1.3647217} {\bibfield
  {journal} {\bibinfo  {journal} {AIP Conf. Proc.}\ }\textbf {\bibinfo {volume}
  {1374}},\ \bibinfo {pages} {621} (\bibinfo {year} {2011})},\ \Eprint
  {https://arxiv.org/abs/1010.0202} {arXiv:1010.0202 [hep-lat]} \BibitemShut
  {NoStop}%
\bibitem [{\citenamefont {Green}\ \emph {et~al.}(2014)\citenamefont {Green},
  \citenamefont {Negele}, \citenamefont {Pochinsky}, \citenamefont {Syritsyn},
  \citenamefont {Engelhardt},\ and\ \citenamefont {Krieg}}]{Green:2014xba}%
  \BibitemOpen
  \bibfield  {author} {\bibinfo {author} {\bibfnamefont {J.~R.}\ \bibnamefont
  {Green}}, \bibinfo {author} {\bibfnamefont {J.~W.}\ \bibnamefont {Negele}},
  \bibinfo {author} {\bibfnamefont {A.~V.}\ \bibnamefont {Pochinsky}}, \bibinfo
  {author} {\bibfnamefont {S.~N.}\ \bibnamefont {Syritsyn}}, \bibinfo {author}
  {\bibfnamefont {M.}~\bibnamefont {Engelhardt}},\ and\ \bibinfo {author}
  {\bibfnamefont {S.}~\bibnamefont {Krieg}},\ }\bibfield  {title} {\bibinfo
  {title} {{Nucleon electromagnetic form factors from lattice QCD using a
  nearly physical pion mass}},\ }\href
  {https://doi.org/10.1103/PhysRevD.90.074507} {\bibfield  {journal} {\bibinfo
  {journal} {Phys. Rev. D}\ }\textbf {\bibinfo {volume} {90}},\ \bibinfo
  {pages} {074507} (\bibinfo {year} {2014})},\ \Eprint
  {https://arxiv.org/abs/1404.4029} {arXiv:1404.4029 [hep-lat]} \BibitemShut
  {NoStop}%
\bibitem [{\citenamefont {Aoki}\ \emph {et~al.}(2017)\citenamefont {Aoki} \emph
  {et~al.}}]{Aoki:2016frl}%
  \BibitemOpen
  \bibfield  {author} {\bibinfo {author} {\bibfnamefont {S.}~\bibnamefont
  {Aoki}} \emph {et~al.},\ }\bibfield  {title} {\bibinfo {title} {{Review of
  lattice results concerning low-energy particle physics}},\ }\href
  {https://doi.org/10.1140/epjc/s10052-016-4509-7} {\bibfield  {journal}
  {\bibinfo  {journal} {Eur. Phys. J. C}\ }\textbf {\bibinfo {volume} {77}},\
  \bibinfo {pages} {112} (\bibinfo {year} {2017})},\ \Eprint
  {https://arxiv.org/abs/1607.00299} {arXiv:1607.00299 [hep-lat]} \BibitemShut
  {NoStop}%
\bibitem [{\citenamefont {Aoki}\ \emph {et~al.}(2022)\citenamefont {Aoki} \emph
  {et~al.}}]{FlavourLatticeAveragingGroupFLAG:2021npn}%
  \BibitemOpen
  \bibfield  {author} {\bibinfo {author} {\bibfnamefont {Y.}~\bibnamefont
  {Aoki}} \emph {et~al.} (\bibinfo {collaboration} {Flavour Lattice Averaging
  Group (FLAG)}),\ }\bibfield  {title} {\bibinfo {title} {{FLAG Review 2021}},\
  }\href {https://doi.org/10.1140/epjc/s10052-022-10536-1} {\bibfield
  {journal} {\bibinfo  {journal} {Eur. Phys. J. C}\ }\textbf {\bibinfo {volume}
  {82}},\ \bibinfo {pages} {869} (\bibinfo {year} {2022})},\ \Eprint
  {https://arxiv.org/abs/2111.09849} {arXiv:2111.09849 [hep-lat]} \BibitemShut
  {NoStop}%
\bibitem [{Note4()}]{Note4}%
  \BibitemOpen
  \bibinfo {note} {We tested that using the CLS continuum value of $t_0$
  instead produces a negligible shift in our final result}\BibitemShut
  {NoStop}%
\bibitem [{Note5()}]{Note5}%
  \BibitemOpen
  \bibinfo {note} {Neglecting the tiny differences in volume and pion mass
  between these two ensembles}\BibitemShut {NoStop}%
\bibitem [{Note6()}]{Note6}%
  \BibitemOpen
  \bibinfo {note} {We chose the references that attempt to quantify the
  uncertainty of the calculation. Reference \cite {Albaladejo:2016ztm} and
  \cite {Yang:2022vdb} use Lattice QCD input and are not fully independent of
  previous lattice results and so we therefore mention them here instead of in
  the overview plot. The results from \cite {Cleven:2010aw} are superseded by
  the results in \cite {Fu:2021wde} and we therefore omit the former from the
  overview plot.}\BibitemShut {Stop}%
\bibitem [{\citenamefont {Dagum}\ and\ \citenamefont {Menon}(1998)}]{660313}%
  \BibitemOpen
  \bibfield  {author} {\bibinfo {author} {\bibfnamefont {L.}~\bibnamefont
  {Dagum}}\ and\ \bibinfo {author} {\bibfnamefont {R.}~\bibnamefont {Menon}},\
  }\bibfield  {title} {\bibinfo {title} {Openmp: an industry standard api for
  shared-memory programming},\ }\href {https://doi.org/10.1109/99.660313}
  {\bibfield  {journal} {\bibinfo  {journal} {IEEE Computational Science and
  Engineering}\ }\textbf {\bibinfo {volume} {5}},\ \bibinfo {pages} {46}
  (\bibinfo {year} {1998})}\BibitemShut {NoStop}%
\bibitem [{\citenamefont {Hudspith}(2015)}]{Hudspith:2014oja}%
  \BibitemOpen
  \bibfield  {author} {\bibinfo {author} {\bibfnamefont {R.~J.}\ \bibnamefont
  {Hudspith}} (\bibinfo {collaboration} {RBC, UKQCD}),\ }\bibfield  {title}
  {\bibinfo {title} {{Fourier Accelerated Conjugate Gradient Lattice Gauge
  Fixing}},\ }\href {https://doi.org/10.1016/j.cpc.2014.10.017} {\bibfield
  {journal} {\bibinfo  {journal} {Comput. Phys. Commun.}\ }\textbf {\bibinfo
  {volume} {187}},\ \bibinfo {pages} {115} (\bibinfo {year} {2015})},\ \Eprint
  {https://arxiv.org/abs/1405.5812} {arXiv:1405.5812 [hep-lat]} \BibitemShut
  {NoStop}%
\end{thebibliography}%

\end{document}